\documentclass[12pt,twoside]{report}

\usepackage{keio-thesis-2e}
\usepackage{hyperref}
\usepackage{ifthen}

\usepackage{times}
\input{epsf}
\epsfverbosetrue
\usepackage{changebar}
\usepackage{graphicx}
\usepackage{color}
\usepackage{dcolumn}
\usepackage{bm}
\usepackage{graphics}
\usepackage{lscape}
\usepackage{subfigure}
\usepackage{amsmath}

\def\comment#1{}

\def\ac{\textsc{ac}}
\def\ntc{\textsc{ntc}}

\def\usec{\mu\text{sec}}

\newenvironment{entry}
  {\begin{list}{}%
      {%
	\setlength{\labelwidth}{50pt}%
	\setlength{\labelsep}{10pt}%
	\setlength{\leftmargin}{\labelwidth}%
      }%
  }%
  {\end{list}}
\newlength{\Mylen}
\newcommand{\Lentrylabel}[1]{%
  \settowidth{\Mylen}{\textsf{#1}}%
  \ifthenelse{\lengthtest{\Mylen > \labelwidth}}%
  {\parbox[b]{\labelwidth}%
    {\makebox[0pt][l]{\textsf{#1}}\\}}%
  {\textsf{#1}}%
  \hfil\relax}
\newenvironment{Lentry}
  {%
    \begin{entry}}
  {\end{entry}}
\newenvironment{chapterquote}
  {\begin{quotation}
      \sf}
  {\end{quotation}}
\def\cq#1#2{\begin{chapterquote}#1\flushright\textbf{#2}\end{chapterquote}}

\title{Architecture of a Quantum Multicomputer Optimized for Shor's
Factoring Algorithm}
\author{Rodney Doyle Van Meter III}
    \principaladviser{Fumio Teraoka}
    \coprincipaladvisor{Kohei M. Itoh}
    \firstreader{Kae Nemoto}
    \secondreader{Hideharu Amano}

\begin{document}


\beforepreface
\prefacesection{Abstract}

Quantum computers exist, and offer tantalizing possibilities of
dramatic increases in computational power, but scaling them up to
solve problems that are classically intractable offers enormous
technical challenges.  Distributed quantum computation offers a way to
surpass the limitations of an individual quantum computer.  I propose
a {\em quantum multicomputer} as a form of distributed quantum
computer.  The quantum multicomputer consists of a large number of
small nodes and a {\em qubus} interconnect for creating entangled
state between the nodes.  The primary metric chosen is the performance
of such a system on Shor's algorithm for factoring large numbers:
specifically, the quantum modular exponentiation step that is the
computational bottleneck.

This dissertation introduces a number of optimizations for the modular
exponentiation, including quantum versions of the classical
carry-select and conditional-sum adders, improvements in the modular
arithmetic, and a means for reducing the amount of expensive,
error-prone quantum computation by increasing the amount of cheaper,
more reliable classical computation.  Parallel implementations of
these circuits are evaluated in detail for two abstract architectural
models, one (called \ac) which supports long-distance communication
between quantum bits, or {\em qubits}, and one which allows only
communication between nearest neighbors in a linear layout (called
\ntc).  My algorithms reduce the latency, or circuit depth, to
complete the modular exponentiation of an $n$-bit number from $O(n^3)$
to $O(n\log^2n)$ for \ac\ and $O(n^2\log n)$ for \ntc.  Including
improvements in the constant factors, calculations show that these
algorithms are one million times and thirteen thousand times faster on
\ac ~and \ntc, respectively, when factoring a 6,000-bit number.  These
circuits also reduce the demands on quantum error correction from
$\sim 210n^4$ to $\sim 12n^3\log_2 n$ for \ac\ and $\sim 3n^4$ for
\ntc, potentially reducing the number of levels of error-correction
encoding or allowing execution on more error-prone hardware.

Extending to the quantum multicomputer, I calculate the performance of
several types of adder circuits for several different hardware
configurations.  Five different qubus interconnect topologies and two
different node sizes are considered, and two forms of carry-ripple
adder are found to be the fastest for a wide range of performance
parameters.  Small nodes (up to five logical qubits) and a linear
interconnection network provide adequate performance; more complex
networks are unnecessary until $n$ reaches several hundred bits.  As
node size grows, it is important that the I/O bandwidth of a node
grow, as well, or performance can actually decline despite the overall
decrease in network activity.  The links in the quantum multicomputer
are serial; parallel links would provide only very modest improvements
in system reliability and performance.  Two levels of the Steane
[[23,1,7]] error correction code will adequately protect our data for
factoring a 1,024-bit number even when the qubit teleportation failure
rate is one percent.



\prefacesection{Acknowledgements}

\cq{I had the good fortune to become acquainted very early with some
characters of very high standing, and to feel the incessant wish that
I could even become what they were.}
{Thomas Jefferson, {\em Autobiography}}

Many, many people have demonstrated a faith in me that can never be
repaid, starting with my parents, who never even suggested, so far as
I recall, that there were any limits to what I could accomplish
(despite sometimes overwhelming evidence to the contrary).  (Though,
at the same time, I have no recollection that they ever suggested I
had a future in the NBA.)

Most especially, I must thank my daughters Sophia and Esther and my
wife Mayumi, who put up with many hours of Daddy being physically
present but mentally elsewhere.  My sisters Sheila and Lera for nearly
forty years have suffered the indignities and logistical difficulties
of a weird, sartorially challenged older brother who lives thousands
of miles away.  To my grandparents, aunts, uncles, cousins and
brother-in-law in a large and close family I also owe an apology for
living so far away.

Every life has its cusps, its turning points that forever change you.
The biggest was Caltech, but joining ISI was an unanticipated stroke
of fortune.  The friends and mentors I made in the Caltech and ISI
days still carry me forward.  Ross Berteig dragged me to the three
most life-changing classes I took at Caltech (Feynman, Ayres, and
Scudder), including the one that led me to ISI.  At ISI, I met Wook,
and my life would never be the same in ways beyond enumerating; I owe
no one a greater debt.  Dale Chase taught me how to be a good employee
and person (and how to play team volleyball).  Andi, Bobo, Brenda,
Daryll, Dave, Dennis, Edie, Gabrielle, Grace, Greg, Harold, Hugo,
Irene, Jessica, John, John, Kevin, Kyu, Liralen, Mimi, Michelle, Min,
Myles, Pam, Rick, Ryuji, Sandy, Steve, Suz, Tiger, Yosufi and the
entire CINC-PAC, WoW, volleyball, Half Moon Bay, Quantum, Nokia, NII,
Keio, and Network Alchemy crowds, just for being there (wherever
``there'' happens to be).

Thanks to Jim Hughes for help on classical cryptography, Reagan Moore
for supercomputing advice, and the rest of the MSSTC EC for years of
companionship and learning.

Without the support and encouragement of Takashi and Nobunori
Shigezaki, Mark Holzbach and the folks at Asaca and ShibaSoku, and the
teaching of Yuko Yamaguchi at Kichijoji Language School and
Misaki-sensei at Keio, Japan would have remained a remote, foreign
land rather than the second home it has become.

When I began working on quantum computing three years ago, I received
important early encouragement from Prof. Kohei Itoh and Eisuke Abe of
Keio University, Prof. Kunihiro of the University of
Electro-Communications, Drs. Kawano and Takahashi of NTT CRL,
Prof. Iwama of Kyoto University, Prof. Yamashita of NAIST, Dave Bacon
of Washington, Mark Oskin of Washington, and Dr. Yamaguchi of
Stanford.  Professors Yoshi Yamamoto, Seth Lloyd, Isaac Chuang, Andrew
Steane, Mio Murao, Hiroshi Imai, Seigo Tarucha and Akira Furusawa, and
Yasunobu Nakamura and J.W. Tsai of NEC, and many researchers at NTT
provided access to their labs and students, without which I would
never understand how to actually build a quantum computer.
Prof. Yamamoto and the others who created and staffed the summer
schools in Okinawa and Kochi not only taught me but brought me into
their community.  I look forward to deepening collaborations with all
of you over the coming years.

Although my name goes on the thesis, my coauthors on the half-dozen
papers that are incorporated deserve much of the credit: Kohei Itoh,
Mark Oskin, Thaddeus Ladd, Kae Nemoto, and Bill Munro.  Thaddeus gets
a special call out for writing advice as well as teaching me physics.
Joe Touch, Ted Faber, Bill Manning and Nick Burke all read (sometimes
awful) drafts of various papers and provided other important support.
Kevin Binkley's stochastic engine, especially the genetic algorithms,
provided insight into optimization problems that remain open.  Suzanne
and Bob Diller get credit for cogent advice on the title and abstract
of my dissertation, as well as providing years of friendship.  Michael
Cohen and Prof. Sagawa of Aizu-Wakamatsu and Prof. Jun Murai, Ryuji
Wakikawa and Shoko Mikawa at the School of Internet ASIA Project
provided teaching opportunities which turned into wonderful learning
experiences for me.  Chip Elliott and others at BBN have provided
important encouragement, as well.

The music of Tatopani, Billy Higgins, John Coltrane, Louis Armstrong,
King Crimson, Kodo, and a plethora of others kept me sane.
Surprisingly, we have no favorite chef in this country (our favorite
anywhere is Jose Luis Ugalde of Cafe Gibraltar), but everyone who has
fed me startling and wonderful meals -- you keep me going.

Without the cooking and baby-sitting of the indomitable Kazuko Arai,
this thesis would have taken a decade to complete, if it ever got done
at all.

This work was supported in part by Ken Adelman and Dave Kashtan under
the Network Alchemy basic research funding plan, and by CREST-JST.
Kae Nemoto also provided funding for travel and a desk at NII.  Karl
and Pattie and Danner and Jenny put me up in Cambridge.

This thesis was created using 100\% free software.  Thanks to the
creators of Linux, X, \TeX and \LaTeX, GNU Emacs, xfig, dia, POVray,
maxima/MACSYMA, octave, gnuplot, bison, flex, gcc, and more --- and,
in some cases, to the researchers on whose work these tools are
founded, though the code base has changed.  And thanks to the creators
of the arXiv, scholar.google.com, researchindex.org, and citebase,
without which I would have missed much important research, and
probably been forced to recreate it poorly and tediously on my own.

I thank Y. Nakamura, T. Yamamoto, D. Wineland, and K. M. Itoh for the
figures.  Thanks to all of the patient physicists who have put up with
my slowness, including Viv Kendon for help with Shor's algorithm.

I have had, in effect, four advisers: Fumio Teraoka in computer
networking, Kohei Itoh in experimental physics, and Kae Nemoto and
Bill Munro in theoretical physics.  All four have been oustanding.
Kae worked harder on reviewing this dissertation than anyone else; it
would not be as clear and correct without her.  Bill is the one person
{\em not} on my committee without whom this technical work could not
have been done.  Thanks.  Thanks to Profs. Amano and Yamasaki, my
other commitee members.  I have also had six godfathers and mentors:
Wook, Darrell Long, Paul Mockapetris, Jun Murai, Rick Carlson, and Bob
Hinden.  Without your advice and encouragement, I wouldn't be
finishing a doctorate in such a challenging and fascinating subject.

\afterpreface

%
%
\thispagestyle{empty}
\newenvironment{dedication}
  {\cleardoublepage \thispagestyle{empty} \vspace*{\stretch{1}} \begin{center} \em}
  {\end{center} \vspace*{\stretch{3}} \clearpage}
\begin{dedication}
For my family
\end{dedication}
\thispagestyle{empty} \cleardoublepage




\chapter{Introduction}

\cq{We are just started on a great venture.}{Dwight Eisenhower,
  November 1942}

\cq{The designer usually finds himself floundering in a sea of
possibilities, unclear about how one choice will limit his freedom to
make other choices, or affect the size and performance of the entire
system. There probably isn't a  `best' way to build the system, or
even any major part of it; much more important is to avoid choosing a
terrible way, and to have clear division of responsibilities among the
parts.\\
\indent I have designed and built a number of computer systems, some
that worked and some that didn't.}{Butler Lampson, ``Hints for
  Computer System Design''~\cite{lampson83:_hints}}

As VLSI features continue to shrink, computers that depend on quantum
mechanical effects to operate are inevitable; indeed, quantum effects
are predicted to affect device behavior within a
decade~\cite{moore65,mead:scaling,bohr98:_silicon_trend_limits,ieong04:_silic_devic_scalin_sub_regim,ITRS2005,bourianoff03:_future_nanocomp}.
The fundamental architectural issue in these future systems is whether
they will attempt to hide this quantum substrate beneath a veneer of
classical digital logic, or will expose quantum effects to the
programmer, opening up the possibilities of dramatic increases in
computational
power~\cite{feynman:_simul_physic_comput,deutsch-jozsa92,deutsch85:_quant_church_turing,benioff82:_qm_turing_machine,bennett:strengths,shor:factor,grover96,aaronson:thesis,lloyd96:_univer_quant_simul,nielsen-chuang:qci}.

Small and unreliable they are, but quantum computers of up to a dozen
nuclear spins~\cite{negrevergne06:_12qubit-bench} and eight
ions~\cite{haeffner05:qubyte} exist.  In these machines, the spin
state of an atomic nucleus or the energy level of an ion can represent
a quantum bit, or {\em qubit}, the smallest unit of quantum
information.  The three most famous quantum algorithms are
Deutsch-Jozsa~\cite{deutsch-jozsa92}, Grover's search~\cite{grover96},
and Shor's factoring~\cite{shor:factor}.  All three of these
algorithms have been experimentally implemented for small-scale
problems~\cite{jones98:_grover-impl,chuang98:_dj-exper,chuang98:_grover_impl,kim00:_implem_deuts_jozsa_nmr,vala02:_exper_deuts_jozsa,vandersypen:thesis,vandersypen:shor-experiment,gulde03:_implem_deuts_jozsa}.
A further extremely broad range of experiments has demonstrated
numerous building
blocks~\cite{walther05:_cluster-exper-nature,baugh05:_exper-hbac,steffen03:_exper_qc_adiab,knill01:_bench,murali02:half-adder-nmr-impl,chiaverini04:_qec-realiz,pittman05:_optical-qec-demo,katz06:_coher_state_evolut_super_qubit}
based on the one- and two-qubit technology demonstrations we will see
in Chapter~\ref{ch:taxonomy}.  Although many theoretical and practical
questions remain open, it seems reasonable to assert that
implementation of quantum computation is on the verge of moving from a
scientific problem to an engineering one.  It is now time to ask what
we {\em can} build, and what we {\em should} build.  Various computer
architecture researchers have begun investigating the former question,
working from the bottom
up~\cite{copsey:q-com-cost,isailovic04:_taco,oskin:quantum-wires,oskin02:_pract_archit_reliab_quant_comput,thaker06:_cqla,isailovic06:_interconnect};
this dissertation and the related papers address the latter question,
working from the top
down~\cite{van-meter04:fast-modexp,van-meter:qarch-impli,van-meter05:_distr_arith_quant_multic,van-meter:qft-topo,van-meter:pay-the-exponential,van-meter:arch-dep-shor}.

\section{Computing Frontiers: Why Study Quantum?}

Why should computer engineers study quantum computation, and why now?
Certainly the field of classical computer architecture is not
moribund, and offers far more immediate impact for much less
intellectual risk.  Work that increases parallelism, reduces power
consumption, improves I/O performance, increases gate speed or reduces
data propagation delays is much more likely to be used in the real
world, and far sooner than quantum technologies.  Intel began sampling
a billion-transistor microprocessor chip in October 2005, a 580
square-millimeter chip built in a 90 nanometer process.  Some
researchers consider integration levels of a trillion transistors per
silicon chip possible~\cite{meindl01:_terascale-si}, though we are
hardly done digesting the implications of a billion transistors on a
chip~\cite{DBLP:journals/computer/PattPEFS97,DBLP:journals/computer/KozyrakisPPAACFGGKTTY97,DBLP:journals/computer/BurgerKMDJLMBMY04}.
Clearly there is room on-chip for many architectural advances.
Ubiquitous computing, sensor networks, augmented reality, and mobile
systems will no doubt be among the most transformative technologies of
the coming decades, relegating today's 3G Internet-connected mobile
phones to the status of Neolithic stone
adzes~\cite{rheingold02:_smart_mobs}.  In ``back end'' systems,
continued research on computational grids and storage are critical.
Among computing exotica, electrical circuits fabricated with
nanotechnology~\cite{zhong03:_nanowire_crossbar-demux,beckman05:_bridg_dimen,martel98:_nanotube-fets,tans98:_swcnt-transistor,rueckes00:_carbon_nanotube-nvram},
DNA computing~\cite{adleman94:_dna-computing}, and amorphous computing
are all other possible fields of
pursuit~\cite{abelson00:_amorp_comput}.  So, why quantum?

Different researchers have different reasons for studying quantum
computing.  Physicists are learning fundamental facts about the
quantum behavior of both individual particles and mesoscopic systems.
Theoretical computer scientists are finding many fascinating new
questions (and answering some of them).  But to a computer systems
person, quantum computation is about one thing: {\em the pursuit of
performance}.  If practical large-scale quantum computers can be
built, we may be able to solve important problems that are classically
intractable.  Potential applications include cryptographically
important functions such as factoring, which appears to offer a
superpolynomial speedup, and scientifically important problems such as
simulations of many-body quantum systems, which may offer exponential
speedup, though recent questions have been raised about whether
exponential speedup is achievable as the desired error bound is
tightened~\cite{brown06:_limit_hamil}.  Quantum computers therefore
hold out the possibility of not just Moore's Law increases in speed,
but a change in computational complexity class and consequent
acceleration on these, and possibly other, problems.

I will not directly address criticisms of the possibility of quantum
computation~\cite{dyakonov02:_QC-skepticism,keyes03:solid-state-qc-problems},
except to note that my response is different from that of Aaronson,
who is excited by the inherent beauty and theoretical importance of
quantum mechanics while searching for the ultimate limits to
computation~\cite{aaronson:thesis}.  I, too, admire these factors, but
more importantly I believe it is inevitable, as silicon devices
continue to scale down in size, that we will have to deal with quantum
effects.  Many researchers are directing their efforts at mitigating
these effects; in my opinion, we will do better by embracing them,
even if ``quantum computing'' ultimately proves to have no
computational advantage over classical.

Studying quantum computing indirectly benefits classical systems, as
well.  Quantum effects are being explored for direct exploitation as
classical logic, for example, the recent work on magnetic quantum dot
cellular automata~\cite{imre06:_qd-majority-gate}.  {\em Plasmonics},
the study of electromagnetic waves propagating in the surface of a
material, is developing rapidly, and might offer improvements in how
we move data within classical chips~\cite{ozbay06:_plasmonics}.  More
broadly, the whole area called {\em spintronics}, directly or
indirectly manipulating the spin of small numbers of electrons, is
already having an impact through the creation of technologies such as
magnetic RAM (MRAM)~\cite{zutic04:_spintronics,wolf05:_spintronics}.
Quantum computers depend on, and have served as an impetus for
developing, thermodynamically reversible computing.  It has been
suggested that classical computers must employ reversible logic to
exceed $10^{22}$ floating point operations per second (10
zettaFLOPS)~\cite{debenedictis05:_reversible}.  Quantum computation
serves as an excellent training ground for engineers destined to work
in these areas, as well as providing both fundamental and practical
results that influence the technological development of these areas.

My analogy is to the field of robotics.  It has been more than eighty
years since the original use of the term {\em robot} to mean an
autonomous, mechanical humanoid (though the idea goes back to
antiquity)~\cite{capek:RUR}, and several decades since the debut of
robotics as a respectable field of inquiry.  Yet the humanoid robots
of science fiction do not roam the streets of Tokyo in the first
decade of the twenty-first century.  This does not mean that robotics
as a field has been barren; indeed, robots dominate many forms of
manufacturing, and related technologies spun off from robotics
research are nearly ubiquitous.  Robotics depends on, and serves as an
impetus for, research as diverse as computer vision, speech
recognition, fuzzy logic, virtual reality, and many mechanical
advances.  The road to development has been long, and the results to
date look nothing like what mid-twentieth century science fiction
writers such as Isaac Asimov anticipated~\cite{asimov50:_i_robot}, but
the results have been extremely valuable nonetheless.  So I expect it
to be with quantum computing.

\section{Defining Quantum Computer Architecture}

\begin{figure}
\centerline{\hbox{
\includegraphics[width=11cm]{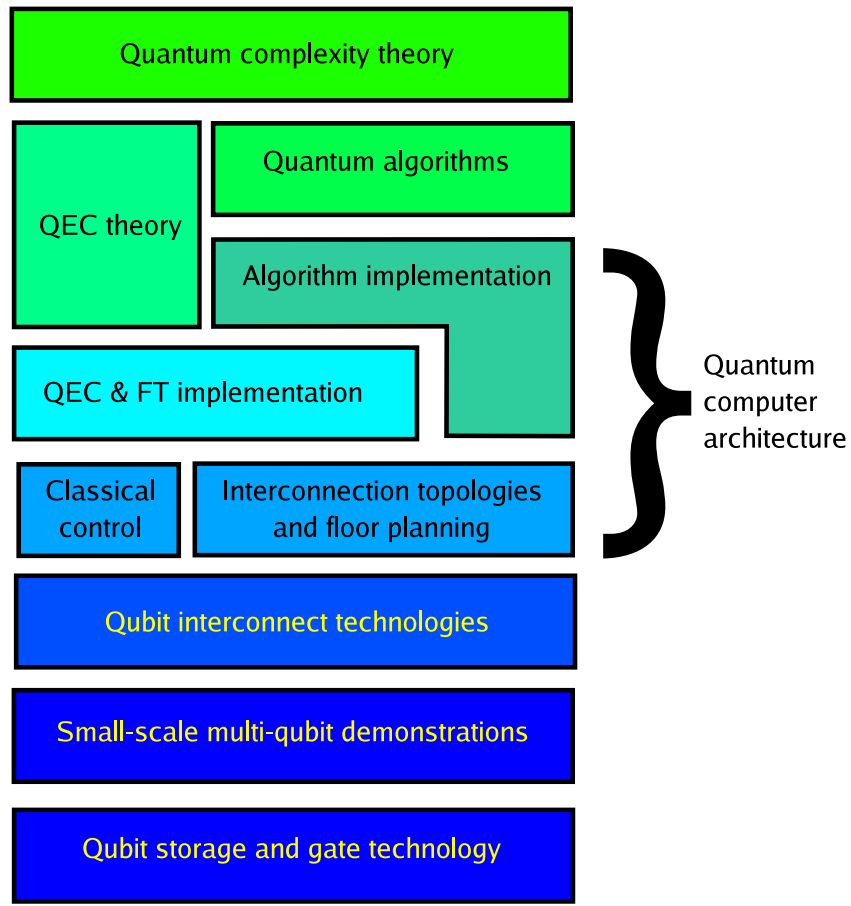}}}
\caption[Quantum computer architecture]{Quantum computer architecture
  among some subfields of quantum computation.}
\label{fig:stack}
\end{figure}

{\em Quantum computer architecture} is an emerging field, spanning the
gap between device physics and algorithms.  If large-scale quantum
computers are to be built, an overall structural plan must be
established; we refer to this plan as the machine architecture of the
quantum computer.  Figure~\ref{fig:stack} shows a representation of
the relationship among some subfields of quantum computing, and which
subfields are part of the broader area of quantum computer
architecture.  I include in this field essentially everything above
device physics up to the design and performance analysis of machines
for specific algorithms.  The component which has (rightly) been the
focus of the most work to date has been quantum error correction,
though effective high-level structures (including physical connection
topologies), control structures, efficient algorithm implementation,
and performance analysis are all receiving increased attention.
Quantum computer architecture can draw heavily on classical computer
architecture, but presents a number of unique challenges.

In most quantum computing technologies, a qubit is the state of a
physical device, more like the state of a flip-flop than a signal
propagating through a circuit.  Qubits that are physically far apart
cannot directly interact, so data must be shuffled from place to place
as they are required to interact with other qubits.  Architects and
compiler writers must cooperate to make this shuffling as efficient as
possible.  In the figure this topic is represented as
``interconnection technologies and topologies''.  Solutions to this
kind of data transport problem form one of the key themes of this
thesis.

Although they are not explicitly represented in the figure, quantum
programming languages and compilers, designed for programming quantum
computers, can be viewed as the interstitial glue that holds the whole
system together~\cite{gay05:_quant_progr_lang}.  Quantum programs are
executed classically, and must be able to manipulate both quantum and
classical data, and make branch and loop decisions based on classical
data.  The ability to look at quantum data during program execution is
extremely limited, as we will see when we discuss measurement in
Section~\ref{sec:measurement}; the operations on the quantum data are
performed more or less blind, without examining the data itself.  In
this sense, programming a quantum computer is like programming a
Connection Machine or systolic array, though the analogy between qubit
and CM processor is weak~\cite{steele86:_connec_machin_lisp}.

Because quantum computer architecture is a young field, many issues
have not yet been addressed in the depth required to evaluate design
choices.  Often clock speed and other architectural features are
ignored as issues in quantum computing devices, assuming that the
quantum speed-up will dominate, making quantum algorithms practical on
any physically realizable quantum computer.  However, this is not
necessarily so.  For example, Shor's factoring algorithm runs in
polynomial time and resources, but the details of the polynomial
matter: what degree is the polynomial, and what are the constant
factors?  How much parallelism can be extracted from both the hardware
and software to reduce the wall-clock time consumed?  All of these
issues are of concern to architects.

Some of these issues are attacked in this thesis.  We will see
others in Section~\ref{sec:future-work}, on future work, at the end of
the dissertation.

\section{The Quantum Multicomputer}

My thesis is the design of a {\em quantum multicomputer}.  Any single,
monolithic quantum computer will have an ultimate limit to its storage
capacity and performance.  Borrowing from classical multicomputer
design and building on the foundations of distributed quantum
computation that have been laid, these limitations can be overcome.
This dissertation describes the architecture of a system suitable for
running highly optimized forms of Shor's factoring algorithm, and
examines the scaling of the performance from sixteen to 1,024 nodes.
This broad range of sizes allows us to see clearly the important
inflection points in behavior as the system scales up, ending at a
performance point well above the capabilities of classical systems.

\begin{figure}
\centerline{\hbox{
\includegraphics[width=11cm]{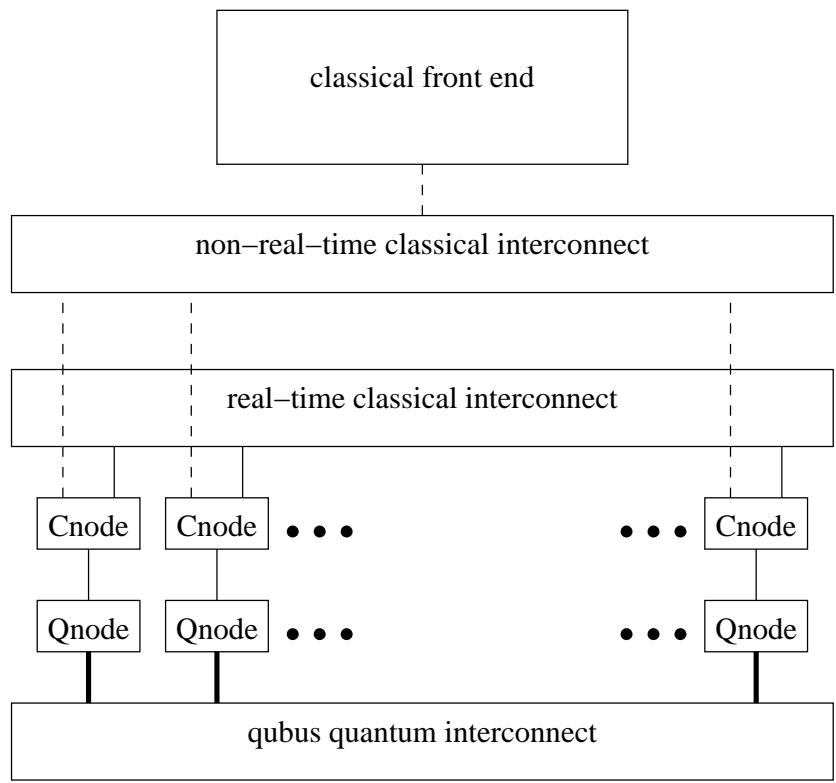}}}
\caption[High-level quantum multicomputer block diagram]{High-level
  quantum multicomputer block diagram.  Dashed lines are non-real-time
  communication; solid lines are real-time communication, either
  classical (thin lines) or quantum (thick lines).  Cnode, classical
  node; Qnode, quantum node.}
\label{fig:multicomp-hw-block}
\end{figure}

A high-level block diagram of the hardware is shown in
Figure~\ref{fig:multicomp-hw-block}.  Like all proposed quantum
computers, it is actually a hybrid quantum-classical system, and to
achieve performance balance the classical portion will be coupled to a
supercomputer-class machine.  The classical front end is responsible
for overall coordination, download of programs and final upload of
data, but has only a loose role in the execution of a program.  The
nodes perform the actual computation.  Each node consists of two
halves, the quantum part (Qnode), which holds the quantum data, and
the classical part (Cnode), which contains the real-time measurement
and control circuitry (including program execution) for the quantum
device.  There are two real-time interconnects, one classical and one
quantum; the quantum interconnect is based on the {\em qubus} approach
for its link technology~\cite{spiller05:_qubus,munro05:_weak}.  These
interconnects may be switched, node-to-node direct, or shared; a major
portion of this thesis is analysis of the traffic on the qubus-based
quantum interconnect for different possible topologies.  We will not
address the classical portions of the system, except that classical
communication and instruction execution are implicitly included in our
timing estimates.

A well-designed architecture can outlive the technological environment
in which it was originally created.  However, some constraints are
necessary as we discuss the initial implementation target, or we are
left adrift on Lampson's Sea.  I have chosen a solid-state qubit
technology, such as Josephson-junction qubits (described in
Sec.~\ref{sec:jj}), as a basis on which to build.  Very, very roughly,
I have chosen to limit the estimated production cost to one hundred
million U.S. dollars, and the size of the system to one hundred meters
square of floor space.

\section{This Dissertation}

\begin{figure}
\centerline{\hbox{
\includegraphics[width=11cm]{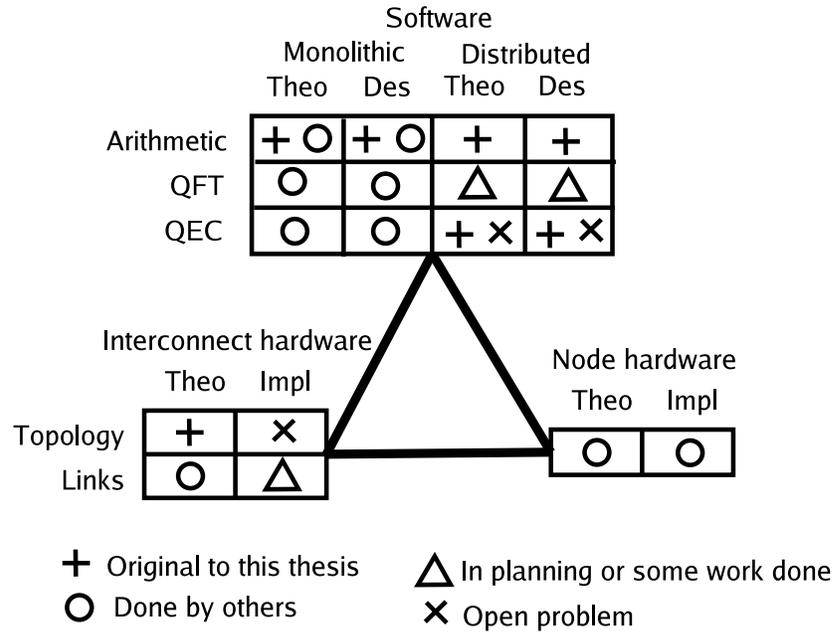}}}
\caption[Current subsystem status]{The status, or relative maturity
  level, of various subsystems within the quantum multicomputer.  QFT,
  quantum Fourier transform; theo, theory; des, design; impl,
  implementation.}
\label{fig:qmc-progress}
\end{figure}

The quantum multicomputer consists of three primary subsystems: the
quantum computational node hardware, the quantum interconnect
hardware, and the software to run on the system.  The status of some
of these subsystems is represented in Figure~\ref{fig:qmc-progress}.
Node hardware is not a primary focus of this thesis; we leave it to
other researchers to meet the hardware requirements outlined in
Chapter~\ref{ch:arch-over}.  Interconnect hardware consists of basic
link technologies and the manner of assembling a complete system,
namely the topology and any necessary lower-level switching
mechanisms; finding an appropriate topology is one of the primary
contributions of this thesis.  Finally, although the arithmetic and
quantum Fourier transform (QFT) algorithms that make up Shor's
factoring algorithm have been described at a high level, we make
significant advances in the former in this thesis.  Although this
thesis makes some progress on distributed quantum error correction
(QEC), I believe this is very much an open problem, so it is marked
with both symbols in the figure.

\subsection{Contributions}

The primary contribution of this thesis is the architecture of a
quantum multicomputer.  To validate design choices, a target workload
of Shor's algorithm for factoring large numbers is used.  This
validation entails analysis and optimization of the performance of
arithmetic, especially adders, on both monolithic and multicomputer
quantum systems.  I have designed new types of reversible adder
circuits, analyzed the parallelism available in Shor's algorithm,
optimized Shor's algorithm, and mapped it to various architectures,
following through with performance analysis for two monolithic machine
types and a variety of adder circuits.  From there, I extend to a
multicomputer.  I define the capabilities necessary for a node.
Detailed analysis shows that the interconnect links may be serial,
rather than parallel, and that a linear network topology will be
adequate into the high hundreds of nodes, when a switched network
becomes more appropriate.  The performance is analyzed assuming nodes
are built on high-speed solid-state qubits, and the performance is
found to be good.  Finally, I investigate very loosely the practical
constraints on the construction of such a system, including cooling,
floor space, packaging, interconnects, control equipment, and
economics.

In summary, the contributions of this thesis are:
\begin{itemize}
\item Fast, architecturally realistic quantum modular exponentiation
  algorithms.
  \begin{itemize}
  \item Based on known and new principles, improvements in both
    asymptotic performance and constant factors in the time required
    for modular exponentiation.  To factor a 6,000-bit number, for
    example, the performance improvement ranges from 13,000 times to
    one million times, depending on architecture, compared to the
    previous best-known algorithm.  The asymptotic performance
    (circuit depth, or latency) improves from $O(n^3)$ to $O(n^2\log
    n)$ or $O(n\log^2 n)$, again depending on architecture.
  \item A classical/quantum tradeoff that reduces the number of
    quantum gates that must be performed.
  \item New square root-depth and logarithmic-depth adder circuits,
    used in some forms of my modular exponentiation algorithms.
  \item Analysis of the demands of arithmetic circuits on the strength
    of quantum error correction, showing that my new algorithms are
    substantially less demanding, and hence have higher probability of
    success and/or can be executing using weaker QEC.
  \item A proposed {\em architectural} taxonomy of qubit technologies,
    complementary to the DiVincenzo criteria that establish minimal
    necessary functionality.
  \item The most detailed architectural performance analysis to date.
  \end{itemize}
\item Architecture of a quantum multicomputer.
  \begin{itemize}
  \item Analysis of performance of adder circuits on various network
    topologies showing that a linear network is adequate up to
    moderately large problem sizes.
  \item Design of link transfer protocols based on quantum
    teleportation and QEC, establishing that serial links perform
    adequately.
  \item Delineation of required traits for the computational nodes.
  \item A high-level analysis of the overall system requirements,
  including floor space and economics, assuming a solid-state qubit
  technology.
  \end{itemize}
\end{itemize}

\subsection{Contents and Structure}

This dissertation is divided into eight chapters.  The first and last
are the overview and conclusions, respectively.
Chapter~\ref{ch:reversible} consists primarily of a review of existing
classical and quantum material. Chapter~\ref{ch:shor} presents Shor's
algorithm.  Chapter~\ref{ch:taxonomy}, the taxonomy of quantum
technologies, reviews the work of experimentalists, but the structure
of the taxonomy is original.  Chapter~\ref{ch:large-perf} describes my
contributions to understanding and improving the performance of the
modular exponentiation for Shor's algorithm, and Chapter~\ref{ch:qmc}
describes the architecture and performance of my quantum
multicomputer.

Sections~\ref{sec:reversible-classical} and \ref{sec:qc-intro}
introduce the fundamental concepts of reversible classical and quantum
computation, including the graphical and mathematical notations used
throughout this dissertation.  Chapter~\ref{ch:shor} describes the
quantum portions of Shor's algorithm for factoring large numbers,
including adder circuits developed by various researchers over the
last decade to support Shor's algorithm.  The taxonomy in
Chapter~\ref{ch:taxonomy} describes existing experimental approaches
to quantum computing developed in many research organizations; I
attempt to extract common themes in these technologies and organize
the information so that it is possible to determine the architectural
promise of each technology.  Chapter~\ref{ch:travel} is a quick sketch
of the mechanisms we need for transferring data in our quantum
multicomputer: the qubus approach to creating entanglement, quantum
teleportation, and the classical concepts of multicomputer networks.

The first section of Chapter~\ref{ch:large-perf} addresses the
practical implications of scalability for large quantum computing
systems, including such mundane issues as economics and floor space.
The rest of the chapter details the mapping of the entire quantum
modular exponentiation necessary for Shor's algorithm to abstract
quantum architectures.  Section~\ref{sec:algo-topo} describes the
management of performance, introducing the \ac\ and \ntc\
architectural models and our performance notation and summarizing the
techniques presented in the following material.
Section~\ref{sec:trading} accelerates the quantum portion of the
algorithm in exchange for more onerous classical computation.
Section~\ref{sec:new-adders} details two new reversible quantum adder
algorithms, the $O(\sqrt{n})$-depth carry-select and $O(\log n)$-depth
conditional sum circuits.  Section~\ref{sec:mono-shor-perf} brings all
of the techniques together and shows overall performance speedups for
both architectural models.

Chapter~\ref{ch:qmc} advances the state of the art in distributed
quantum computation by creating specific hardware models and
performance estimates for the quantum multicomputer, starting with a
system overview.  Section~\ref{sec:qec-qubus} covers the distributed
form of quantum error correction and its impact on link design, and
shows that serial links are acceptable.  Finally,
Section~\ref{sec:dist-shor} brings us to the goal of analyzing the
behavior of Shor's algorithm on realistic hardware models.

A small glossary is provided as Appendix~\ref{ch:glossary}.

\subsection{How to Read This Dissertation}

The primary target audience of this dissertation is computer systems
researchers with little or no prior background in quantum computing.
As such, the mathematics are limited and informal, but heavy on
examples.  Systems researchers will probably benefit most from reading
the dissertation linearly from beginning to end.

Physicists who are already familiar with quantum computing may want to
skip most of Chapters~\ref{ch:reversible} and \ref{ch:shor}, though
they may find enough new tidbits in
Section~\ref{sec:reversible-classical} to repay the time invested.
Such readers may be less familiar with some of the concepts in
Section~\ref{sec:qc-adders} and Chapter~\ref{ch:travel}, and are
encouraged to skim Chapter~\ref{ch:taxonomy} for some insight into the
technology issues that matter to a system architect.

For those readers interested in only the major results, besides the
overview and conclusion chapters, the most important sections
are~\ref{sec:mono-shor-perf}, \ref{sec:new-adders}, \ref{sec:trading},
and especially \ref{sec:dist-shor}.

\section{What We're Not Going to Talk About}


Quantum information processing (QIP), despite its youth, is already a
very broad field, and there are many important and fascinating topics
that I am {\em not} going to present in this dissertation.  This
section merely identifies a few for completeness, and provides some
pointers to further literature for those whose curiosity is piqued by
this dissertation.  Readers interested in more depth are referred
first to popular~\cite{williams:_ultim_zero_one,nielsen:sciam} and
technical~\cite{nielsen-chuang:qci,kitaev:cqc,galindo02:rmp-qc,preskill98:_lecture-notes}
texts on the subject.

Probably the most important area not addressed is computational
complexity.  Computer science theorists are rapidly advancing our
understanding of what quantum computers are, and are not, capable of
computing
efficiently~\cite{bennett:strengths,yao93quantum,bernstein97:_quant_complexity,aaronson:_compl_zoo,aaronson:thesis}.
This research is also advancing our knowledge of classical
computational complexity, and has the potential to ultimately shed
light on the fundamental $P \stackrel{?}{=} NP$ question.

Feynman originally conceived of a quantum computer as a device for
quantum
simulation~\cite{feynman:_simul_physic_comput,lloyd96:_univer_quant_simul,boghosian98:_quant-simul,abrams97:_qc_fermi_simul,byrnes06:_simul,brown06:_limit_hamil,aspuru-guzik05:_simul_quant_comput_molec_energ}.
Quantum simulation may very well be the first production use of
quantum computing technology.  However, it bears less resemblance to a
general-purpose, programmable machine derived from known classical
architectural principles, which is my goal in this thesis.

Other important algorithms besides Shor's factoring algorithm have
been developed.  The first quantum algorithm invented was
Deutsch-Jozsa, which can determine whether a function is {\em
constant} (returns the same value for all inputs) or {\em balanced}
(returns zero for half of its inputs and one for the other half),
using only a single call to the function~\cite{deutsch-jozsa92}.
Grover's search algorithm can search an unstructured space of $N$
possibilities in $O(\sqrt{N})$ time.  It is sometimes referred to as
{\em amplitude amplification} and has been found to be useful for
quantum counting, and as a wrapper for other
algorithms~\cite{grover96,grover05:_fixed_point_quant_search,brassard98:_quant_counting}.
Although they are important, we will not delve into Simon's
algorithm~\cite{simon:_power_quant_comput},
Hallgren's~\cite{hallgren:_quantum-pell}, or the fascinating topic of
quantum random
walks~\cite{aharonov93:_quant_random_walks,kempe03:_quant-random-walks}.

Quantum networking, especially as typified by quantum key
distribution, is a vital and fascinating area, and the only area of
QIP in which products are already
available~\cite{elliott:qkd-net,paterson04:why-qkd,bennett:bb84}.
Dense coding is also a clever and important idea by Bennett and
Wiesner~\cite{mattle96:_dense_coding_exper,bennett92:_dense-coding}
which essentially allows one system to ``presend'' half of the bits in
a message to its partner {\em before} computing the data.  Many
researchers have worked on various aspects of quantum information
theory, including quantum channel capacities analogous to Shannon's
capacity for a classical channel.  The last third of Nielsen and
Chuang deals with this topic, including derivation of quantum error
correction from this point of view~\cite{nielsen-chuang:qci}.

Perhaps the most interesting advance in quantum computing theory in
recent years is the development of {\em cluster state computing}, or
{\em one-way
computing}~\cite{raussendorf03:_cluster-state-qc,nielsen:cluster-review}.
We refer to cluster state occasionally in this dissertation, but will
not have the space to deal seriously with it.

Researchers have begun designing programming languages for quantum
computers~\cite{oemer:qcl}, and several workshops have been held.
Gay's survey and extensive bibliography is a good place to start
studying this topic~\cite{gay05:_quant_progr_lang}.

All of the quantum computers being seriously discussed today are
essentially hybrid computers: some of the data is quantum, but other
data and all of the program are classical.  We will confine ourselves
to such systems for this thesis, though some researchers have
investigated the next advance in quantum computer architecture: true
{\em quantum} programs, leading to a quantum instruction set
architecture
(ISA)~\cite{nielsen97:_pqga,hillery:prob-prog-qc,paz03:_quant-gate-arrays,rovsko03:_gen-meas}.

Quantum games~\cite{meyer00:_quant-games,eisert99:_quant-games},
quantum computing through wormholes~\cite{bacon04:_qc_wormholes} and
relativistically accelerated
devices~\cite{ralph05:_time_displaced_qc}, and the amount of
computation that can be performed by given amounts of
matter~\cite{lloyd00:_ultimate} or even the Universe as a
whole~\cite{lloyd02:_comput_capac_univer} are mind-boggling ideas.  We
are not discussing quantum cellular automata (QCA) or quantum Turing
machines~\cite{benioff82:_qm_turing_machine,deutsch85:_quant_church_turing,galindo02:rmp-qc},
despite their importance (quantum wires and the original Lloyd model
of a quantum computer are forms of
QCA~\cite{lloyd:model,oskin:quantum-wires}).  We are not going into
any significant detail on entanglement theory.  We are also not going
to discuss qutrits, or continuous quantum variables (qunats).

And, of course, even in a work the length of a thesis it is impossible
to go into any topic in the depth it truly deserves; the device
technologies we discuss in Chapter~\ref{ch:taxonomy} are but a few of
the dozens of proposed and even instantiated types.  In addition to the
taxonomy and references in this dissertation, I recommend the ARDA
road map for its breadth~\cite{arda:qc-roadmap-v2} and Chapter 7 of
Nielsen and Chuang for its clarity of
exposition~\cite{nielsen-chuang:qci}.

\section{Summary}

The fundamental principles of small-scale quantum computing have been
demonstrated experimentally, and matching theory is progressing
nicely, though both have plenty of challenges ahead.  What has been
much less clear is whether truly scalable systems can be built;
indeed, the real-world feasibility of creating entanglement across
thousands of qubits remains very much open to question.  Distributed
quantum computation is one possible way to overcome the limitations of
an individual quantum computer.  The basic idea of distributed quantum
computation is straightforward, but detailed analysis of its
implementation has been lacking: what hardware will it run well on,
under what conditions is it robust, and can it bring improvements in
both qubit storage capacity and algorithmic performance?  This thesis
clarifies these issues.  The quantum multicomputer framework, like a
good classical architecture, has the potential to far outlive the
technological environment in which it was originally conceived.  Ladd
has speculated that production quantum computers are likely to be
built on technologies which have not yet been invented; the principles
outlined here will apply even in that eventuality.

Before we can demonstrate that the quantum multicomputer has
acceptable performance and reliability for large but finite problems,
we must evaluate and optimize the proposed workload.  Prior even to
that, we begin by investigating the foundations of classical
reversible and quantum computation.  The road to a working, useful,
reliable, economically viable quantum computer is long, dangerous, and
in large measure unknown, but, like Hokusai's stages of the Tokaido,
the sights and stops along the way are beautiful, fascinating and
important.  In the next chapter, we take the first step.

  \chapter{Reversible and Quantum Computation}
\label{ch:reversible}


\cq{``[A civilized man] can go up against gravitation in a balloon, and why
should he not hope that ultimately he may be able to stop or
accelerate his drift along the Time-Dimension, or even turn about and
travel the other way?''}
{The Time Traveler, in H.G. Wells' {\em The Time Machine}, 1895}

In good time, as it were, we will come to our performance analysis of
the arithmetic necessary to run Shor's algorithm for factoring large
numbers, and our {\em quantum multicomputer} architecture designed to
run the algorithm.  Let us begin prior to the genesis of quantum
computation, with the development of {\em reversible computing}.
Gates in quantum computation depend on concepts developed for
reversible classical computing, which is sometimes also called
``conservative logic''.  Once we understand the basics of reversible
classical computation, it will be easier to understand the circuits
and algorithms for quantum computation presented in the second and
third parts of this chapter, first the basic principles of quantum
computing then the major topic of quantum error correction.

\section{Reversible Classical Computation}
\label{sec:reversible-classical}

In a reversible computation, it is possible to recover the complete
initial state of the system having only the final state.  A
\textsc{not} gate, for example, is reversible; applying a second
\textsc{not} gate recovers the initial state with no loss of
information.  An AND gate is not reversible; from the single output
bit it is not always possible to determine the input state
unambiguously.  If the output is 1, we know that the input was 11, but
if the output is 0, we can't tell whether the input was 00, 01, or 10.
Similarly, an OR gate is not reversible; if the output is 1, we don't
know whether the input state was 10, 01, or 11.  A single bit of
output is insufficient to discriminate among the possible states of
multiple bits of input.  These examples suggest an important rule:
\begin{quotation}
Reversible gates must have the same number of outputs as inputs, and
the mapping of input to output states must be $1:1$.
\end{quotation}

First, we briefly discuss the history and importance of reversible
computation, then show the important two-bit reversible gate, followed
by three-bit gates and the emulation of Boolean logic.  We finish by
presenting ancilla management techniques without which the space
required for most interesting computations would grow unacceptably.
We do not discuss the thermodynamics of computation in any detail
here; interested readers will find this topic covered in the papers
referenced here.

\subsection{History and Importance}

Reversible computation was developed in the early 1970s by Charles
Bennett~\cite{bennett73:reversible}, acting on inspiration from
Landauer's discovery that the {\em erasure} of information requires an
increase in
entropy~\cite{landauer61:_irreversibility,bennett88:_notes}.  In
traditional logic, erasing information may involve, for example,
discharging a capacitor, which dissipates energy.  At first glance
this appears to be an implementation-dependent fact, but Landauer
proved that it is in fact fundamental.  Bennett initially proposed
reversible Turing machines, and discussed reversibility in the context
of the contents of several tapes.  We shall discuss reversibility in
the form of circuits and gates, rather than Turing machines, in this
thesis.  In order to be computationally complete, single-bit and even
two-bit gates are not enough; at least one three-bit operation is
necessary.  Fredkin and Toffoli invented the two most commonly used
three-bit reversible gates, discussed
below~\cite{fredkin82:_conserv_logic}.

Studying reversible computation is interesting in its own
right~\cite{feynman:lect-computation}: Kerntopf has identified more
than sixty research papers on the topic, including a variety of basic
logic gates that we will not detail here~\cite{kerntopf02:rev-synth}.
Perhaps the most famous classical example of reversible computing is
the billiard ball computer developed by Fredkin, Feynman, and others,
in which colliding billiard balls compute functions~\footnote{Ross
Berteig, Takako Matoba and I implemented a small-scale circuit based
on these principles in 1985, when taking Feynman's class on
``Potentialities and Limitations of Computing Machines''.}.  Such a
system is easier to design when conserving billiard balls, making
reversible logic the obvious choice.  For more practical circuits,
Bruce et al.  recently designed reversible carry-ripple and carry-skip
adders using Fredkin gates, intended to be implemented in
silicon~\cite{bruce02:_revers-adders}.  Hall designed a reversible
instruction set equivalent to a PDP-10~\cite{bell78:_evolut_dec10}
more than a decade ago, before quantum computation became a hot
research topic~\cite{hall94reversible}.  More recently, Vieri, Frank
and others, working in the Tom Knight group at MIT, designed and
fabricated a reversible microprocessor known as
Pendulum~\cite{vieri99:_thesis,vieri98:fully}.  They developed not
only the microprocessor, but also a small compiler.  Frank's thesis
discusses in detail topics such as options for subroutine call and
branch structure, and operating systems for reversible computers; as
reversible and quantum computer architectures advance, this thesis
will be a valuable resource~\cite{frank99:_thesis}.

Reversible computation benefits the thermodynamics of a system.  The
minimum amount of energy that a circuit must dissipate is proportional
to the number of bits of information that are {\em erased}.  Although
the minimum amount of energy to erase a bit is very small, this factor
eventually must be addressed in classical systems.  Athas, Koller and
their collaborators have investigated its importance for lowering
power consumption in adiabatic CMOS and found that power distribution
and clocking issues are manageable, but that the increase in chip area
required is significant~\cite{athas:reversible,koller:_adiabatic}.
They suggest occasionally relaxing the constraints on reversibility,
discarding a few intermediate results to reduce the area consumed.
Their chips operate far above the theoretical minimum for irreversible
logic, but take advantage of adiabatic charging and discharging of
capacitors to reduce power consumption.  DeBenedictis has argued that
building a high-performance computer system capable of exceeding $\sim
10^{26}$ logic gates per second or 10 zettaFLOPS ($10^{22}$ floating
point operations per second), roughly 6-7 decimal orders of magnitude
more than the current most powerful systems, within a realistic power
budget (750 kilowatts to the active logic components) will require the
use of reversible logic~\cite{debenedictis05:_reversible}.

\subsection{Two-Bit Gates}

Classically, the only important one-bit gate is the \textsc{not} gate, and, as
noted, it is reversible.  For two-bit gates, we have the \textsc{cnot} and \textsc{swap},
and construct \textsc{fanout}.

First, let us look at the controlled-\textsc{not} gate, or \textsc{cnot}.  One variable
(or input) is designated as the control line, and the other as the
target.  If the control bit is one, a \textsc{not} gate is performed on the
target bit; if the control bit is zero, the target bit is left
unchanged.  The output is the exclusive OR (XOR) of the two bits, and
one of the input bits: $(a,b)\rightarrow(a,a\oplus b)$.
Table~\ref{tab:cnot} shows the truth table for a \textsc{cnot} with A as the
control bit and B as the target bit.  Applying a \textsc{cnot} gate twice to
the same bits returns to the system to its original state,
$(a,b)\rightarrow(a,a\oplus b)\rightarrow(a,a\oplus b\oplus b) = (a,b)$.

\begin{table}
\centerline{
\begin{tabular}{cc|cc}
\multicolumn{2}{c|}{input} & \multicolumn{2}{c}{output} \\\hline
A & B & A & B \\
0 & 0 & 0 & 0 \\
0 & 1 & 0 & 1 \\
1 & 0 & 1 & 1 \\
1 & 1 & 1 & 0 \\
\hline
\end{tabular}
}
\caption{\textsc{cnot} truth table.}
\label{tab:cnot}
\end{table}

Swapping two bits is an important capability.  Physically, if data
signals are propagating through a circuit, routing of wires may
accomplish the swap.  However, if two register bits are to be swapped,
and no temporary storage location is available, we need a different
approach.  In standard logic, three consecutive XORs will swap two
bits or two entire registers without the use of intermediate,
temporary variables~\cite{beeler72:_hakmem}.  A similar trick, using
three \textsc{cnot}s, can be done in reversible computation, as shown in
Figure~\ref{fig:reversible-gates} on
page~\pageref{fig:reversible-gates}.

In reversible notation, we must explicitly specify the fanout of a
signal, an operation generally done implicitly with a wire in
irreversible logic.  A \textsc{cnot} performed with the variable to be copied
as the control and a zero in the target bit accomplishes this task for
us.

\subsection{Three-Bit Gates: Toffoli and Fredkin}

The two seminal reversible three-bit gates are the Toffoli and Fredkin
gates.  Table~\ref{tab:ccnot} shows the truth table for the
control-control-\textsc{not} (\textsc{ccnot}), or Toffoli gate.  If both control lines,
A and B, are one, then a \textsc{not} gate is performed on the target bit, C,
otherwise, no action is performed.  Table~\ref{tab:fredkin} shows the
control-\textsc{swap}, or Fredkin, gate.  This gate has one control line (A)
and two target lines (B and C).  If the control is one, the two
targets have their values swapped; if the control is zero, the targets
are unaffected.  Either of these gates is adequate to perform
universal computation; any computable circuit or equation can be
reduced to a set of Toffoli gates or a set of Fredkin gates.  Smaller
gates, such as the \textsc{cnot} and \textsc{not}, can of course be simulated by setting
one or two of the inputs to the gate to zero or one, as appropriate.

\begin{figure}
\input{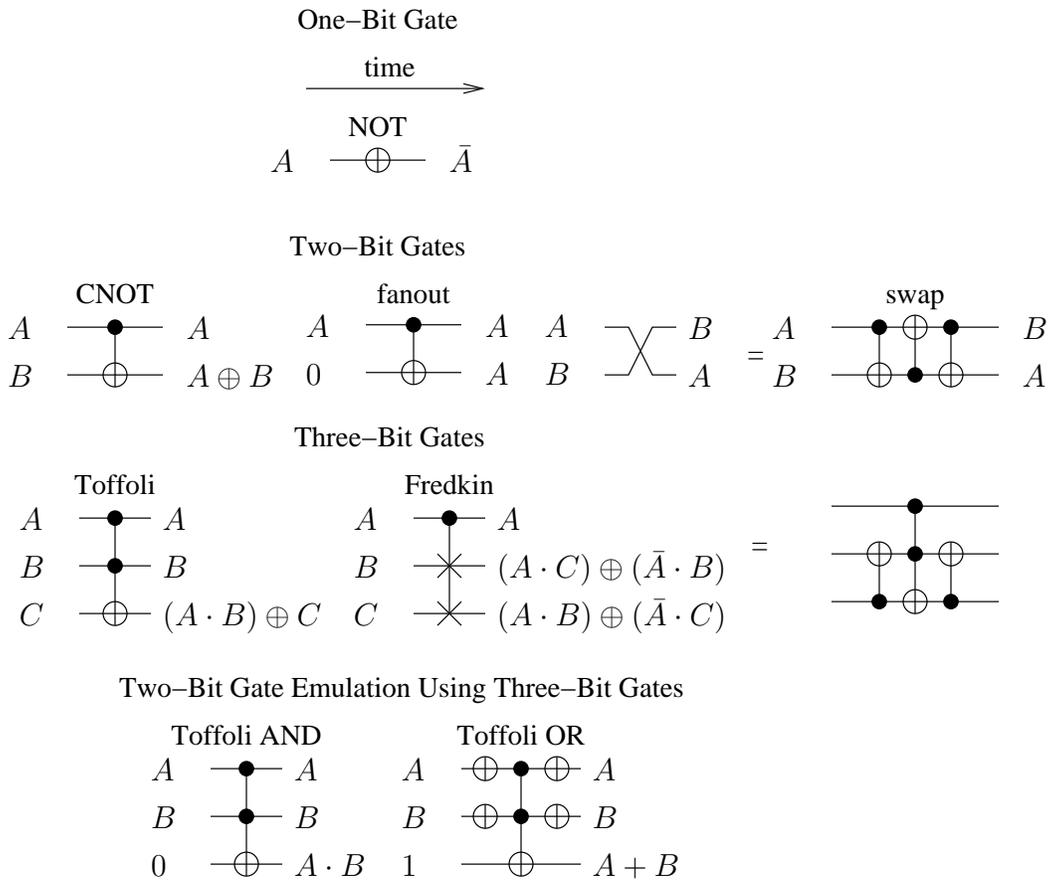}
\caption[Reversible gates]{Reversible gates: \textsc{cnot}, \textsc{ccnot} (Toffoli),
  control-\textsc{swap} (Fredkin), \textsc{not}, fanout and swap, and emulation of
  Boolean AND and OR using the Toffoli gate.}
\label{fig:reversible-gates}
\end{figure}

Graphic symbols for these gates are shown in
Figure~\ref{fig:reversible-gates}.  In all circuit diagrams in this
thesis, time flows left to right, a horizontal line represents a
single bit through time, and vertical line segments represent gates.
A filled dot indicates a control variable, while an open circle
represent a \textsc{not} gate on that variable -- the target of the gate, for a
\textsc{cnot} or \textsc{ccnot}.

\begin{table}
\centerline{
\begin{tabular}{ccc|ccc}
\multicolumn{3}{c|}{input} & \multicolumn{3}{c}{output} \\\hline
A & B & C & A & B & C \\
0 & 0 & 0 & 0 & 0 & 0 \\
0 & 0 & 1 & 0 & 0 & 1 \\
0 & 1 & 0 & 0 & 1 & 0 \\
0 & 1 & 1 & 0 & 1 & 1 \\
1 & 0 & 0 & 1 & 0 & 0 \\
1 & 0 & 1 & 1 & 0 & 1 \\
1 & 1 & 0 & 1 & 1 & 1 \\
1 & 1 & 1 & 1 & 1 & 0 \\
\hline
\end{tabular}
}
\caption{\textsc{ccnot} (Toffoli gate) truth table.}
\label{tab:ccnot}
\end{table}

\begin{table}
\centerline{
\begin{tabular}{ccc|ccc}
\multicolumn{3}{c|}{input} & \multicolumn{3}{c}{output} \\\hline
A & B & C & A & B & C \\
0 & 0 & 0 & 0 & 0 & 0 \\
0 & 0 & 1 & 0 & 0 & 1 \\
0 & 1 & 0 & 0 & 1 & 0 \\
0 & 1 & 1 & 0 & 1 & 1 \\
1 & 0 & 0 & 1 & 0 & 0 \\
1 & 0 & 1 & 1 & 1 & 0 \\
1 & 1 & 0 & 1 & 0 & 1 \\
1 & 1 & 1 & 1 & 1 & 1 \\
\hline
\end{tabular}
}
\caption{Control-\textsc{swap} (Fredkin gate) truth table.}
\label{tab:fredkin}
\end{table}

\subsection{Ancilla Management}

Every temporary variable created --- every term in a logical
expression --- consumes a bit.  For example, in the simple expression
$(A\wedge B)\vee(C\wedge D)$, the terms $(A\wedge B)$ and $(C\wedge
D)$ each require a temporary bit during the calculation of the final
result.  These temporary variables, in reversible logic terminology,
are {\em ancillae}.  Without a method for recovering these ancillae,
the space required for a computation would grow in direct proportion
to the length of the computation.  Of course, since we are using only
reversible gates in this computation, we could clean our ancillae
(collect our garbage) by applying the exact same set of gates in the
reverse order.  Unfortunately, that would return the state of the
entire system to the initial state, including resetting our desired
output to zero.  We need a way to keep the output but clean up the
garbage, and maybe even ``delete'' the input if what we really want to
keep is just the output.

Bennett discovered a method for cleaning ancillae while retaining the
important results bits.  He originally constructed this method for
Turing machines; we will describe it in terms of circuits and
registers.  We will illustrate the computation in terms of three
registers used in the computation itself (the INPUT, TEMPVARS for
intermediate variables, and TEMPOUT, which holds the result
immediately after completing the computation), though in practice the
roles assigned to bits may not be that clearly delineated.  A fourth
register, OUTPUT, gets the final result.  The computation is run
forward (step 1), then the results are ``copied'' out to the OUTPUT
register (step 2)~\footnote{In this thesis, we use the term ``copy''
to mean the fanout operation described above.}, then the
ancillae are returned to their initial (generally, zero) state by
reversing the computation (step 3).  This is illustrated in
Table~\ref{tab:erasing-ancillae}.  Bennett also defined a seven-step
method for doing in-place computation (erasing the input state,
leaving only the output), and Feynman stated that he had a method for
doing a $2n$-step irreversible computation reversibly in only $3n$
steps, though as far as I can tell he did not publish this result and
it has never been replicated~\cite{feynman:_future_computers}.

\begin{table}
\centerline{
\begin{tabular}{llllll}
Step & action & INPUT & TEMPVARS & TEMPOUT & OUTPUT \\\hline
0. & initial state & input & 0 & 0 & 0 \\
1. & forward computation & input & garbage & result & 0 \\
2. & ``copy'' using \textsc{cnot} fanout & input & garbage & result & result \\
3. & reverse computation & input & 0 & 0 & result \\
\hline
\end{tabular}
}
\caption{Erasing ancillae.}
\label{tab:erasing-ancillae}
\end{table}

\section{Introduction to Quantum Computing}
\label{sec:qc-intro}

\begin{chapterquote}
Alice laughed. ``There's no use trying,'' she said: ``one can't
 believe impossible things.''

``I daresay you haven't had much practice,'' said the Queen. ``When I
    was your age, I always did it for half-an-hour a day. Why,
    sometimes I've believed as many as six impossible things before
    breakfast.''\\
\flushright{\textbf{Lewis Carroll}, {\em Through the Looking Glass},
    1871}
\end{chapterquote}

A {\em quantum computer} is a device that takes advantage of quantum
mechanical effects to perform certain computations faster than a
purely classical machine can.  It relies on {\em quantum parallelism},
using physical phenomena that can be held, like Schr\"odinger's cat,
in more than one state at once, allowing us to compute on all of those
states at the same time, using a single operation.  Quantum
parallelism is best understood in the context of the concepts of {\em
superposition}, {\em entanglement} and {\em measurement}; of course,
we must also learn how quantum data is represented and manipulated.  A
quantum computer performs, in principle, exponentially many
computations simultaneously; however, exponentially many {\em results}
of those computations cannot be read out, leaving us with the
fascinating problem of how to use such a machine to accelerate
computations that interest us.  The most famous result in quantum
computing to date, Shor's algorithm for factoring large numbers (which
we will discuss in more detail in the next chapter), appears to offer
superpolynomial speedup, but no general method for finding quantum
analogs to classical algorithms is known.

This section reviews the basics of quantum computing. We begin with
quantum mechanics, presenting Dirac's {\em ket} notation, with a few
notes on linear algebra, then Schr\"odinger's equation and Hamiltonian
dynamics.  We then informally define a qubit, discuss its state-vector
and Bloch sphere representations and corresponding manipulations.
Two-qubit gates and their relationship to the reversible gates
presented above are explained, along with constructions for the
Toffoli gate.  Once we have begun to understand these fundamentals, we
can discuss DiVincenzo's criteria for physical realization of quantum
computation.  We end the chapter with a discussion of distributed
quantum computation, which is the purpose of our proposed quantum
multicomputer.  Readers are also referred to both
popular~\cite{nielsen:sciam,williams:_ultim_zero_one} and
technical~\cite{kitaev:cqc,nielsen-chuang:qci} texts on the topic for
more breadth and depth.

\subsection{Notation and a Few Linear Algebra Notes}

First, let us introduce the notation commonly used in quantum
computing.  We will not give rigorous definitions, instead limiting
ourselves to a few of the practical matters that a working engineer
needs to understand.

$|\psi\rangle$ is Dirac's {\em ket} notation for vectors, and this can
be referred to as the state-vector representation of a qubit.
$\langle\psi|$ is the {\em bra} corresponding to the ket.  The bra is
a complex-conjugate row vector and the ket is a column vector.
$\langle\psi_1|\psi_2\rangle$ is the dot product of the two vectors
$\psi_1$ and $\psi_2$, and $|\psi_1\rangle\!\langle\psi_2|$ is their
outer product.

For a single qubit, $|0\rangle$ is the zero state, and $|1\rangle$ is
the one state.  For a multiple-qubit register, we will often write the
binary expansion of the state as e.g. $|0111\rangle$ (a four-qubit state
with the value seven).  This state can also be written
$|0\rangle|1\rangle|1\rangle|1\rangle$ or
$|0\rangle\otimes|1\rangle\otimes|1\rangle\otimes|1\rangle$,
emphasizing that it is the tensor product of four separate two-level
systems.  Sometimes, we will write $|7\rangle$ as the state of the set
of qubits.  Although the number may be written base ten for
convenience, it is represented in binary in the quantum register (many
physical phenomena, such as the energy levels of an atom, may have
more than two levels and therefore may use e.g. $|2\rangle$ to
represent the third level, but we will confine ourselves to two-level
qubits in this thesis).  The size of the register will usually be
understood from context, and if the integer is small the high-order
bits are of course understood to be zero.  Occasionally, it may be
necessary to write $|0\rangle^{\otimes k}$ to indicate a set of $k$
qubits all in the zero state.

We describe an arbitrary $n$-qubit quantum gate via the $2^n\times2^n$
matrix $U$, which must be a {\em unitary} transform.  A unitary matrix
obeys the equation $U^\dagger U = UU^\dagger = I$, where $U^\dagger$
is the adjoint of $U$.  In keeping with normal matrix multiplication
rules, a series of gates or transforms applied to a register can be
written
\begin{equation}
U_k\cdots U_3 U_2 U_1 |\psi\rangle
\end{equation}
where $U_1$ is the first gate applied, $U_2$ is the second, etc.  This
can be confusing, as we draw circuit diagrams with time flowing left
to right.  We introduced the graphical notation for reversible gates
in Chapter~\ref{ch:reversible}; we extend that to quantum gates in
Section~\ref{sec:qgate}, and larger circuits will appear in later
chapters.

\subsection{Schr\"odinger's Equation}

Schr\"odinger's equation
\begin{equation}
i\hbar\frac{\partial|\psi\rangle}{\partial t} = H|\psi\rangle
\end{equation}
describes the dynamics of a quantum system.  Solutions describing the
time evolution of the system are of the form
\begin{equation}
|\psi\rangle \rightarrow e^{-iHt/\hbar}|\psi\rangle = U|\psi\rangle.
\end{equation}
$H$, in this equation, is an operator (represented as a matrix) known
as the {\em Hamiltonian} of the system, and $U$ is the corresponding
unitary transform.  Solutions to the Schr\"odinger equation can be
weighted, linear combinations of any of the possible solutions, such
that the weights all add up to 1.  Experimentalists usually describe
the behavior of the system in terms of its Hamiltonian to emphasize
the temporal nature of the evolution, but we are interested in
specific types of behavior achieved by using fixed time intervals, so
it will be easiest for us to use the unitary operators.  Unitary
operators can, in turn, be expressed as gates, which we will use
throughout this thesis.

\subsection{Qubits}

\subsubsection{What's a Qubit?}

A {\em qubit} is either a true two-level system, such as the direction
of polarization of a photon or the direction of spin of an electron,
or a pseudo-two-level system, such as two energy levels of an atom
that can be treated as a two-level system.  We will see more examples
in Chapter~\ref{ch:taxonomy}.  Of course, an electron spins in either
the ``up'' or ``down'' direction, not zero and one, so we chose to
label the two states as our zero and one states, much as we choose
e.g. $+5$ volts to be a logical one and ground to be a logical zero in
classical circuits.  The difference between a classical bit and a
qubit is that a qubit can be in a {\em superposition} of the two
states; it can be partially zero and partially one.  The state of a
qubit can be written as
\begin{equation}
|\psi\rangle = \alpha|0\rangle + \beta|1\rangle
\end{equation}
where $\alpha$ and $\beta$ are complex numbers, $|\alpha|^2$ is the
probability of finding the qubit in the state 0, and $|\alpha|^2 +
|\beta|^2 = 1$: the qubit must be found to be in one state or the
other.

The above expression can also be written
\begin{equation}
|\psi\rangle = \left[\begin{array}{c}\alpha \\ \beta \end{array} \right]
\end{equation}
showing the same probabilities for finding the states 0 and 1,
implicit in the position within the vector.  The top element of the
vector corresponds to the zero state, and the bottom element to the
one state.  Technically, the 0 and 1 inside the ket are labels for the
states; we could choose to represent any two basis vectors by
$|0\rangle$ and $|1\rangle$, but in this dissertation we will always
use the convention that
\begin{equation}
|0\rangle\equiv\left[\begin{array}{c} 1 \\ 0 \end{array} \right],
|1\rangle\equiv\left[\begin{array}{c} 0 \\ 1 \end{array} \right].
\end{equation}

\begin{figure*}
\centerline{\hbox{
\input{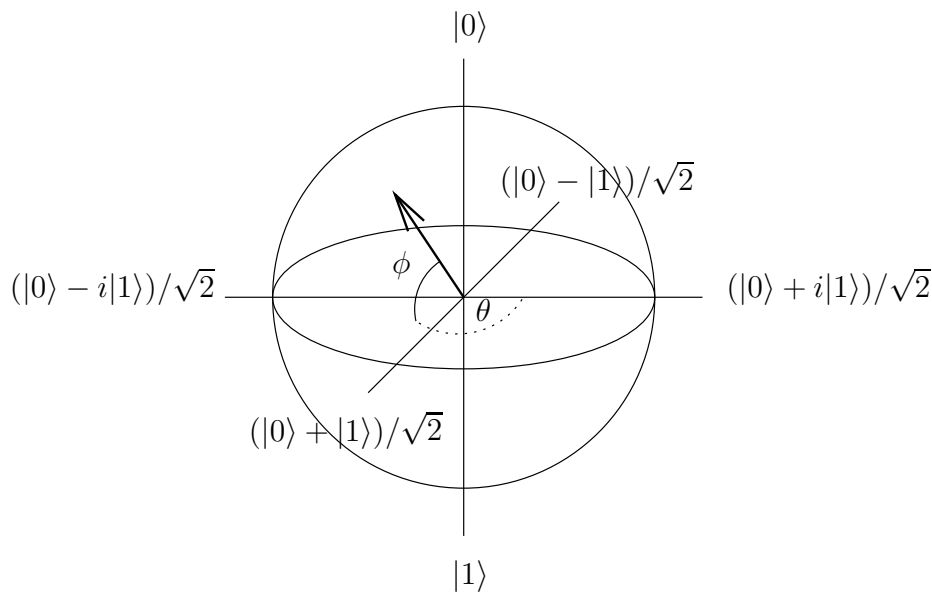}}}
\caption{The Bloch sphere.}
\label{fig:bloch-sphere}
\end{figure*}

The state of a single qubit is often thought of in terms of the {\em
Bloch sphere} representation, in which the state of a qubit is a unit
vector, as shown in Figure~\ref{fig:bloch-sphere} (this sphere is
often called the {Poincar\'e sphere} by researchers working in
optics).  If the vector points at the north pole, our qubit is in the
$|0\rangle$ state, and if it points at the south pole, the qubit is in
the $|1\rangle$ state.  The north-south axis is the $Z$ axis, the
positive $X$ axis is toward the reader (out of the page or screen, for
a 2-D representation), and the $Y$ axis is right-left.  When the unit
vector points toward you, that is the $(|0\rangle +
|1\rangle)/\sqrt{2}$ state, when it points away from you that is the
$(|0\rangle - |1\rangle)/\sqrt{2}$ state.  The positive $Y$ axis is
$(|0\rangle+i|1\rangle)/\sqrt{2}$, and the negative $Y$ axis is
$(|0\rangle-i|1\rangle)/\sqrt{2}$.  The {\em phase} is the position of
our vector about the $Z$ axis (the angle $\theta$ in the figure).

Physicists, especially theorists, occasionally refer to a large
unitary transform as a {\em quantum gate}, but in this dissertation we
will restrict the use of the term to smaller units, which for most
proposed implementations will be more physically realistic.  Our gates
will be one-, two-, and three-qubit transforms only.

\subsubsection{Quantum Registers and Weighted Probabilities}
\label{sec:q-reg}

We will refer to a related set of two or more qubits as a {\em quantum
 register}.  Two classical bits can be in any of the four states $00$,
 $01$, $10$, and $11$.  Two qubits can be in a weighted combination of
 all four states at the same time.  For two qubits, we can write
\begin{equation}
|\psi\rangle = \alpha|00\rangle + \beta|01\rangle + \gamma|10\rangle + \delta|11\rangle
\end{equation}
where $|\alpha|^2 + |\beta|^2 + |\gamma|^2 + |\delta|^2 = 1$.  For
example, if $\alpha = \delta = 1/\sqrt{2}$ and $\beta = \gamma = 0$,
we have a fifty percent probability of finding $|00\rangle$ and a
fifty percent probability of finding $|11\rangle$, but no chance of
finding the other states.

Similarly, three qubits can be in eight states, and $n$ qubits can be
in all $2^n$ possible states at once,
\begin{equation}
|\psi\rangle = \sum_{i = 0}^{2^n-1}\alpha_i|i\rangle,
\label{eq:psi-full}
\end{equation}
subject to the constraint that their total
weights $\alpha_i$ must sum to 1,
\begin{equation}
\sum|\alpha_i|^2 = 1.
\end{equation}
Of course, some of the $\alpha_i$ may be zero.

\subsubsection{Entanglement}
\label{sec:entanglement}

Two quanta can be in a shared state in which operations on one affect
the other.  The quanta are said to be {\em entangled}.  One
consequence is that the probabilities of two entangled qubits are not
independent (but see Section~\ref{sec:pure-mix} below for an important
caveat).  If the state of the system is e.g. $(|00\rangle +
|11\rangle)/\sqrt{2}$ ($\alpha = \delta = 1/\sqrt{2}$, in the above
notation), when we measure the system, we will find either that both
qubits are zero, or that both qubits are one.  Although each qubit has
a 50\% probability of being zero and a 50\% probability of being one,
their state is not independent.  Starting from this state, we will
never find one qubit to be zero and the other qubit to be one.

Entanglement is a continuous phenomenon, not discrete.  There are
numerous measures of the amount of entanglement present in a system,
but they all use a scale running from zero to one, where zero is
completely unentangled and one is fully entangled (see Munro et
al. and references therein~\cite{munro01:_bell_inequal}.)  For the
purposes of this thesis, our primary interest will be in
fully-entangled and fully-unentangled pairs of qubits, though the
process of purifying a set of partially entangled pairs of qubits into
fully-entangled pairs will figure into the qubus network protocol
(Chapter~\ref{ch:travel} and
Section~\ref{sec:qec-qubus})~\cite{cirac97:_distr_quant_comput_noisy_chann,bennett95:_concen,spiller05:_qubus}.

\subsubsection{Decoherence}
\label{sec:decoherence}

Quantum states are very fragile: excited atoms decay and spins of
electrons and atomic nuclei spontaneously flip.  Any quantum system
can be affected by interacting with its environment, leaking
information about its state out into the environment where we cannot
recover or use the information.  We call this gradual decay of the
state of a system {\em decoherence}.  When decoherence sets in,
measurement of the system probably will not produce the desired
results, causing the failure of our quantum algorithm.  The two key
measures of decoherence are the $T_1$ and $T_2$ times.  $T_1$ is the
energy relaxation time, and $T_2$ is the phase relaxation time.  Both
processes are memoryless, with probabilistic behavior.  The amount of
time we can count on the state of a qubit remaining in a usable state
is the minimum of $T_1$ and $T_2$.  Researchers determine these values
experimentally, and an important area of device research is extending
these times by careful engineering of the environment and control
system.

\subsubsection{Pure and Mixed States and the Density Matrix}
\label{sec:pure-mix}

Quantum states can be either {\em pure} or {\em mixed}.  So far, we
have discussed only pure states.  ``Pure'' does not mean that the
superposition, when written out in state-vector form, contains only
one term; pure means that it is {\em possible} to write the state in
state-vector form.  For example, $|\psi\rangle = |0\rangle$ and
$|\psi\rangle = (|0\rangle + |1\rangle)/\sqrt{2}$ are both pure
states.  However, not all quantum states can be written out completely
in the state-vector form.  Experimentalists often prefer to write the
state using the $2^n\times2^n$ {\em density matrix} form, which can
represent a more complex state of the system.  In particular, the
density matrix representation allows us to write down a representation
of the state of the system when the complete state cannot be known,
such as when part of the information in the quantum state has leaked
out into the environment.  Using the example of our basic entangled
state, $|\psi\rangle = (|00\rangle + |11\rangle)/\sqrt{2}$, our
density matrix is
\begin{equation}
\rho = |\psi\rangle\!\langle\psi| = \frac{1}{2}|00\rangle\!\langle 00| + 
\frac{1}{2}|00\rangle\!\langle 11| + \frac{1}{2}|11\rangle\!\langle 00| +
\frac{1}{2}|11\rangle\!\langle 11|
= \left[\begin{array}{cccc}
\frac{1}{2} & 0 & 0 & \frac{1}{2} \\
0 & 0 & 0 & 0 \\
0 & 0 & 0 & 0 \\
 \frac{1}{2} & 0 & 0 & \frac{1}{2} \\
\end{array} \right].
\label{eq:dens-mat}
\end{equation}
The entries along the diagonal of the density matrix correspond to the
probability of finding the system in a particular state.  To be a
valid density matrix, the trace (the sum of the diagonal) must be one,
written $\operatorname{Tr}(\rho) = 1$.  The trace must be one because,
when measured, the system will be found to be in {\em some} state.
For pure states, the square of the density matrix also has trace one,
$\operatorname{Tr}(\rho^2) = 1$.  If the density matrix is
diagonalized (achieved via an appropriate change of basis), a pure
state will have only a single non-zero element.  The eigenvector
corresponding to this eigenstate is the state of the system.  The
Bloch sphere can be used to visualize mixed states of a single qubit
as points inside the sphere; the closer the state is to pure, the
closer the length of the vector is to unity.

In Section~\ref{sec:entanglement} above, we referred to a caveat on
our definition of entanglement; with this understanding of the
difference between pure and mixed states we are now ready to discuss
it.  The state of two qubits can, in fact, be dependent, without being
entangled, if the state is mixed.  In contrast to the state in
Equation~\ref{eq:dens-mat}, we can also have the state
\begin{equation}
\frac{1}{2}(|00\rangle\!\langle 00| + |11\rangle\!\langle 11|)
= \frac{1}{2}\left[\begin{array}{cccc}
1 & 0 & 0 & 0 \\
0 & 0 & 0 & 0 \\
0 & 0 & 0 & 0 \\
0 & 0 & 0 & 1 \\
\end{array} \right].
\label{eq:dens-mat-unentangled}
\end{equation}
In this mixed state, the state of the two qubits is not independent,
but they are not entangled; actions on one qubit cannot affect the
state of the other.  In this particular case, the density matrix now
represents classical dependent probabilities.

\subsubsection{Measurement}
\label{sec:measurement}

{\em Measurement} of a qubit causes the collapse of the wave function,
forcing the state of the system into just one term of the
superposition.  In the famous thought experiment of Schr\"odinger,
measurement is opening the box containing his cat and finding out if
the cat is dead or alive.  Until measurement takes place, the state of
the system can be in the superposition state, with various histories
and outcomes only determined probabilistically.  When we measure the
system, the state and history pick one consistent ``storyline'' that
the system must have followed, in effect choosing among possible pasts
based on their relative probabilities.  If we measure such that more
than one history is possible, the system remains in a state that is
consistent with all of them, as in the double-slit quantum
interference experiment (see, for example, V. I Ch. 37 of the Feynman
Lectures~\cite{feynman63:_lect_physics}).

In our basic example of $|\psi\rangle = |0\rangle$, we know the system
is 100\% in the zero state.  Measurement of the qubit's state will
definitely produce a zero~\footnote{Assuming the measurement is
performed along the Z (0/1) axis; we will not deal with measurements
in other bases in this dissertation.}.  For $|\psi\rangle = (|0\rangle
+ |1\rangle)/\sqrt{2}$, zero and one each have a fifty percent
probability of being found.  Once our measurement determines the state
(e.g., 0), the entire system will be forced to a state consistent with
the idea that our qubit has been zero all along.

For two or more qubits, we can measure either the entire system, or
only part.  Measuring a single qubit can alter the state of the
system.  For example, consider our two-qubit state $|\psi\rangle =
(|00\rangle + |11\rangle)/\sqrt{2}$.  If we measure the low-order
bit (the right-hand one of our pair), we have a fifty percent
probability of each outcome, and our result will force the system to a
matching state.  We can write the measurement outcome and the
resulting state as
\begin{align}
0:&\quad|\psi\rangle\rightarrow|0\rangle\\
1:&\quad|\psi\rangle\rightarrow|1\rangle.
\end{align}
In this case, measuring one qubit has determined the state of the
other.  For the state $|\psi\rangle = (|00\rangle +
|10\rangle)/\sqrt{2}$, we can factor the state as $|\psi\rangle =
(|0\rangle+|1\rangle)|0\rangle/\sqrt{2}$.  Measuring the low-order
qubit will clearly always yield the result 0.  The state of the system
then moves to $(|0\rangle+|1\rangle)/\sqrt{2}$; the high-order qubit
(now our only qubit) has not changed.  We can say that two qubits were
{\em separable}; there was no entanglement between them.

Measurement is a complex and sometimes counter-intuitive topic.  It is
important and deep enough that books and conferences are devoted to
it~\cite{alter01:_quant_measur_singl_system}.  One good place to start
studying this topic is Preskill's lecture
notes~\cite{preskill98:_lecture-notes}.  We will see an example of how
to use measurement in the discussion of quantum error correction in
Section~\ref{ch:err-mgmt}.

\subsubsection{The Partial Trace}
\label{sec:partial-trace}

We are now ready to discuss the {\em partial trace} of a system.  We
use the partial trace for various purposes, including expressing the
loss of a photon in optical quantum computing or the ``leaking'' of
information about the state out into the environment.

We can discuss the state of a system in terms of the {\em system} and
the {\em reservoir}, where system in this case refers to the qubits we
are interested in and have control over, and reservoir refers to the
rest of the world.  Initially, the system and the reservoir are not
entangled; that is, they are separable, and the state can be written
\begin{equation}
\rho = \rho_S \otimes \rho_R
\end{equation}
where $\rho$ is our overall state, $\rho_S$ is the state of the
quantum system, and $\rho_R$ is the state of the reservoir (which we
can never know fully).  Over time, information leaks out of the
quantum system into the larger world, or the reservoir.  If $\rho(t)$
is the state at time $t$,
\begin{equation}
\rho = \rho_S \otimes \rho_R.
\rho_S(t) = \operatorname{Tr}_R(\rho(t))
\end{equation}
where $\operatorname{Tr}_R$ is the partial trace with respect to the
reservoir.

For a two-qubit system, numbering our qubits 0 and 1, in keeping with
normal computer architecture convention, we will let $\rho^0$ be the
density matrix for the system traced out over qubit 1, and $\rho^1$ be
traced out over qubit 0.  Defining the partial trace as
\begin{equation}
\rho^0 = \operatorname{Tr}_1(\rho) = 
\langle{_1}0|\rho|0_1\rangle + \langle{_1}1|\rho|1_1\rangle,
\end{equation}
where $|0_1\rangle$ is the basis vector for the zero state for qubit
one.  Noting that $\langle 0|0\rangle = \langle 1|1\rangle = 1$ and
$\langle 0|1\rangle = \langle 1|0\rangle = 0$, and that the trace is
linear, the partial trace for the example in
equation~\ref{eq:dens-mat} is
\begin{equation}
\begin{split}
\rho^0 =& \operatorname{Tr}_1(\rho) =
\frac{1}{2}\operatorname{Tr}_1(|00\rangle\!\langle 00|) +
\frac{1}{2}\operatorname{Tr}_1(|11\rangle\!\langle 00|) +
\frac{1}{2}\operatorname{Tr}_1(|00\rangle\!\langle 11|) +
\frac{1}{2}\operatorname{Tr}_1(|11\rangle\!\langle 11|) \\
=& \frac{1}{2}\langle{_1}0|00\rangle\!\langle 00|0_1\rangle + 
 \frac{1}{2}\langle{_1}0|11\rangle\!\langle 00|0_1\rangle + 
 \frac{1}{2}\langle{_1}0|00\rangle\!\langle 11|0_1\rangle + 
 \frac{1}{2}\langle{_1}0|11\rangle\!\langle 11|0_1\rangle \\
&+ \frac{1}{2}\langle{_1}1|00\rangle\!\langle 00|1_1\rangle + 
 \frac{1}{2}\langle{_1}1|11\rangle\!\langle 00|1_1\rangle + 
 \frac{1}{2}\langle{_1}1|00\rangle\!\langle 11|1_1\rangle + 
 \frac{1}{2}\langle{_1}1|11\rangle\!\langle 11|1_1\rangle \\
=& \frac{1}{2}|0\rangle\!\langle 0| + \frac{1}{2}|1\rangle\!\langle 1| \\
=& \left[\begin{array}{cc}
\frac{1}{2} & 0 \\
0 & \frac{1}{2} \\
\end{array} \right].
\end{split}
\end{equation}
$\operatorname{Tr}((\rho^0)^2) = 1/2$, indicating that our state is
now a mixed state.  Our pure state has become mixed with the
environment, and we can no longer write down a definitive description
of the quantum register alone.  

\subsubsection{Interference}
\label{sec:interference}

The state of a quantum system is a wave function that matches
Schr\"odinger's equation.  As with classical wave mechanics, two waves
can {\em interfere}, depending on the relative phases of the waves.
That interference can be positive, enhancing the amplitude (hence,
probability) of a particular state, or negative, decreasing the
probability.  Since the phase of a state is actually complex, the
addition of phases is also complex.

As a simple example, consider the state created by application of a
Hadamard gate (which we will define below) to the $|0\rangle$ state,
\begin{equation}
|\psi\rangle = \frac{|0\rangle+|1\rangle}{\sqrt{2}} = \frac{1}{\sqrt{2}}\left[\begin{array}{c} 1 \\ 1 \end{array} \right].
\end{equation}
The state now consists of two terms, a superposition of two states.
Applying a second Hadamard gate will return the system to its original
state by interfering the two terms,
\begin{equation}
H|\psi\rangle = \frac{1}{2}\left[\begin{array}{rr} 1 & 1 \\ 1 & -1
  \end{array}\right]\left[\begin{array}{c} 1 \\ 1 \end{array} \right]
= \frac{1}{2}\left[\begin{array}{c} 1 + 1 \\ 1 - 1 \end{array}\right]
= \left[\begin{array}{c} 1 \\ 0 \end{array}\right]
= |0\rangle.
\end{equation}
The top element in the array exhibits positive interference ($1+1$),
and the bottom element shows negative interference ($1-1$).

\subsection{Manipulating Qubits}

Quantum computation proceeds by taking a set of qubits, modifying
their state such that a ``computation'' of some interest is performed,
and reading out the result so that we learn what happened.  Feynman
originally conceived of quantum computers as systems designed to
simulate the physical behavior of many-body systems, which are hard to
examine experimentally or in classical simulation, solving quantum
mechanical problems directly in an analog fashion rather than via
numerical calculation of properties of the wave
function~\cite{feynman:_simul_physic_comput,lloyd96:_univer_quant_simul,abrams97:_qc_fermi_simul,byrnes06:_simul}.
This approach is similar to e.g. simulating a set of mechanical
resonators using a set of electrical resonators, as is done in analog
computing~\cite{korn56:_elect_analog_comput,gilbert64:_desig_use_elect_analog_comput,mead89:_analog_vlsi}.
However, this is not the only way to use quantum phenomena to solve
problems.  A quantum computation can be defined as a circuit, in which
the system is built and programmed and behaves roughly analogously to
a classical digital computer.  Recent advances include adiabatic
quantum
computing~\cite{farhi01:_quant_adiab_science,steffen03:_exper_qc_adiab,aharonov:_adiab_qc_equiv}
and cluster-state
computing~\cite{raussendorf03:_cluster-state-qc,nielsen:cluster-review,walther05:_cluster-exper-nature}.
All of these are equivalent in computational power, but are believed
to be very different in how useful algorithms are found.  In this
dissertation, we will deal almost exclusively in terms of the circuit
model, which is the basis for Shor's factoring algorithm and most of
the other important quantum algorithms discovered to date.

\subsubsection{What's a Quantum Gate?}
\label{sec:qgate}

In the circuit model, quantum computations are decomposed into
separate gates, and can be organized more or less along the lines of
classical circuits.  These gates are based on the concepts of
reversible computing discussed in the last section, extended to
accommodate the Bloch sphere.  In order for our computational
capabilities to be ``universal'', we must be able to reach any point
on the Bloch sphere for a single qubit, and we must be able to
entangle two qubits.  First we discuss the individual gates that
compose a quantum computation, and in the next subsection we discuss
larger circuits in more detail.

\subsubsection{Single-Qubit Gates and the Bloch Sphere}

Only one interesting single-bit operation, the \textsc{not} gate, exists in the
classical world (ignoring setting and resetting the bit).  In the
quantum world, a single-qubit operation can be any rotation on the
Bloch sphere.  Rotations about the axes of the Bloch sphere can be
described in terms of the {\em Pauli matrices}.  The transforms for
$180^\circ$ rotations are
\begin{equation}
X = \sigma_x = \left[\begin{array}{rr}0 & 1 \\ 1 & 0 \end{array} \right]
\end{equation}
\begin{equation}
Y = \sigma_y = \left[\begin{array}{rr}0 & -i \\ i & 0 \end{array} \right]
\end{equation}
\begin{equation}
Z = \sigma_z = \left[\begin{array}{rr}1 & 0 \\ 0 & -1 \end{array} \right].
\end{equation}
For rotation of an angle $\theta$ about each axis, the transforms
(modulo a global phase factor we will ignore) are (from Nielsen \&
Chuang~\cite{nielsen-chuang:qci}):
\begin{equation}
R_x(\theta) = e^{-i\theta X/2} = \left[\begin{array}{rr}
\cos\frac{\theta}{2} & -i\sin\frac{\theta}{2} \\
-i\sin\frac{\theta}{2} & \cos\frac{\theta}{2} \end{array} \right]
\end{equation}
\begin{equation}
R_y(\theta) = e^{-i\theta Y/2} = \left[\begin{array}{rr}
\cos\frac{\theta}{2} & -\sin\frac{\theta}{2} \\
\sin\frac{\theta}{2}  & \cos\frac{\theta}{2} \end{array} \right]
\end{equation}
\begin{equation}
R_z(\theta) = e^{-i\theta Z/2} = \left[\begin{array}{cc}
e^{-i\theta/2} & 0 \\ 0 & e^{i\theta/2} \end{array} \right]
\end{equation}
which we will need only for the quantum Fourier transform and for our
decomposition of the Toffoli gate.

\begin{figure*}
\centerline{\hbox{
\input{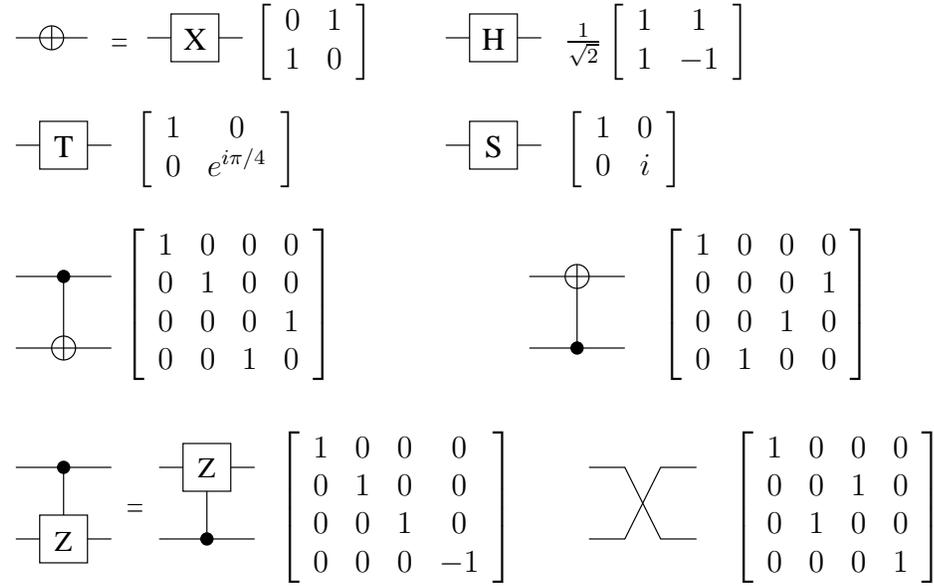}}}
\caption[Basic quantum gates]{Basic one-qubit \textsc{not} (X), Hadamard (H),
  $\pi/8$ (T), and phase (S) gates (top two rows), and two-qubit
  \textsc{cnot}s, control-phase, and swap gates (bottom two rows).}
\label{fig:basic-gates}
\end{figure*}

Universal quantum computation requires that we be able to reach any
location on the Bloch sphere starting from any other.  Naturally, we
do not need arbitrary rotations about all three axes in order to
achieve this; two will do.  Moreover, arbitrary rotations can be
approximated using a small set of fixed rotations.
Figure~\ref{fig:basic-gates} shows one such set of gates, with their
graphic representations and unitary transform matrices.  The
particular set shown is technically redundant; the control-Z and swap
gates can be constructed from the others.

As a simple example, consider the state created by application of a
Hadamard gate to the $|0\rangle$ state,
\begin{equation}
|\psi\rangle = H|0\rangle \rightarrow \frac{1}{\sqrt{2}}\left[\begin{array}{rr} 1 & 1
 \\ 1 & -1 \end{array} \right]\left[\begin{array}{r} 1
 \\ 0 \end{array} \right] = \frac{1}{\sqrt{2}}\left[\begin{array}{r} 1
 \\ 1 \end{array} \right] \rightarrow \frac{|0\rangle+|1\rangle}{\sqrt{2}}.
\end{equation}
The state now consists of two terms, a superposition of two states.
Likewise, applying the Hadamard to the $|1\rangle$ state, we have
\begin{equation}
|\psi\rangle = H|1\rangle \rightarrow \frac{1}{\sqrt{2}}\left[\begin{array}{rr} 1 & 1
 \\ 1 & -1 \end{array} \right]\left[\begin{array}{r} 0
 \\ 1 \end{array} \right] = \frac{1}{\sqrt{2}}\left[\begin{array}{r} 1
 \\ -1 \end{array} \right] \rightarrow \frac{|0\rangle-|1\rangle}{\sqrt{2}}.
\end{equation}
Geometrically, we visualize the Hadamard gate as a $180^\circ$
($\pi$) rotation about the Z axis, followed by a $90^\circ$
($\pi/2$) rotation about the Y axis.  The rotation about the Z axis
does not directly affect the probability of finding either a 0 or a 1
if the state is measured right away, but this two-step manipulation
shows clearly the importance of the phase (angle about the Z axis).

Unfortunately, visualizing the state of more than one qubit is more
complicated than a set of spheres, one per qubit.  If it were that
easy, there would be no exponential growth in the complexity of our
states, and quantum computation would be uninteresting.  It is
possible to visualize the state of more than one qubit as a {\em set}
of points on the Bloch sphere, in what is called the {\em Majorana
representation}.  Its utility is limited to pure states; there are not
enough degrees of freedom to represent mixed
states~\cite{markham04:_thesis}.

\subsubsection{Two-Qubit Gates}

In Chapter~\ref{ch:reversible}, we discussed classical reversible
computation using control-\textsc{not} (\textsc{cnot}) gates as our primary two-qubit
gate.  The \textsc{cnot} is an extremely useful gate in quantum computation, as
well, and will figure prominently in our quantum arithmetic.  However,
the \textsc{cnot} is not the only type of two-qubit quantum gate.  As with the
one-qubit gates, we must consider the phase of the system, resulting
in analog gates equivalent to the rotations about the axes we saw for
single-qubit gates.  We can create a ``control-$U$'' two-qubit gate,
where $U$ is any single-qubit unitary gate.

First, let us look at the unitary transforms for single-qubit gates
applied to two-qubit systems, so we can see the form the matrices
take.  For operations on multi-qubit registers, we will let $U_i$ be
the single-qubit unitary operation $U$ on the $i$th qubit in the
register.  We will number qubits from zero, with qubit zero being the
``low order'' qubit in the system.  Qubit $i$ then corresponds to the
value $2^i$ in the binary expansion (note that this is in keeping with
common computer architecture practice, but physicists usual number
from qubit 1, starting at the left, or high-order, bit).  In circuit
diagrams, the low-order qubit will be the bottom qubit.  The transform
for a Hadamard gate on the low-order qubit is
\begin{equation}
H_0\equiv I\otimes H = \frac{1}{\sqrt{2}}\left[\begin{array}{rrrr}
1 & 1 & 0 & 0 \\
1 & -1 & 0 & 0 \\
0 & 0 & 1 & 1 \\
0 & 0 & 1 & -1 \end{array}\right]
\end{equation}
and for one on the high-order qubit is
\begin{equation}
H_1 \equiv H \otimes I = \frac{1}{\sqrt{2}}\left[\begin{array}{rrrr}
1 & 0 & 1 & 0 \\
0 & 1 & 0 & 1 \\
1 & 0 & -1 & 0 \\
0 & 1 & 0 & -1 \end{array}\right]
\end{equation}
where $I_i$ is the identity operation on qubit $i$ and $H_i$ is the
Hadamard on qubit $i$.  Because the two gates operate on independent
qubits, the order in which we compose the larger unitary in does not
matter,
\begin{equation}
H_0H_1 = H_1H_0 = \frac{1}{2}\left[\begin{array}{rrrr}
1 & 1 & 1 & 1 \\
1 & -1 & 1 & -1 \\
1 & 1 & -1 & -1 \\
1 & -1 & -1 & 1 \end{array}\right].
\label{eq:h0h1}
\end{equation}
The two-qubit swap gate has a very simple transform,
\begin{equation}
SWAP = \left[\begin{array}{cccc}
1 & 0 & 0 & 0 \\
0 & 0 & 1 & 0 \\
0 & 1 & 0 & 0 \\
0 & 0 & 0 & 1 \end{array}\right].
\end{equation}
When we write a \textsc{cnot} gate, occasionally it will be necessary to
distinguish which qubit is which.  In that case, the first subscript
will be the control qubit and the second subscript the target qubit,
e.g.,
\begin{equation}
CNOT_{1,0} = \left[\begin{array}{cccc}
1 & 0 & 0 & 0 \\
0 & 1 & 0 & 0 \\
0 & 0 & 0 & 1 \\
0 & 0 & 1 & 0 \end{array}\right]
\end{equation}
and
\begin{equation}
CNOT_{0,1} = \left[\begin{array}{cccc}
1 & 0 & 0 & 0 \\
0 & 0 & 0 & 1 \\
0 & 0 & 1 & 0 \\
0 & 1 & 0 & 0 \end{array}\right].
\end{equation}

In some physical implementations, a control-phase gate is the natural
Hamiltonian.  The control-phase or control-Z unitary is
\begin{equation}
CZ_{1,0} = \left[\begin{array}{rrrr}
1 & 0 & 0 & 0 \\
0 & 1 & 0 & 0 \\
0 & 0 & 1 & 0 \\
0 & 0 & 0 & -1 \end{array}\right],
\end{equation}
or, more generally, for an arbitrary rotation by an angle $\theta$
about the Z axis,
\begin{equation}
CZ_{1,0}(\theta) = \left[\begin{array}{cccc}
1 & 0 & 0 & 0 \\
0 & 1 & 0 & 0 \\
0 & 0 & 1 & 0 \\
0 & 0 & 0 & e^{i\theta} \end{array}\right],
\end{equation}
which is not quite what we need for most logic.  However, we can
construct a \textsc{cnot} gate from CZ easily, by wrapping the CZ in a
pair of Hadamards on the target qubit:
\begin{equation}
H_0 CZ_{1,0} H_0 = CNOT_{1,0}.
\end{equation}
DiVincenzo described other related constructions in an early
paper~\cite{divincenzo98:_gates-and-circuits}.  The control-phase gate
is actually symmetric; it does not matter which of the two qubits we
treat as the control and which we treat as the target.  The change in
the system state is the same.  This fact is illustrated in
Figure~\ref{fig:basic-gates} on page~\pageref{fig:basic-gates} with
the control-Z gate both ``right side up'' and ``upside down''.  This
feature can result in unwanted error propagation, as discussed in
Section~\ref{ch:err-mgmt}.

\subsubsection{Three-Qubit Gates}
\label{sec:3qb-gates}

\begin{figure}
\centerline{\hbox{
\includegraphics[height=2cm]{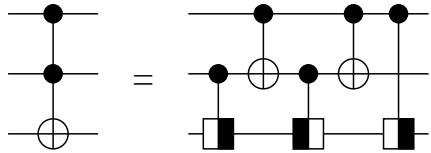}}}
\caption[Our \textsc{ccnot} construction]{Our \textsc{ccnot} construction.  The box with
  the bar on the right represents the square root of $X$, and the box
  with the bar on the left its adjoint.}
\label{fig:5gcc-only}
\end{figure}

We have already discussed the importance of the Toffoli and Fredkin
gates in classical reversible computation.  They form the two most
important three-qubit gates in the quantum domain, as well.  Most
quantum algorithms are defined using Toffoli gates.  The transform for
the Toffoli \textsc{ccnot} gate with the low-order qubit being the target is
\begin{equation}
CCNOT = \left[\begin{array}{cccccccc}
1 & 0 & 0 & 0 & 0 & 0 & 0 & 0 \\
0 & 1 & 0 & 0 & 0 & 0 & 0 & 0 \\
0 & 0 & 1 & 0 & 0 & 0 & 0 & 0 \\
0 & 0 & 0 & 1 & 0 & 0 & 0 & 0 \\
0 & 0 & 0 & 0 & 1 & 0 & 0 & 0 \\
0 & 0 & 0 & 0 & 0 & 1 & 0 & 0 \\
0 & 0 & 0 & 0 & 0 & 0 & 0 & 1 \\
0 & 0 & 0 & 0 & 0 & 0 & 1 & 0 \end{array}\right],
\end{equation}
and the transform for the Fredkin control-\textsc{swap} gate with the
high-order bit being the control is
\begin{equation}
CSWAP = \left[\begin{array}{cccccccc}
1 & 0 & 0 & 0 & 0 & 0 & 0 & 0 \\
0 & 1 & 0 & 0 & 0 & 0 & 0 & 0 \\
0 & 0 & 1 & 0 & 0 & 0 & 0 & 0 \\
0 & 0 & 0 & 1 & 0 & 0 & 0 & 0 \\
0 & 0 & 0 & 0 & 1 & 0 & 0 & 0 \\
0 & 0 & 0 & 0 & 0 & 0 & 1 & 0 \\
0 & 0 & 0 & 0 & 0 & 1 & 0 & 0 \\
0 & 0 & 0 & 0 & 0 & 0 & 0 & 1 \end{array}\right].
\end{equation}
The \textsc{ccnot} cannot be implemented directly on most quantum technologies,
so we need a breakdown into two-qubit gates.  The breakdown we choose
uses a two-qubit gate which we will call the ``square root of $X$'',
\begin{equation}
\sqrt{X} = \frac{1}{2}\left[\begin{array}{cccc} 1+i & 1-i \\ 1-i & 1+i
\end{array}\right]
\end{equation}
and its adjoint
\begin{equation}
\sqrt{X}^\dagger = \frac{1}{2}\left[\begin{array}{cccc} 1-i & 1+i \\ 1+i & 1-i
\end{array}\right].
\end{equation}
Our graphic representation is shown in Figure~\ref{fig:5gcc-only}.  We
will use this construction and an additional variant in
Section~\ref{sec:algo-topo}, when we discuss the interaction of
architecture and gates in more detail.

\subsubsection{Quantum Circuits}

A quantum computation, in the abstract, is a unitary transformation on
the initial state of the system, creating a desired output.  The
complete unitary transform on $n$ qubits, of course, is a $2^n\times
2^n$ matrix, so direct construction of the unitary to implement a
complex function of more than a few qubits is difficult.  The physical
phenomena used for quantum computation do not, in general, lend
themselves well to direct implementation of complex transforms.
Moreover, human beings are not good at imagining such large systems,
but are very good at composing large systems from smaller components.
Thus, the abstraction of a {\em quantum circuit} is important.  A
quantum circuit effects the overall transform via a series of smaller
gates (generally, one- to three-qubit gates) applied in a prescribed
order on the appropriate qubits.

Researchers have found several methods for decomposing a specific
unitary transform into a series of small gates or operations that we
know how to implement.  Some methods find optimal evolution paths (not
necessarily composed of discrete gates) but are highly theoretical,
and it is not immediately clear how to compile a large program by
employing these
methods~\cite{nielsen06:_quant_comput_geomet,carlini06:_time_optim_quant_evolut}.
Using the most general method, the number of gates grows exponentially
as the size of the problem increases, negating any advantage in
computational complexity that quantum computing appears to
offer~\cite{schuch03:_progr}.  Most of the work on quantum programming
languages and tools for them essentially defers the decomposition
problem to the
programmer~\cite{gay05:_quant_progr_lang,oemer:qcl,aho:palindromes,svore06:_sw-arch-computer}.
Fortunately, many quantum algorithms depend on a few basic building
blocks that have known efficient decompositions (such as the quantum
Fourier transform), or on ideas translated directly from classical
analogues (such as arithmetic).

\subsection{DiVincenzo's Criteria}
\label{sec:divincenzo}

DiVincenzo~\cite{divincenzo95:qc} enumerated five abilities which are
necessary for real-world quantum computing devices.  A quantum
computer must:

\begin{enumerate}
\addtolength{\itemsep}{-2mm}
\item Be a scalable physical system with well-defined qubits;
\item Be initializable to a known state prior to computation;
\item Have adequately long decoherence times;
\item Have a universal set of quantum gates; and
\item Permit high efficiency quantum measurements.
\end{enumerate}

Two additional criteria focus on moving quantum information between
two different quantum computers.  A viable quantum communications
technology must:

\begin{enumerate}
\addtolength{\itemsep}{-2mm}
\item[6.] Be able to convert between physical realizations of qubits that are
  stationary and moving; and
\item[7.] Be able to faithfully transmit a physical realization of a qubit
  between specified locations.
\end{enumerate}

The first criterion means there must be some physical entity, such as
energy levels of an ion, polarization of a photon, or spin of an
electron, that is the actual carrier of the qubit.  It must meet basic
criteria of quantum behavior and support two distinct states which can
be treated as zero and one.  Item 1 also refers to ``scalability'',
which means different things in different contexts; we will explore
its system aspects beginning in Chapter~\ref{ch:scalability-defn}.

The second item may seem obvious, but some qubits, especially nuclear
spins, are difficult to ``reset'' to zero.  Schulman and Vazirani
developed a method for taking a poorly-initialized machine and
improving the state, ``cooling'' the system
algorithmically~\cite{schulman01:cooling-qubits}.

The third item, decoherence, has important implications for quantum
computer architecture.  In order to fault tolerantly compute on a
quantum computer, the native error rate must be below a certain
threshold.  Aharonov and Ben-Or initially calculated the threshold
(``errors per quantum gate'') to be
$10^{-6}$~\cite{aharonov99:_threshold}.  However, this factor is
architecture dependent, with real architectures requiring
substantially lower thresholds.  Furthermore, in order to not have
undue overhead from error correction processes as the size of the
computation scales, quantum technologies really need to achieve error
rates well below this critical
threshold~\cite{steane02:ft-qec-overhead}.

The fourth criterion requires that a quantum computer be able to
compute any quantum function.  It must be able to rotate a single
qubit by any angle, and must be able to entangle a pair of qubits.
The single-qubit rotations may be difficult to achieve, so a small
number of ``universal'' gates that can be used to synthesize larger,
more complex gates serves as an alternative, at polynomial
cost~\cite{barenco:elementary,kitaev97:_quant,harrow02:_effic-discrete-gates,fowler05:_const-arb-ft-gates}.
This is equivalent to saying that a classical computing technology
should be able to perform at least a \textsc{nor} or \textsc{nand}
operation.  For quantum computers, one such set of universal gates is
\textsc{x}, \textsc{h}, \textsc{t}, and \textsc{cnot}, the gates we
have already introduced in Figure~\ref{fig:basic-gates} on
page~\pageref{fig:basic-gates}.  \textsc{x, h} and \textsc{cnot} are
relatively simple to make fault tolerant, while \textsc{t} requires a
more complex circuit; nearly one hundred gates in one
construction~\cite{fowler05:_const-arb-ft-gates}.

Item 5 is the measurement we discussed above; there must be a reliable
way to read out the state of a qubit.  However, as noted, measurement
is far more important than retrieving results at the end of a
computation; it occurs almost continuously as part of quantum error
correction and the fault-tolerant execution of gates on encoded
bits~\cite{shor:qecc,calderbank96:_good-qec-exists,steane02:ft-qec-overhead,gottesman99:ft-local,steane02:_quant-entropy-arch}.

Items 6 and 7 deal specifically with moving quantum information across
long distances for purposes of computation.  Criterion 6 only applies
to systems that compute complex quantum algorithms via shared state.
It does not apply to other uses of quantum effects, such as quantum
cryptography~\cite{bennett:bb84,elliott:qkd-net} and basic
demonstrations of quantum
teleportation~\cite{bennett:teleportation,furusawa98} (though
teleportation may be used in quantum computer
architectures~\cite{gottsman99:universal_teleport,grover97:_quant_telec}).

These criteria have been used as a basis for evaluation of quantum
computing {\em
technologies}~\cite{nielsen-chuang:qci,spiller:qip-intro,arda:qc-roadmap-v2}.
They are a necessary set of capabilities, but not sufficient to
understand the difficulty of building a quantum computer or its speed
and utility once built.  Ladd has suggested that DiVincenzo's five
criteria can be restated as three~\cite{ladd05:_thesis}.  A
complementary set of criteria for quantum computer {\em systems} is
discussed in Chapter~\ref{ch:taxonomy}.

\subsection{Quantum Algorithms}

We observed in Section~\ref{sec:q-reg} that an $n$-qubit quantum
register can be in a superposition of all possible $2^n$ states
$|0\rangle$ to $|2^n-1\rangle$ at the same time.  Usually, quantum
algorithms begin by placing one input register in this superposition.
This effect allows a quantum computer to calculate a function on all
possible inputs at the same time, in a single pass.  The hard part is
getting a useful answer out.  At the end of the calculation, the
result register is a superposition of all of the results, one for each
of the $2^n$ possible inputs.  However, we can't directly read out all
of those results.  If we measure the result register to get our
answer, the superposition collapses into a single state with a
probability according to the weights discussed above.  Then we have
only a single value; our end result is no better than if we had used a
classical computer to compute the function for one possible input
chosen at random.  How do we structure a quantum algorithm so that
useful results come out, taking advantage of these quantum effects to
accelerate computation?  We must find a way to drive the system toward
the state where the weights $\alpha_i$ from Equation~\ref{eq:psi-full}
of undesirable states are zero and desirable states (the solutions to
our problem) have large weights.

Deutsch discovered the key to a quantum
algorithm~\cite{deutsch-jozsa92}: use quantum interference to increase
the probability that a useful state is found when the quantum register
is measured.  Deutsch's algorithm, later refined in collaboration with
Jozsa, classifies an unknown function as one of two types.  One type
of function will create interference so that the register cannot read
0; the other type of function creates interference so that all of the
non-zero values cancel, leaving only the state 0.  This is perhaps the
most profound observation in all of quantum computing: we can take
advantage of the wave nature of particles to achieve computation.

What we colloquially call quantum algorithms are, in reality, hybrid
algorithms with both classical and quantum components.  Moreover, the
quantum portion of many algorithms is probabilistic, often
necessitating multiple runs to get the desired result (even ignoring
the physical issues of decoherence).  The complete cycle of a
``quantum'' computation is as follows:
\begin{enumerate}
\item Pre-calculate certain classical factors.
\item Repeat:
  \begin{enumerate}
    \item Initialize quantum computer.
    \item Prepare input state.
    \item Execute quantum portion of the algorithm.
    \item Measure output register.
    \item Post-process output to determine if desired result achieved.
    \item Exit if desired result.
  \end{enumerate}
\item Finish post-processing.
\end{enumerate}

We will see in Section~\ref{ch:err-mgmt} that this process is applied
recursively in the implementation of quantum error correction.  The
quantum computer can be initialized starting from a
partially-initialized state using quantum algorithms, as well, using
this procedure for step 2.a~\cite{schulman01:cooling-qubits}.

\subsection{Distributed Quantum Computation}
\label{sec:intro-dist-qc}

Distributed quantum computation (DQC) is the cooperative use of
multiple, independent quantum computers working to solve a single
problem.  The theoretical foundations of DQC have been laid, but very
little work on designing a machine to run DQC has been done.  Early
suggestions of distributed quantum computation include
Grover~\cite{grover97:_quant_telec}, Cirac et
al.~\cite{cirac97:_distr_quant_comput_noisy_chann}, and Steane and
Lucas~\cite{steane:ion-atom-light}.  A recent paper has proposed
combining the cluster state
model~\cite{raussendorf03:_cluster-state-qc,nielsen:cluster-review}
with distributed computation~\cite{lim05:_repeat_until_success}.
D'Hondt has done work on formal models of distributed quantum
computation, drawing on formal classical
techniques~\cite{dhondt05:_dist-qc}; D'Hondt and Tani et al. have
worked on the leader election problem, one of the few true distributed
quantum algorithms~\cite{tani05:_quant_leader_elect}.  A distributed
system generally requires the capability of transferring qubit state
from one physical representation to another, such as nuclear spin
$\leftrightarrow$ electron spin $\leftrightarrow$ photon, as in
DiVincenzo's seventh
criterion~\cite{mehring03:_entan-electron-nuke,jelezko04:_observ,childress05:_ft-quant-repeater}.

Yepez distinguished between distributed computation using entanglement
between nodes, which he called type I, and without inter-node
entanglement (i.e., classical communication only), which he called
type II~\cite{yepez01:_type_ii}.  Our quantum multicomputer is a type
I quantum computer.  Jozsa and Linden showed that Shor's algorithm
requires entanglement across the full set of qubits, concluding that a
type II quantum computer cannot achieve exponential
speedup~\cite{jozsa03:entangle-speedup,love05:_type_ii_quant_algo}.
Much of the work on our multicomputer involves creation and management
of that shared entanglement.

Yimsiriwattana and Lomonaco have discussed a distributed version of
Shor's algorithm~\cite{yimsiriwattana04:dist-shor}, based on one form
of the Beckman-Chari-Devabhaktuni-Preskill modular exponentiation
algorithm~\cite{beckman96:eff-net-quant-fact}.  The form they use
depends on complex individual gates, with many control variables,
inducing a large performance penalty compared to using only two- and
three-qubit gates.  Their approach is similar to our telegate
(Sec.~\ref{sec:telegate}), which we show to be slower than teledata
(Sec.~\ref{sec:teledata}).  They do not consider differences in
network topology, and analyze only circuit complexity, not depth (time
performance).

\section{Error Management in Quantum Computers}
\label{ch:err-mgmt}

\cq{By failing to prepare, you are preparing to fail.}
{Benjamin Franklin}

\cq{There are no mistakes, save one: the failure to learn from a mistake.}
{Robert Fripp}

\cq{\noindent O throw away the worser part of it,\\
And live the purer with the other half.}
{Shakespeare's {\em Hamlet}, quoted by Lampson}

A bewildering array of error processes bedevil quantum computing
technologies.  There are normal, independent errors of decay ($T_1$
and $T_2$ memoryless processes) that affect a single qubit only,
correlated error processes caused by environmental defects that affect
more than one qubit, unwanted interactions between qubits, stochastic
gate errors, propagation of errors by gates, ``hot'' and ``cold''
gates, accidental measurement of qubits, leakage of information into
the environment creating mixed states, and finally, loss of the qubits
themselves (photons or, occasionally, ions).

Error management in quantum computers is accordingly a rich and
complex field.  In this section, we provide a general introduction to
quantum error correction (QEC), including a look at how QEC helps
reinforce the digital nature of quantum computing, and briefly present
the notion of a {\em threshold}.  We then skim over other error
control techniques such as decoherence-free subspaces and composite
gate sequences, very different from error-correcting codes and more
tightly bound to the quantum nature of the data we are protecting.
Our goal in this section is not to cover the mathematics of quantum
errors or to provide complete coverage of the topic, but to give
computer architects a feel for the nature of the problems and the
solutions.  For a more thorough understanding, see Chapter~10 of
Nielsen \& Chuang~\cite{nielsen-chuang:qci} (which runs seventy-five
pages) and the many papers referenced both there and in this chapter.
Keyes' paper is a good introduction to some of the physical concerns
associated with solid-state
systems~\cite{keyes03:solid-state-qc-problems}.  In my opinion, the
single most important paper for engineers to read and understand, for
the practicality of its results, is one by
Steane~\cite{steane02:ft-qec-overhead}.  This topic alone would easily
warrant development of a full book.

\subsection{Error Models}

As suggested above, there are many ways in which quantum data can be
damaged.  Error processes also operate at many time scales: errors may
occur at fabrication time, over the course of many gates, or over the
course of a single gate.  Atoms are identical, but fabricated
structures are not, and the resulting differences may alter
e.g. oscillation frequencies, affecting gate time and coupling of
qubits.  Temperatures drift over time, influencing behavior.  Atoms
may vary their position relative to a laser beam or optical cavity,
altering the ideal gate time on a moment-by-moment basis.  Stray
magnetic fields may influence large groups of qubits.

This plethora of problems suggests that we should look for
similarities and simplifying abstractions.  The first models of errors
in quantum computation assumed that independent errors occurred before
or after the execution of logical gates.  If we assume independent,
random errors (an assumption we will gradually relax), it can be shown
that all errors can be reduced to $X$ or $Z$ gate errors on individual
qubits.  \comment{What about the failure of the entangling operation,
  such as the \textsc{cnot}, itself?  Is this adequately covered?}

\subsubsection{Error Propagation}
\label{sec:err-prop}

In classical circuits, whether analog or digital, we are accustomed to
errors propagating from source to target; an error in an AND gate
creates an incorrect result, but does not affect its inputs.  In the
quantum world, we have the same kind of errors, but additionally have
errors that propagate in counter-intuitive fashion.

\begin{figure}
\centerline{\hbox{
\includegraphics{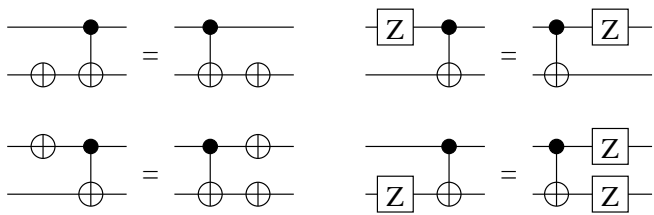}}}
\caption{Error propagation in two-qubit gates.}
\label{fig:err-prop}
\end{figure}

In Figure~\ref{fig:err-prop}, we show how errors propagate through
quantum gates.  An $X$ error (a \textsc{not} error, drawn as $\oplus$ in the
figures) on the target qubit of a \textsc{cnot} gate behaves the same before or
after the qubit.  An $X$ error on the control bit before the gate
execution, in contrast, propagates the error to both the control and
target qubits at the output; our single error has become two errors.
Worse, a $Z$ error (drawn as a box with a $Z$ in it in the figure) on
the target qubit of a \textsc{cnot} prior to the gate propagates a $Z$ back to
the control, as well; our intuition about the flow of errors in the
system fails us in this case.  This effect affects our ability to
correctly execute quantum error correction itself, which we will see
below.

\subsubsection{Steane's Error Models}
\label{sec:steane-models}

The basic model introduced above correctly models the {\em logic} of
errors as single-qubit gates that occur before or after the execution
of logic gates.  For accurately assessing the {\em probability} of
errors, it is somewhat simplistic; we will see in
Section~\ref{sec:qc-adders} that, for many calculations, many of the
qubits sit idle for long periods of time.  A better model will
therefore take into account memory errors and gate-induced errors.
Steane introduced just such a model, which we will call the $KQ$ model
or the {\em space-time} model~\cite{steane02:ft-qec-overhead}.
Letting $K$ be the number of logical qubits in the computation and $Q$
be the number of time steps to complete the computation, then the
accuracy of our logical operations must be related to the inverse of
the space-time product, $\sim 1/KQ$.  In this remarkable paper, Steane
went further and discussed the difference between the gate error
probability, which he labeled $\gamma$, and memory error probability,
which he labeled $\epsilon$, and produced numerical values for the
size of computations ($KQ$) achievable for various system
characteristics.

\subsection{Quantum Error Correction Codes}

Until the advent of quantum error correction, many researchers
believed that these problems were
insurmountable~\cite{keyes03:solid-state-qc-problems,dyakonov02:_QC-skepticism,unruh95:_maintaining-coherence}
or at least limited the range of problems to which quantum computing
can be
applied~\cite{chuang95:_factoring-decoherence,barenco:approx-qft}.
However, in 1995, almost simultaneously, several researchers
discovered and developed mechanisms for applying classical error
correction codes, such as Reed-Solomon codes, to quantum
data~\cite{shor:qecc,calderbank96:_good-qec-exists,steane96:_qec}.
The most important class of quantum error correction (QEC) codes is
now called the Calderbank-Shor-Steane codes, after its inventors, and
includes quantum analogs of Hamming, Golay, and other types of
classical error correcting codes.

In classical systems, we often use multiple levels of error
correction.  The same principle can be applied in quantum systems, in
a manner called {\em concatenation}.  In a concatenated system,
physical qubits are grouped to encode a logical qubit, and a group of
logical qubits is further encoded (using the same or a different code)
to provide greater protection against errors.  We discuss
concatenation in Section~\ref{sec:threshold}.

\begin{figure*}
\centerline{\hbox{
\input{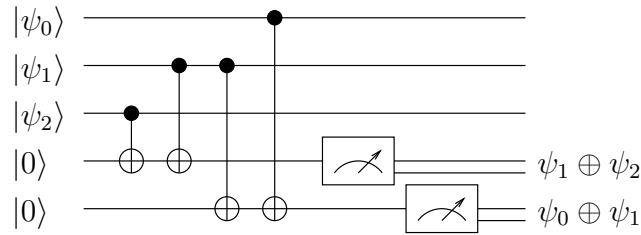}}}
\caption{Parity measurement for quantum error correction.}
\label{fig:qec-parity}
\end{figure*}

First, let us examine how to correct bit-flip errors in a quantum
state.  The CSS codes, like classical codes, redundantly encode
information so that an error in one component qubit can be detected by
comparing to the other qubits, and the error corrected.  In the
simplest example, one qubit is encoded by making two fanout
``copies''~\footnote{Again, be careful that when we use the term
``copy'', we are referring to a fanout, rather than an independent,
cloned copy of the state, which we know is
impossible~\cite{wootters:no-cloning}.} of the qubit.  Three ones will
be our logical one, and three zeroes will be our logical zero, i.e.
\begin{align}
|0\rangle&\rightarrow|0_L\rangle\equiv|000\rangle\\
|1\rangle&\rightarrow|1_L\rangle\equiv|111\rangle.
\end{align}
Our canonical unknown single-bit state then becomes
\begin{equation}
|\psi\rangle = \alpha|0\rangle+\beta|1\rangle\rightarrow
|\psi_L\rangle = \alpha|0_L\rangle+\beta|1_L\rangle 
= \alpha|000\rangle + \beta|111\rangle.
\end{equation}
Now that we have our proposed logical states, how do we execute gates,
and how do we perform our actual error correction?  Taking the second
question first, error correction is done by a series of parity
calculations and measurements.  Letting $|\psi_j\rangle, j=2,1,0$ be
the three qubits in our logical state $|\psi_L\rangle$, we want to
calculate the parity of the 0-1 pair and the 1-2 pair. If the state is
still unmarred, both calculations will return zero (even parity).
However, if we find, for example, that the 1-2 pair is even but the
0-1 pair is odd, we can infer that bit 0 is in error, and needs to be
corrected.  If both pairs are odd, we can infer that bit 1 is in
error.  The basic circuit for these parity measurements is shown in
Figure~\ref{fig:qec-parity}.

Although it is not immediately obvious, this parity measurement will
{\em not} disrupt our qubit state, causing the collapse of the wave
function and ruining our computation.  We saw in
Section~\ref{sec:partial-trace} that measurement of a single qubit in
a superposition takes out one qubit, shrinking the entangled state of
the system.  Intuitively, this is reasoned by considering what we {\em
learn} from the measurement.  By doing a parity measurement, we learn
only whether the two qubits are the same, not whether they are one or
zero.  When the states are correct, both bits will be one, or both
bits will be zero, in accordance with the usual behavior of entangled
qubits.

The error case works the same.  Our error model assumes a bit {\em
flip}, not a qubit being {\em set} to either one or zero.  Thus, an
error on bit 1, for example, would lead to the state
$\alpha|010\rangle+\beta|101\rangle$.  Parity measurement of the 0-1
pair produces a 1 (odd parity).  Writing out both the correct case and
the case of an error on bit 1, adding a parity qubit whose state is
created using the circuit in Figure~\ref{fig:qec-parity}, we have
\begin{align}
\alpha|000\rangle + \beta|111\rangle&\rightarrow
\alpha|0000\rangle + \beta|1110\rangle = 
(\alpha|000\rangle + \beta|111\rangle)|0\rangle \\
\alpha|010\rangle + \beta|101\rangle&\rightarrow
\alpha|0101\rangle + \beta|1011\rangle = 
(\alpha|010\rangle + \beta|101\rangle)|1\rangle
\end{align}
where the right-hand factoring makes it explicit that measuring the
last qubit will not affect the prior state, neither collapsing the
superposition nor altering the values of $\alpha$ and $\beta$.

Once the parity has been calculated and measured, we know whether or
not an error occurred, and if so, on which qubit.  Assuming we found
an error on qubit 1, we correct by applying an $X$ gate,
\begin{equation}
X_1|\psi_L\rangle = X_1(\alpha|010\rangle+\beta|101\rangle) =
\alpha|000\rangle+\beta|111\rangle
\end{equation}
and our desired state is restored.


The second type of error we must correct is phase errors.  When a
phase error occurs on our three-bit encoded state,
\begin{equation}
\alpha|0_L\rangle + \beta|1_L\rangle \rightarrow \alpha|0_L\rangle -
\beta|1_L\rangle
\end{equation}
regardless of which qubit the phase error affected.  Obviously, our
three-bit code does not detect such errors.  However, if we apply a
Hadamard to the three-qubit state before an error occurs, then we
shift into a space where a phase error will show up as a bit error
when the parities are calculated.  Combining a three-bit code for
protecting against bit flips and a three-bit code for protecting
against phase flips, we have a nine-bit encoding for a single logical
qubit known as the {\em Shor nine-bit code}~\cite{shor:qecc}.

QEC traditionally depends on interleaving measurement and logic gates,
and there has been recent experimental progress on this
front~\cite{roos04:3qubit-entangled}.  However, it is possible to
perform QEC without measurement, at a cost of a number of ancillae
that grows with the number of applications of error
correction~\cite{nielsen-chuang:qci}; this approach is not supportable
in a large computation, but may be applied in short sequences.

QEC builds on concepts from classical error correcting codes.  {\em
Stabilizer codes} represent an important advance in the mathematical
representation of QEC, providing a more compact representation of the
code word states and simplifying construction of fault-tolerant
operations~\cite{gottesman97:_thesis}.  

QEC demands to be taken into account when designing a quantum
computer.  Indeed, Steane has referred to a quantum computer as a
machine whose purpose is to execute error correction; computation is a
side effect~\cite{steane02:_quant-entropy-arch}.  Currently, some
researchers are analyzing the behavior of QEC on proposed
architectures and attempting to design machines that are well-adapted
to performing
QEC~\cite{svore05:_local-ft,cross05:masters-thesis,fowler04:_qec_lnn,steane02:_quant-entropy-arch,copsey02:_quant-mem-hier,oskin02:_pract_archit_reliab_quant_comput,burkard99:_qec-opt,devitt04:_shor_qec_simul,metodiev:ion-trap-sim-prelim,metodi05:qla},
or exploring the interaction of QEC with {\em cluster state
computing}~\cite{nielsen-2004}.  Others are demonstrating QEC and
decoherence-free subspaces (DFS, described below) experimentally,
either partially or completely, on NMR~\cite{knill01:_bench},
optical~\cite{pittman05:_optical-qec-demo}, Josephson
junction~\cite{katz06:_coher_state_evolut_super_qubit}, or ion trap
systems~\cite{haeffner05:_robust,roos04:3qubit-entangled,chiaverini04:_qec-realiz}.
Knill et al. have even suggested that the ability to run QEC be used
as a reliability benchmark for quantum computing
technologies~\cite{knill01:_bench}.

\subsubsection{CSS Codes and Larger Blocks}

Now that we understand the basics of the error correction processes,
surely we will want more efficient codes than the Shor nine-bit code.
To discuss the efficiency of the encoding of various schemes, we need
a notation.  We will describe a quantum error correcting code using
the notation [[$n$,$k$,$d$]], where $n$ is the number of physical
bits, $k$ is the number of logical bits encoded, and $d$ is the
Hamming distance ($(d-1)/2$ errors can be corrected by the code).  In
this notation, the nine-bit Shor code is [[9,1,3]].  Nine physical
qubits encode a single logical qubit, and can correct any single-qubit
error, whether bit flip or phase flip (or both).

\begin{figure*}
\centerline{\hbox{
\input{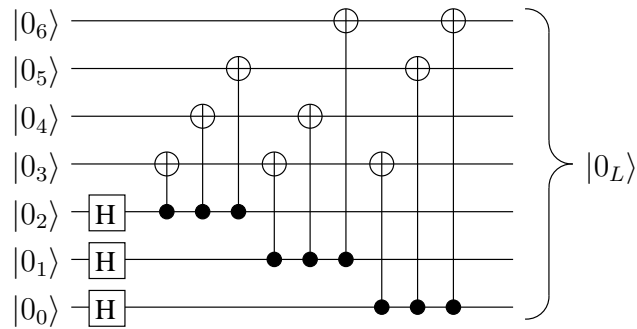}}}
\caption{Circuit to create the $|0_L\rangle$ state for the Steane
  [[7,1,3]] code.}
\label{fig:713-zero}
\end{figure*}

More efficient encodings for a single qubit are known.  The most
commonly used example is the [[7,1,3]] Steane
code~\cite{steane96:_qec}.  Thus, for the Steane 7-bit code, we encode
each logical qubit in seven physical qubits, and this state can
correct a single error.  In this code, logical zero and logical one
are~\cite{calderbank96:_good-qec-exists}
\begin{equation}
\begin{split}
|0_L\rangle &= \frac{1}{\sqrt{8}}
(|0000\underline{000}\rangle + |1101\underline{001}\rangle
+ |1011\underline{010}\rangle + |0110\underline{011}\rangle \\
&+ |0111\underline{100}\rangle + |1010\underline{101}\rangle
+ |1100\underline{110}\rangle + |0001\underline{111}\rangle) \\
|1_L\rangle &= \overline{X}|0_L\rangle.
\end{split}
\label{eq:713-zero}
\end{equation}
In the equation, we have underlined the last three bits and ordered
the terms in the superposition to emphasize that all of the binary
values 0 to 7 appear there.  Figure~\ref{fig:713-zero} shows a circuit
that can create the logical zero state; the Hadamards on the bottom
three qubits give us our superposition of 0 to 7 from which the rest
of the state is built.  The subscripts in the figure are the bit
number in the QEC block, with qubit 6 being the leftmost bit in the
state as written in Equation~\ref{eq:713-zero}. The quality of the
state must be verified after creation and before use.

This seven-bit code is still not the limit for a single qubit; within
months of the elucidation of the basic concepts of quantum error
correction, two groups had discovered a [[5,1,3]] code, which was
demonstrated experimentally on an NMR system in
2001~\cite{laflamme96:_5qubit-qec,bennett96:_5qubit-qec,knill01:_bench}.
However, this code is difficult to work with; executing many kinds of
logical gates on the logical states for this code requires long
sequences of physical gates.

As with classical error correction, we can encode more than a single
qubit into a block that is collectively protected.  In classical
systems, even with strong codes, in practice the overhead is rarely
more than 30\%.  Unfortunately, in the quantum world, even with
modest-sized blocks, the overhead runs to a factor of three or so.
Steane described codes as efficient as [[63,39,5]], with an overhead
of only 1.6, but this one can correct only two errors in the entire
block, and the other efficient codes likewise trade protection for
space.  Steane recommends a [[23,1,7]] code based on a classical Golay
code as giving higher error tolerance (a larger possible
application-level $KQ$) for a given overhead in storage, when multiple
layers of QEC are concatenated.  For a concatenated code, he
recommends $k=1$ for the lowest level(s) of the system, it being much
easier to construct higher-level codes in this
case~\cite{steane02:ft-qec-overhead}.

\subsection{Fault Tolerance}

{\em Fault tolerance}, as the term is usually applied in quantum
computing, means that dynamic errors in our state do not propagate
uncontrollably throughout the system.  The system can tolerate
individual errors and still successfully compute.  Thus, fault
tolerance is primarily a set of techniques for controlling error
propagation.  Fault tolerance does not mean, as the term is often used
in classical systems, that the quantum computer is prepared to deal
with near-permanent failure of large hardware subsystems.

As we saw above, errors can propagate from one qubit to another as
gates are executed.  For this reason, errors are especially dangerous
to QEC blocks that contain more than one logical qubit.  A code that
can correct only a single error across multiple qubits can never be
robust against a logical gate error when the gate is applied between
two qubits in the same QEC code block.  Therefore, to execute a gate
between qubits in the same block, one logical qubit must first be
swapped out.  Next, we apply the gate laterally between blocks and
perform error correction separately in each block, after which the
qubit can be swapped back in to its original location, if desired.

Figure~\ref{fig:qec-parity} on page~\pageref{fig:qec-parity} shows a
simple, ideal circuit for calculating the error syndromes.  To prevent
the kind of error propagation described in Section~\ref{sec:err-prop},
we cannot use this circuit directly; we must have a scheme which
prevents phase error propagation.  Steane described an algorithm for
this, based on earlier work by himself, Shor, Zalka, and
others~\cite{steane02:ft-qec-overhead}.  Figure~\ref{fig:ft-syndrome}
shows Steane's algorithm, slightly reformulated.  In actual
implementation, the creation of the logical $|0_L\rangle$ state will
be decoupled from the syndrome measurement part of the subroutine.
The syndrome measurement will draw from a pool of logical zeroes that
is refilled continuously, tuned to guarantee that logical zeroes are
available when necessary, and as fresh as possible, while minimizing
the number of qubits required.

\begin{figure}
\begin{verbatim}
subroutine get_one_syndrome:
    repeat
        prepare n qubits in state |0>
        apply circuit to create logical |0>
        verify logical |0> state
    until logical |0> state is good
    couple |0> to data block
    Hadamard transform |0>
    measure
    return result
endsubroutine

routine syndrome:
    get_one_syndrome
    if syndrome = 0 then 
        return 0
    else
        repeat r-1 times
            get_one_syndrome
        if more than chosen limit of r syndromes agree then
            return syndrome
        else
            fail
endroutine
\end{verbatim}
\caption{Fault-tolerant error syndrome measurement algorithm.}
\label{fig:ft-syndrome}
\end{figure}

Many researchers have studied fault tolerance, including the
composition of fault-tolerant logical
gates~\cite{steane03:ft-css-codes,burkard99:_qec-opt,fowler05:_const-arb-ft-gates}.  We will not
delve further into this topic here.

\subsection{Threshold Calculations and Concatenation}
\label{sec:threshold}

Error correction only improves the quality of the state of our system
if, on average, it repairs more errors than it introduces.  If the
resulting error rate is still inadequate, we can concatenate multiple
levels of QEC, pushing down the net error rate to the necessary
level.

In an $h$-level concatenated encoding, the effective error probability
is $(cp)^{2^h}/c$, where $c$ is the {\em threshold} value and $p$ is
the error probability (which is assumed to be the same at each level,
for the moment).  The threshold, in this equation, is the number of
operations required to execute a single level of error correction.  If
$cp < 1$, then each level of encoding we add to the system decreases
our net probability of failure.  If an encoding level uses $n$ qubits
from the level below to encode a single qubit, our total cost per
logical qubit is $n^h$ physical qubits.  In two-level concatenated
QEC, with different inner and outer codes, [[$n_i$,1,$d_i$]] and
[[$n_o$,$k$,$d_o$]], respectively, we use $n_i n_o$ physical qubits to
represent $k$ logical qubits.

Aharonov and Ben-Or were among the first to calculate a numerical
value for a threshold~\cite{aharonov99:_threshold}.  They found a
value of $\sim 10^{-6}$ for a particular set of assumptions.  That is,
if more than one gate out of a million fails (a level {\em well}
beyond experimental capabilities for all quantum technologies at the
moment), using fault tolerant techniques actual makes the state of the
system worse rather than better.  If less than one in a million gates
fails, fault tolerance makes the state of the system better, and via
repeated application of fault tolerance we can reach an arbitrary
level of reliability.  Aharonov and Ben-Or also proved (without
providing a numerical figure) that a threshold exists even when the
qubits are arranged in a linear nearest neighbor-only topology, which
we will see in Section~\ref{sec:algo-topo}.  Thresholds have been
calculated many times for different sets of physical assumptions and
error correcting codes, with answers varying by several orders of
magnitude in both
directions~\cite{knill98:_resil_quant_comput,dawson-2005,fowler04:_qec_lnn,gottesman99:ft-local,knill00:loqc-threshold,szkopek04:_thres_error_penalty,gottesman97:_thesis,preskill98:_reliab_quant_comput};
Knill has suggested that, under some conditions, error rates as high
as 1\% might be acceptable~\cite{knill04:very-noisy-qc}.  In this
dissertation, we will work with the Steane algorithm and memory/gate
error assumptions described above, working toward a finite computation
of a particular size and ignoring the issues around thresholds.

\subsection{Why QEC Suppresses Over-Rotation Errors}

One counter-intuitive aspect of operating on encoded states is the
suppression of over-rotating gates (gates running ``hot'') or
under-rotating gates (gates running
``cold'')~\cite{miquel97:_overrotation}~\footnote{This is a key factor
in the ``quantum computation is not analog computation'' argument.}.
It is easy to see that QEC corrects a single gate error, but if all of
the physical gates comprising a logical gate over-rotate by similar
amounts, can that be corrected?

Examining the three-bit encoding once again,
\begin{equation}
|\psi_L\rangle = \alpha|000\rangle + \beta|111\rangle
\end{equation}
an $X$ gate that runs hot on a single qubit will actually perform the
gate
\begin{equation}
X_\epsilon = R_X((1+\epsilon)\pi) = 
\left[\begin{array}{cc}
\sin\frac{\epsilon}{2} & \cos\frac{\epsilon}{2} \\
\cos\frac{\epsilon}{2} & \sin\frac{\epsilon}{2}
\end{array}\right].
\end{equation}
The logical $X$ gate for this encoding is $\overline{X} = XXX$ ($X$
gates on all three component qubits), where the over-line indicates a
logical operation.  A mis-rotation at the {\em logical} level is
\begin{equation}
\overline{X_\epsilon}|\psi\rangle =
(\alpha\sin\frac{\epsilon}{2}+\beta\cos\frac{\epsilon}{2})|000\rangle
+ (\alpha\cos\frac{\epsilon}{2} + \beta\sin\frac{\epsilon}{2})|111\rangle
\end{equation}
{\em but} $\overline{X_\epsilon} \neq X_\epsilon X_\epsilon
X_\epsilon$!  It is easy to be confused about how the system
distinguishes between a deliberate attempt to rotate by $\pi$ and
$1.1\pi$.  The answer is that this construction $X_\epsilon X_\epsilon
X_\epsilon$ suppresses the (apparent) over-rotation and {\em forces}
$X_\epsilon X_\epsilon X_\epsilon \sim \overline{X}$.  This fact can
be seen by doing the vectors explicitly.
\begin{equation}
X_\epsilon X_\epsilon X_\epsilon |\psi\rangle =
\left[\begin{array}{c}
\alpha\sin^3\frac{\epsilon}{2} + \beta\cos^3\frac{\epsilon}{2} \\
\alpha\sin^2\frac{\epsilon}{2}\cos\frac{\epsilon}{2} + \beta\cos^2\frac{\epsilon}{2}\sin\frac{\epsilon}{2} \\
\alpha\sin^2\frac{\epsilon}{2}\cos\frac{\epsilon}{2} + \beta\cos^2\frac{\epsilon}{2}\sin\frac{\epsilon}{2} \\
\alpha\sin\frac{\epsilon}{2}\cos^2\frac{\epsilon}{2} + \beta\cos\frac{\epsilon}{2}\sin^2\frac{\epsilon}{2} \\
\alpha\sin^2\frac{\epsilon}{2}\cos\frac{\epsilon}{2} + \beta\cos^2\frac{\epsilon}{2}\sin\frac{\epsilon}{2} \\
\alpha\sin\frac{\epsilon}{2}\cos^2\frac{\epsilon}{2} + \beta\cos\frac{\epsilon}{2}\sin^2\frac{\epsilon}{2} \\
\alpha\sin\frac{\epsilon}{2}\cos^2\frac{\epsilon}{2} + \beta\cos\frac{\epsilon}{2}\sin^2\frac{\epsilon}{2} \\
\alpha\cos^3\frac{\epsilon}{2} + \beta\sin^3\frac{\epsilon}{2}
\end{array}\right]
\end{equation}
{\em before} applying the error correction.  This encoding suppresses
the angular error to $O(\sin^2\frac{\epsilon}{2}) = O(\epsilon^2)$, even without
going through the QEC correction step, but it's easier to see once
we've applied the QEC.  Assuming perfect QEC, the final result is
\begin{eqnarray}
|\psi'_L\rangle &=& (\alpha\sin^3\frac{\epsilon}{2} + \beta\cos^3\frac{\epsilon}{2} \nonumber\\
&&+ 3\alpha\sin^2\frac{\epsilon}{2}\cos\frac{\epsilon}{2} + 3\beta\cos^2\frac{\epsilon}{2}\sin\frac{\epsilon}{2})|000\rangle \nonumber\\
&&+ (\alpha\cos^3\frac{\epsilon}{2} + \beta\sin^3\frac{\epsilon}{2} \nonumber\\
&&+ 3\alpha\cos^2\frac{\epsilon}{2}\sin\frac{\epsilon}{2} 
+ 3\beta\sin^2\frac{\epsilon}{2}\cos\frac{\epsilon}{2})|111\rangle
\end{eqnarray}
and we see that the angular rotation error is in the $\sin^2$ terms.

Mathematicians would say, of the deliberate attempt to rotate by
$1.1\pi$, that ``it's not in the Clifford group,'' or ``it's not in
the normalizer.''  The importance of this mathematical distinction is
that there are only a few gates, such as the $X$ gate, that are easily
constructed by applying the same gate in a transverse fashion to all
elements of our logical qubit.  It is not possible to (easily)
construct a deliberate rotation by $1.1\pi$ on the logical state.  We
will not delve further into these mathematical issues or terminology,
though they affect compilation of efficient programs and cluster-state
computing as well as quantum error correction, and are influenced by
the natural gate for a specific
technology~\cite{harrow02:_effic-discrete-gates,fowler05:_const-arb-ft-gates}.

\subsection{Other Error-Suppression Techniques}

Other forms of error management techniques exist, some based on deep
theoretical insights.  One particularly intriguing one, from a
theoretical point of view, is {\em topological quantum memory}, in
which a 2-D array or torus of qubits is entangled in various patterns
to make a logical qubit~\cite{dennis:topo-memory}.  The state is
stable because it is the patterns of the connections, rather than the
value or phase of any single qubit, that determines the logical state.
The resources required are large, and it is not immediately clear how
to implement this scheme on a physical system.

QEC works best on systems with uncorrelated errors on separate
qubits.  When error processes are more likely to affect groups of
nearby qubits, a technique known as {\em decoherence free subspaces}
(DFS) helps to mitigate these
problems~\cite{lidar03:dfs-review,haeffner05:_robust,lidar98:dfs}.  In
a DFS, the logical value is encoded in the relative, rather than
absolute, state of a group of qubits.  A stray magnetic field that
caused them all to flip, for example, would not affect the logical
state.

In optical systems, the principal source of error is loss of photons.
In this case, {\em erasure codes} (in contrast to {\em error
correcting codes}) work well~\cite{knill00:loqc-threshold}.  Erasure
codes can be as simple as a parity check.  Reconstruction of the state
is straightforward when the position of the missing qubit is known.
Erasure codes are used in RAID arrays, where the position of the disk
spindle that has failed plus a simple parity check provide enough
information to reconstruct the original data~\cite{patterson:raid}.

As we noted above, individual gates can run hot or cold, over- or
under-rotating compared to the intended angle.  Besides using
(digital) quantum error correction, analog techniques for improving
the accuracy of gates have been developed.  Composite pulses break
down a rotation into a series of steps designed so that similar errors
in each step
cancel~\cite{vandersypen04:_nmr_techn,collin04:_nmr-like}.  As a
simple example, a rotation from the north pole to the south pole can
be broken down into a $90^\circ$ rotation about the $X$ axis, then a
$180^\circ$ rotation about the $Y$ axis, then another $90^\circ$
rotation about $X$.  If the $X$ rotations both under-rotate, the $Y$
rotation will compensate by mirroring the position about the equator
between $X$ rotations.  Realistic sequences for arbitrary (and
unknown) starting positions and gates are substantially more complex
but valuable.  Some sequences can reduce an error of $\epsilon$ in
each step of the process to an error $O(\epsilon^6)$ in the final
outcome.

\section{Summary}

Reversible computation allows us to reverse the arrow of time and
return to the starting point of a computation, recovering all inputs
to the system.  This is possible because information is conserved,
rather than destroyed, as in common Boolean logic; each gate has an
inverse that undoes its operation.  In reversible classical logic, the
inverse of a gate is the same gate, but in quantum that is not
necessarily so, as we saw in Section~\ref{sec:3qb-gates}.  In
reversible classical logic, we need a three-bit gate in order to have
universal computation; we have also seen that in quantum computation
we can construct the three-bit gates from many types of two-bit gates.

When Bennett, Feynman, Fredkin, Toffoli and others originally
developed the concepts behind reversible computing in the 1970s, they
were searching for the ultimate limits to the energy consumption of a
computation, as well as playing with remarkable intellectual facets of
information.  They probably had no notion that beginning just a few
years later they would help to found the fields of quantum
computation, quantum information theory and quantum communication, and
that their names would be indelibly linked with those fields.
Feynman, Benioff and Deutsch conceived of quantum computing in the
1980s as utilizing quantum effects to, potentially, dramatically
accelerate computation of certain
functions~\cite{benioff82:_qm_turing_machine,feynman:_simul_physic_comput,deutsch-jozsa92}.

\comment{I don't like this paragraph...}
Quantum computing must be contrasted with classical computation
performed using quantum phenomena.  Of course, the behavior of
semiconductors can be viewed as an analog quantum phenomenon, but
transistors currently use large numbers of charge carriers, allowing
us to treat transistors as classical digital devices.  As device size
continues to decrease according to Moore's Law, we will soon move into
the range where individual electrons are used~\cite{moore65,ITRS2005}.
Other approaches involve using quantum cellular automata as logic
gates, or more directly manipulating the spin of small numbers of
electrons for e.g. magnetic RAM and logic devices, in a field broadly
called {\em
spintronics}~\cite{imre06:_qd-majority-gate,zutic04:_spintronics,wolf05:_spintronics}.
Although the physics of the devices and the technology for
manipulating such states have much in common with the experimental
techniques for quantum computation, there is a key difference.  In
what we refer to as quantum computation, we are attempting to take
direct advantage of the key aspects of superposition and entanglement,
whereas in quantum-executed classical computation, the goal is to
suppress these effects as unwanted, and maintain a clear binary state.

A quantum bit, or qubit, can be in a superposition of states, rather
than the definite zero or one state of a classical bit.  In this
chapter, we have presented the basic concepts of qubit state, starting
with the relationship between the wave function and the probability of
getting certain results.  We discussed representing the state of a
single qubit as a point on the Bloch sphere; visualization of the
state of multiple qubits is much harder, and if the qubits are
entangled they cannot be represented independently.  We discussed the
basic principles of quantum superposition, entanglement, measurement,
and decoherence.  We can entangle multiple qubits and interfere the
terms in the superposition, driving the system toward our desired
states.  Measuring the system will produce values that would be
difficult to calculate using only classical computers, in some cases,
exponentially more difficult.  Designing algorithms that generate
superpositions with useful speedups has proved to be a difficult
problem.

We have outlined some of the coherence and computational accuracy
problems inherent in quantum computing devices, and shown a variety of
ways of mitigating these problems.  In particular, we focused on
quantum error correction (QEC).  Besides simple bit errors, QEC must
be able to correct phase errors as well.  This fact results in
substantially less efficient codes than in the classical case.  The
counter-intuitive propagation of phase errors also forces complex
fault-tolerance mechanisms.  The state of a qubit is something of an
analog phenomenon, with a continuum of states for the phase and
probabilities of different states; fortunately, as we have seen, QEC
helps to suppress analog errors, at the expense of requiring more
complex processes to effect many logical qubit rotations.

This chapter has described the building blocks of quantum computation.
The material presented so far gives only the vaguest notion how these
concepts cooperate to give us the power of quantum computation.  We
will gradually elaborate on these topics, beginning in the next
chapter with Shor's algorithm for factoring large numbers.

  \chapter{Shor's Algorithm for Factoring Large Numbers}
\label{ch:shor}

\begin{chapterquote}
``I am fairly familiar with all forms of secret writings, and am
myself the author of a trifling monograph upon the subject, in which I
analyze one hundred and sixty separate ciphers, but I confess that
this is entirely new to me. The object of those who invented the
system has apparently been to conceal that these characters convey a
message, and to give the idea that they are the mere random sketches
of children.''\\
\textbf{Sherlock Holmes in ``The Adventure of the Dancing
Men,'' Sir Arthur Conan Doyle, 1903.}
\end{chapterquote}

Before we can design a computer, we have to understand how it will be
used.  Characterizing the workload of a proposed system is the first
important task in the design process.  For our quantum multicomputer
design, we have chosen Shor's algorithm as our primary target
application~\cite{shor:factor,ekert96:_quant-shor}.  Shor's algorithm
requires arithmetic and the quantum Fourier transform (QFT), both of
which are considered fundamental building blocks of other algorithms.
Moreover, Shor's algorithm is a famous and important result in its own
right.  This chapter presents an informal overview of the algorithm.
Our discussion does not detail the theoretical mathematics of the
algorithm, instead covering the importance and structure of the
algorithm, and its relationship to the quantum mathematical building
blocks which are the primary focus of this thesis.  The chapter begins
by discussing the factoring problem, then presents the QFT, followed
by arithmetic algorithms for reversible and quantum addition and
modular exponentiation, then combines the parts into Shor's overall
algorithm.

\section{The Importance of Factoring}

Authentication of identity is one of the key factors in computer
security.  To authenticate yourself, you prove in some fashion that
you are who you claim to be (or, at least, have rights that you claim
to have).  Authentication is often said to depend on something you
have, something you are, or something you know (but that is not known
to other people).  A door key, for example, is one way to authenticate
that you are allowed to pass through the corresponding door; it is
something you have.  Biometric sensors, such as fingerprint or iris
readers, are canonical examples of ``something you are''
authentication.  A computer password is something you know.

The RSA algorithm (Rivest-Shamir-Adelman, named for its developers) is
the most important authentication mechanism on the Internet
today~\cite{rivest78:_rsa,schneier96:_applied_crypto}.  RSA is a
classic example of a {\em public key}, or {\em asymmetric}, encryption
algorithm.  RSA is used primarily for authentication, rather than
encryption of bulk data, because it is expensive to calculate relative
to other encryption algorithms.  In RSA, a cryptographic key has two
parts, the public key and the private key.  The public key can be
disclosed to anyone, and should be made available via some trustworthy
means.  This trustworthy publication of the public key is beyond the
scope of our discussion, but can be recursive use of the same
authentication mechanism leading back to a trusted source such as a
friend or the RSA Corporation, or an out-of-band trust mechanism such
as publication in the New York Times.  The private key is used to
calculate a function whose result can be disclosed publicly.  Using
the result and the previously-announced public key, any party can then
verify that the function result was calculated by the holder of the
private key, thereby authenticating the identity of the creator.

Factoring a large integer into its components would seem to be a
rather esoteric problem, but in fact, it is directly relevant to this
issue of authentication.  The difficulty of cracking RSA is known to
be related to the difficulty of factoring a large, composite number
into its prime factors.  Letting $C$ be the ciphertext and $M$ be the
original message, the function calculated in RSA is
\begin{align}
C &= M^e \bmod n \\
M &= C^d \bmod n.
\end{align}
The {\em encryption key}, or public key, is $(e,n)$, and the {\em
decryption key}, or private key, is $(d,n)$.  $n$ is chosen to be a
simple composite number, the product of two primes, $n = pq$.  $d$ is
a large, random number which is relatively prime to $(p-1)(q-1)$. $e$
must then be the multiplicative inverse of $d$, modulo $(p-1)(q-1)$,
such that
\begin{align}
ed = 1 \bmod (p-1)(q-1).
\end{align}
From this, we can easily see that the ability to factor $n$ into $p$
and $q$ would allow the encryption scheme to be broken.  Thus, the
security of RSA is said to depend on the computational difficulty of
the factoring problem.

\section{Historical Progress in Factoring}

The factoring problem has never been {\em proved} to be impossible to
solve classically in polynomial time, though many researchers strongly
believe it to be impossible.  The best known classical algorithm, the
general Number Field Sieve (NFS), consumes total resources that are
superpolynomial in the length of the number~\cite{knuth:v2-3rd}.  Its
asymptotic computational complexity on large numbers is
\begin{equation}
\label{eq:nfs}
O(e^{(nk \log^2 n)^{1/3}})
\end{equation}
where $n$ is the length of the number, in bits, and $k =
\frac{64}{9}\log 2$.

\begin{figure}
\centerline{\hbox{
\includegraphics[width=12cm]{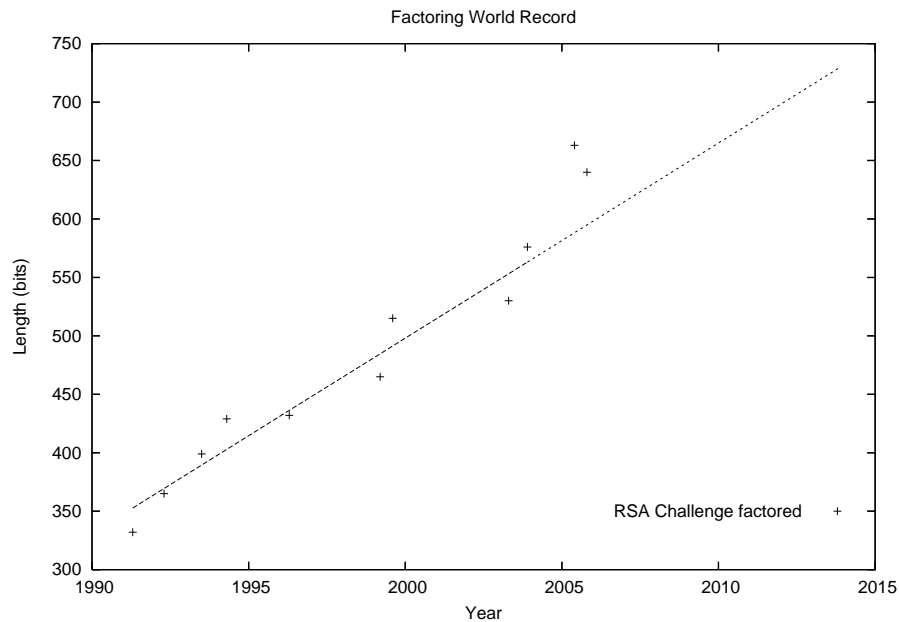}}}
\caption[Length of factored RSA Challenge numbers]{Length of RSA
  Challenge numbers successfully factored, in bits, plotted versus
  date accomplished.}
\label{fig:factoring-record}
\end{figure}

RSA, the company founded by the inventors of the RSA algorithm, which
owns the (now expired) patents on the RSA algorithm and much related
software, issues an ongoing series of public challenges to the
factoring community, in the form of numbers to be factored.  These
challenges carry with them cash prizes that are currently modest but
grow into the hundreds of thousands dollars for longer
numbers~\cite{rsa04-factoring}.  Figure~\ref{fig:factoring-record}
shows the progress of the RSA Challenge factoring records since 1991.

RSA places no restrictions on the amount or type of computing power to
be used in the challenge.  At a constant dollar value of computing
power used, in the current range of $\sim 600$ bits, Moore's Law
applied to CPU power alone (ignoring memory and I/O, and software
improvements) suggests that the longest number factorable using NFS
should be growing at about 18 bits per year.  In the data through
2003, we see roughly this trend.  The line on the plot is a
least-squares fit to the records through 2003.  The current world
record for factoring is 663 bits; a German team (Bahr, Boehm, Franke,
and Keinjung) announced the factoring of the RSA-200 challenge number
in May, 2005.  This data point appears to be an anomalously large
leap; whether it represents a shift in the long-term trend remains to
be seen.  Cavallar et al. estimated in 2000 that a 768-bit RSA key
will be factored by 2010, and a 1024-bit one by
2018~\cite{cavallar00:_factor_bit_rsa_modul}; progress appears to be on
track to meet those predictions.  Lenstra et al.  have also noted that
NFS scales well to large numbers of parallel processors and is
amenable to custom hardware acceleration; they suggest that a machine
that could factor a 1,024-bit number in one year could be built for
US\$10M using 2003 technology~\cite{lenstra03:_factor_rsa}.  It may be
possible to use an Internet-scale distributed system, such as the
Berkeley Open Infrastructure for Network Computing (BOINC), to attack
this problem~\cite{hughes06:nfs-private,anderson05:boinc-potential}.
BOINC, upon which SETI@home is based, has the potential to manage
100,000 or more nodes simultaneously attacking the same problem, a
1,000-fold increase over the size of systems deployed to date on
factoring problems.  We can infer that, at this point, moderately
large jumps in factoring records are primarily a matter of commitment
of resources.

\begin{figure}
\centerline{\hbox{
\includegraphics[width=12cm]{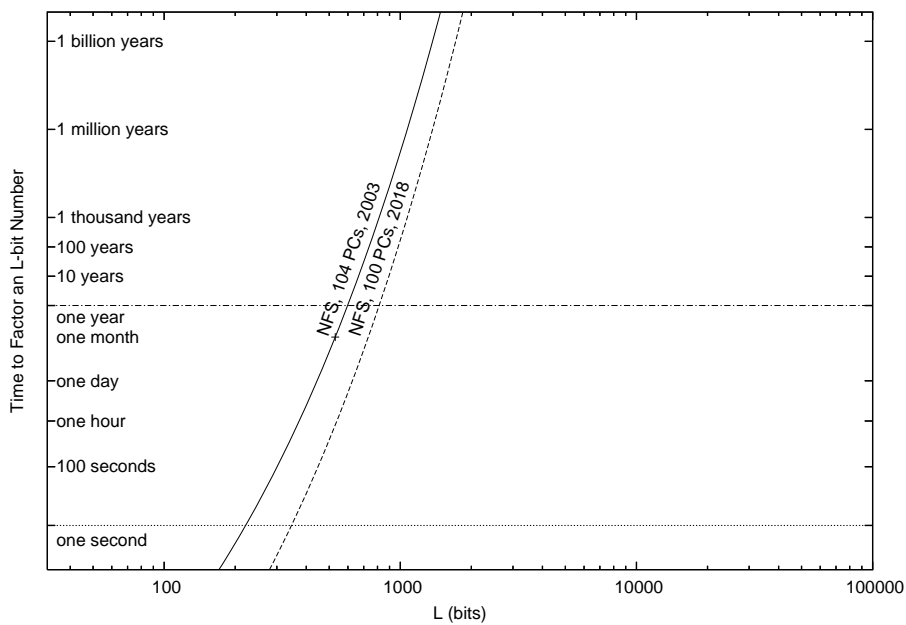}}}
\caption[Scaling of number field sieve]{Scaling of number field sieve
  (NFS) on classical computers.  Both horizontal and vertical axes are
  log scale.  The horizontal axis is the size of the number being
  factored, in bits.}
\label{fig:scaling-nfs}
\end{figure}

The execution time to factor a number using NFS an a set of on
classical computers is shown in Figure~\ref{fig:scaling-nfs}.  The
left curve is extrapolated performance based on the previous world
record, factoring a 530-bit number in one month, established using 104
PCs and workstations made in 2003~\cite{rsa04-factoring}.  The right
curve is speculative performance using 1,000 times as much computing
power.  This could be 100,000 PCs in 2003, or, based on Moore's law,
100 PCs in 2018.  From these curves it is easy to see that Moore's law
has only a modest effect on our ability to factor large numbers.
Factoring a 1,000-bit number is only a matter of time, but a 2,000-bit
number awaits either some theoretical advance or the advent of
large-scale quantum computers.

\section{The Quantum Fourier Transform}

We have noted several times that quantum parallelism effectively
calculates exponentially many functions at the same time, but that the
difficulty lies in extracting useful information from the
superposition of results.  Shor's remarkable insight showed the path
to creating a desirable superposition by interfering periodic
elements.  Some problems exhibit periodicity in their results, but
with a changing offset from zero.  Classically, one method for finding
a period in such an environment is to Fourier transform the data,
which eliminates phase (the offset) and leaves the frequency (or
period) information.

The quantum Fourier transform (QFT) transforms each individual
basis state in the following way:
\begin{equation}
|j\rangle\stackrel{\operatorname{QFT}}{\longrightarrow}\frac{1}{\sqrt{L}}\sum_{j=0}^{L-1}e^{2\pi ijk/L}|k\rangle
\end{equation}
where $L$ is $2^l$, and $l$ is the length of our state in bits.
Writing out the entire transform for $l=3$ and letting $\omega =
e^{2\pi i/8} = \sqrt{i}$, we have
\begin{equation}
\frac{1}{\sqrt{8}}
\left[\begin{array}{cccccccc}
1 & 1 & 1 & 1 & 1 & 1 & 1 & 1 \\
1 & \omega & \omega^2 & \omega^3 & \omega^4 & \omega^5 & \omega^6 &
\omega^7 \\
1 & \omega^2 & \omega^4 & \omega^6 & 1 & \omega^2 & \omega^4 &
\omega^6 \\
1 & \omega^3 & \omega^6 & \omega & \omega^4 & \omega^7 & \omega^2 &
\omega^5 \\
1 & \omega^4 & 1 & \omega^4 & 1 & \omega^4 & 1 &
\omega^4 \\
1 & \omega^5 & \omega^2 & \omega^7 & \omega^4 & \omega & \omega^6 &
\omega^3 \\
1 & \omega^6 & \omega^4 & \omega^2 & 1 & \omega^6 & \omega^4 &
\omega^2 \\
1 & \omega^7 & \omega^6 & \omega^5 & \omega^4 & \omega^3 & \omega^2 &
\omega
\end{array}
\right].
\end{equation}

Let us look at the input and output of the QFT in more
detail~\footnote{These examples are borrowed from Lieven Vandersypen's
thesis~\cite{vandersypen:thesis}.}.  In Table~\ref{tab:qft-values},
$\alpha_j$ are the coefficients of the values $j$ in the input
superposition $\sum\alpha_j|j\rangle$.  $\beta_k$ are the coefficients
in the output superposition.  The top left entry, for example, has a
one in the leftmost $\alpha_j$ column, corresponding to the state
$|0\rangle$.  The next line includes $|0\rangle$ and $|4\rangle$,
corresponding to the two ones.  $r$ is the period of repetition, that
is, how often ones appear in the fully-written-out superposition.
\begin{table}
\centerline{
\begin{tabular}{|c|rccccccc||rccccccc|c|}
\hline
r & \multicolumn{8}{c||}{input $\alpha_j$} &
\multicolumn{8}{c|}{output $\beta_k$} &
$L/r$ \\
\hline
 & $j=0$ & 1 & 2 & 3 & 4 & 5 & 6 & 7   & $k=0$ & 1 & 2 & 3 & 4 & 5 & 6
& 7 & \\
\hline
8  & 1 & 0 & 0 & 0 & 0 & 0 & 0 & 0   & 1 & 1 & 1 & 1 & 1 & 1 & 1 & 1
& 1 \\
4  & 1 & 0 & 0 & 0 & 1 & 0 & 0 & 0   & 1 & 0 & 1 & 0 & 1 & 0 & 1 & 0
& 2 \\
2  & 1 & 0 & 1 & 0 & 1 & 0 & 1 & 0   & 1 & 0 & 0 & 0 & 1 & 0 & 0 & 0
& 4 \\
1  & 1 & 1 & 1 & 1 & 1 & 1 & 1 & 1   & 1 & 0 & 0 & 0 & 0 & 0 & 0 & 0
& 8 \\
\hline
\end{tabular}
}
\caption{Transform values of the coefficients in the QFT.}
\label{tab:qft-values}
\end{table}
\begin{table}
\centerline{
\begin{tabular}{|cccccccc||cccccccc|}
\hline
\multicolumn{8}{|c||}{input $\alpha_j$} & \multicolumn{8}{c|}{output $\beta_k$} \\
\hline
$j=0$ & 1 & 2 & 3 & 4 & 5 & 6 & 7   & $k=0$ & 1 & 2 & 3 & 4 & 5 & 6
& 7 \\
\hline
1 & 0 & 0 & 0 & 1 & 0 & 0 & 0   & 1 & 0 & 1 & 0 & 1 & 0 & 1 & 0 \\
0 & 1 & 0 & 0 & 0 & 1 & 0 & 0   & 1 & 0 & $i$ & 0 & -1 & 0 & $-i$ & 0 \\
0 & 0 & 1 & 0 & 0 & 0 & 1 & 0   & 1 & 0 & -1 & 0 & 1 & 0 & -1 & 0 \\
0 & 0 & 0 & 1 & 0 & 0 & 0 & 1   & 1 & 0 & $-i$ & 0 & -1 & 0 & $i$ & 0 \\
\hline
\end{tabular}
}
\caption{Transform of different offsets into phase via the QFT.}
\label{tab:qft-phase}
\end{table}
The table can be used, for example, to see the following
transformation:
\begin{equation}
\frac{1}{\sqrt{2}}(|0\rangle+|4\rangle)\stackrel{\operatorname{QFT}}{\longrightarrow} \frac{1}{2}(|0\rangle+|2\rangle+|4\rangle+|6\rangle)
\end{equation}
What happens if the values in the superposition are period four, but not
$|0\rangle$ and $|4\rangle$, perhaps being $|1\rangle$ and $|5\rangle$
instead?  Such an offset difference shows up in a difference in the
{\em phase} of the output, as shown in Table~\ref{tab:qft-phase},
giving e.g.
\begin{equation}
\frac{1}{\sqrt{2}}(|1\rangle+|5\rangle)\stackrel{\operatorname{QFT}}{\longrightarrow} \frac{1}{2}(|0\rangle+i|2\rangle-|4\rangle-i|6\rangle).
\end{equation}
After the transform, {\em all} of the period four superpositions will
have an equal chance of returning 0, 2, 4, or 6 when the register is
measured, regardless of their original input values (this discarding
of offset or phase is a characteristic of the classical Fourier
transform, as well).

Thus, when we have an unknown superposition that we suspect consists
of some terms $|j\rangle$ where the $j$s have a periodic relationship,
the quantum Fourier transform will allow us to extract that period.
Shor has used quantum interference to cause undesirable terms to
cancel when transformed.  This remarkable result concentrates portions
of our total probability into superposition terms that tell us
something useful about the entire superposition when measured, holding
out the tantalizing possibility of an exponential increase in
computational power.

Shor built on work by Simon to develop his
algorithm~\cite{simon:_power_quant_comput}.  Many researchers have
examined the QFT in more detail, including describing how to implement
it, and discussing the necessity of exponentially small rotations in
the low-order
bits~\cite{barenco:approx-qft,chiaverini05:qft-impl,cleve:qft,coppersmith:approx-qft,hales:improv-qft,van-meter:qft-topo,fowler04:_limited-rotation-shor}.
We will leave off discussing the QFT, and move on to arithmetic, which
we also need for Shor's algorithm.

\section{Prior Art in Quantum Adders}
\label{sec:qc-adders}

%

Shor's factoring algorithm depends on the creation of a superposition
consisting of the modular integer exponentiation of a randomly-chosen
number $x$ raised to all powers 0 to $2^{2n}-1$, for an $n$-bit
number.  Exponentiation, of course, depends on integer multiplication,
which in turn depends on addition.  In this section we will review
several types of quantum adders developed by other researchers, which
will be used to construct the complete modular exponentiation in the
following section.

Classically, engineers have found many ways of building adders and
multipliers; choosing the correct one is a technology-dependent
exercise~\cite{ercegovac-lang:dig-arith}.  The performance of an adder
depends primarily on how quickly the information about the carry can
propagate from bit to bit.  The most obvious methods result in latency
that is linear in the number of bits to be added, but more complex
techniques can reduce that to $O(\sqrt{n})$ or even $O(\log n)$.
Classical multipliers are usually built by deferring the carry
calculation, allowing the $n$ additions necessary for a multiplication
to be completed in much less than $n$ times the latency of an
individual adder; we will see below that this is less attractive for
quantum arithmetic.  Only a few of these classical techniques have
been explored for quantum computation.  We review these circuits in
this chapter.  For our purposes, we need only unsigned integer
arithmetic, so the standard unsigned integer representation is used.

We begin by explaining our notation for performance, then analyze
progressively faster types of adders developed by other researchers,
saving the presentation of my new adder types for
Section~\ref{sec:new-adders}.  Rather than the details of why these
circuits work, we are more interested in how to implement them and
evaluate their performance.

\subsection{Arithmetic Performance Notation}
\label{sec:arith-nota}

We express the circuit cost using the notation $(\textsc{ccnot}s; \textsc{cnot}s;
NOTs)$ or \\
$(\textsc{cnot}s; \textsc{not}s)$.  The values may be total gates or circuit
depth (latency), depending on context.  The notation is sometimes
enhanced to show required concurrency and space,
$(\textsc{ccnot}s; \textsc{cnot}s; \textsc{not}s) \#(concurrency; space)$.

$t$ is time, or latency to execute an algorithm, and $S$ is space,
subscripted with the name of the algorithm or circuit subroutine.
When $t$ or $S$ is superscripted with \ac\ or \ntc, the values are for
the latency of the construct on that architecture, as described in
Section~\ref{sec:arch-models}.  Equations without superscripts are for
an abstract machine assuming no concurrency.  $R$ is the number of
calls to a subroutine, subscripted with the name of the routine.

\subsection{Linear-Time Adders}

The two most commonly cited modular exponentiation algorithms are
those of Vedral, Barenco, and Ekert~\cite{vedral:quant-arith}, which
we will refer to as VBE, and Beckman, Chari, Devabhaktuni, and
Preskill~\cite{beckman96:eff-net-quant-fact}, which we will refer to
as BCDP.  Both the BCDP and VBE algorithms build multipliers from
variants of carry-ripple adders, the simplest but slowest method.
Draper designed an adder that acts in the Fourier transform space
whose principal advantage is its smaller
size~\cite{draper00:quant-addition}.  Cuccaro, Draper, Kutin and
Moulton have more recently shown the design of a smaller, faster
carry-ripple adder, which we call
(CDKM)~\cite{cuccaro04:new-quant-ripple}, which appears to make the
Fourier adder obsolete.

\subsubsection{VBE Carry-Ripple}
\label{sec:vbe}

We use the VBE adder in several of our algorithmic variants described
in Chapter~\ref{ch:large-perf}.  In this algorithm, the values to
be added in (the convolution partial products of $x^a$, in the overall
modular exponentiation) are programmed into a temporary register
(combined with a superposition of $|0\rangle$ as necessary) based on a
control line and a data bit via appropriate \textsc{ccnot} gates.  Here we
examine just the adder itself.

\begin{figure}
\centerline{\hbox{
\includegraphics[width=14cm]{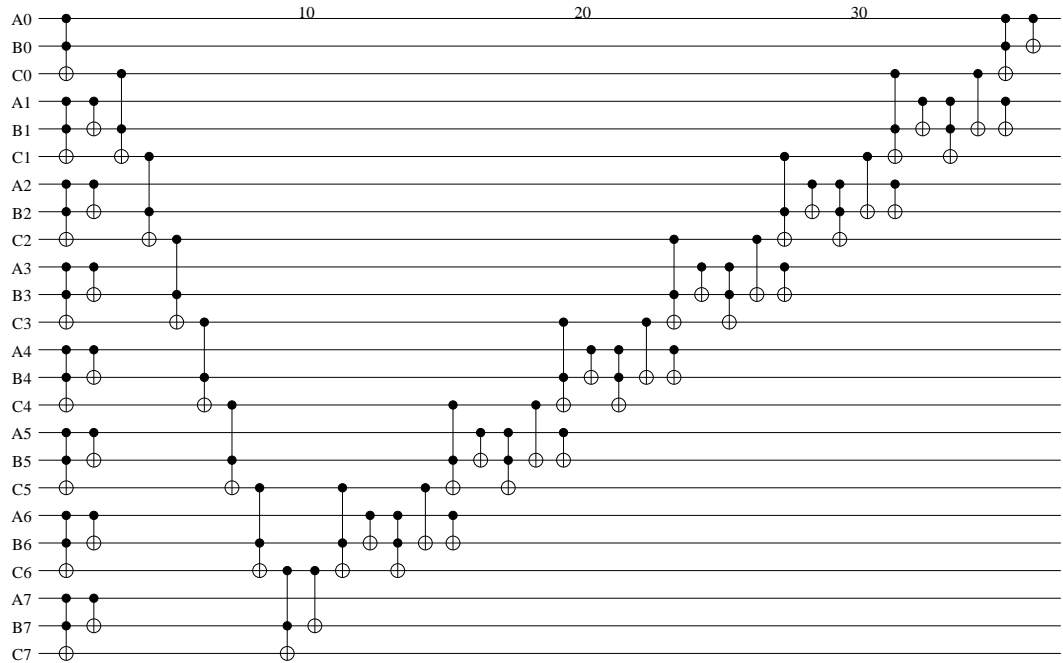}}}
\caption{An eight-bit VBE adder.}
\label{fig:vbe8}
\end{figure}

The latency of ADDER~\footnote{When we write ADDER in all capital
letters, we mean the complete VBE $n$-bit construction, with the
necessary undo; when we write adder in small letters, we are usually
referring to a smaller or generic circuit block.}, assuming {\em no}
concurrent gate execution, is
\begin{equation}
\label{eq:adder}
t_{ADD} = (4n-4; 4n-3; 0)\#(1; 3n)
\end{equation}
that is, $4n-4$ \textsc{ccnot} times plus $4n-3$ \textsc{cnot} times
and zero NOT times, executing only one gate at a time and using $3n$
qubits.  Since we are assuming no concurrent gate operations, this
value is the same as the total number of gates in the circuit.  In
Figure~\ref{fig:vbe8}, we have drawn the circuit with multiple gates
being executed in some time slots; the actual expression for the
performance of the circuit as drawn is
\begin{equation}
t_{ADD}^{AC} = (3n-3;\thinspace{}2n-3;\thinspace{}0)\#(3;3n)
\label{eq:adderac}
\end{equation}
It requires that at least 3 gates can be executed concurrently in
order to meet the performance specified, and uses $3n$ qubits during
the calculation.  These numbers are calculated assuming that gates on
independent qubits can be executed concurrently, and that \textsc{ccnot}s take
longer to execute than \textsc{cnot}s.

Figure~\ref{fig:vbe8} shows the circuit for an eight-bit VBE adder,
adding the $A$ and $B$ registers, with the $C$ register used as
temporary variables that begin in the zero state and must be returned
to that state at the end.  The graphical notation used for quantum
circuits is a superset of the classical reversible notation introduced
in Figure~\ref{fig:reversible-gates} on
page~\pageref{fig:reversible-gates}; we will introduce new gates as
necessary.  The structure of the circuit is straightforward.  Along
the left-hand edge, all of the partial sums are computed concurrently
(as drawn, the concurrency used is $n$, but it is easy to see that
doing the partial sums in a ``just in time'' fashion would result in a
concurrency of 3).  Next, descending from the top edge, we see a chain
of \textsc{ccnot} gates; these propagate the carry from one bit to the next.
The entire latter two-thirds of the circuit cleans up the ancillae we
have used, leaving the $A$ register in its original state and the $B$
register containing the eight-bit value $A+B$, with $C7$ the output
carry.  The numbers across the top of the diagram are clock cycles.
These numbers are counted assuming that all gates require the same
amount of time, which is not the case in most systems, so the numbers
should be treated as a guideline rather than an actual performance
figure.

Murali et al. experimentally demonstrated a half-adder subunit
of the VBE carry-ripple on an NMR
system~\cite{murali02:half-adder-nmr-impl}.  This experiment and the
NMR implementation of Shor's algorithm to factor the number
fifteen~\cite{vandersypen:shor-experiment} are, to the best of my
knowledge, the only experimental demonstrations of quantum arithmetic
circuits.

\subsubsection{BCDP Carry-Ripple}

The BCDP algorithm is also based on a carry-ripple adder.  It differs
from VBE in that it more aggressively takes advantage of classical
computation, adding a classical number into the register conditional
on a quantum enable bit.  However, for our purposes, this makes it
harder to use some of the optimization techniques presented in later
chapters.  Beckman et al. present several optimizations and
tradeoffs of space and time, slightly complicating the analysis.  The
latency of their adder is
\begin{equation}
\label{eq:oaddn}
t_{OADDN} = (6n-2; 2n; 2)
\end{equation}
which, assuming \textsc{ccnot} gates are slower than \textsc{cnot}s, is slower than the
VBE adder.

\subsubsection{Gossett Carry-Ripple}
\label{sec:gossett-ripple}

Shortly after the publication of the VBE and BCDP algorithms, Gossett
realized that it is possible to do much better than carry-ripple
arithmetic, drawing on the important classical Boolean techniques of
{\em carry-save arithmetic}~\cite{gossett98:q-carry-save}.  Gossett
does not provide a full modular exponentiation circuit, only adders,
multipliers, and a modular adder.  Carry-save arithmetic is
particularly well suited to incorporation into a larger multiplier
structure, but in this case a large penalty in the number of qubits
required must be paid.  Unfortunately, the paper's secondary
contribution, Gossett's carry-ripple adder, as drawn in his figure 7,
seems to be incorrect.  Once fixed, his circuit optimizes to be
similar to VBE.

\subsubsection{Draper QFT-based Adder}

Draper developed a clever method for doing addition on
Fourier-transformed representations of
numbers~\cite{draper00:quant-addition}.  It uses only $2n$ qubits, but
it requires $n$ concurrent gates.  Moreover, the comparison operations
necessary for modular arithmetic are difficult in the Fourier space,
necessitating frequent transformation of the representation between
integer and Fourier forms.  The accuracy required in the gate
rotations is very high, which may be difficult to achieve.  Finally,
although the latency is $O(n)$, I believe the constant factors to
actually implementing this circuit on encoded logical states will be
large, making it ultimately an unattractive option for most purposes.

\subsubsection{CDKM Carry-Ripple}
\label{sec:cdkm}

Cuccaro et al. have recently introduced a carry-ripple circuit, which
we will call CDKM, which uses only a single ancilla
qubit~\cite{cuccaro04:new-quant-ripple}.  The authors do not present a
complete modular exponentiation circuit; we will use their adder in
our algorithms {\bf F} and {\bf G} (Section~\ref{sec:mono-shor-perf}).
This adder, we will see in section~\ref{sec:algo-g}, is the most
efficient known for some architectures.

\begin{figure}
\centerline{\hbox{
\includegraphics[width=12cm]{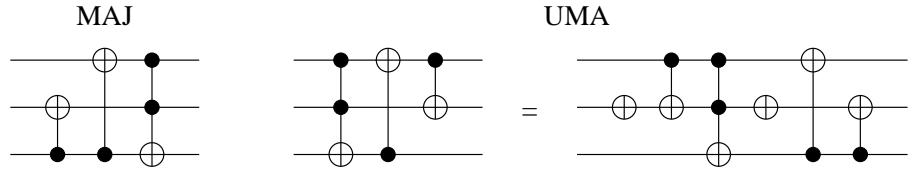}}}
\caption{Building blocks for the CDKM adder.}
\label{fig:cdkm-blocks}
\end{figure}

\begin{figure}
\centerline{\hbox{
\includegraphics[width=12cm]{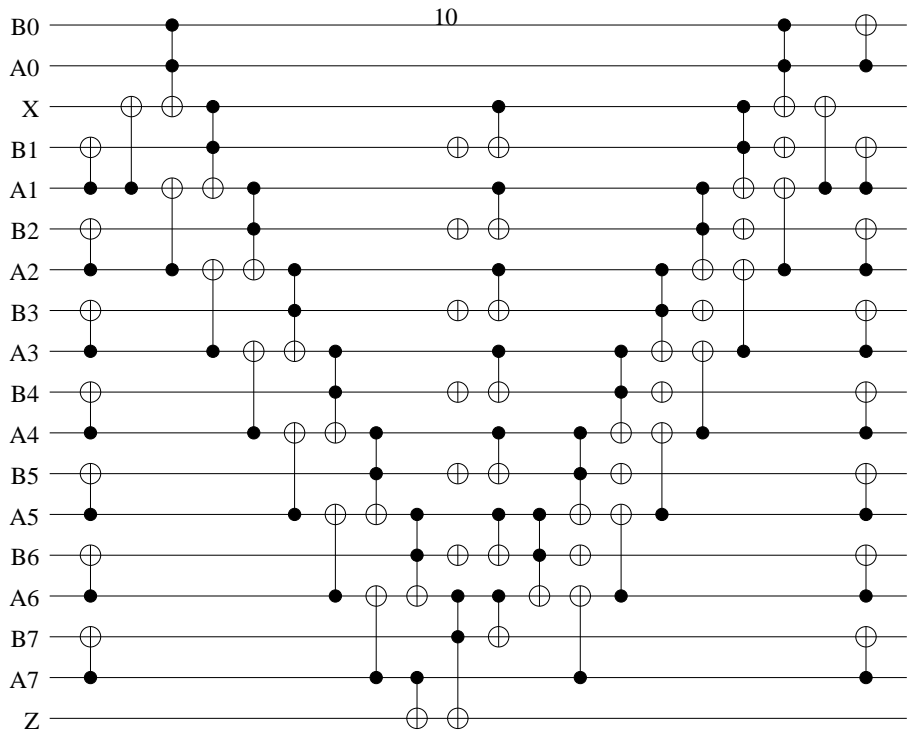}}}
\caption[An eight-bit CDKM adder.]{An eight-bit CDKM adder.  X is a
temporary variable, and Z is the carry out.}
\label{fig:cdkm8}
\end{figure}

Figure~\ref{fig:cdkm-blocks} shows the building blocks of the CDKM
adder.  MAJ is the majority function; the bottom qubit winds up
holding zero if two or three of the bits are zero, and one if two or
three of the bits are one.  It is the basis of the carry calculation
chain.  UMA is unmajority and add, undoing the MAJ calculation while
turning the middle bit into the correct, carry-adjusted final sum.
Two ways to construct the UMA function are shown.  A full adder
circuit is illustrated in Figure~\ref{fig:cdkm8}, using the right-hand
construct for UMA, which is more gates than the left-hand construct
but can be pipelined more effectively, overlapping the execution of
multiple gates and reducing the total latency.

The latency of their adder is
\begin{equation}
t_{CDKM} = (2n-1;\thinspace{}5;\thinspace{}0)\#(6;2n+2).
\end{equation}
This circuit uses only $2n+2$ qubits and runs perhaps one and a half
times as fast as the VBE adder (again, depending on implementation
details), but requires higher concurrency in gate operations.  This
factor affects the performance of the distributed forms of our
algorithms, presented in Section~\ref{sec:dist-shor}.

\subsection{$O(\log n)$ Adders}

Carry-save, carry-lookahead and conditional-sum (see
Sec.~\ref{sec:csum}) are all adder types that reach $O(\log n)$
performance by deferring carry computation or by communicating the
carry to distant parts of the circuit more rapidly.

\subsubsection{Gossett Carry-Save}
\label{sec:gossett}

Gossett's arithmetic is pure quantum, as opposed to the mixed
classical-quantum of BCDP.  Gossett's carry-save
adder~\cite{gossett98:q-carry-save}, the primary contribution of the
paper, can run in $O(\log n)$ time.  More importantly, carry-save
adders are designed to combine well into fast multiplier circuits.
However, such a circuit will remain impractical for the foreseeable
future due to the large number of qubits required; Gossett estimates
$8n^2$ qubits for a full multiplier, which would run in $O(\log^2 n)$
time.  It bears further analysis because of its high speed and
resemblance to standard fast classical multipliers.

\subsubsection{Carry-Lookahead}
\label{sec:qcla}

\begin{figure}
\centerline{\hbox{
\includegraphics[width=12cm]{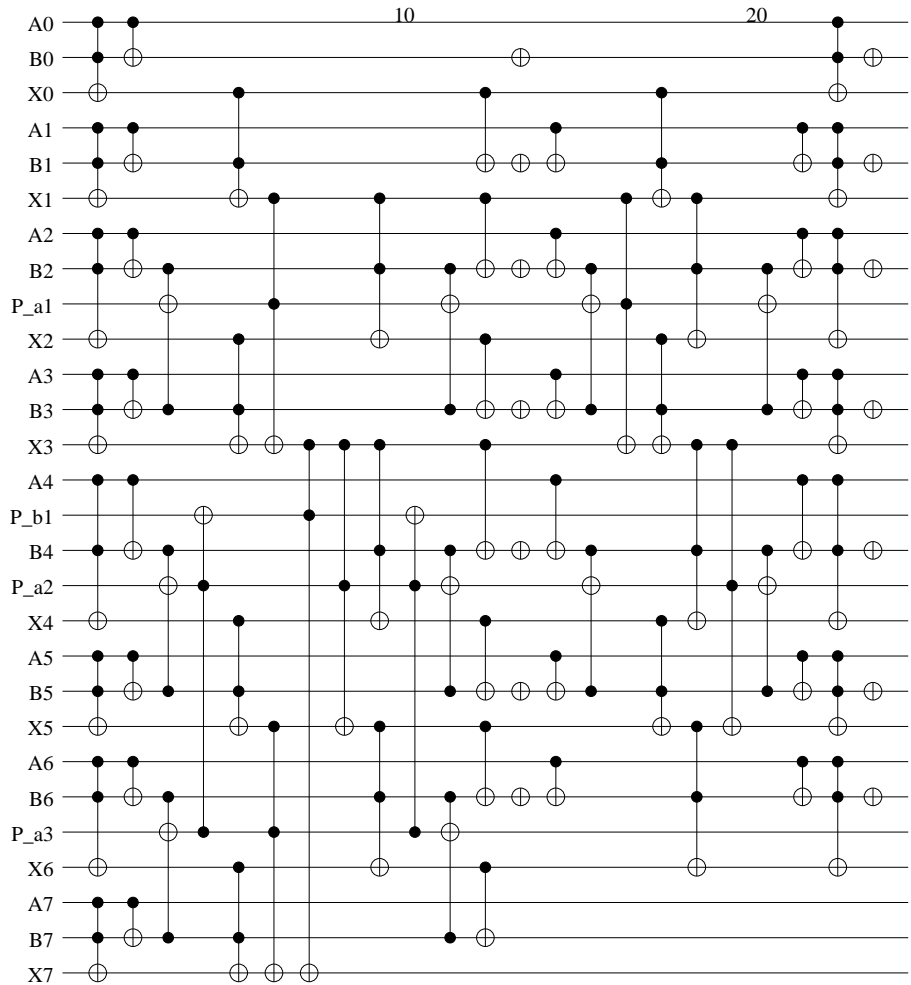}}}
\caption{An eight-bit carry-lookahead adder.}
\label{fig:look8}
\end{figure}

Draper, Kutin, Rains, and Svore have recently designed a
carry-lookahead adder, which we call
QCLA~\cite{draper04:quant-carry-lookahead}.  This method allows the
latency of an adder to drop to $O(\log n)$.  The latency and storage
of their adder is
\begin{equation}
t_{LA} = (4\log_2 n + 3;\thinspace{}4;\thinspace{}2)
\#(n;\thinspace4n-\log n-1).
\end{equation}
This circuit is illustrated in Figure~\ref{fig:look8}.  Although an
eight-bit carry-lookahead adder is not faster than a CDKM carry-ripple
adder, the logarithmic advantage quickly becomes apparent as $n$
grows.  When looking at this figure, it is immediately obvious that
the circuit is denser than the carry-ripple adders.  All quantum
carry-ripple adders exhibit a ``V'' shape in which many of the qubits
sit idle for long periods while the carry propagates down and back the
length of the register.  In the carry-lookahead adder, various carry
signals leapfrog up and down the register, with the overall state
gradually converging on the correct value.  In the figure, this
leapfrogging is illustrated by gates that stretch is much as half the
height of the total circuit.  We will see shortly that such gates are
not always practical, and that this issue will place limits on our
achievable performance.

\subsection{Ultimate Limits on Performance of Addition}
\label{sec:adder-ultimate}

The performance of any circuit must be specified with respect to a
particular architecture.  Architectural assumptions are implied in the
numbers provided throughout this chapter; we will detail these more
carefully in Section~\ref{sec:mono-shor-perf}.

Engineers tend to use the $O(\cdot)$ notation more loosely than
theorists.  The behavior of an algorithm is generally understood to
hold only for a particular range of problem size, or as long as a
certain set of assumptions holds.  In particular, signal propagation
times are often approximated to be zero, an assumption which clearly
does not hold indefinitely.  {\em All} algorithms which require any
signal to propagate to all parts of a computation are ultimately
limited to $O(\sqrt[3]{n})$ for any system in which bits occupy a
finite volume, as the signal propagation is constrained to the finite
speed of light and bits can only be packed in three dimensions.  This
constraint holds for addition; our assertion above that certain adders
can reach $O(\log n)$ performance holds only until signal propagation
effects come into play.  We will present the system behavior for more
realistic conditions when we discuss both monolithic and distributed
computation.

\subsection{Summary}

Recent focus on quantum arithmetic has provided a bounty of new
reversible addition algorithms.  With the exception of Draper's
quantum Fourier transform-based adder, all of the adder circuits we
have just presented will benefit classical reversible logic, as well.
In Boolean logic, the carry-ripple adder is so straightforward that
there are not many distinctions to be made.  In the reversible and
quantum arenas, we now have the VBE, BCDP, and CDKM carry-ripple
circuits, using different numbers of ancillae qubits and having
different performance characteristics.  We also have various more
complex adder circuits that reach square-root or logarithmic depth
instead of the linear depth of carry-ripple.  These faster circuits
include the carry-save adder, the carry-lookahead adder, and my two
circuits, the conditional-sum and carry-select adders, which we will
see in Section~\ref{sec:new-adders}.  All of these adders except the
carry-ripple ones require qubits that are some distance apart to
interact.  Classically, the choice of adder circuit in modern systems
is made not based on actual gate count, but on the time and space
required for the wiring to connect the bits; this approach will
inevitably be necessary in quantum computing, as well.

Integer arithmetic, of course, is the foundation of all computer
arithmetic, but has been extended in many ways to make more complex
functions, including integer multiplication and floating-point
arithmetic.  Research into these areas for reversible logic remains
very basic.  The next section introduces two methods for composing the
complete quantum modular exponentiation, and several optimizations,
but multiplication is still created by serial execution of addition.


\section{Quantum Modular Exponentiation}
\label{sec:modexp}

We now come to the part of the algorithm most relevant to this
thesis.  The modular exponentiation of a random integer is the most
computationally intensive portion of Shor's algorithm, and is our
benchmark for the behavior of our quantum multicomputer.  These
algorithms are introduced here and improved throughout
Chapter~\ref{ch:large-perf}.

To factor the number $N$ using Shor's algorithm~\cite{shor:factor}, a
quantum computing device must evolve to hold the state
\begin{equation}
\frac{1}{2^n}\sum_{a=0}^{2^{2n}-1}|a\rangle|x^a \bmod N\rangle.
\end{equation}
for a randomly chosen, fixed $x$, where $n$ is the bit length of $N$.
$|a\rangle$ is the register that holds the superposition of all values
$0..2^{2n}-1$, created by applying a Hadamard gate to each qubit in
$|a\rangle$.  Depending on the algorithm chosen for modular
exponentiation, $x$ may appear explicitly in a register in the quantum
computer, or may appear only implicitly in the choice of instructions
to be executed.

In general, quantum modular exponentiation algorithms are created from
building blocks that do modular multiplication,
\begin{equation}
|\alpha\rangle|0\rangle \rightarrow |\alpha\rangle|\alpha\beta \bmod
 N\rangle
\end{equation}
where $\beta$ and $N$ may or may not appear explicitly in quantum
registers.  This modular multiplication is built from blocks that
perform modular addition,
\begin{equation}
|\alpha\rangle|0\rangle \rightarrow |\alpha\rangle|\alpha + \beta \bmod
 N\rangle
\end{equation}
which, in turn, are usually built from blocks that perform addition
and comparison.

In most modular exponentiation algorithms, the multiplication step is
performed $2n$ times, once for each bit in the register
$|a\rangle$~\cite{vedral:quant-arith,beckman96:eff-net-quant-fact}.
The running product is multiplied by a value held in a quantum
register.  That value is either 1, if the corresponding bit of
$|a\rangle$ is zero, or $x^{2^i}$, if the corresponding bit is one.
Let $d_i = x^{2^i}$, and $a_{n-1}a_{n-2}..a_0$ be the binary expansion
of $a$.  The $d_i$ can be calculated classically, but $|a\rangle$ is a
quantum register.  The value $x^a \bmod N$ can be
rewritten~\cite{kunihiro:factoring-time,vedral:quant-arith} as
\begin{equation}
\prod_{j=0}^{2n} d_j^{a_j} \bmod N.
\end{equation}

Fundamentally, quantum modular exponentiation is $O(n^3)$; that is,
the number of quantum gates or operations scales with the cube of the
length in bits of the number to be
factored~\cite{shor:factor,vedral:quant-arith,beckman96:eff-net-quant-fact}.
It consists of $2n$ modular multiplications, each of which consists of
$O(n)$ additions, each of which requires $O(n)$ operations.  However,
$O(n^3)$ {\em operations} do not necessarily require $O(n^3)$ {\em
time steps}.  On an abstract machine, it is relatively straightforward
to see how to reduce each of those three layers to $O(\log n)$ time
steps, in exchange for more space and more {\em total} gates, giving a
total running time of $O(\log^3 n)$ if $O(n^3)$ qubits are available
and an arbitrary number of gates can be executed concurrently on
separate qubits.  Such large numbers of qubits are not expected to be
practical for the foreseeable future, so much interesting engineering
lies in optimizing for a given set of constraints.

\subsection{VBE, BCDP and Others}
\label{sec:vbe-etal}

Both the VBE and BCDP algorithms construct modular multiplication from
a straightforward series of modular additions.  Each modular addition
is performed by adding in the chosen number, comparing to $N$ to see
if the result has overflowed, and subtracting $N$ if so.  This method
results in a large number of additions and subtractions, which can
easily be reduced, as will be demonstrated in
Chapter~\ref{ch:large-perf}.

The VBE algorithm~\cite{vedral:quant-arith} builds full modular
exponentiation from smaller building blocks.  The bulk of the time is
spent in $20n^2-5n$ calls to ADDER.  The full circuit requires $7n+1$
qubits of storage: $2n+1$ for $a$, $n$ for the other multiplicand, $n$
for a running sum, $n$ for the convolution products, $n$ for a copy of
$N$, and $n$ for carries.

In this algorithm, the values to be added in, the convolution partial
products of $x^a$, are programmed into a temporary register (combined
with a superposition of $|0\rangle$ as necessary) based on a control
line and a data bit via appropriate \textsc{ccnot} gates.  The latency $t_{V}$
of the complete VBE algorithm is
\begin{eqnarray}
\label{eq:vbe}
t_{V} &=& (20n^2-5n)t_{ADD}\nonumber\\
&=& (80n^3-100n^2+20n;\thinspace{}96n^3-84n^2+15n;  \nonumber\\
&& \thinspace{}8n^2-2n+1).
\end{eqnarray}

The BCDP algorithm is similar in structure to VBE, but uses more
complicated gates and presents numerous engineering
tradeoffs. Borrowing from their equation 6.23, the latency $t_{B}$ of
the complete BCDP algorithm is
\begin{eqnarray}
t_{B} &=& (54n^3-127n^2+108n-29; \nonumber\\
&&\thinspace{}10n^3+15n^2-38n+14;\nonumber\\
&&\thinspace{}20n^3-38n^2+22n-4).
\end{eqnarray}
The exact sequence of gates to be applied is also dependent on the
input values of $N$ and $x$, saving space but making it less suitable
for hardware implementation with fixed gates (e.g., in an optical
system).  In the form we analyze, it requires $5n+3$ qubits, including
$2n+1$ for $|a\rangle$.

Beauregard has designed a circuit for doing modular exponentiation in
only $2n+3$ qubits of space~\cite{beauregard03:small-shor}, based on
Draper's clever method for doing addition on Fourier-transformed
representations of numbers~\cite{draper00:quant-addition}.  The depth
of Beauregard's circuit is $O(n^3)$, the same as VBE and BCDP.
However, we believe the constant factors on this circuit are very
large; every modulo addition consists of four Fourier transforms and
five Fourier additions.  Moreover, its primary advantage, reduction of
the scratch space used in addition, has been partially nullified by
the development of a carry-ripple adder that likewise uses only $2n+1$
qubits~\cite{cuccaro04:new-quant-ripple}.

Fowler, Devitt, and Hollenberg have simulated Shor's algorithm using
Beauregard's algorithm, for a class of machine they call {\em linear
nearest neighbor}
(LNN)~\cite{fowler04:_shor_implem,devitt04:_shor_qec_simul}.  LNN
corresponds approximately to our \ntc.  In their implementation of
the algorithm, they found no significant change in the computational
complexity of the algorithm on LNN or an \ac -like abstract
architecture, suggesting that the performance of Draper's adder, like
a carry-ripple adder, is essentially architecture-independent.

\subsection{Cleve-Watrous Parallel Multiplication}
\label{sec:cw-parallel-mult}

Modular exponentiation is often drawn as a string of modular
multiplications, but Cleve and Watrous pointed out that these can
easily be parallelized, at linear cost in space~\cite{cleve:qft}.
We always have to execute $2n$ multiplications; the goal is to do
them in as few time-delays as possible.

\begin{figure}
\centerline{\hbox{
\includegraphics[width=8.6cm]{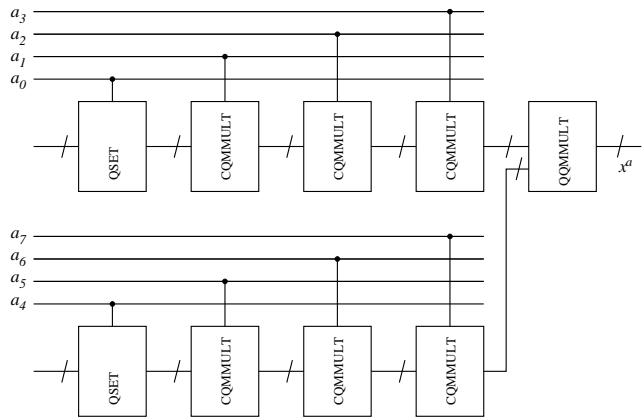}}}
\caption[Concurrent modular multiplication]{Concurrent modular
  multiplication in modular exponentiation using two multipliers.
  QSET simply sets the sum register to the appropriate value.}
\label{fig:pmodexp}
\end{figure}

\begin{figure}
\centerline{\hbox{
\includegraphics[width=12cm]{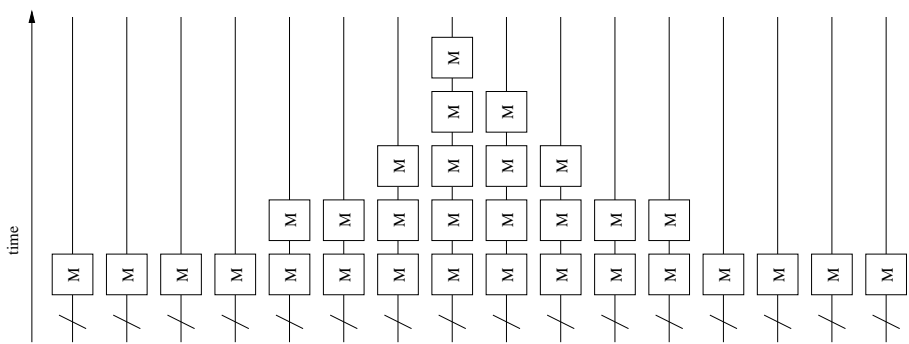}}}
\caption[Cleve-Watrous parallel multiplication]{Cleve-Watrous parallel
  multiplication (rotated ninety degrees relative to other graphs,
  with time flowing bottom to top).}
\label{fig:cw}
\end{figure}

To go (almost) twice as fast, use two multipliers.  For four times,
use four.  Naturally, this can be built up to $n$ multipliers to
multiply the necessary $2n+1$ numbers, in which case a tree
recombining the partial results requires $\log_2 n$ quantum-quantum
(Q-Q) multiplier latency times, as shown in Figure~\ref{fig:cw}.  We
will analyze this method in more detail in Section~\ref{sec:conc-exp}.

\subsection{Sch\"onhage-Strassen}

The Sch\"onhage-Strassen multiplication algorithm is often quoted in
quantum computing research as being $O(n \log n \log \log n)$ in
complexity for a single
multiplication~\cite{zalka98:_fast_shor,knuth:v2-3rd}.  However,
simply citing Sch\"onhage-Strassen without further qualification is
misleading for several reasons.  Most importantly, the constant
factors matter.  Shor noted this in his original paper, without
explicitly specifying a bound.  Quantum modular exponentiation based
on Sch\"onhage-Strassen is only faster than basic $O(n^3)$ algorithms
for more than approximately 32 {\em kilobits}~\footnote{Zalka found
that his approach would be faster for 8kilobits, using a slightly
different set of assumptions.}.  In this thesis, we will concentrate
on smaller problem sizes, and exact, rather than $O(\cdot)$,
performance.  Note also that this bound is for a Turing machine; a
random-access machine can reach $O(n\log n)$ using
Sch\"onhage-Strassen.

\section{Shor's Algorithm}

Finally, we come to Shor's factoring algorithm itself.  The algorithm
consists of both classical and quantum portions, with the quantum
portion being a period-finding method based on the QFT and arithmetic
to calculate the modular exponentiation of two integers.  The
period-finding method operates on two quantum registers, the control
register and the function result register; in the end, we will
actually measure the {\em control} register to find the period of the
function (this is perhaps the most counter-intuitive feature of the
algorithm).

To factor a number $N$ whose length is $n$ bits, we begin by checking
that the number is not even and determining that it not an integer
power, $a^b$, for $a \ge 1$ and $b > 2$.  Efficient classical methods
are known for this calculation and for finding the greatest common
divisor ($\operatorname{gcd}$), which we will not present.  Next,
choose an integer $2 < x < N$, and check that $\operatorname{gcd}(x,N)
= 1$; if not, return $\operatorname{gcd}(x,N)$.  The value of $x$ need
not be strictly random, but is not important except that repeating the
algorithm after a failure sometimes requires that $x$ be changed.

Next, use the quantum period-finding method to determine the order $r$
of $x$ modulo $N$.  If $r$ is even and $x^{r/2} \ne -1 \bmod N$,
calculate $\operatorname{gcd}(x^{r/2}-1,N)$ and
$\operatorname{gcd}(x^{r/2}-1,N)$.  One of these should be a factor of
$N$.  If not, or if $r$ is odd, repeat the algorithm, choosing a
different $x$.

The order of $x$ modulo $N$ is found by noting that we can calculate
the modular exponentiation $x^a \bmod N$ for all $a$.  We use two
quantum registers, which will hold, respectively, $a$ and $x^a \bmod
N$.  The register for $a$ must be $2n$ qubits long.  Starting from the
state
\begin{equation}
\frac{1}{2^L}\sum_{a=0}^{2^{2L}-1}|a\rangle|1\rangle
\end{equation}
in which all of the qubits are disentangled, the modular
exponentiation then produces the state
\begin{equation}
\frac{1}{2^L}\sum_{a=0}^{2^{2L}-1}|a\rangle|x^a \bmod N\rangle.
\end{equation}
Once we have that entangled state~\cite{kendon04:_entan-shor}, we
apply the QFT {\em to the first register}, measure both registers, and
use the value in the first register (discarding the second) to find
the order of $x$ modulo $N$, and from there the factors of $N$.


How the QFT creates a state that can tell us the order of the function
is mysterious, almost spooky, and certainly difficult to grasp.  To
make this more concrete, let's look at an example.  15 is the smallest
number upon which Shor's algorithm works properly, and we will choose
$x=7$ as a good example.  For reasons we won't go into here, we really
need at least one bit more in our $a$ register than the length of $N$
itself, but we will restrict ourselves to four bits for $a$ to keep
the size of the example manageable.  This gives us
\begin{equation}
\begin{split}
\frac{1}{4}\sum_{a=0}^{15}|a\rangle|x^a\bmod N\rangle = &
\frac{1}{4}(|0\rangle|1\rangle+|1\rangle|7\rangle+|2\rangle|4\rangle+|3\rangle|13\rangle\\
& + |4\rangle|1\rangle+|5\rangle|7\rangle+|6\rangle|4\rangle+|7\rangle|13\rangle\\
& + |8\rangle|1\rangle+|9\rangle|7\rangle+|10\rangle|4\rangle+|11\rangle|13\rangle\\
& + |12\rangle|1\rangle+|13\rangle|7\rangle+|14\rangle|4\rangle+|15\rangle|13\rangle)\\
= & \frac{1}{4}((|0\rangle+|4\rangle+|8\rangle+|12\rangle)|1\rangle\\
& + (|1\rangle+|5\rangle+|9\rangle+|13\rangle)|7\rangle\\
& + (|2\rangle+|6\rangle+|10\rangle+|14\rangle)|4\rangle\\
& + (|3\rangle+|7\rangle+|11\rangle+|15\rangle)|13\rangle).
\end{split}
\label{eq:period}
\end{equation}
The second form makes it clear that what we have accomplished so far
is to {\em group the values of $a$ based on $x^a\bmod N$}.  Each of
these groups -- 0-4-8-12, 1-5-9-13, etc. -- has elements that skip
four values, but with an offset that differs from group to group.
This information -- the length of that stride between elements of the
superposition in each group -- is what will allow us to find the
order.  But how can we extract that piece of information?

If we were to apply the QFT to our original raw $a$ register
$\frac{1}{4}\sum_{a=0}^{15}|a\rangle$, the result would simply be
$|0\rangle$.  The grouping created by the modular exponentiation now
creates sets of elements that can effectively be Fourier transformed
independently.  The Fourier transform, as noted, eliminates the
offset, ``hiding'' it in the phase of the elements of the
superposition and leaving the frequency components in the numeric
values.  The QFT of Equation~\ref{eq:period} is
\begin{equation}
\begin{split}
\operatorname{QFT}(&\frac{1}{4}((|0\rangle+|4\rangle+|8\rangle+|12\rangle)|1\rangle\\
& + (|1\rangle+|5\rangle+|9\rangle+|13\rangle)|7\rangle\\
& + (|2\rangle+|6\rangle+|10\rangle+|14\rangle)|4\rangle\\
& + (|3\rangle+|7\rangle+|11\rangle+|15\rangle)|13\rangle))\\
=& (\frac{1}{4}((|0\rangle+|4\rangle+|8\rangle+|12\rangle)|1\rangle\\
& + (|0\rangle+i|4\rangle-|8\rangle-i|12\rangle)|7\rangle\\
& + (|0\rangle-|4\rangle+|8\rangle-|12\rangle)|4\rangle\\
& + (|0\rangle-i|4\rangle-|8\rangle+i|12\rangle)|13\rangle)).
\end{split}
\end{equation}
Now, when we measure the two registers, we will always find one of 0,
4, 8, or 12 in the first register, with equal probability.  If we find
0, the algorithm has failed and we must repeat.  Otherwise, we use the
number found as $r$, and apply Euclid's algorithm for finding greatest
common denominators to find the GCD of $N$ and $x^{r/2}-1$, and of $N$
and $x^{r/2}+1$, as described above.

\section{Summary}

In this chapter, we have introduced Shor's algorithm for factoring
large numbers, and discussed its significance.  The creation of a
machine that executes Shor's algorithm would have implications for
security on the Internet, breaking the widely-used RSA public-key
crypto system.  Most of the tasks assigned to RSA can be accomplished
via other mechanisms, including symmetric, private-key encryption, but
such solutions may be less efficient in using resources both locally
and globally~\cite{schneier96:_applied_crypto}.

Shor's algorithm rests on the breakthrough insight that certain
functions produce the same results for inputs that are separated by a
specific period, and that the quantum Fourier transform can extract
that period efficiently.  For factoring large composite integers, the
function of interest is the exponentiation of a random number modulo
$N$, the number to be factored.  The modular exponentiation is
constructed in a straightforward fashion from integer addition and
comparison, and we saw various circuits for addition.  We will see in
later chapters how to implement these operations efficiently; we
turn next to a taxonomy of quantum computing technologies which might
used to build systems on which Shor's algorithm can be run.

  \chapter[Taxonomy of Quantum Computing Technologies]{A Taxonomy of Quantum Computing Technologies}
\label{ch:taxonomy}

In this chapter we present a classification scheme for quantum
computing technologies, based on the characteristics most relevant to
computer systems architecture, and apply it to analyze several
candidate technologies.  This taxonomy is complementary to the
DiVincenzo criteria introduced in Section~\ref{sec:qc-intro}.  Whereas
the DiVincenzo criteria help define whether or not it is {\em
possible} to build a quantum computer based on the specified
technology, in our taxonomy we are concerned with whether or not it is
{\em practical}.  This taxonomy will be used in our definition of a
scalable system (Section~\ref{ch:scalability-defn}), and the
performance-relevant portions will affect our analysis of systems
throughout the remainder of this thesis.  We will describe each
criterion as well as some of its high-level architectural
implications.  In the last section, we will use this taxonomy to
evaluate several proposed computing technologies.

\section{Taxonomy Framework}

\subsection{Basic Features}

\paragraph{Stationary, flying and mobile:}
Quantum computing technologies can be divided into two categories:
those in which the qubits are represented by constantly moving
phenomena (photons) and those in which qubits are represented by
static phenomena (nuclear or electron spins).  For phenomena that
move, gates are physical devices which affect qubits as they flow
through the gate.  These are called ``flying qubits''.  Optical
implementations generally fall into this category, where photons are
qubits and e.g. beam splitters serve as gates.  For ``stationary''
phenomena, qubits occupy a physical place and gate operations from an
application are applied to them.  The ``stationary'' notion applies
\textit{only} during gate operation.  Some stationary technologies,
such as the proposed scalable ion trap~\cite{kielpinski:large-scale},
permit the physical qubit carrier to be moved prior to application of
a gate; we will call these ``mobile'' qubits.

The key reason to make the distinction between stationary and flying
implementations is dynamic control.  In a flying qubit device, the
order and type of gates must typically be fixed in advance, often at
device construction time; different program execution is achieved by
classical control of switches that route qubits through different
portions of the circuit.  A stationary qubit device has more
flexibility to reconfigure gates.  In this sense, using stationary
devices is like classical programming, while flying qubit designs are
more like classical circuit design~\cite{yao93quantum}.

\paragraph{Single system versus ensemble:}
A significant distinction in quantum computing technologies is the
choice of {\em ensemble} computing or {\em singleton} computing.  In
ensemble computing, generally implemented on stationary qubit systems,
there are many identical quantum computers, all receiving the same
operators and executing the same program on the same data (except for
noise).  Singleton systems have the ability to directly control a
single physical entity that is used to represent the qubit.

From a technology perspective, ensemble systems are easier to
experiment with, as techniques for manipulating and measuring large
numbers of atoms or molecules are well understood.  Hence, the largest
quantum computing system demonstrations to date have all been on
bulk-spin NMR~\cite{vandersypen:shor-experiment,boulant03:_exper_nmr},
which uses an ensemble of molecules to compute.

\paragraph{Quantum I/O:}
There are a variety of reasons why we may want to move quantum data
from one place to another: we may simply be aggregating multiple
devices into a larger machine, or the far node may provide different
computational capabilities (e.g. long-term storage) or have access to
different data.  In some cases, we may wish to move quantum data
between devices of different
technologies~\cite{matsukevich:matter-light-xfer}.  In our quantum
multicomputer, we will be aggregating homogeneous nodes into a larger
system using the qubus protocol described in Chapter~\ref{ch:travel}.

Quantum I/O (QIO) is a very error-prone process.  Therefore, it is
done by first using QIO on ``empty'' qubits, which we will call QIO
sites or transceiver qubits, creating an entangled state between a
pair of devices.  Once the existence of the entangled state is
confirmed through a process called
purification~\cite{bennett95:_concen,cirac97:_distr_quant_comput_noisy_chann,pan03:_exper-purification},
it can be used to transfer any desired quantum state by using quantum
teleportation (Chapter~\ref{ch:travel}).

Question marks appear in the QIO entries in
table~\ref{tab:qubit-tech-char} because experimental demonstration in
structures similar to those expected to be used in quantum computers
has not yet been done, or because adequate fidelity has not been
shown.  In some cases, basic experimental confirmation or proposals
backed by relatively solid analysis exist; in others, only a few
sentences in a longer paper.

\paragraph{Measurement:}
In Section~\ref{sec:measurement} we discussed measurement in the
abstract, and in Chapter~\ref{ch:err-mgmt} we saw its importance for
quantum error correction.  Four architectural features characterize
different measurement schemes: (1) Can measurements of multiple
quantum bits be performed in parallel or must they be serialized?  (2)
Does measurement of a quantum bit require interaction with another
``clean'' qubit in order to produce a result?  (3) Is the speed of
measurement about as fast (in the same order of magnitude) as
performing an operation?  (4) Can measurement be performed almost
anywhere, or must the physical entities that are used to represent the
qubits be moved to specialized measurement sites?

Reliably computing on a quantum system will mean that many of the
total quantum operations will be measurements, as we discussed in the
last chapter.  From an architectural perspective, if measurements must
be performed serially, or are inordinately slow, then Amdahl's
Law~\cite{amdahl67} will apply and measurement will be the bottleneck
in computation.  Furthermore, if additional ancillae qubits are
required for measurement to take place, then we must plan for the
initialization of those qubits to occur frequently.  Similarly, if
technologies restrict where measurement can occur, then those
restrictions will need to be designed into the architecture and
algorithms.

\subsection{Algorithmic Efficiency Features}
\label{sec:algo-eff}

Many features of the various quantum computing proposals will have
profound implications for the execution of quantum algorithms on
realistic architectures.

\paragraph{Concurrency (control parallelism):}

The most fundamental feature required to accelerate quantum
computation is concurrent execution of gates.  This is useful at the
algorithmic (logical) level, but critical at the physical level, where
concurrent operation is required in order to execute quantum error
correction frequently enough to prevent decoherence of large numbers
of qubits.

Despite the advantages in computational complexity class that some
quantum algorithms promise, it is still important to extract
parallelism from quantum algorithms.  If all operations had to be
sequentialized, then on some proposals, such as silicon
NMR~\cite{skinner03:_hydrog_spin}, it would still require significant
time to factor large values.  For example,
Kunihiro~\cite{kunihiro05:_exact_ieice} has estimated the sequential
running time of Shor's algorithm factoring a 530-bit number at 1.18
years for a 1kHz device (approximately NMR speeds), 10 hours for a
1MHz device, or 37 seconds for a 1GHz device.

Fortunately, there is significant parallelism
available~\cite{moore98:_parallel_quantum} in quantum software (error
correction~\cite{steane02:ft-qec-overhead} and
factoring~\cite{cleve:qft,van-meter04:fast-modexp}).  The ability to
exploit this parallelism, however, requires technologies with parallel
control.  This parallel control will require significant classical
support circuitry.  If this circuitry cannot be located ``on chip''
near the qubits then a high-bandwidth interface between a classical
device generating control pulses and a quantum device containing the
actual qubits will be required.  This may be a control line per qubit,
or may be multiplexed across the wire, reducing the need for I/O pads
at the cost of reduced concurrent operation (and longer times between
QEC cycles).  Thaker et al. have designed a large-scale ion trap with
separate storage and gate action sites (see
below), and investigated the use of the carry-lookahead adder on this
system, finding that performance grows only linearly due to limited
application-level concurrency~\cite{thaker06:_cqla}.

\paragraph{Total available qubits:}
The feature with the single largest impact on the scalability,
usefulness, and reliability of the computer is the actual number of
physical qubits available.  Clearly, too few qubits and the ability to
execute on large data sizes will be inhibited.  Additional qubits can
be utilized for increased reliability via error correction, as well as
algorithmic parallelism.

All entries in table~\ref{tab:qubit-tech-topo} are followed by
question marks because of very high uncertainty; in some cases, even
which factors will prove to be the practical limits are not yet
clear.  As most researchers are still focusing on very small numbers
of qubits, they have not yet attempted to circumscribe this upper
limit.

\paragraph{Wiring topologies:}
Optimization of the architecture to support the data movement of a
useful class of algorithms is one of the key areas where computer
architects can contribute.  In many proposed technologies, only
neighboring qubits are allowed to perform two-qubit gates.  Either the
physical entities representing the qubits (using a control
process~\cite{kielpinski:large-scale}) or just the state (using
quantum wires~\cite{oskin:quantum-wires}) must be moved around the
machine to support computation.  In some cases, technological
constraints limit the interconnection topology to a one-dimensional
line; in others, a loose two-dimensional lattice, full 2-D mesh, or
even 3-D structure have been proposed~\cite{lloyd:model}.  A few
proposals support long-distance gates with various tradeoffs, such as
limited concurrency~\cite{you:scalable}.

\paragraph{Addressability:}
In some systems, addressing specific qubits is difficult, because
localization of the classical control required (e.g.,
microwave-frequency electromagnetic field) to just the small region
the qubit occupies is difficult.  One solution, the original Lloyd
model, proposes forming small groups of qubits into cellular
automata~\cite{lloyd:model}.  One suggested implementation is long
molecular chains with a repeating pattern in which each unit is a C.A.
Each qubit position in the automaton can be addressed via a specific
electromagnetic frequency.  Each automaton follows the same program,
effected by electromagnetic radiation blanketing the whole device,
which is, in effect, a fully concurrent SIMD machine.  One technique
for turning a cellular automata into a more-easily-controlled serial
machine is to include in the cellular automata a token that is passed
from automaton to automaton; only the automaton holding the token
performs the indicated action.  We expect that designing architectures
and software systems for technologies without the ability to address
and operate on specific qubits will be difficult.

\paragraph{Operations on all qubits:}
In most physical implementations, all qubits are identical; any qubit
can have any operation performed on it during any clock cycle.  A few
technologies, however, notably the scalable ion trap, separate storage
and action locations, so that qubits (e.g., individual atoms) must be
physically moved from a storage location to an action location before
a gate can be executed on the qubit.

\subsection{Time and Gate Characteristics}
\label{sec:time-gate}

\paragraph{Decoherence time:}
We discussed decoherence in Section~\ref{sec:decoherence}.  The upside
to good isolation from environmental effects is long {\em coherence
time}, or the time which a qubit can be ``kept''.  As a broad
generalization, those technologies relying upon electrons to maintain
quantum state have short coherence times because electrons are fairly
mobile and tend to interact with their surrounding environment.
Technologies that utilize nuclear effects are more stable.  However,
the downside to good isolation from environment effects is relatively
slow operation times for two-qubit gates.  Across the technologies we
examine, the gate speed and decoherence time vary over eight orders of
magnitude or more~\cite{ladd:coherence-time}.  Coherence time is an
especially important research area and will be subject to potentially
large advances as QC technology progresses.  Gate operation time,
however, is often tied directly to physical processes with limited
flexibility in engineering parameters.

\paragraph{Measurement time:}
How long does it take to accurately measure the state of a qubit?
For many technologies the measurement time is longer than a gate time,
dominating the time for a quantum error correction cycle and hence the
logical clock speed.

\paragraph{Single-qubit and two-qubit gate clock speeds:}
In some cases, the time it takes to perform a one-qubit gate can be
vastly different from the time for a two-qubit gate, so we must
specify both.

\paragraph{Natural two-qubit gate:}
Various sets of gates have been shown to form elementary basis
sets~\cite{barenco:elementary,divincenzo94:twobit}.  The standard set
of universal gates presented in Section~\ref{sec:divincenzo}
(\textsc{x, h, t, cnot}) is just one example, and all serious
proposals for quantum computing technologies include enough operations
to provide this or an equivalent universal set.  Beyond universality,
however, are three important characteristics.  (1) Does the technology
provide an arbitrary single qubit rotation, or must it be synthesized
from \textsc{x, h} and \textsc{t}; (2) How complex are the syntheses
for a \textsc{cnot} and three qubit controlled-controlled-not (a
\textsc{toffoli} gate), which is commonly used in quantum
algorithms~\cite{barenco:elementary}; (3) Do specific gates have
unwanted effects on qubits that are \textit{not} the intended operands
(that is, are other qubits being implicitly manipulated)?  We will
discuss these in more detail below.

Several of most common physical interactions result in a controllable
exchange (\textsc{SWAP}), the $J$ coupling~\cite{vandersypen:thesis},
and a controlled phase shift, which, when applied for the appropriate
amounts of time, give us possible two-qubit natural gates with these
unitary transforms:
\begin{align}
\sqrt{\text{SWAP}} &= \left[\begin{array}{cccc}
1 & 0 & 0 & 0 \\
0 & \frac{1}{\sqrt{2}} & \frac{1}{\sqrt{2}} & 0 \\
0 & \frac{1}{\sqrt{2}} & -\frac{1}{\sqrt{2}} & 0 \\
0 & 0 & 0 & 1 \end{array}\right]
\end{align}
\begin{align}
J &= \left[\begin{array}{cccc}
-i & 0 & 0 & 0 \\
0 & i & 0 & 0 \\
0 & 0 & i & 0 \\
0 & 0 & 0 & -i \end{array}\right]
\end{align}
\begin{align}
\text{CZ} &= \left[
\begin{array}{cccc}
1 & 0 & 0 & 0 \\
0 & 1 & 0 & 0 \\
0 & 0 & 1 & 0 \\
0 & 0 & 0 & -1 \end{array}\right].
\end{align}
From these three possible entangling two-qubit gates, we can construct
a \textsc{CNOT} with only a few single-qubit rotations on the two qubits.

In stationary qubit devices such as ion traps or NMR systems, several
electromagnetic pulses are generally required to implement each gate.
A typical number is five or six, though the exact number and timing
are dependent on the gate to be executed.  One side effect in NMR
systems is that nearby qubits are affected by these pulses and are
implicitly operated on by them.  To overcome this, additional control
sequences called {\em decoupling pulses} are
required~\cite{beckman96:eff-net-quant-fact,leung99:_effic_hadam}.

\subsection{Other Features}

\paragraph{Logical Encoding:}
Quantum algorithms are written to manipulate abstract, logical qubits.
Logical qubits, however, are not always represented by a single
physical phenomenon such as a single ion or photon.  We call the
entities that software manipulates ``logical qubits'' (or ``encoded
qubits'' when quantum error correction is involved) and the entities
that technologies use to implement them ``elementary qubits'' or
``physical qubits''.  This is not the same as the ensemble /
singleton distinction outlined above.

In some technologies, such as electron count (charge) in quantum dots,
a ``dual rail'' encoding is used.  Similarly, a single photon may take
either the left or right path through a circuit, corresponding to
logical different quantum states (i.e. 0 or 1).  In both of these
technologies, it is possible to talk about a single quantum dot (or
path) as a single qubit, but we arrange computation and measurement to
take place on the encoded pair.

\paragraph{Gate-Level Timing Control:}
Because the state of an individual qubit is something of an analog
phenomenon, precise timing of gates is critical.  What will limit our
ability to achieve the necessary precision?  And, in the case of
photons or other flying qubits, how do we dynamically adjust their
arrival times so that multiple qubits can be in the right place at the
same time?  Most qubits oscillate; how do we keep the relative phases
of multiple qubits right?

\paragraph{Scalability Limits:}
Scaling to large numbers of qubits is, for most architectures, a
function of all of the above factors and more.  Other factors not yet
described are technology specific.  For example, in lithography-based
systems, they include I/O pads on the chip, the supporting
infrastructure such as rack-mount microwave generators, and the
practical challenge of simply providing enough control wires to such a
small device.  Few of the proposals suggest that an actual numerical
upper bound exists because of any of these factors, yet they are
critical to the success of building systems.  In the next section we
will highlight what the primary scalability limit is perceived to be
for each technology.

\subsection{Manufacturing and Operating Environment}
\label{sec:manufop}

At the moment, all scalable quantum computing technologies are
proposals and significant advances in manufacturing will be required
to bring them to reality.  Nevertheless, some proposals have less
onerous technological hurdles in front of them than others.
Furthermore, certain proposed technologies integrate better with
existing classical silicon-based computing.

\paragraph{Fabrication challenges:}
To what extent do the proposed technologies rely on
difficult-to-achieve advances in manufacturing?  For example, the Kane
silicon-based NMR technology relies upon the ability to dope silicon
with precisely placed individual phosphorus atoms, and to align those
with overlaid structures created using standard VLSI
lithography~\cite{kane:nature-si-qc}.  All of the solid-state circuit
techniques require classical control lines (e.g.,
~\cite{fujisawa98:double-dot,nakamura99:_coher_cooper}), which may
benefit from expected improvement in VLSI feature sizes following
Moore's Law~\cite{moore65,ITRS2005}.  In our taxonomy we will
highlight the major technological challenges facing each quantum
computing proposal and discuss the latest advances in overcoming them.

\paragraph{Operating temperature:}
In order to control noise, most proposals call for extremely low
temperatures achievable only with liquid helium.  Others, such as
superconducting qubits and quantum dot qubits, require still colder
\textit{millikelvin} temperatures achieved through a dilution
refrigerator.  A dilution refrigerator, or dil fridge, uses the
different condensation characteristics of helium-3 and helium-4 to
cool things down to millikelvin
temperatures~\cite{craig04:_hguide_dil_fridge}.

Although there are numerous models, the dil fridges made by Oxford
Instruments seem to be popular.  The most commonly used ones are
almost two meters tall and a little under a meter in diameter.  The
researcher loads the test sample in from the top on a long insert, so
another two meters' clearance above (plus a small winch) are
required. The lowest temperature a dil fridge can reach is limited in
theory to approximately 7 millikelvin, and in practice to higher
values depending on model. A dil fridge can typically extract only a
few hundred microwatts of heat from the device under test, which is
limited to a few cubic centimeters.  This thermal limit will limit the
number of devices per chip and the operating speed of the devices,
imposing an important constraint on scalability.  These low
temperatures are not only operationally challenging, but also affect
the ability of classical circuits to operate, complicating the design
of the control process~\cite{oskin:quantum-wires}.

The atom chip~\cite{folman00:_atom-chip} and ion
trap~\cite{cirac95:_cold_trapped_ions} operate by cooling individual
atoms to extremely low temperatures using lasers and electrical and
magnetic control fields, but the devices themselves are kept at room
temperature and no elaborate cooling mechanisms are required.

\paragraph{Supporting equipment:}
Some technologies require complex supporting equipment, notably
high-frequency microwave and voltage signal generators and
high-precision lasers.  One or more of these per qubit may be needed;
as systems scale, switching or sharing of this equipment or direct
integration into on-chip systems are likely to be required.

\section{Quantum Technologies}
\label{sec:survey}

In this section we survey a variety of proposed quantum computing
technologies using the taxonomy framework described in the last
section.  We have chosen to focus on eight technologies: Si-NMR,
P-NMR, solution NMR, quantum dot charge, scalable ion traps, Josephson
junction charge, linear optics, and optical lattice.  This selection
should by no means be interpreted as exhaustive; several dozen viable
proposals
exist~\cite{folman00:_atom-chip,shahriar02:_spectral-holes,pellizzari95:_cavity-QED,childress05:_ft-quant-repeater}.
These systems were chosen for their near and long term
implementability, and/or scalability and/or pedagogical interest.  It
is also worth noting that the fundamental technology, in some case,
can lead to several possible qubit representations, such as spin,
energy level, or particle count.  The information is summarized in
Tables~\ref{tab:qubit-tech-char}-\ref{tab:manuf}.  Below we will
briefly discuss each technology and its architectural implications.

\begin{table}
\centerline{
\begin{tabular}{|p{1.2in}|c|c|c|p{1.25in}|p{0.8in}|}\hline
technology & stationary/    & single/  & QIO? & measurement & references \\ 
           & flying/mobile & ensemble & & & \\ 
\hline\hline
Si NMR		& stationary    & ensemble & N & mechanical vibration, concurrent, frequency analysis & ~\cite{ladd:si-nmr-qc}\\
\hline
solution NMR 	& stationary    & ensemble      & N & concurrent, frequency analysis & ~\cite{vandersypen:shor-experiment}\\
\hline
quantum dot charge & stationary & single        & Y? & concurrent, on-chip auxiliary structures, similar to quantum dots in size and structure & ~\cite{loss:qdot-comp} \\
\hline
scalable ion trap & mobile        & single        & Y? & limited concurrent, optically induced fluorescence & ~\cite{cirac95:_cold_trapped_ions,kielpinski:large-scale}\\
\hline
JJ charge 	& stationary    & single        & Y?  & concurrent, on-chip charge probe & ~\cite{pashkin:oscill,you:many-qubits}\\
\hline
Kane model 	& stationary    & single        & N?  & concurrent, single-electron spin measurement & ~\cite{kane:nature-si-qc}\\
\hline
LOQC	 	& flying        & single        & Y   & single qubit
polarization via single photon number resolving optical detectors & ~\cite{knill01:klm}\\
\hline
optical lattice	& stationary	& single	& N?  & fluorescence,
but resolution of individual atoms difficult & ~\cite{brennen99:_optical_lattice,jaksch99:_cold_collisions,treutlein06:_qip_optic_lattic_magnet_microt}\\
\hline
\end{tabular}
}
\caption[Qubit technology basic characteristics]{Qubit technology
  basic characteristics.  Question marks under QIO indicate that
  experimental verification has not yet been shown.  JJ: Josephson
  junction, LOQC: linear optics quantum computing}
\label{tab:qubit-tech-char}
\end{table}

\begin{table}
\centerline{
\begin{tabular}{|p{0.75in}|p{1.1in}|c|p{0.9in}|p{1.2in}|p{0.4in}|} \hline
technology & concurrency & max qubits & wiring topologies & addressability & ops on all qubits? \\ 
\hline\hline
Si NMR 	& limited by ability to suppress activity of uninvolved qubits
& hundreds?	 	& linear nearest neighbor  & by frequency, all independent                         & Y \\
\hline
solution NMR 	& limited by ability to suppress activity of uninvolved qubits
& low tens? 	& linear nearest neighbor, limited non-neighbor 	     & by frequency, all independent & Y \\
\hline
quantum dot charge & limited by control mechanism 
& large? 	& linear nearest neighbor    & localized, independent
control via on-chip systems   & Y \\
\hline
scalable ion trap & limited by \# of action sites with lasers
& large? 	& open, irregular, up to 2-D? 
& individual ions and chains moved from addressable storage to action sites & N \\
\hline
JJ charge 	& limited by coupling mechanism
& large? 	& 1-D, 2-D?, long-distance possible?	   &
localized, independent control via on-chip systems & Y \\
\hline
Kane model 	& limited by control mechanism 
& large? 	& 1-D or 2-D? 		   & localized, independent control via on-chip systems & Y \\
\hline
LOQC 	& unlimited? 
& large?	& physical routing, essentially unlimited     & physical position & Y \\
\hline
optical lattice & mandatory & thousands? & 1-, 2-, or 3-D neighbors &
none & Y \\
\hline
\end{tabular}
}
\caption[Features affecting algorithm efficiency]{Features affecting
  algorithm efficiency on specific qubit technologies.  The maximum
  number of qubits in all technologies remains undetermined with any
  reliability.  Question marks in topologies indicate that the natural
  area for layout is 2-D, but practical engineering constraints may
  limit full 2-D layout.}
\label{tab:qubit-tech-topo}
\end{table}

\begin{table}
\centerline{
\begin{tabular}{|p{0.75in}|p{0.8in}|p{0.85in}|p{0.9in}|p{0.75in}|p{0.75in}|} \hline
technology & decoherence time & measurement time & single-qubit gate clock speed &
two-qubit gate clock speed & natural two-qubit gate \\
\hline\hline
Si NMR 	& 25s & long & 40kHz & 400Hz & $J$ coupling \\
\hline
solution NMR 	& seconds & long & 50kHz & 50Hz & $J$ coupling \\
\hline
quantum dot charge & a few ns & 10-100ns~\cite{fujisawa98:double-dot,loss:qdot-comp} & 10GHz & 10GHz &
exchange \cite{loss:qdot-comp} \\
\hline
scalable ion trap & 1ms-20s &
100$\mu$s~\cite{metodiev:ion-trap-sim-prelim} to
10msec~\cite{schmidt-kaler03:ion-cnot} & can trade off speed for
  gate fidelity in the range of 14kHz to 100kHz; also limited by ion
  movement times to $\sim 20$kHz & & conditional phase shift \\
\hline
JJ charge 	& a few ns & 10ns & 10GHz & 10GHz & conditional phase shift
\\
\hline
Kane model 	& long? & long & 75kHz & 75kHz & $J$ coupling \\
\hline
LOQC	 	& limited by scattering and absorption & 5-10ns &
$<1$ns & limited by detector time & several possibilities, including conditional phase shift \\
\hline
optical lattice & seconds? & N/A & 160kHz & 5kHz & conditional phase
shift \\
\hline
\end{tabular}
}
\caption{Clock speed and gate characteristics}
\label{tab:qubit-gates}
\end{table}

\begin{table}
\centerline{
\begin{tabular}{|p{1in}|p{0.9in}|p{1.1in}|p{2in}|} \hline
technology & logical: elementary encoding & gate-level timing control &
scalability limit \\
\hline\hline
Si NMR 	& 1:1 & slow gates make precise timing feasible &
quality of initialization (no more than
$1/n$ copies may be mis-polarized for large $n$, to achieve adequate
SNR), precision of placement in static magnetic field, area of
high-quality magnetic field \\
\hline
solution NMR 	& 1:1 & slow gates make precise timing
feasible & SNR falls exponentially in $n$ \\
\hline
quantum dot charge & 1:3 & gates must be precise, but jitter is not a problem & external wiring/control \\
\hline
scalable ion trap & 1:1 & recommends use of decoherence-free subspace to reduce jitter & 
probably ability to accurately track large numbers of individual ions, and their movement times
\\
\hline
JJ charge 	& 1:1 & active control of phases &
cross-qubit interference; inductance of
Josephson junctions; large numbers of rack-mount microwave
generators and getting wires into the dilution refrigerator
\\
\hline
Kane model 	& 1:1 & & manufacturing complexity \\
\hline
LOQC 	& 1:1 but many auxiliary photons used & ``stopped'' light~\cite{fleischhauer00:stopped-light} & 
skew and jitter in both input
generation and gates; single-photon photodetector efficiencies of
$\sim 0.9$ will scale poorly when used for large numbers of
independent qubits; deep circuits subject to loss
\\
\hline
optical lattice & 1:1 & slow gates make precise timing feasible &
region of high-quality lattice tens of sites per side?\\
\hline
\end{tabular}
}
\caption{Other Features}
\label{tab:qubit-misc}
\end{table}

\begin{table}
\begin{tabular}{|p{1in}|p{1.3in}|p{1.5in}|p{1.6in}|} \hline
technology & fabrication & operating environment & supporting equipment \\ 
\hline\hline
Si NMR 	& Si micromachining & 4 K, 7 T magnetic field & r.f. signal generator \\\hline
solution NMR 	& chemical & room temperature, 11 T magnetic field & r.f. signal generator \\\hline
1-D quantum dot charge 	& GaAs lithography & 20 mK & GHz voltage pulse generator (per qubit?) \\\hline
scalable ion trap 	& macroscopic electromechanical assembly &
supercooled ions in room temperature vacuum & multiple lasers (gates and measurement), electronic signal generators (ion movement control), CCD cameras (state detection) \\\hline
JJ charge 	& Si lithography & 30 mK & GHz voltage pulse generator (per qubit?) \\\hline
Kane model 	& P implanted in Si lithography & 1.5 K, 2 T magnetic field &  \\\hline
LOQC	 	& macroscopic electromechanical assembly & dependent
on optical detectors; liquid helium to room temperature & high speed
optical switches, atomic clocks \\
\hline
optical lattice & vacuum chamber, lasers, macroscopic
electromechanical assembly & ultracold atoms in room temp. vacuum &
multiple lasers\\
\hline
\end{tabular}
\caption[Manufacturing and operating environment]{Manufacturing and
operating environment.  K, degrees Kelvin; mK, millikelvin.}
\label{tab:manuf}
\end{table}


\subsection{Solution NMR}

Probably the most complete demonstrations of quantum computation to
date are the solution NMR
experiments~\cite{vandersypen:shor-experiment,boulant03:_exper_nmr,knill00:_algo-bench}.
In an NMR system, the qubit is represented by the spin of the nucleus
of an atom.  When placed in a magnetic field, that spin precesses, and
the spin can be manipulated via microwave radiation.  In solution NMR,
a carefully-designed molecule is used.  Some of the atoms in the
molecule have nuclear spins, and the frequency of radiation to which
they are susceptible varies depending on their position in the
molecule, so that different qubits are addressed by frequency.  In
some cases, isotopic composition must be carefully controlled.  Many
copies of the molecule are held in a liquid solution; each molecule is
a separate quantum computer, run independently, with the large numbers
providing adequate signal strength for readout.  This is the canonical
ensemble system.  Solution NMR has been used to factor the number 15
using Shor's algorithm, which required 720
milliseconds~\cite{vandersypen:shor-experiment}.  The largest
demonstration to date is 12
qubits~\cite{negrevergne06:_12qubit-bench}.

No special cooling apparatus is required for this ensemble system.
However, its scalability is believed to be quite limited due to
falling signal/noise ratio as the number of qubits increases.

\begin{itemize}
\addtolength{\itemsep}{-2mm}
\item {\bf strengths:} good decoherence time, room temperature
  operation, advanced experimental verification
\item {\bf weaknesses:} slow gates, poor scalability, difficult
  concurrent operations
\end{itemize}

\subsection{Josephson Junction}
\label{sec:jj}

\begin{figure}
\includegraphics[width=16cm]{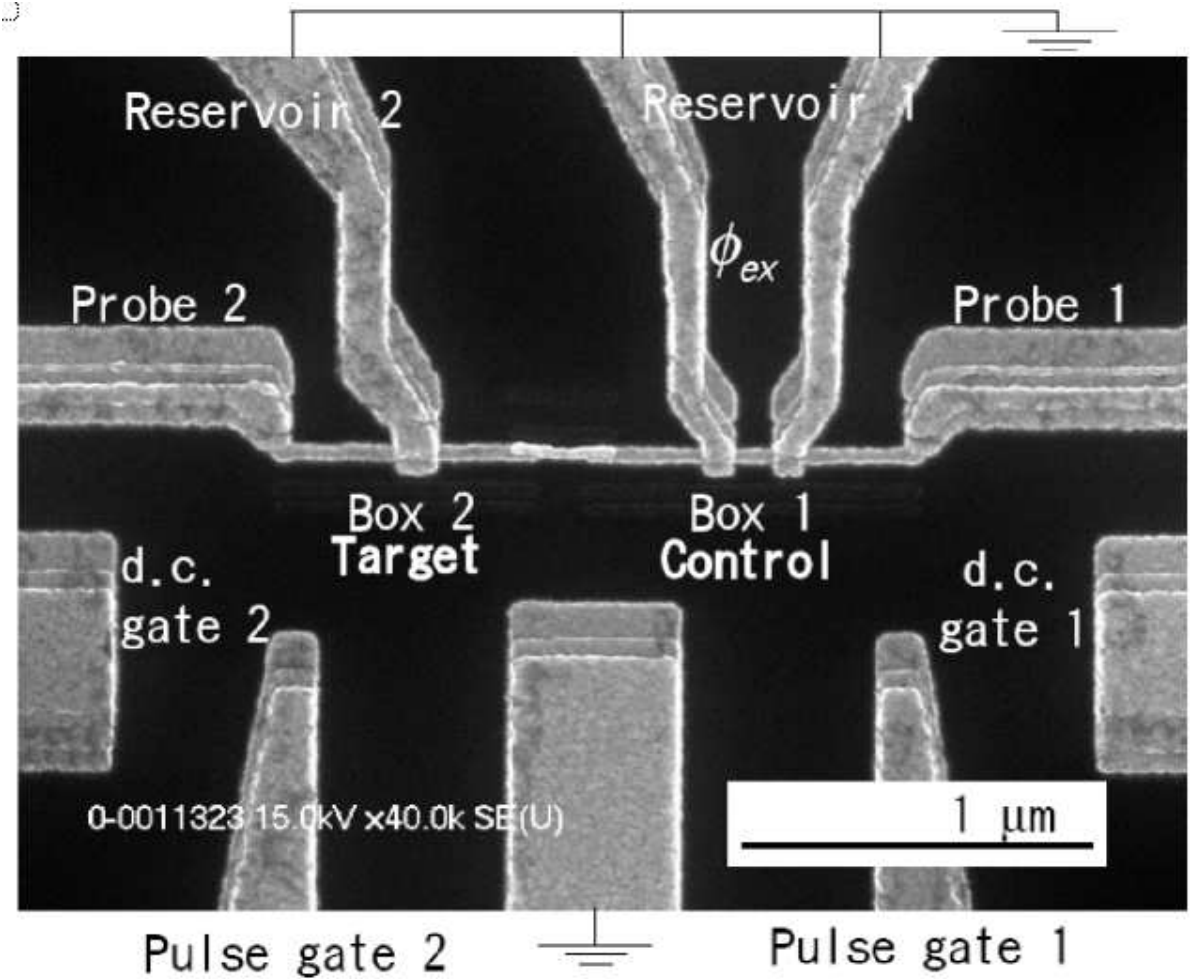}
\caption[A pair of coupled Josephson-junction charge qubits]{A pair of
  coupled Josephson-junction charge qubits (labeled Box 1 and Box 2).
  This device is designed to execute a two-qubit gate between the
  qubit labeled ``Control'' and the one labeled ``Target''.  The
  coupling between the two qubits is fixed in hardware in this device.
  Image courtesy of Y. Nakamura and T. Yamamoto, NEC.}
\label{fig:nec-charge}
\end{figure}

Josephson junction-based (JJ) quantum computing devices are
superconducting systems~\cite{shnirman97:_jj_qubit}.  They come in
four flavors: those that represent qubits using charge (such as the
device shown in
figure~\ref{fig:nec-charge})~\cite{nakamura99:_coher_cooper,pashkin:oscill},
those that use
flux~\cite{mooij:jj-qubit,chiorescu04:_coher,plourde05:_flux-qubits},
those that use
phase~\cite{yu:temp-oscil-phase,martinis02:_rabi_oscil_phase}, and a
recently-designed high-temperature form~\cite{bauch06:_d-wave-jj};
most of the information in the tables applies to all but the latter.
Fabrication is done using conventional electron-beam lithography and
shadow evaporation of Al onto an SiN$_x$ insulating substrate.  In the
JJ charge qubit, a sub-micron size superconducting box (essentially, a
small capacitor) is coupled to a larger superconducting reservoir.  In
a superconductor, electrons move in pairs known as Cooper pairs.  The
qubit representation is the number of Cooper pairs in the box,
controlled to be either zero or one, or a superposition of both.
Similarly, for the flux qubit, Cooper pairs are introduced into a
superconducting ring, where they circulate and induce a quantized
magnetic flux.  Because the flux qubit has slower gate times but a
relatively even longer coherence time, experimental efforts appear to
be shifting toward the flux qubit approach.

Josephson junction technologies can couple qubits in a variety of
ways~\cite{bertet06:_param,brink05:_mediat,cosmelli04:_contr_flux_coupl_integ_flux_qubit,niskanen05:_tunab,liu05:_contr-coupling,liu05:_optical_select_rules,plourde04:_entan-flux}.
In one proposed scalable form of the charge qubit, neighboring qubits
are linked in a one-dimensional structure that supports only
nearest-neighbor gates, but concurrent gates on independent qubits may
be allowed~\cite{lantz04:jj-switching}.  In another proposal, it is
possible to address any two qubits and couple them through a shared
inductance~\cite{you:scalable}.  In this case, the restriction of
operations involving only neighboring qubits in a linear array is
removed, but execution is limited to one gate at a time.  Rigetti et
al. have proposed a scheme that borrows ideas from NMR to couple
neighboring qubits of either flux or charge
type~\cite{rigetti05:_jj-protocol}; their proposal has the benefit
that slight differences in fabrication between qubits are a help
rather than a hindrance.

The high-temperature JJ device requires complex fabrication and
careful alignment to crystal lattice axes.  ``High temperature'', in
this case, refers to the materials potentially being superconductors
at liquid nitrogen temperatures, but the experiments described are
conducted at 15mK to minimize other sources of decoherence.

\begin{itemize}
\addtolength{\itemsep}{-2mm}
\item {\bf strengths:} very fast gates, advanced experimental
  demonstration, straightforward fabrication (for all but the
  high-temperature device)
\item {\bf weaknesses:} low coherence time relative to measurement
time, sensitivity to background charge fluctuations and local magnetic
fields
\end{itemize}

\subsection{All-Silicon NMR}

\begin{figure}
\centerline{\hbox{
\input{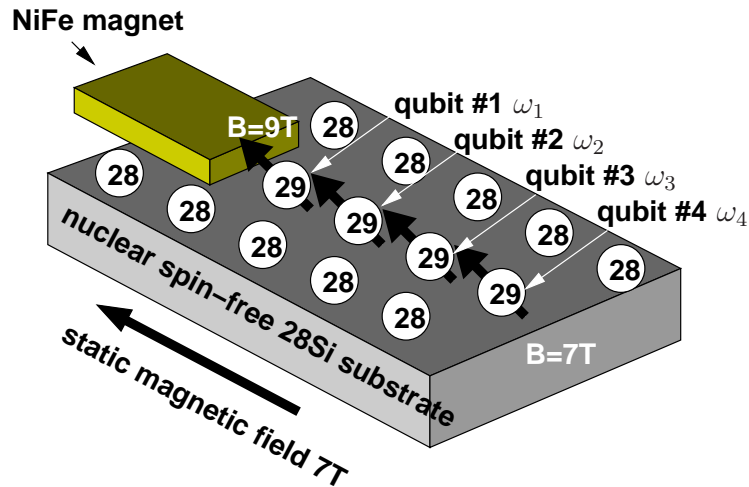}}}
\caption[Schematic of the all-silicon NMR computer]{Schematic of the
  all-silicon NMR computer.  Qubits are the spin of $^{29}$Si nuclei
  on a spin-free base of $^{28}$Si.  Distance from the micromagnet
  determines oscillation frequency $\omega_i$ and provides individual
  qubit addressability.  Image courtesy of K. M. Itoh, Keio
  University.}
\label{fig:all-si}
\end{figure}

Ladd et al. have proposed an all-silicon NMR-based quantum computer
which stores qubits in the nuclear spin of a chain of $^{29}$Si (spin
$1/2$ nucleus) in a substrate of spin 0 nuclei ($^{28}$Si and
$^{30}$Si).  In one form, the $^{29}$Si atoms are laid down in a line
across a micromechanical bridge~\cite{ladd:si-nmr-qc}.  Readout is
done via magnetic resonance force microscopy (MRFM), reading
oscillations of the bridge.  Other measurement schemes for the same
basic architecture are being pursued, as well~\cite{itoh05:_nmr-qc}.
This is an ensemble system; $10^5$ copies are required to get an
adequate signal for measurement.  One form of the system is
illustrated in Figure~\ref{fig:all-si}.  Only one chain of $^{29}$Si
is shown.  Initialization is done via electrons whose spin is set with
polarized light (optical pumping).  Operations are done via microwave
radiation directed at the device.  A micromagnet provides a high field
gradient, allowing individual atoms to be addressed by frequency.  The
device is fabricated via near-atomically precise machining, then
refined by passing electrical current through it in a carefully
controlled
fashion~\cite{sekiguchi05:_self_paral_atomic_wires,yoshida05:_atomic_straight_steps,wang05:_magnet_mesa}.

\begin{itemize}
\addtolength{\itemsep}{-2mm}
\item {\bf strengths:} longest known decoherence time
\item {\bf weaknesses:} slow gates, no QIO, measurement still being
designed, difficult fabrication
\end{itemize}

\subsection{Scalable Ion Trap}

\begin{figure}
\centerline{\hbox{
\includegraphics[width=12cm]{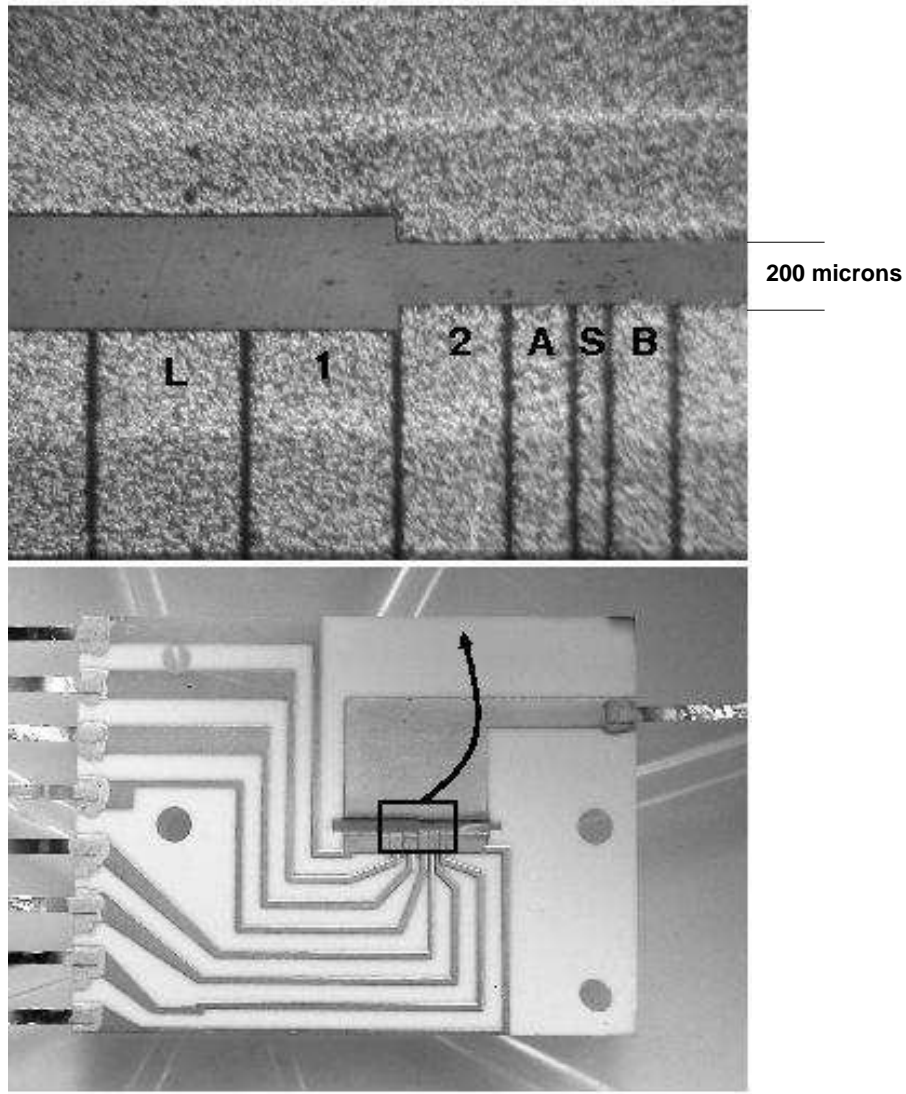}}}
\caption[A six-zone ion trap]{A six-zone ion trap capable of moving
  individual ions.  Ions are inserted in the landing zone L, and
  manipulated in the zones A, S, and B.  Image courtesy of
  D. Wineland, NIST.}
\label{fig:nist-trap}
\end{figure}

One of the few systems which explicitly separates storage areas from
interaction areas is the scalable ion
trap~\cite{kielpinski:large-scale,wineland05:_quant,kim05:_system,metodiev:ion-trap-sim-prelim,steane04:_gop-qc,balensiefer:isca05,thaker06:_cqla,acton:_near_perfect_meas,hensinger06:_t-junction}.
Initially designed and built at NIST, this is a proposal to scale up
an ion trap quantum
computer~\cite{cirac95:_cold_trapped_ions,steane97:ion-trap,soerensen00:_entan,schmidt-kaler03:ion-cnot}.
In ion trap systems, qubits are usually stored in the energy levels of
individual ions.  In early ion trap experiments, small numbers of ions
were held in a single trap known as an RF Paul trap.  In the scalable
trap system, which is a large system of interconnected, individually
controllable traps, the ions are kept suspended in a vacuum in a
channel in the device and are literally moved around using magnetic
fields until they reach locations in the system designated for
operations, as shown in Figure~\ref{fig:nist-trap}.  Small numbers of
ions are brought together and formed into chains to execute
multi-qubit gates.  Gates are effected by laser pulses; readout is
also accomplished by laser pulses creating fluorescence (interpreted
as a 1) or not (0).  Gate times are moderate, but overall system
performance will likely be driven by ion movement times (which
naturally depend on distance and topology), times for creating and
splitting chains of atoms, time to cool atoms heated by the movement
process, and multiplexing of gate operations.  For both gates and
measurement in scalable ion trap systems, many laser beams must excite
many ions.  Complex optics and photon detectors may be required to
read out the state of many qubits at once; CCD cameras involve a
direct tradeoff of speed versus noise, while avalanche photodiodes are
difficult to integrate and photon counters require cryogenic
operation~\cite{kim05:_system}.

The Monroe group has recently shown the ability to move ions around
corners, a fundamental engineering advance in control of individual
atoms~\cite{hensinger06:_t-junction}.  As noted above, the efficiency
of algorithms implemented on ion traps will depend on realizable
concurrency, and on the time to move and cool ions.

In Table~\ref{tab:qubit-gates}, we list the decoherence time of ions
as a range of 1 millisecond to 20 seconds.  The lifetime of individual
ions has been shown to be in the millisecond range, but H\"affner et
al., in the Blatt group in Austria, recently encoded a state on a pair
of ions using a decoherence free subspace and experimentally measured
a lifetime of 20 seconds~\cite{haeffner05:_robust}.  Other experiments
from both the Blatt group and the Wineland group at NIST have recently
confirmed the existence of entangled groups of 6, 7, and 8 ions,
prompting the coining of the term
``qubyte''~\cite{haeffner05:qubyte,leibfried05:_six-atom-cat}.  While
these accomplishments do not yet surpass the size of the Cory group's
12 qubit NMR system, researchers are excited because ion trap
technology is viewed as a strong candidate for a scalable system.  It
will be interesting to see when it becomes possible to draw a
``Moore's Law'' parallel for the size of an entangled system, graphing
the doubling time of the largest entanglement demonstrated in ion
traps.

\begin{itemize}
\addtolength{\itemsep}{-2mm}
\item {\bf strengths:} scalability of storage
\item {\bf weaknesses:} slow gates~\cite{steane00:_speed}; limitations
on concurrent operations and measurements
\end{itemize}

\subsection{All-Optical}

All-optical systems come in two flavors: those that depend on
non-linear effects to execute gates, and
those in which the only necessary non-linearity is measurement, known
as {\em LOQC} (linear optics quantum computation)~\cite{knill01:klm}.
Research on all-optical systems has focused on photon sources capable
of generating precise numbers of photons with the necessary timing
precision~\cite{santori02:_indist-photon-src}, gates based on
measurement~\cite{knill01:klm,scheel03:_meas-nonlin,knill03:_prob-bound,browne05:_cluster-resource,yoran03:_deter_linear_optical-qc},
and high-quality single-photon
detectors~\cite{miller03:_infra-detector,waks03:_high-eff-detector,james02:atomic-vapor-detector}.

Measurement-based gates are inherently probabilistic in nature, though
it has been shown that these gates can be built into a scalable
feed-forward
network~\cite{knill01:klm,ralph05:_loss_toler_optic_qubit}.  Much of
the current experimental work is focusing on this approach, and
individual gates have been shown to
work~\cite{pittman02:_demon,o-brien:optical-cnot,pittman03:_exper_not,gasparoni04:_realiz_not,sanaka04:_exper_nonlin}.

Jitter and skew are likely to be managed by ``stopped light'', created by
electromagnetically induced
transparency~\cite{fleischhauer00:stopped-light,harris97:_elect_induc_trans},
which has also recently been shown to be useful for creating and
managing single photons both directly~\cite{eisaman05:_eit-single} and
in combination with other techniques~\cite{chaneliere05:_storage}.

\begin{itemize}
\addtolength{\itemsep}{-2mm}
\item {\bf strengths:} well-understood physics and easy fabrication
\item {\bf weaknesses:} photon losses; for non-linear systems, weak
non-linear effects give poor gate quality; high resource requirements
for probabilistic gates; large physical size of systems
\end{itemize}

\subsection{Quantum Dot}

A ``quantum dot'', as used in quantum information processing, is a
lithographically-defined structure that confines electrons at the
boundary layer between two materials, creating a two-dimensional
electron gas (2DEG).  By varying the surrounding electrical potential,
individual electrons can be confined to a small area, called the
quantum dot.  A qubit can be defined based on the number of electrons
in a quantum dot or the spin or energy levels of a single electron
held in a quantum dot.

Several quantum dot devices are under development; one experimentally
advanced approach uses a pair of quantum dots as a dual-rail encoded
logical qubit, with a single electron in the left dot representing a
logical 0, and the electron in the right dot representing a logical
1~\cite{fujisawa98:double-dot,taylor05:_ft-qdot-arch}.  Another
approach uses a linear array of single-electron quantum dots, and
encodes the qubit in the spin of the excess
electron~\cite{loss:qdot-comp}.

In a third approach, DiVincenzo et al. proposed that the only
operation needed is an exchange between two neighboring qubits,
accomplished by lowering the electrical potential and allowing the
electrons to
tunnel~\cite{divincenzo:universal-exchange,loss:qdot-comp,myrgren:exchange-only-feas}.
This is easier to accomplish than precise control of a magnetic field,
which would be required in order to effect gates on specifically
addressable bits.  Perhaps the biggest drawback of this approach is
that exchange-only computation requires encoding a single logical
qubit onto multiple physical qubits.  A \textsc{CNOT}, for example,
requires each logical qubit to be encoded in three physical qubits,
and the exchange times must be controlled fairly precisely.  The
\textsc{CNOT} on neighboring logical qubits requires 19 exchange
operations~\cite{divincenzo:universal-exchange}, though Myrgren and
Whaley have found interesting optimizations that allow non-neighbor
operations to be effected in 28\% fewer total operations than the
obvious formulation of repeated use of the 19-exchange
\textsc{CNOT}~\cite{myrgren:exchange-only-feas}.  Continued compiler
work may reduce the encoded execution time penalty further, though the
important storage penalty remains.

\begin{itemize}
\addtolength{\itemsep}{-2mm}
\item {\bf strengths:} advanced fabrication
\item {\bf weaknesses:} low coherence time
\end{itemize}

\subsection{Kane Solid-State NMR}

Kane has proposed a solid-state NMR system with excellent scalability,
built on VLSI techniques for control~\cite{kane:nature-si-qc}, and
Clark et al. have made progress in
fabrication~\cite{clark03:kane-progress}.  In this system, individual
phosphorus atoms are embedded in a silicon substrate, and standard
photolithography techniques are used to build control structures on
the surface.  The qubit is held in the spin of the phosphorus nucleus,
and interactions between neighboring qubits are mediated by electrons
coupled to the nuclei via hyperfine interactions.  The shape of the
electron wave function is controlled via the control structures built
on the Si surface; the distance between neighboring P atoms and the
accuracy of aligning the control gates to the P impurities will
determine the quality of qubit interactions.  Some Si isotopes have a
nuclear spin; the presence of atoms of these isotopes could
potentially disrupt the operation of the Kane structure.  Abe et
al. have studied the behavior of such a system as the isotopic
composition of the Si substrate is
varied~\cite{abe04:_elect,abe06:_thesis}.  Oskin, Copsey et al.  have
performed engineering studies, suggesting that teleportation may be
required to move qubits long distances even for error correction, and
that matching the pitch of the necessary lithographically-created
control structures to the desirable atomic spacing is
difficult~\cite{oskin:quantum-wires,copsey:q-com-cost}.

\begin{itemize}
\addtolength{\itemsep}{-2mm}
\item {\bf strengths:} long coherence time
\item {\bf weaknesses:} difficult fabrication, creating adequate
  overlap in electron wave functions
\end{itemize}

\subsection{Optical Lattice}

In an optical lattice, qubits are the internal states of individual
atoms~\cite{jaksch99:_cold_collisions,brennen99:_optical_lattice,treutlein06:_qip_optic_lattic_magnet_microt}.
The optical lattice itself is a set of standing waves of light,
creating magnetic fields that hold individual atoms in place in an
array, suspended in a vacuum.  Two-qubit gates are executed by
adjusting the positions of the peaks and troughs of the light waves so
that neighboring atoms collide.  This basic approach is similar to
trapping of ions, but since the atoms are neutral rather than charged,
they do not interact with the environment as strongly, and hence have
the potential to have much longer lifetimes.  The lifetime of a
Bose-Einstein condensate (a coherent quantum state rather different
from qubits) has been measured in seconds in a
lattice~\cite{greiner01:_be_coherence}.  The lattice may work well in
multiple dimensions.  The principal drawbacks to this approach are
that individual addressing and readout of atoms have not been shown.
Each pair of atoms in the lattice acts exactly the same, and the
spacing between the atoms is too small for optical resolution for
fluorescent readout.  The ``atom chip'' approach uses similar physics
for the qubits and gates, but is a dramatically different engineering
approach, using lithographically created structures to move individual
atoms at will, something like the scalable ion
trap~\cite{folman00:_atom-chip,treutlein06:_qip_optic_lattic_magnet_microt,kishimoto06:_atom-chip}. 

\begin{itemize}
\addtolength{\itemsep}{-2mm}
\item {\bf strengths:} long coherence time, easy fabrication
\item {\bf weaknesses:} no individual addressability for gates or readout
\end{itemize}

\section{Summary}

DiVincenzo laid down the defining characteristics of a viable quantum
computing technology~\cite{divincenzo95:qc}.  Many engineering factors
extend beyond the DiVincenzo criteria to determine how practical it is
to build a machine based on a given
technology~\cite{van-meter:qarch-impli}.  These factors include such
basic issues as possible measurement schemes, the difficulty of
building and operating large-capacity devices, and several issues
affecting performance, notably clock speed, the qubit-to-qubit layout
topology and possible concurrent operation.  For our purposes, some
quantum I/O mechanism is necessary; without one, we cannot build a
quantum multicomputer, and the system's scalability with respect to
number of qubits (and possibly concurrent operation) will be quite
limited.  In the next chapter, we will develop the {\em qubus}
mechanism and accompanying {\em teleportation} techniques that we will
use to connect quantum computers together.

This chapter organizes information about quantum computers in a way
that specifically focuses on scalability, implementability, and
architectural implications.  The evaluation criteria we have laid out
should make it possible to compare technologies and determine which
will be useful in different roles of a system, and how application
algorithms can be mapped to and compiled for various architectures.

Each of the technologies discussed here has its own particular set of
technological hurdles to overcome before it can be considered
practical.  NMR-based systems have slow gate times, but have good
coherence times; if a QIO mechanism can be
designed~\cite{wallraff04:_strong-coupling}, they will make excellent
storage devices, but pure NMR systems are unlikely to make adequate
factoring machines.  Josephson-junction devices and quantum dots have
extremely fast gate times, but have poor coherence times.  Both of
these systems have yet to demonstrate scalability in implementation
and addressing of qubits, though both have been designed.  Pure
optical systems need more efficient single-photon detectors.  Ion
traps have many desirable features that make them scalable
architecturally.

The complex tradeoffs in controlling a quantum computer include
trading speed for coherence time.  The quantum wiring and classical
control are under investigation in both technology-dependent and
-independent fashions, but many scaling questions remain.  Work on
both programming language design to support quantum computation and
back-end optimization for specific architectural characteristics has
just
begun~\cite{oemer:qcl,aho:palindromes,nakajima-2005,kawano04:compiler}.
The mapping of algorithms to these architectures will determine the
performance and practicality of particular architectures.

  \chapter{Networking}
\label{ch:travel}

\cq{True and serious traveling is no pastime, but is as serious as the
grave.}{Henry David Thoreau}

Our quantum multicomputer will require a quantum network, as
illustrated in Figure~\ref{fig:multicomp-hw-block} on
page~\pageref{fig:multicomp-hw-block}.  The physical layer of the
network must be quantum, of course, but the techniques for describing
and understanding classical networks can be applied easily to quantum
networks.  In this chapter, we take a quick look at the qubus physical
layer for creating entangled pairs, and the classical ways of
describing network topologies and their performance.

\section[Qubus Entanglement Protocols]{Weak Nonlinearity and Qubus
  Entanglement\\ Protocols}
\label{sec:qubus}


{\em EPR pairs}, or Einstein-Podolsky-Rosen pairs, are pairs of
particles or qubits which are entangled so that actions on one affect
the state of the other, such as the state
$(|00\rangle+|11\rangle)/\sqrt{2}$ (which can also be called a Bell
pair).  EPR pairs can be created in a variety of ways, including
reactions that simultaneously emit pairs of photons whose
characteristics are related and many quantum gates on two qubits.  For
an ion trap system, for example, two ions can be moved together, an
entangling operation performed, and the ions separated.  As long as
the quantum state remains coherent, the ions can be separated by any
physical distance and their state will remain related.  In the next
section, we will see how to use EPR pairs both to move data and to
execute gates remotely, via a process known as quantum teleportation.
In this section, we present our mechanism for making the EPR pairs.
Technically, an EPR pair is a maximally entangled pair; that is,
operations on one qubit have the strongest possible influence on the
other.  In this thesis, we use the term somewhat more loosely,
including pairs whose entanglement has decayed somewhat from the
maximum, or whose entangling operations failed to produce a perfect
pair.

Our approach to creating EPR pairs contains no direct qubit-qubit
interactions and does not require the use of single photons, as
e.g. Kimble's team has recently demonstrated~\cite{chou05:_measur}.
We use the invention of Munro, Nemoto and Spiller, which uses laser or
microwave pulses as a {\em probe
beam}~\cite{nemoto04:_nearly_deter,munro05:_weak}.  Two qubits are
entangled indirectly through the interaction of qubits with a common
quantum field mode created by the probe beam -- a continuous quantum
variable -- which can be thought of as a communication bus, or
``qubus''~\cite{spiller05:_qubus}.  We call this process the qubus
entanglement protocol (QEP).

\begin{figure}
\centerline{\hbox{
\includegraphics{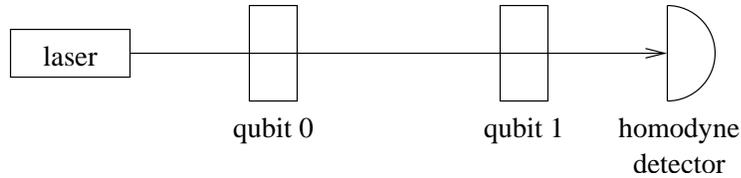}}}
\caption{Physical configuration of a qubus.}
\label{fig:qep-phy}
\end{figure}

Physically, the qubus consists of a laser or microwave source, a pair
of qubits and some means of interacting them with the probe beam, and
a {\em homodyne detector}~\cite{armen02:_adaptive-homodyne}, as shown
in Figure~\ref{fig:qep-phy}.  The distance between the qubits can be
arbitrarily large, limited only by losses in the probe beam.  The
probe beam consists of a large number of photons, each of which
interacts minutely with the qubits.  If the qubits are single photons,
this is accomplished using a type of crystal with a property known as
a {\em cross-Kerr nonlinearity}.

\begin{figure}
\centerline{\hbox{
\input{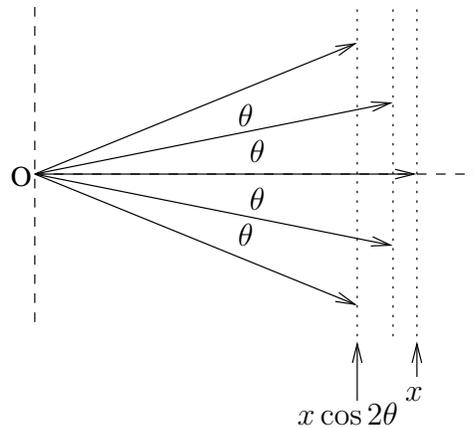}}}
\caption{Phase space diagram of the qubus entanglement protocol.}
\label{fig:qep-phase}
\end{figure}

For some solid state qubit systems, we can put the qubits in a
microwave resonant cavity and use a microwave pulse to create the
qubus effect.  The interaction with a bus mode takes the effective
form of a cross-Kerr nonlinearity, analogous to that for optical
systems, described by an interaction Hamiltonian of the form
\begin{equation}
\label{Hintck}
H_{int} = \hbar \chi \sigma_z a^{\dagger} a.
\end{equation}
In this equation, $a^{\dagger}$ and $a$ are, respectively, the
creation and annihilation operators, representing the raising or
lowering of the number of photons present in the probe beam.  When
acting for a time $t$ on a qubit-bus system where the nonlinear
interaction is of strength $\chi$, this interaction causes a rotation
in phase space by an angle $\pm \theta$ on a bus coherent state, where
$\theta = \chi t$ and the sign depends on the qubit computational
basis amplitude.  In a phase space diagram, the horizontal and
vertical axes correspond to the quadrature amplitudes of two
variables.  They are commonly referred to as position ($x$) and
momentum ($p$), respectively, due to mathematical similarities in
their behavior, but they do not physically represent these quantities.
The diagram for this interaction is shown in
Figure~\ref{fig:qep-phase}.  By interacting the probe beam with the
qubit, the probe beam picks up a $\theta$ phase shift if it is in one
basis state (e.g., $|0\rangle$) and a $-\theta$ phase shift if it is
in the other (e.g., $|1\rangle$).  If the same probe beam interacts
with two qubits, it is straightforward to see that the probe beam
acting on the two-qubit states $|0\rangle |1\rangle$ and $|1\rangle
|0\rangle$ picks up no net phase shift because the opposite-sign
shifts cancel, while the probe beam acting on the states $|0\rangle
|0\rangle$ and $|1\rangle |1\rangle$ picks up phase shift $\pm
2\theta$.

\begin{figure}
\centerline{\hbox{
\input{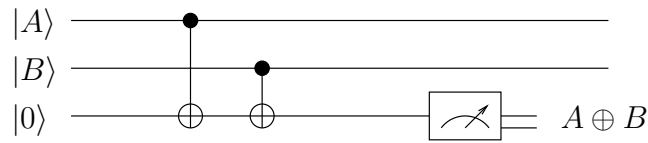}}}
\caption{Logical equivalent of the qubus entanglement protocol.}
\label{fig:qep-logical}
\end{figure}

The homodyne measurement projects the point in phase space onto the
$x$ axis (position).  This projection determines whether the probe
beam has been phase shifted (in effect taking the absolute value of
the angular shift), projecting the qubits into either an even parity
state or an odd parity state.  The measurement shows only the parity
of the qubits, not the actual values, leaving them in an entangled
state.  If the homodyne measurement returns $x\cos 2\theta$, we know
that the state is either $|00\rangle$ or $|11\rangle$.  If the
measurement returns $x$, we know that the state is either $|01\rangle$
or $|10\rangle$.  In the latter case, we can apply a NOT gate to
either qubit, moving the state into $|00\rangle$ or $|11\rangle$.
Figure~\ref{fig:qep-logical} shows a circuit that is logically similar
to QEP, differing only in its possible error propagation
characteristics, which we will not detail.

Although the qubus is physically asymmetric, with a probe beam source
and homodyne detector at opposite ends of the physical layout and a
definite ordering of qubits along the bus, this layout does not
influence the logic of the qubus.  The qubus is used to create EPR
pairs, which are symmetric.  Each teleportation operation, as we will
see in the next section, consumes one EPR pair to send a qubit from
node to node.  We can schedule use of the bus as if it is a {\em
half-duplex} bus.

This procedure is general, and can be applied to any pair of qubits to
determine their parity.  If all of the terms of the superposition have
the same parity, the state of the superposition is not affected by the
parity measurement, beyond a small phase change which can be corrected
with single-qubit gates.  If we start with both qubits in the state
$(|0\rangle+|1\rangle)/\sqrt{2}$, we are left with the state
$(|00\rangle+|11\rangle)/\sqrt{2}$, which is a good state for
beginning the teleportation protocols described in the next section.

\section{Teleportation}
\label{sec:teleportation}

Teleportation, discovered by Bennett and his collaborators, transfers
the state of one quantum to another by using EPR pairs.  Teleportation
of quantum states has been known for more than a
decade~\cite{bennett:teleportation}.  It has been demonstrated
experimentally~\cite{furusawa98,bouwmeester:exp-teleport}, and has
been suggested as being necessary for moving data long distances
within a single quantum
computer~\cite{oskin:quantum-wires,metodi05:qla}.  Teleportation can
also form part of the process of transferring quantum state from one
physical representation to another.

For our quantum multicomputer, we propose using the qubus entanglement
protocol (QEP), described in the last section.  Entanglement is a
continuous, not discrete, phenomenon, and several weakly entangled
pairs can be used to make one strongly entangled pair using a process
known as {\em
purification}~\cite{bennett95:_concen,cirac97:_distr_quant_comput_noisy_chann}.
Purification starts with EPR pairs in a known (but possibly degraded)
state, then essentially performs an error correction protocol that is
specific to that state.  This is more efficient than full-fledged
quantum error correction.

\comment{Thaddeus says D\"ur, Briegel, PRA 59, 169 (1999), Quantum
Repeaters Based on Entanglement Purification, is the right reference
for purification.  Isailovic also cites Bennett, PRL 76:722, 1996;
Deutsch, PRL 77:2818, 1996 as their two teleportation protocols.
Isailovic says purification only at the end of a series of hops is the
best, which seems to be in contrast to what others have said...}

\subsection{Teleporting Data}
\label{sec:teledata}

\begin{figure}
\centerline{\hbox{
\input{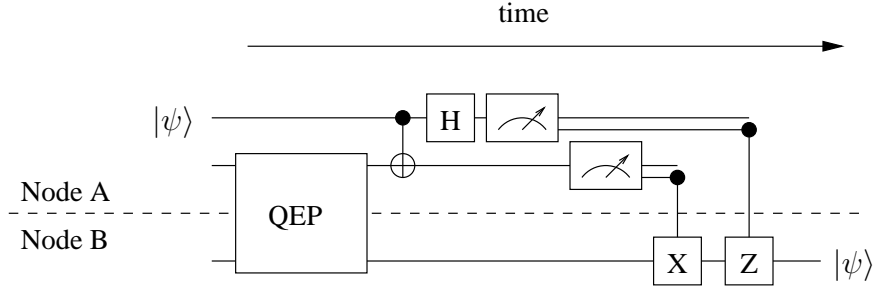}}}
\caption{Teleporting a single qubit.}
\label{fig:teleport}
\end{figure}

Figure~\ref{fig:teleport} shows the basic teleportation circuit to
move a single qubit from one location to another.  The box labeled QEP
is the qubus entanglement protocol; the output of the box is the EPR
pair.  The near and far ends of the teleportation each hold one member
of the entangled pair.  To teleport the qubit $|\psi\rangle =
\alpha|0\rangle+\beta|1\rangle$, the first step is to perform a CNOT
at the source between the qubit and the source-side EPR member,
causing the change
\begin{equation}
|\psi\rangle\frac{|00\rangle+|11\rangle}{\sqrt{2}}\rightarrow
\frac{\alpha}{\sqrt{2}}|000\rangle+\frac{\alpha}{\sqrt{2}}|011\rangle+
\frac{\beta}{\sqrt{2}}|110\rangle+\frac{\beta}{\sqrt{2}}|101\rangle
\end{equation}
where the qubits in our written representation correspond top to
bottom to the qubits in the figure.  That is, the left-most qubit in
our notation is the original qubit, the middle one is the source-side
EPR pair member, and the right-most qubit is the member of the EPR
pair at the destination.  We then apply a Hadamard gate to the
original qubit, moving to the state
\begin{multline}
\frac{\alpha}{2}|000\rangle+\frac{\alpha}{2}|100\rangle+
\frac{\alpha}{2}|011\rangle+\frac{\alpha}{2}|111\rangle
-\frac{\beta}{2}|101\rangle-\frac{\beta}{2}|110\rangle+
\frac{\beta}{2}|001\rangle+\frac{\beta}{2}|010\rangle \\
= \frac{1}{2}(|00\rangle(\alpha|0\rangle+\beta|1\rangle)+
|01\rangle(\beta|0\rangle+\alpha|1\rangle)+
|10\rangle(\alpha|0\rangle-\beta|1\rangle)+
|11\rangle(-\beta|0\rangle+\alpha|1\rangle)).
\end{multline}
The last representation makes it clear that the destination qubit now
has some relationship to the state of the original qubit.  In the
first term, if the first two qubits are zero, then the last qubit
holds the state of our original qubit,
$\alpha|0\rangle+\beta|1\rangle$.  In the other three terms, the state
of the last qubit is a simple permutation of the original qubit, which
can be recovered via an $X$ gate, a $Z$ gate, or both.  The four terms
correspond to the states 00, 01, 10, and 11 in the first two qubits.
Thus, if we force the state of the system into one of those four
states, we can determine which gates to apply to ``fix'' the
destination qubit, so that it ends in the starting state of the qubit
we wanted to send, $|\psi\rangle$.

In the figure, this is shown by the measurements, followed by
``control'' $X$ and $Z$ gates.  Of course, the outcomes of the
measurements are classical bits, so our control, in this case, is a
classical choice to apply an $X$ gate or not, depending on the
measured bit.  After the measurements but before the control gates,
the original qubit and the source-side EPR pair member have both been
``destroyed'' (the physical carriers of the qubits likely still exist,
but we no longer have a useful quantum state, as the superposition has
collapsed).

As an example, assume that the node A bits are measured, and produce
the value 11.  This value is then transmitted via classical means to
node B. At node B, we now know that the state of the destination qubit
is $-\beta|0\rangle+\alpha|1\rangle$.  We apply both $X$ and $Z$
gates, and the state shifts to $\alpha|0\rangle+\beta|1\rangle$,
recovering the original qubit $|\psi\rangle$ at the destination.

The ``spooky action at a distance'' of entangled pairs of particles
was one of Einstein's concerns about quantum mechanics, especially
because it appears to violate relativity.  Part of the answer to his
concern is that {\em information} cannot travel faster than the speed
of light.  Thus, although the state of the qubit at the destination
may change ``instantaneously'' as we perform the measurements at the
source, the state of the qubit remains in the indeterminate state
until we receive the classical, relativity-limited information telling
us which gates to apply to recover the pure state we are teleporting.

\subsection{Teleporting Gates}
\label{sec:telegate}

\begin{figure}
\centerline{\hbox{
\input{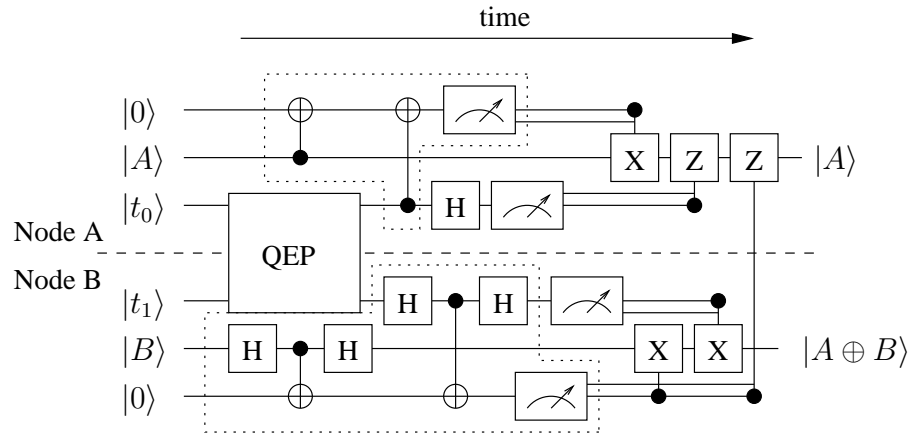}}}
\caption{A teleported control-NOT (CNOT) gate.}
\label{fig:telegate}
\end{figure}

So far, we have discussed the teleportation of data.  It is also
possible to teleport gates.  Gottesman and Chuang showed that
teleportation can be used to construct a control-NOT (CNOT)
gate~\cite{gottsman99:universal_teleport}.  Their original teleported
gate requires two EPR pairs.  We use an approach based on parity gates
that consumes only one EPR pair, as shown in
figure~\ref{fig:telegate}~\cite{munro05:_weak}.  Locally, the parity
gates can be implemented with two CNOT gates and a measurement
(outlined with dotted lines in the figure).  Double lines are
classical values that are the output of the measurements; when used as
a control line, we decide classically whether or not to execute the
quantum gate, based on the measurement value.  The last gate involves
classical communication of the measurement result between nodes.  As
shown, this construction is not fault tolerant; it must be built over
fault-tolerant gates.  Alternatively, the qubus approach can be used
as the node-internal interconnect.  Its natural gate is the parity
gate, and is fault tolerant; this is the approach we will use when we
come to distributed computing in Chapter~\ref{ch:qmc}.

\section{Multicomputer Networks}
\label{sec:multi-net}

The theme of this dissertation is the design of a quantum
multicomputer, a collection of smaller quantum computers connected via
a message-passing network so as to collaborate to solve a single
problem~\cite{athas:multicomputer}.  A multicomputer is a
distributed-memory multiprocessor, in which processing units run
programs independently, and cannot directly access the memory of other
processing units.  All shared computation is accomplished by
exchanging messages through an interconnection network.  In this
section, we take a very brief, technology-independent look at the
interconnection networks that turn a group of individual computers
into a multicomputer.  In Chapter~\ref{ch:qmc}, we will apply these
principles to our quantum system, designing an interconnect network to
create EPR pairs.

Networks consist of {\em nodes} and {\em links}.  A node is a
computational element, where data is stored and manipulated.  A link
transfers messages from one node to another.  A link may be serial,
with one data line, or parallel, with several.  A serial link requires
only a single {\em transceiver}, whereas a parallel link requires one
per wire, or the {\em bus width}.  The current trend in local-area
networks and peripheral buses (such as Fibre Channel, USB, and serial
ATA) is serial links, which allow tighter packaging, lower power
requirements, simpler cabling, etc.  The savings in those areas offset
the cost of a single higher-speed transceiver, generally meaning that
serial networks wind up being roughly as fast as the parallel ones
they replace.

For multicomputer
networks~\cite{hennessy-patterson:arch-quant3ed,hwang:capp,dally04:_interconnects},
as with all networks, we have most of what we need to know about the
topology when we know four characteristics:

\begin{itemize}
\item{\bf degree}  The number of links from each node.
\item{\bf diameter} The maximum distance across the network, measured
  in hops.
\item{\bf average distance} The average distance between any two nodes.
\item{\bf bisection} The minimum number of links you must cut to chop the
  machine in half.
\end{itemize}

This assumes, generally, a regular network, though the same principles
apply for arbitrary topologies.  For a link, we also need to know the
link latency, bandwidth, and protocol and processing overhead; we will
mostly ignore those issues and express our results in units of a
single transfer, or EPR pair creation.  We also include aggregate
system bandwidth in our analysis.

These characteristics give us some guidelines and hint at the
generality (or lack thereof) of a particular network.  What ultimately
matters, of course, is how long it takes to execute the application
algorithm(s) that comprise our workload.  In most cases, this is a
function of both the network topology and the message-passing pattern
of the algorithm.  ``Incast'' problems (two nodes trying to send to
the same destination at the same time) inevitably cause contention
(competition for access to resources); we will see some of the effects
of contention in Section~\ref{sec:dist-shor}.

\begin{table*}
\begin{tabular}{|l|c|c|c|c|c|}\hline
Topology & degree & diameter & avg. dist. & bisection & tot. BW (links) \\
\hline
Bus & 1 & 1 & 1 & 1 & 1 \\
Line & 2 & $N-1$ & $N/2$ & 1 & $N-1$ \\
2D Mesh & 4 & $2(\sqrt{N}-1)$ & $2\sqrt{N}/3$ & $\sqrt{N}$ &
$2N-2\sqrt{N}$ \\
Hypercube(2-cube) & $\log_2 N$ & $\log_2 N$ & $(\log_2 N)/2$ & $N/2$ &
$(N\log_2 N)/2$ \\
Fully Connected & $N-1$ & 1 & 1 & $N^2/4$ & $N(N-1)/2$ \\
\hline
\end{tabular}
\caption[Some common interconnect topologies.]{Some common
interconnect topologies.  $N$, number of total nodes.}
\label{tab:topos}
\end{table*}

\begin{figure}
\centerline{\hbox{
\includegraphics{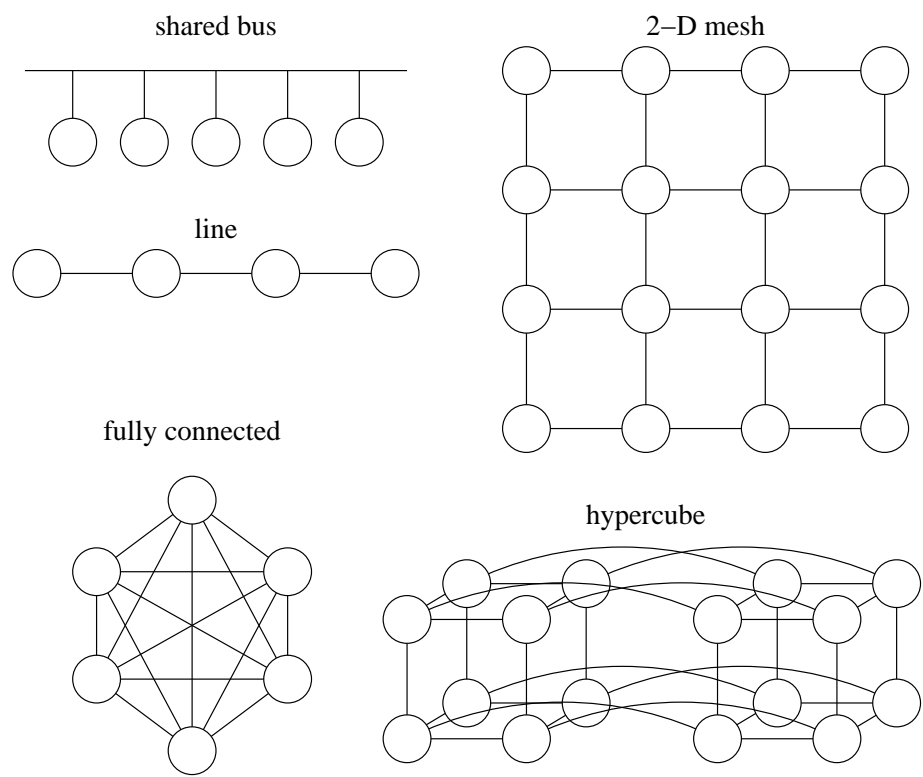}}}
\caption{Five important interconnect network topologies.}
\label{fig:c-topos}
\end{figure}

Table~\ref{tab:topos} and Figure~\ref{fig:c-topos} show five
topologies.  The {\em bus} is a single, shared medium on which any
node can send a message directly to any other node, but only one pair
can be communicating at a time; this configuration roughly corresponds
to the original Ethernet scheme and most computer buses.  In a {\em
line} configuration, each node has a neighbor to the left and a
neighbor to the right, and can exchange messages with both of them
simultaneously.  In a {\em 2D mesh}, each node has four neighbors,
north, east, south and west, and the nodes are laid out in a
two-dimensional grid; the Intel Touchstone Delta and other large-scale
systems found this topology to be a good choice.  The original Caltech
Cosmic Cube was a hypercube, with each of the 64 nodes connected to
$\log_2 64 = 6$ neighbors.  Scaling this system up is difficult, as
each doubling of the number of nodes requires adding a link to each
one of the existing nodes; packaging constraints quickly become a
problem.  In a fully-connected network, each node can communicate
directly with each other node.  Given that this requires $O(N^2)$
links, it is clearly impractical, but serves as a theoretical upper
bound.

All of these topologies are {\em direct network}, also sometimes
called {\em distributed switch}, topologies, where the hardware to
route messages from location to location resides with the compute
nodes.  It is also possible to use {\em indirect network}, also called
{\em centralized switch} topologies, such as crossbars and fat trees.
In indirect networks, packets must pass through switching nodes in the
middle of the network whose sole purpose is routing packets.  For
reasons that will become clear in later chapters, we ignore indirect
network systems.

The performance of a system depends on several factors besides the
topology.  Although a hypercube offers excellent theoretical
properties, with no node more than $\log_2 N$ hops away, if each hop
is slow, the overall system suffers.  The most straightforward
implementation, {\em store and forward}, requires waiting for an
entire message to arrive at a node before beginning the retransmission
along the next hop.  Based on this experience, 2D meshes such as the
Intel Touchstone Delta were implemented with {\em wormhole
  routing}, allowing the start of a message to begin transmitting
while the tail is still arriving, giving excellent overall performance
with more scalable hardware.  These issues matter less in our
environment.

For most of the 1980s and 1990s, with fine-grained parallelism and
many processors attempting to send messages at the same time, careful
matching of applications to network topologies and management of
resources (principally, access to the network) were required.  In
recent years, the availability of fast, cheap, general-purpose
networking hardware and improving software tools for larger-grained
parallel systems, such as Beowulf, MPI, and BOINC, have largely
decoupled parallel applications from the need for such
hardware-specific
tuning~\cite{sterling95:beowulf,dongarra95:_mpi,anderson05:boinc-potential}.

The field of interconnection networks for distributed, parallel
computation is a vast one; here we have hardly begun to even hint at
the
scope~\cite{dally04:_interconnects,hennessy-patterson:arch-quant3ed}.
Our current needs for a quantum multicomputer are modest, so this
level of analysis will suffice.

\section{Summary}

In this chapter, we have introduced the disparate concepts needed to
build a quantum multicomputer: the fundamental qubus technology we
intend to use to create entangled pairs of qubits (EPR pairs), the
teleportation of both quantum data and quantum gates that will use
EPR pairs to effect distributed quantum computation, and the
principles of store-and-forward multicomputer networks that will
determine how efficient the system can be.

We now come to the end of not only the chapter on the qubus, but of
the entire first part of this thesis, covering the fundamentals of
quantum computation.  We have studied the basic ideas of quantum
computation, seen Shor's algorithm for factoring large numbers, which
we will use as our target application, explained how to control
errors, and discussed many different quantum computing technologies.
And finally, we presented quantum teleportation and the qubus protocol
upon which we will build our quantum multicomputer.

We now set aside the distributed nature of our system for a while, and
move into the detailed analysis of the performance and limitations of
a monolithic quantum computer.  Once that analysis is complete, we
will return to the quantum multicomputer in Chapter~\ref{ch:qmc}.

  \chapter{Performance of Large-Scale Systems}
\label{ch:large-perf}

\cq{[T]he period matters little until the acceleration itself is
admitted. The subject is even more amusing in the seventeenth than in
the eighteenth century, because Galileo and Kepler, Descartes,
Huygens, and Isaac Newton took vast pains to fix the laws of
acceleration for moving bodies, while Lord Bacon and William Harvey
were content with showing experimentally the fact of acceleration in
knowledge...}{Henry Adams, ``A Law of Acceleration,'' 1905}

We are now prepared to design the architecture of a quantum computer
and evaluate its performance.  Up to this point, we have examined what
it means to do quantum computation, discussed what a quantum computer
could be used for, and analyzed the technologies available to build
such a system.  In Section~\ref{sec:divincenzo}, we saw DiVincenzo's
five criteria which must be met by any useful quantum computing
technology~\cite{divincenzo95:qc}.  In addition to these criteria, a
useful quantum computing technology must also support a quantum
computer {\em system architecture} which can run one or more quantum
algorithms in a usefully short time.  This observation subsumes into
one requirement several issues which, while not strictly necessary to
build a quantum computer, will have a strong impact on the possibility
of engineering a practical, useful system; we presented our analysis
of those requirements in Chapter~\ref{ch:taxonomy}.

The process of adapting abstract algorithms to quantum computers
naturally depends on the architecture, but the application of
classical computer architecture principles to quantum computers has
only just begun, making it difficult to definitively pronounce that a
certain quantum computer will be ``useful'' in solving real-world
problems.  In this chapter, my aim is to advance our understanding of
this design process, including designing some specific algorithmic
subroutines that are appropriate for certain architectures.  I analyze
and optimize the performance of the modular exponentiation that forms
the largest part of Shor's factoring algorithm, based on the
Vedral-Barenco-Ekert algorithm as discussed in
Section~\ref{sec:modexp}.  We have found ways to improve the scaling
of performance with respect to the length of the number being
factored; the acceleration is thousands of times for important problem
sizes, reaching one million times when factoring a 6,000-bit number.
We show that this acceleration depends on the architecture of the
system, and how to optimize for certain constraints.  We also show
that the faster modular exponentiation algorithms reduce the demands
on the error management subsystems and increase the fidelity of our
calculation.

The first section of this chapter provides a brief overview of the
techniques we use to accelerate arithmetic, then discusses the impact
of architecture on quantum error correction, and presents our
architectural models and notation.  The next two sections explain the
tradeoff between classical and quantum computation and present our new
adder designs, the carry-select and conditional-sum adders.
Section~\ref{sec:mono-shor-perf} closes this chapter with our major
analytical and numerical results for the complete modular
exponentiation algorithm.  The material presented here should help
other researchers analyze the performance of systems they design, both
large and small; in the next part of this dissertation, I use these
techniques to analyze the behavior of a quantum multicomputer based on
an overall structure I propose.

\section{Managing Performance}
\label{sec:algo-topo}


The realized performance of a system is a product of both the
underlying technology and the architecture imposed above it.  In
Sections~\ref{sec:algo-eff}, \ref{sec:time-gate} and
\ref{sec:manufop}, we introduced the technological factors that affect
performance of the system: physical and logical clock speed,
concurrency or parallelism, the number of available qubits, the
ability of qubits to communicate with each other (the ``wiring
topology''), addressability of individual qubits, and the
decomposition of logical gates into physical ones.  From this point
forward in the dissertation, we will ignore addressability and assume
individual control over qubits.  For our purposes (primarily
arithmetic circuits), the issue of direct or polynomial approximation
of arbitrary rotations only concerns us as described below, in the
breakdown of \textsc{ccnot}.  The ability of a system to retire application
instructions as quickly as possible derives from more than the clock
speed; extracting parallelism and moving data as efficiently as
possible strongly impact behavior, and these issues drive much of the
rest of this dissertation.

Concurrent quantum computation is the execution of more than one
quantum gate on independent qubits at the same time.  We generally use
the term {\em concurrency} rather than parallelism, to avoid confusion
with the concept of quantum parallelism.  Utilizing concurrency, the
latency, or circuit depth, to execute a number of gates can be smaller
than the number of gates.  We discussed parallel multipliers in
Section~\ref{sec:cw-parallel-mult}.  Circuit depth is also explicitly
considered in Cleve and Watrous' parallel implementation of the
quantum Fourier transform~\cite{cleve:qft}, various types of
arithmetic~\cite{cuccaro04:new-quant-ripple,draper04:quant-carry-lookahead,van-meter04:fast-modexp,gossett98:q-carry-save},
and Zalka's Sch\"onhage-Strassen-based implementation of modular
exponentiation~\cite{zalka98:_fast_shor}.  Moore and Nilsson define
the computational complexity class {\bf QNC} to describe certain
parallelizable circuits, and show which gates can be performed
concurrently, proving that any circuit composed exclusively of
Control-\textsc{not}s (\textsc{cnot}s) can be parallelized to be of depth $O(\log n)$
using $O(n^2)$ ancillae on an abstract
machine~\cite{moore98:_parallel_quantum}.  In
Chapter~\ref{ch:taxonomy}, we discussed the capability of different
technologies to perform concurrent gates; in this part of the thesis,
we combine the theoretical and practical concerns to analyze the
demands of the algorithms.

Here we summarize the techniques which are detailed in following
sections.  Our fast modular exponentiation circuit is built using
the following optimizations:

\begin{itemize}
\item Trade classical for quantum computation, to reduce the length of
  the expensive and difficult quantum portions
  (Section~\ref{sec:trading}).
\item Move to better adders; our algorithms concentrate on the use of
  the conditional-sum adder (Section~\ref{sec:csum}), carry-lookahead
  adder (Section~\ref{sec:qcla}), and CDKM carry-ripple adder
  (Section~\ref{sec:cdkm}).
\item Look for concurrency within addition; our concurrent version of
  VBE forms our baseline case, and the other adder circuits are
  defined with concurrency in mind.
\item Do multiplications concurrently, using Cleve-Watrous
  (Sections~\ref{sec:modexp} and \ref{sec:conc-exp}).
\item Reduce modulo comparisons, only subtract $N$ on overflow;
  this incurs a small space penalty and requires some cleanup at the
  end, in exchange for a nearly $5\times$ reduction in the number of
  calls to the adder routine (Section~\ref{sec:cut-mod}).
\item Select correct qubit layout and subsequences to implement gates,
  then hand
  optimize~\cite{vandersypen:thesis,aho:palindromes,kawano04:compiler,kunihiro:new-req,takahashi:_const_depth,yao93quantum,ahokas:_qft_complexity}.
\end{itemize}

\subsection{Error Correction, Architecture, and Clock Speed}
\label{sec:ecc-arch-clock}

A basic understanding of the pressures that quantum error correction
and fault tolerance place on architecture is critical.  As we saw in
Chapter~\ref{ch:err-mgmt}, QEC and FT demand the continuous
preparation and measurement of a set of ancillae (temporary work
qubits), and raise the overall cost of quantum computation by as much
as four orders of magnitude for {\em each} level of QEC built into the
system -- and it appears that two or more levels may be necessary.
The logical clock speed of the system will correspond roughly to the
QEC cycle time, and is correspondingly slower than the physical clock
speed, though the exact ratio will depend on both technology- and
machine-dependent details.

QEC codes encode one or more qubits into a code word.  The error
syndromes on this code word are continuously calculated and measured,
and corrective actions applied to the code word.  The measurement of
the syndrome actually effects a key portion of the error control
process; it forces (``projects'') the state either back into a good
state (with high probability) and returns a zero (no error) syndrome,
or an error state (with low probability) and returns a non-zero
syndrome.  When the syndrome is non-zero, one or two corrective gates
are indicated and applied.  Unfortunately, this syndrome calculation
and measurement process may also introduce errors.  Technologies that
support nearest-neighbor-only interactions require swapping of qubits
in order to calculate the error syndrome, with the swap gates possibly
introducing errors themselves, making the threshold requirements for
effective error correction more stringent; in some studies, as much as
175 times
worse~\cite{svore05:_local-ft,steane02:ft-qec-overhead,aharonov99:_threshold,fowler04:_qec_lnn,szkopek04:_thres_error_penalty}.
The parity calculations necessary to retrieve the error syndrome
cannot be carried out directly, but must operate indirectly using a
logical zero ($|0_L\rangle$) state to defend against propagation of
errors.  That state preparation requires as many qubits as the code
word itself, and may be the driver of the cycle time for QEC.
Measurement of qubit state on some technologies is slow compared to
the gate time, so this also figures prominently into the cycle time.

As qubits are subject to error processes when idle, as well as while
being used, the total amount of error correction in the system is
dependent on the size of the machine, as well as the number of logical
gates being executed.  If each qubit must be ``refreshed'' at
one-tenth the QEC cycle rate, for example, then we must build a system
in which one-tenth of the qubits can all be undergoing QEC at the same
time.  Longer waits for correction increase the probability of error;
this must be balanced against the number of levels of QEC and the
engineering difficulties of initialization and measurement.  Quantum
dots and superconducting qubits require additional on-chip structures
to perform measurement~\cite{petta05:_coher_manip}, limiting layout
flexibility and consuming die space.  If possible, it will be
desirable to perform entire QEC sequences on-chip; however, in the
short run, it may be necessary to use off-chip signal generators and
control circuitry, requiring a wide, high-bandwidth I/O interface from
the chip itself.

To manage errors effectively, then, we can say that a technology must
support large numbers of concurrent qubit state preparations, gates,
and measurements.  As the required operations are much more complex
than a DRAM refresh cycle, and are close to the universal gate set,
a large-scale difference in structure akin to the CPU/RAM dichotomy is
unlikely.  However, at the small scale, systems which store qubits in
nuclear spins while idle and shift to electron spins for active gates
have been
proposed~\cite{steane:ion-atom-light,kane:nature-si-qc,mehring03:_entan-electron-nuke,jelezko04:_observ,childress05:_ft-quant-repeater}.

\subsection{\ac\ and \ntc\ Architectural Models}
\label{sec:arch-models}

This dissertation analyzes two separate architectures, still abstract
but with some important features that help us understand performance.
For both architectures, we assume any qubit can be the control or
target for only one gate at a time.  The first, the \ac, or {\em
Abstract Concurrent}, architecture, is our more abstract model.  It
supports \textsc{ccnot} (the three-qubit Toffoli gate, or Control-Control-\textsc{not}),
arbitrary concurrency, and gate operands any distance apart without
penalty.  It does not support arbitrary control strings on control
operations, only \textsc{ccnot} with two ones as control.  \ac\ corresponds to
the machine we have implicitly assumed to this point.  The second, the
\ntc, or {\em Neighbor-only, Two-qubit-gate, Concurrent} architecture,
is similar but does not support \textsc{ccnot}, only two-qubit gates, and
assumes the qubits are laid out in a one-dimensional line, and only
neighboring qubits can interact.  The 1D layout will have the highest
communications costs among possible physical topologies.

The \ntc\ model is a reasonable description of several important
experimental approaches, including a one-dimensional chain of quantum
dots~\cite{loss:qdot-comp}, the original Kane
proposal~\cite{kane:nature-si-qc}, and the all-silicon NMR
device~\cite{ladd:si-nmr-qc}.  Superconducting
qubits~\cite{pashkin:oscill,you:scalable} may map to \ntc, depending
on the details of the qubit interconnection.

\begin{figure}
\centerline{\hbox{
\includegraphics[height=2cm]{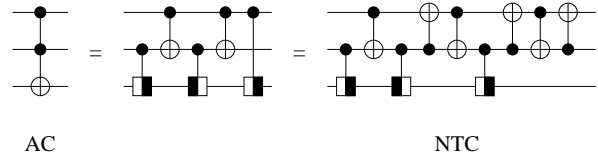}}}
\caption[\textsc{ccnot} constructions for our architectures \ac\ and
  \ntc]{\textsc{ccnot} constructions for our architectures \ac\ and \ntc.
  The box with the bar on the right represents the square root of $X$,
  and the box with the bar on the left its adjoint.}
\label{fig:5gcc}
\end{figure}

For \ntc, which does not support \textsc{ccnot} directly, we compose \textsc{ccnot}
from a set of five two-qubit gates~\cite{barenco:elementary}, as shown
in figure~\ref{fig:5gcc}.  The box with the bar on the right
represents the square root of $X$,
$\sqrt{X} = \frac{1}{2}\left[\begin{array}{cccc} 1+i & 1-i \\ 1-i & 1+i
\end{array}\right]$
and the box with the bar on the left its adjoint.  We assume that this
gate requires the same execution time as a \textsc{cnot}.

The difference between \ac\ and \ntc\ is critical; beyond the
important constant factors as nearby qubits shuffle, we will see in
Section~\ref{sec:mono-shor-perf} that \ac\ can achieve $O(\log{n})$
performance on addition where \ntc\ is limited to $O(n)$.  Most real,
scalable architectures will have constraints with this flavor, if
different details, so \ac\ and \ntc\ can be viewed as bounds within
which many real architectures will fall.  The layout of variables on
this structure has a large impact on performance; what is presented
here is the best we have discovered to date, but we do not claim it is
optimal.

\subsection{Notation}
\label{sec:notation}

In the rest of this dissertation, as in Chapter~\ref{ch:shor}, where
we introduced Shor's factoring algorithm, we will use $N$ as the
number to be factored, and $n$ to represent its length in bits.  For
convenience, we will assume that $n$ is a power of two, and that the
high bit of $N$ is one.  $x$ is the random value, smaller than $N$, to
be exponentiated, and $|a\rangle$ is our superposition of exponents,
with $a < 2N^2$ so that the length of $a$ is $2n+1$ bits.

As described in Section~\ref{sec:arith-nota}, when discussing circuit
cost, the notation we use is $(\textsc{ccnot}s; \textsc{cnot}s;
\textsc{not}s)$ or $(\textsc{cnot}s; \textsc{not}s)$.  The values will
usually be circuit depth (latency), but may be total gate count,
depending on context.  The notation is sometimes enhanced to show
required concurrency and space,\\
$(\textsc{ccnot}s; \textsc{cnot}s; \textsc{not}s) \#(concurrency; space)$.

$t$ is time, or latency to execute an algorithm, and $S$ is space,
subscripted with the name of the algorithm or circuit subroutine.
When $t$ is superscripted with \ac\ or \ntc, the values are for the
latency of the construct on that architecture.  Equations without
superscripts are for an abstract machine assuming no concurrency,
equivalent to a total gate count for the \ac\ architecture.  $R$ is
the number of calls to a subroutine, subscripted with the name of the
routine.

$m$, $g$, $f$, $p$, $b$, and $s$ are parameters that determine the
behavior of portions of our modular exponentiation algorithm.  $m$,
$g$, and $f$ are part of our carry-select/conditional-sum adder
(Section~\ref{sec:new-adders}).  $p$ and $b$ are used in our
indirection scheme (Section~\ref{sec:trading}).  $s$ is the number of
multiplier blocks we can fit into a chosen amount of space
(Section~\ref{sec:conc-exp}).

\section{Trading Classical for Quantum Computation}
\label{sec:trading}

\cq{Any software problem can be solved by adding another layer of
indirection.}%
{David Wheeler}

This section discusses balancing the overall {\em system} performance.
With a classical computer as much as $10^{15}$ times as fast as
quantum computer~\footnote{Very, very roughly, a modern microprocessor
has $10^9$ transistors, of which perhaps 10\% are involved in a
``gate'' in a clock cycle, of which there are $10^9$ per second,
yielding some $10^{17}$ gates/second.  In contrast, the slowest
quantum devices (liquid NMR) may run at only a few tens of gates per
second, {\em before} applying quantum error correction.  Note that
this ignores both parallel classical computation and faster quantum
devices, but the point is still valid.}, we can afford to trade many
classical operations for a single quantum
one~\cite{van-meter:pay-the-exponential}.  The same principle applies
if the metric of interest is economic cost, rather than time
performance; quantum gates will remain many orders of magnitude more
expensive than classical ones for the foreseeable future.

As we saw in earlier chapters, modular exponentiation is the most
expensive portion of Shor's algorithm, consisting of $2n$
multiplication operations to exponentiate an $n$-bit number.  Here, I
show that it is possible to reduce the number of quantum modular
multiplications necessary by a factor of $w$, at a cost of performing
$2^w$ times as many classical modular multiplications and adding
temporary storage space and associated machinery for a table of $2^w$
entries.  The storage space may be quantum-addressable classical
memory, pure quantum memory, or pure classical memory.  Values of $w$
from 2 to 30 seem attractive; physically feasible values depend on the
implementation of the memory.

\subsection{Introduction}

To factor the number $N$ using Shor's algorithm~\cite{shor:factor}, a
quantum computing device must evolve to hold the state
\begin{equation}
\frac{1}{2^n}\sum_{a=0}^{2^{2n}-1}|a\rangle|x^a \bmod N\rangle.
\end{equation}
This is the {\em modular exponentiation} step discussed in
Section~\ref{sec:modexp}, the first major quantum step in the
order-finding process.  We also saw that the value $x^a \bmod N$ can
be rewritten~\cite{kunihiro:factoring-time,vedral:quant-arith} as
\begin{equation}
\prod_{j=0}^{n-1} d_j^{a_j} \bmod N
\end{equation}
where $d_i = x^{2^i}$, and $a_{n-1}a_{n-2}..a_0$ is the binary
expansion of $a$.  The $d_i$ can be calculated classically, but
$|a\rangle$ must be a quantum register.

This approach treats $|a\rangle$ as a sequence of {\em bits}; my
approach to reducing the number of multiplications is to treat
$|a\rangle$ as a series of short {\em words}.  Dividing $|a\rangle$ up
into $l$ words of length $w$, let $|t_k(a)\rangle$, the $k$th word in
$|a\rangle$, be $|t_k(a)\rangle =
|a_{w(k+1)-1}a_{w(k+1)-2}..a_{wk+1}a_{wk}\rangle$ for $0 \leq k < l, l
= \lceil n/w\rceil$.  $|t_k(a)\rangle$, as part of $|a\rangle$, will
hold a superposition of all values $0$ to $2^w-1$.

We can reduce the $2n$ quantum multiplications to $l$ by iterating
over the words in $|a\rangle$, using the superposition
$|t_k(a)\rangle$ as a quantum index into a memory array holding the
$2^w$ $n$-bit entries with values $b_{m,k} = x^{m2^{wk}} \bmod N$,
where $m$ is the index into the array and $k$ is the iteration number,
0 to $l-1$.
The superposition of values retrieved from the memory is multiplied
with the current value, giving
\begin{equation}
\frac{1}{2^n}\sum_{a=0}^{2^{2n}-1}|x^a \bmod N\rangle =
\frac{1}{2^n}\sum_{a=0}^{2^{2n}-1} |\prod_{j=0}^{l-1} b_{t_j(a),j}
\bmod N\rangle.
\end{equation}
A total of $lw2^w = n2^{w+1}$ classical and $l$ quantum modular
multiplications must be performed, compared with $2n$ classical and
$2n$ quantum modular multiplications using Vedral's
formulation~\cite{vedral:quant-arith}.

\subsection{Indirection}

In a computer, arguments to an instruction (or function) can be passed
{\em by value} or {\em by reference}.  By value arguments appear
directly in the bits of the instruction.  When accessing arguments by
reference, the address of the argument is held in the instruction; the
actual value must be retrieved from memory before the function can be
evaluated. The address is called a {\em pointer} or an {\em
index}. {\em Indirection} is a generalization of by reference, in
which the value retrieved from memory may itself be a pointer which
must in turn be dereferenced.

In the straightforward, bit-based implementation of quantum modular
exponentiation, the $d_i$ values are classical values programmed into
a register with a superposition of 0, based on the matching bit in the
superposition $|a\rangle$.  In the word-oriented approach, the
$b_{m,k}$ values are held in a table.  Logically, a portion of the
$|a\rangle$ superposition is used as an index into that table,
fetching one of the values to use as the multiplicand (more correctly,
fetching a superposition of the $b_{m,k}$ values to use as the
multiplicand).  That is, we are accessing the arguments for our
multiplication through a single level of indirection.

\subsection{The $b$ Array}

The $b$ array is our bridge from classical computation to quantum.
Each entry is $n$ bits.  We must compute $2^w$ values for the table,
requiring $w$ classical modular multiplications each, before each of
the $l$ quantum multiplications.  Then, we must figure out how to get
$b_{m,k}$ values into the multiplicand register, in superposition.  We
can use quantum memory, classical memory, or a type of mixed device to
hold the data.

\subsubsection{Quantum-Addressable Classical Memory}

\begin{figure*}
\centerline{\hbox{
\input{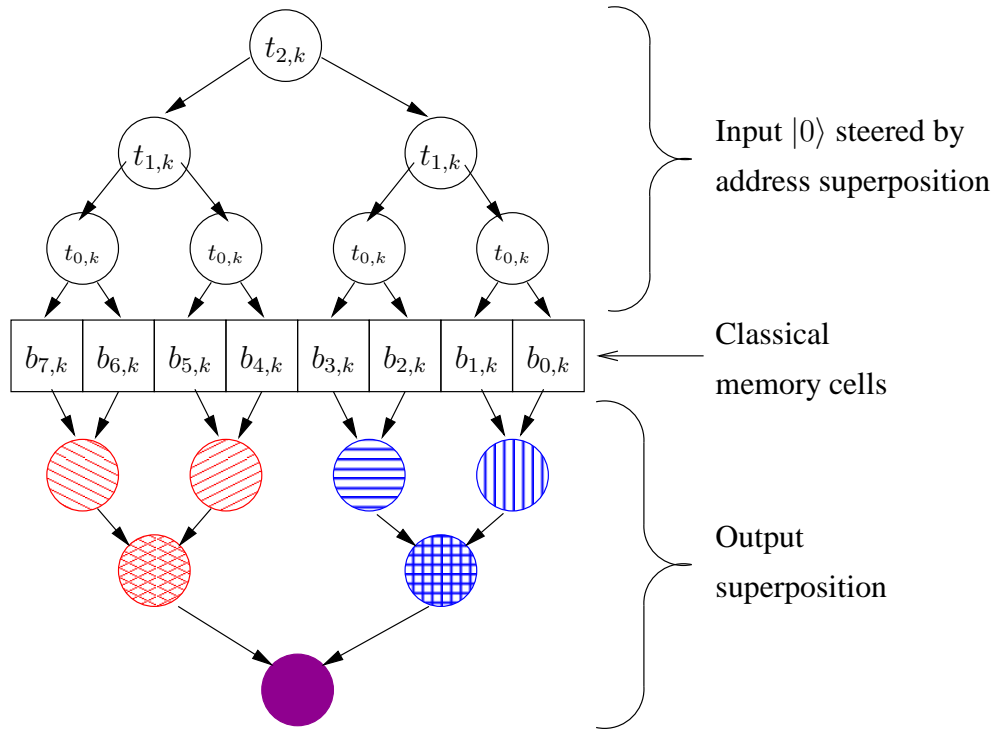}}}
\caption{Quantum-Addressable Classical Memory (QACM)}
\label{fig:qacm}
\end{figure*}

The array can be held using a quantum addressable classical memory
(QACM)~\cite{nielsen-chuang:qci-qacm}.  In such a device, memory cells
(the modular exponentiation values) are classical, but a quantum
superposition is used as an address, and the read out value is a
superposition of each classical value in proportion to the ``amount''
of its address present in the address superposition.  One such
possible device is an optical plate, with photons steered through the
various cells according to the value of specific address bits.
Figure~\ref{fig:qacm} shows a 3-bit example.  At the top, the input
(generally $|0\rangle$) is steered left or right according to the
high-order bit of the address superposition (carried on a control line
not shown in the figure).  Subsequent circles steer left or right
according to their address bits, to reach the appropriate classical
data memory cells.  The values retrieved from the memory are combined
to give the full output superposition, in weights according to the
address superposition.

\subsubsection{Pure Quantum Memory}

\begin{figure*}
\centerline{\hbox{
\input{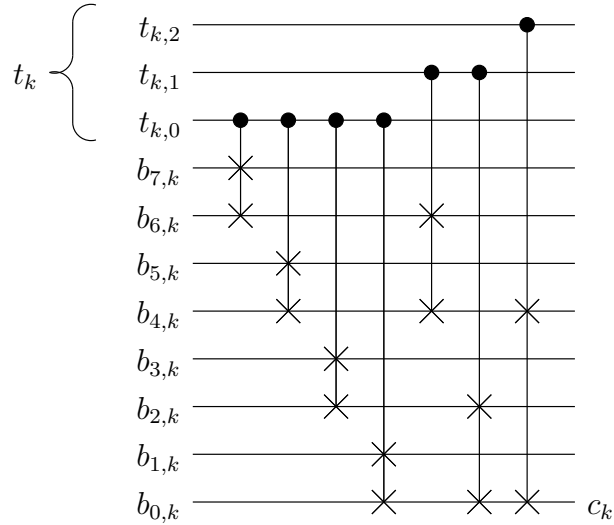}}}
\caption{3-bit Quantum Select Circuit (Q-SEL)}
\label{fig:3b-qsel}
\end{figure*}

An equivalent array of qubits can be used in place of the QACM.
However, in that case, the cost of filling the table must be accounted
for, and our limitation will be the number of available qubits.
Figure~\ref{fig:3b-qsel} shows a 3-bit select circuit composed of
Fredkin gates which will choose from among the 8 possible arguments
for the modular multiplier.  The desired value $c_k = b_{t_k,k}$
occupies the location as shown on the right of the figure; it is then
used as the argument to the modular multiplier.  This select circuit
can be reversed following the multiplication to restore the original
locations of the $b_{j,k}$ values.

\subsubsection{Classically-Driven Setting of Multiplicand Register}

\begin{figure}
\centerline{\hbox{
\input{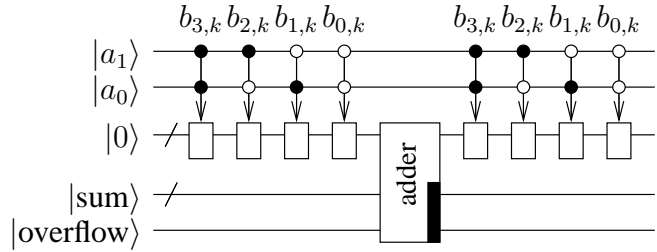}}}
\caption[Implicit indirection using a classical memory]{Implicit
  indirection using a classical memory.  The arrows pointing to blocks
  indicate the setting of the multiplicand register to the value
  above, based on the control lines.}
\label{fig:imp-ind}
\end{figure}

In the VBE algorithm, the multiplicand register is filled using \textsc{cnot}s,
with the appropriate bit of $|a\rangle$ as the control.  For our
word-oriented approach, we can implicitly perform the lookup by
choosing which gates to apply while setting the argument.  In
Figure~\ref{fig:imp-ind}, we show the setting and resetting of the
argument for $w = 2$, where the arrows indicate \textsc{ccnot}s to set the
appropriate bits of the 0 register to 1.  The $x^i$ values are
classically calculated and stored; we are setting the $|0\rangle$
register to a superposition of the $b$ values.  The actual
implementation can use a calculated enable bit to reduce the \textsc{ccnot}s to
\textsc{cnot}s.  Only one of the values $x^0$, $x^1$, $x^2$, or $x^3$ will be
enabled, based on the value of $|a_{1}a_{0}\rangle$.

The setting of this input register may require propagating $|a\rangle$
or the enable bit across the entire register.  Use of a few extra
qubits ($2^{w-1}$) will allow the several setting operations to
propagate in a tree.  The cost of setting the argument is
\begin{equation}
t_{ARG} = 
\begin{cases}
2^w(1;0;1) = (4; 0; 4) w = 2 \\
2^w(3;0;1) w = 3,4 \end{cases}.
\end{equation}

\begin{figure}
\centerline{\hbox{
\includegraphics{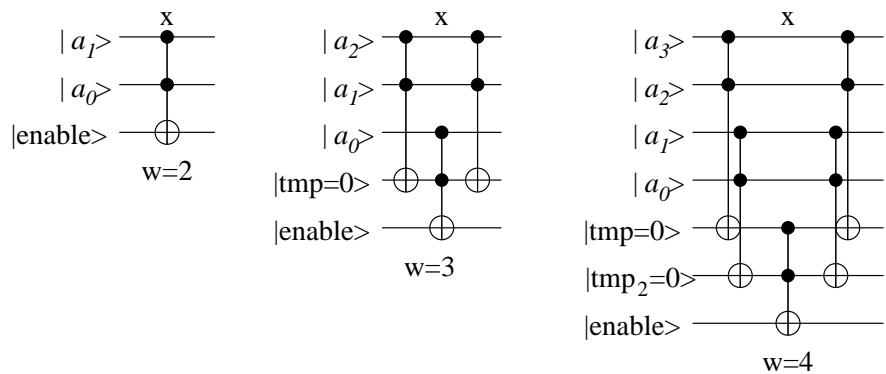}}}
\caption[Argument setting for indirection for different values of
$w$]{Argument setting for indirection for different values of $w$.
For the $w=4$ case, the two \textsc{ccnot}s on the left can be executed
concurrently, as can the two on the right, for a total latency of 3.}
\label{fig:arg-setting}
\end{figure}

For $w = 2$ and $w = 3$, we calculate that setting the argument adds
$(4;0;4)\#(4,5)$ and $(24;0;8)\#(8,9)$, respectively, to the latency,
concurrency and storage of each adder.  We create separate
enable signals for each of the $2^w$ possible arguments and pipeline
flowing them across the register to set the addend bits.  We consider
this cost only when using indirection.  Figure~\ref{fig:arg-setting}
shows circuits for $w=2,3,4$.

\subsection{The Algorithm}
\label{sec:alg}

In essence, the algorithm involves moving from a bit-oriented
breakdown of the multiplications to a word-oriented breakdown.  The
algorithm consists of two main parts: classically calculating the $b$
array values, and calculating their products in the quantum domain.
We pay the classical cost in step~\ref{a:cc} in the algorithm below,
and the quantum cost in step~\ref{a:qc}.

The cost of setting up to use the $k$th iteration of the $b$ array is
technology dependent; only one of steps \ref{a:pu} and \ref{a:up} is
necessary.  $O(n2^w)$ gates may be required to set a quantum memory,
or only the change of a single pointer or position if a QACM is large
enough to hold the entire $b$ array at once.

\begin{enumerate}
\item Calculate the $b$ array elements:
\begin{enumerate}
\item Classically calculate $b_{j,0} = x^j$ for all $j, 0 \le j < 2^w$
\item\label{a:cc} For $k$ from $1$ to $l-1$, classically square
  (modulo $N$) all $2^w$ elements $b_{j,k-1}$ $w$ times to create $b_{j,k}$
\item\label{a:pu} (Store all $b_{j,k}$ into QACM)
\end{enumerate}
\item Initialize $|p\rangle$ to 1
\item For $k$ from $0$ to $l-1$, do
\begin{enumerate}
\item\label{a:up} (Set up to use $b_{j,k}$ values: store into QACM or
quantum memory)
\item In quantum domain, use $|t_k(a)\rangle$ as index into $b$,
  $|c_k\rangle = |b_{t_k(a),k}\rangle$
\item\label{a:qc} $|p\rangle = |c_kp \bmod N\rangle$
\end{enumerate}
\end{enumerate}

\begin{figure*}
\centerline{\hbox{
\input{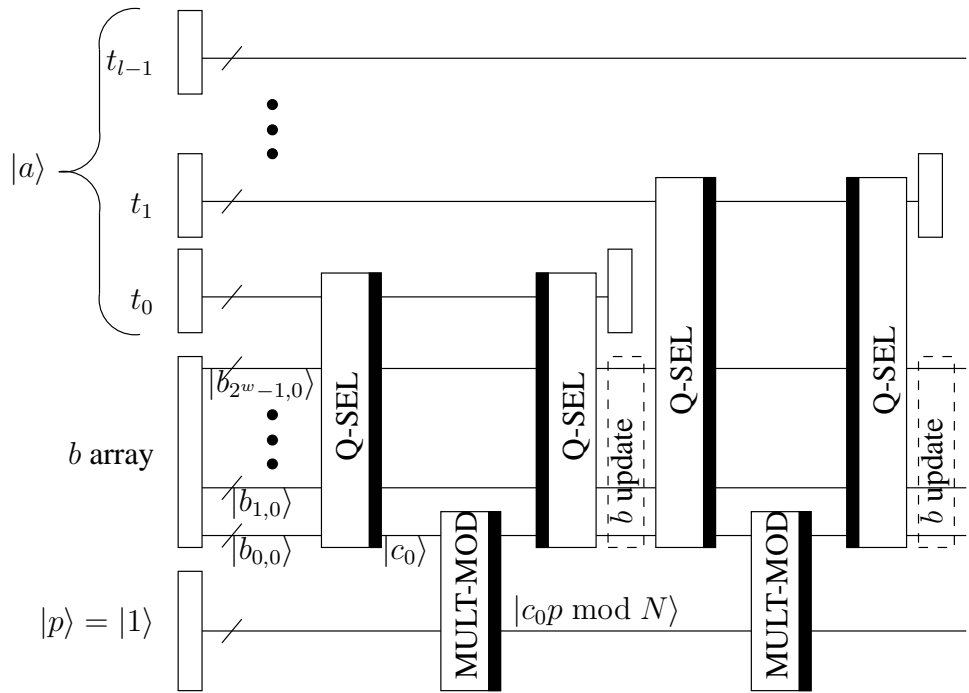}}}
\caption{Multiplication Using Indirection, Based on Vedral's Circuit}
\label{fig:ind-mult}
\end{figure*}

Figure~\ref{fig:ind-mult} shows a portion of a modified form of
Vedral's circuit using indirection.  The dashed box indicates where
update of the $b$ array takes place, if necessary; only one-qubit
gates are required. Note also that Q-SEL and its reverse are used,
but, unlike Vedral's circuit, we do not need the
reverse of multiplication to free up our argument.  The degenerate
case of $w = 1$ is therefore faster than Vedral's circuit.

\subsection{Evaluating Cost and Selecting Word Length}

\begin{figure*}
\def\epsfsize#1#2{.8\hsize}
\centerline{\hbox{
\epsfbox{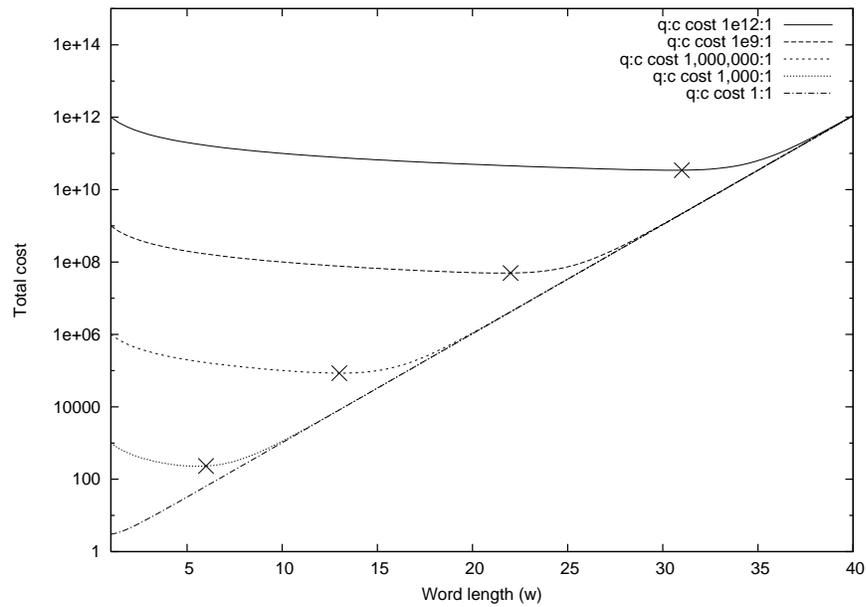}}}
\caption{Total Cost}
\label{fig:cost-plot}
\end{figure*}

The goal of this work is to minimize the cost of executing Shor's
algorithm, for some metric of cost important to the user.  In
Figure~\ref{fig:cost-plot} we show the total cost of calculating the
modular exponentiation, as a function of word length $w$.  ``Cost'' in
this graph is an arbitrary metric; it may be wall clock time, total
time on parallel machines, price tag, or some other economic cost of
quantum and classical machines.  Perhaps the easiest cost to consider
is simply time to perform a modular multiplication.  The five curves
represent total cost at different ratios of quantum:classical cost,
ranging from 1:1 to $10^{12}$:1.  The 'x' marks on each curve are the
nearest integer value of $w$ to the minimum.  This recommended word
length increases by approximately eight bits for each factor of one
thousand the relative quantum cost increases.

This graph is somewhat simplified, in that the cost ratio is
treated as fixed.  In reality, the QACM cost will almost certainly
depend on the word length.

Commodity microprocessors may be as much as $10^{15}$ times as fast as
quantum computing devices, even before accounting for quantum error
correction.  Faster technologies, ranging up to gigahertz clock rates,
still leave several orders of magnitude difference between classical
and quantum aggregate gate rates.  Combined with the success
probability, it is clear that the limitation on $w$ will be the
practical size of the $b$ array rather than computational cost.

This section has shown that the standard computer science technique of
indirection can be used in the quantum domain to accelerate the
modular exponentiation that is the primary cost of Shor's algorithm.
This technique reduces the number of multiplications necessary, and is
independent of the multiplication algorithm chosen.  The price we pay
for this is a large classical/quantum tradeoff; we perform $2^w$ more
multiplications in the classical domain in exchange for reducing the
quantum multiplications by a factor of $w$.  This basic technique will
likely apply to other algorithms, as well.

\section{New Adder Types}
\label{sec:new-adders}

\begin{chapterquote}
``I only took the regular course...Reeling and Writhing, of course,
  to begin with, and then the different branches of Arithmetic --
  Ambition, Distraction, Uglification and Derision.''\\
the Mock Turtle, in \textbf{Lewis Carroll's {\em Alice's Adventures in
  Wonderland}, 1865}
\end{chapterquote}

Quantum versions of the classical carry-select and conditional-sum
adders deepen the toolbox of arithmetic routines available for
matching software to
hardware~\cite{ercegovac-lang:dig-arith,van-meter04:fast-modexp}.  The
basic carry-select adder concurrently calculates two possible results
without knowing the value of the carry in, one assuming that the carry
in will be zero, one assuming that the carry in will be one.  Once the
carry in becomes available, the correct output value is selected using
a multiplexer (MUX).  The type of MUX determines whether the latency
of the circuit is $O(\sqrt{n})$ (called a carry-select adder) or
$O(\log n)$ (called a conditional-sum adder).

Zalka has proposed a carry-select adder, without calling it by
name~\cite{zalka98:_fast_shor}.  He did not present a full circuit,
making it difficult to reproduce his results, and my circuit produces
slightly different numbers than his.

\subsection{Basic Carry-Select Adder}
\label{sec:csla}

First, we present the basic carry-select adder, then show the MUX
structure that completes the circuit.  To add two $n$-bit numbers, we
will divide the numbers into groups and run an adder for each group.
The bits are divided into $g$ groups of $m$ bits each, $n = gm$.  The
first group may have a different size, $f$, than $m$, since it will be
faster, but for the moment we assume they are the same.  The
carry-select adder for a single group we will call CSLA.

\subsubsection{VBE-Based Adder}

\begin{figure}
\centerline{\hbox{
\includegraphics[width=8.6cm]{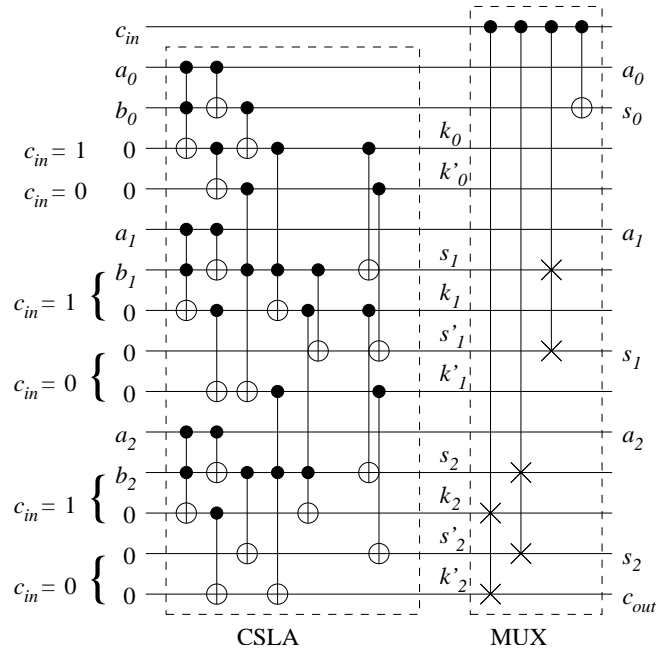}}}
\caption[Three-bit carry-select adder (CSLA) with multiplexer
  (MUX)]{Three-bit carry-select adder (CSLA) with multiplexer (MUX).
  $a_i$ and $b_i$ are addends.  The control-SWAP gates in the MUX
  select either the qubits marked $c_{in} = 1$ or $c_{in} = 0$
  depending on the state of the carry in qubit $c_{in}$.  $s_i$ qubits
  are the output sum and $k_i$ are internal carries.}
\label{fig:csel3}
\end{figure}

Figure~\ref{fig:csel3} shows a three-bit carry-select adder, CSLA,
plus an example MUX.  This generates two possible results, assuming
that the carry in will be zero or one.  All of the outputs without
labels are ancillae to be garbage collected.  The circuit shown here
is based on the Vedral-Barenco-Ekert (VBE) carry-ripple adder
described in Section~\ref{sec:vbe}.  As drawn, a full
carry-select circuit requires $5m-1$ qubits to speculatively add two
$m$-bit numbers.  The MUX can be implementing using the optimization
of the Fredkin gate shown in Figure~\ref{fig:basic-gates} on
page~\pageref{fig:basic-gates}.

A larger $m$-bit carry-select adder can be constructed so that its
internal delay, as in a normal carry-ripple adder, is one additional
\textsc{ccnot} for each bit, although the total number of gates increases
(because we are essentially running two additions at the same time)
and the distance between gate operands increases.  The latency for the
CSLA block is
\begin{equation}
\label{eq:cslaac}
t_{CS} = (m;\thinspace{}2;\thinspace{}0).
\end{equation}
Note that this is not a ``clean'' adder; we still have ancillae to
return to the initial state.

\subsubsection{CDKM-Based Adder}

\begin{figure}
\centerline{\hbox{
\input{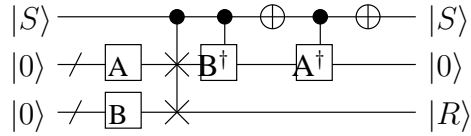}}}
\caption[In-place post-select circuit]{In-place circuit and MUX to
  post-select either $R = A|0\rangle$ or $R = B|0\rangle$, based on
  the select signal $S$.}
\label{fig:inplace-mux}
\end{figure}

It is possible that a design optimized for space could reduce the
number of qubits required, perhaps by utilizing the
Cuccaro-Draper-Kutin-Moulton (CDKM) carry-ripple adder
(Section~\ref{sec:cdkm}), which is more space-efficient.  The CDKM
adder uses only $2n+2$ bits to add two $n$-bit numbers (including the
carry out).  By simply fanning out a ``copy'' of both the $A$ and $B$
input registers and running separate adders in parallel, it is easy to
reduce the $5m-1$ qubits required above to $4m$, a noticeable savings
in space.  Figure~\ref{fig:inplace-mux} outlines one approach to
performing the demultiplexing in place; this approach results in {\em
very} fast availability of the result, but the ancillae garbage
collection is slow.  The circuit in the figure is general; applying it
to carry-select addition, $A$ and $B$ are almost identical, but
disentangling the carry in signals slows down the total circuit.  I am
still in the process of designing this adder, and expect to report on
its performance at a later date.

\subsection{$O(\sqrt{n})$ Carry-Select Adder}

\begin{figure}
\centerline{\hbox{
\includegraphics{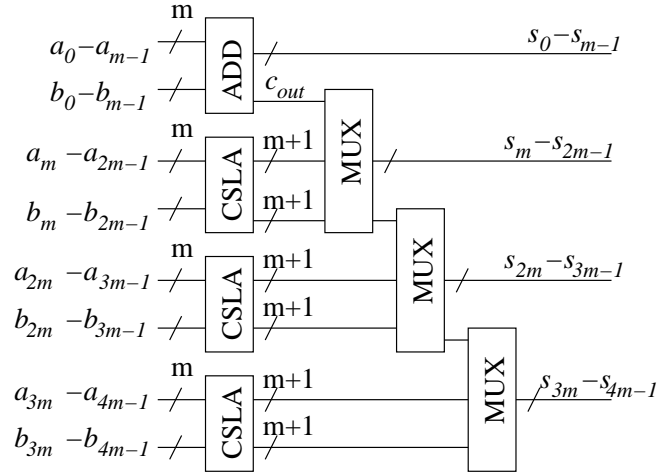}}}
\caption[Block-level diagram of four-group carry-select
  adder]{Block-level diagram of four-group carry-select adder.  $a_i$
  and $b_i$ are addends and $s_i$ is the sum.  Additional ancillae not
  shown.}
\label{fig:csel4b}
\end{figure}
The right-hand portion of Figure~\ref{fig:csel3} is the MUX which
selects the output to use; it is constructed from Fredkin gates using
the carry in as the control bit.  Notice that the carry in is not used
until after all of the adder blocks have completed.  This feature
allows the parallelism that makes the carry-select adder structure
fast.  One CSLA for each of the $g$ groups is used; all of the CSLAs
are executed concurrently, then the output MUXes are cascaded, as
shown in Figure~\ref{fig:csel4b}.

The most difficult implementation problem will be creating an
efficient MUX.  Figure~\ref{fig:csel4b} makes it clear that the total
carry-select adder is only faster than the carry-ripple adder if the
latency of MUX is substantially less than the latency of the full
carry-ripple adder.  The delay of the initial part of the VBE adder
for a group of $m$ qubits would be $(m; 0; 0)$.  If the carry out from
the MUX requires less than $m$ \textsc{ccnot} times, it may be faster.  The
carry out can be generated in a constant number of time steps by
prioritizing the last bit in the addition as the first to be MUXed
out.  The latency of the carry ripple from MUX to MUX (not qubit to
qubit) can be arranged to give a total MUX cost of $(4g+2m-6;0;2g-2)$.

Within the block, it is certainly easy to see how the MUX can use a
fanout tree consisting of more ancillae and \textsc{cnot} gates to distribute
the carry in signal, as suggested by
Moore~\cite{moore98:_parallel_quantum}, allowing all MUX Fredkin gates
to be executed concurrently.  A full fanout requires an extra $m$
qubits in each adder.  For intermediate values of $m$, we will use a
fanout of 4, reducing the MUX latency to $(4g+m/2-6;2;2g-2)$ in
exchange for 3 extra qubits in each group.  The space used for the
full, clean, VBE-based adder is $(6m-1)(g-1)+3f+4g$ when using a
fanout of 4.

The total latency of the CSLA, MUX, and the CSLA undo is
\begin{eqnarray}
\label{eq:cslamuac}
t_{SEM} &=& 2t_{CS}+t_{MUX} \nonumber\\
&=& (4g+5m/2-6;\thinspace{}6;\thinspace{}2g-2).
\end{eqnarray}
Optimizing, based on equation~\ref{eq:cslamuac}, the delay will be the
minimum when $m \sim \sqrt{8n/5}$, giving asymptotic performance
$O(\sqrt{n})$.

\subsection{$O(\log n)$ Conditional Sum Adder}
\label{sec:csum}

\begin{figure}
\centerline{\hbox{
\includegraphics[height=10cm]{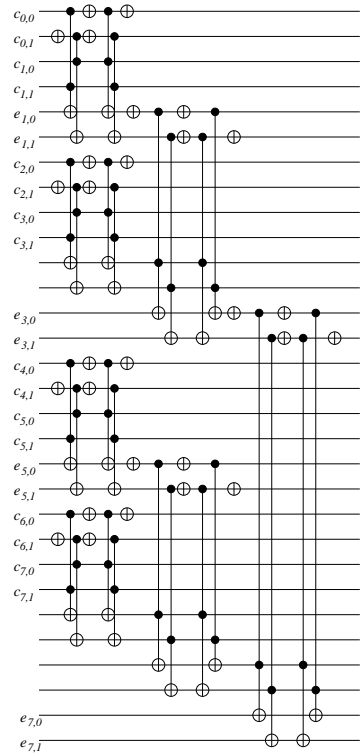}}}
\caption[$O(\log n)$ MUX for conditional-sum adder]{$O(\log n)$ MUX
  for conditional-sum adder, for $g=9$ (the first group is not shown).
  Only the $c_{i,j}$ carry out lines from each $m$-qubit block are
  shown, where $i$ is the block number and $j$ is the carry in value.
  At each stage, the span of correct effective swap control lines
  $e_{i,j}$ doubles.  After using the swap control lines, all but the
  last must be cleaned by reversing the circuit.  Unlabeled lines are
  ancillae to be cleaned.}
\label{fig:csla-new-mux}
\end{figure}

To reach $O(\log n)$ performance, we must add a multi-level MUX to our
carry-select adder.  This structure is called a conditional sum adder,
which we will label CSUM.  Rather than repeatedly choosing bits at
each level of the MUX, we will create a multi-level distribution of
MUX select signals, then apply them once at the end.
Figure~\ref{fig:csla-new-mux} shows only the carry signals for eight
CSLA groups.  The $e$ signals in the figure are our effective swap
control signals.  They are combined with a carry in signal to control
the actual swap of variables.  In a full circuit, a ninth group, the
first group, will be a carry-ripple adder and will create the carry in
to the rest of our tree; that carry in will be distributed
concurrently in a separate tree.

The total adder latency will be
\begin{eqnarray}
t_{CSUM} &=& 2t_{CS} +\nonumber\\
&& (2\lceil \log_2(g-1)\rceil - 1)
\times(2;\thinspace{}0;\thinspace{}2)\nonumber\\
&& + (4;\thinspace{}0;\thinspace{}4)\nonumber\\
&=&(2m + 4\lceil\log_2 (g-1)\rceil + 2;\thinspace{}4;\nonumber\\
&&\thinspace{}4\lceil\log_2 (g-1)\rceil+2)
\end{eqnarray}
where $\lceil x\rceil$ indicates the smallest integer not smaller than
$x$.

For large $n$, this generally reaches a minimum for small $m$, which
gives asymptotic behavior $\sim4\log_2 n$, the same as the
carry-lookahead adder from Section~\ref{sec:qcla}.  CSUM is noticeably
faster for small $n$, but requires more space.  The MUX uses $\lceil
3(g-1)/2\rceil - 2$ qubits in addition to the internal carries and the
tree for dispersing the carry in.  Our space used for the full, clean
adder is $(6m-1)(g-1)+3f+\lceil 3(g-1)/2 -2+ (n-f)/2\rceil\approx 6n$.
Section~\ref{sec:mono-shor-perf} details the tradeoffs in overall
system design caused by the extra space required.

Maslov et al. have recently improved on the performance of this MUX by
reducing the pair of \textsc{ccnot}s to one \textsc{ccnot} and two
\textsc{cnot}s, using the breakdown of the Fredkin gate from
Figure~\ref{fig:reversible-gates}.

\subsection{Summary}

Carry-select addition speculatively executes two additions in
parallel, one assuming a carry in of zero, and one assuming a carry in
of one.  After completion of the addition, when the input carry
becomes available, one result is chosen and the other discarded, in
direct analog to the speculative execution of instructions in modern
microprocessors.  The basic concept of a carry-select addition process
is a flexible framework allowing different choices of group size,
inner adder type, and multiplexer structure.  This structure can even,
in theory, be applied to other operations besides addition, by using
the general circuit in Figure~\ref{fig:inplace-mux}.  The adders I
have designed have latency of $O(\log n)$ or $O(\sqrt{n})$ to add two
$n$-bit numbers, when evaluated for the abstract \ac architecture.  We
turn next to the mapping of these and other algorithms to specific
sets of hardware constraints, primarily restrictions on the distance
of gate operands on the \ntc\ architecture.

\section[Monolithic Shor Performance]{Performance of Shor's Algorithm on a Monolithic Quantum Computer}
\label{sec:mono-shor-perf}

\begin{figure}
\centerline{\hbox{
\includegraphics[width=12cm]{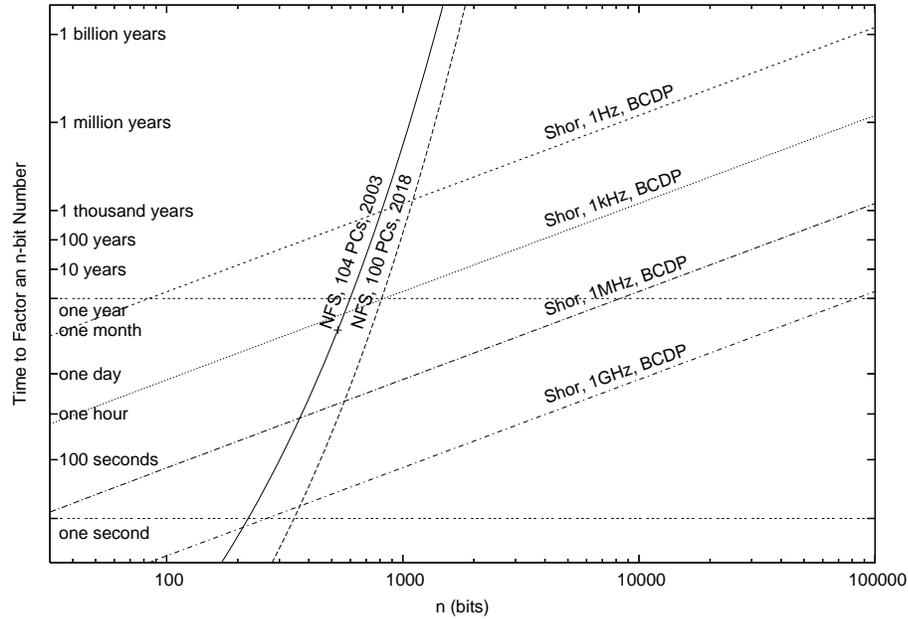}}}
\caption[Scaling of number field sieve]{Scaling of number field sieve
  (NFS) on classical computers and Shor's algorithm for factoring on a
  quantum computer, using BCDP modular exponentiation with various
  clock rates.  Both horizontal and vertical axes are log scale.  The
  horizontal axis is the size of the number being factored, in bits.}
\label{fig:scaling-bcdp}
\end{figure}

In Chapter~\ref{ch:shor}, particularly Figure~\ref{fig:scaling-nfs}
and Section~\ref{sec:vbe-etal}, we introduced the performance of
factoring on classical machines and quantum computers, but left that
analysis incomplete.  We know that Shor's algorithm is polynomial in
the length of the number being factored, which will be a straight line
on a log-log plot, but where should it fall on the graph?  We were
missing a key piece of information, namely, the logical clock speed of
the quantum system, as discussed in Section~\ref{sec:algo-topo}.  A
comparison of the execution time to factor a number on classical and
quantum computers is shown in Figure~\ref{fig:scaling-bcdp}.  It
compares the performance of Shor's algorithm on a quantum computer
using the Beckman-Chari-Devabhaktuni-Preskill (BCDP) modular
exponentiation algorithm~\cite{beckman96:eff-net-quant-fact} to
classical computers running the general Number Field Sieve.  The steep
curves are for NFS on a set of classical computers.  The shallower
curves on the figure are predictions of the performance of a quantum
computer running Shor's algorithm, using the BCDP modular
exponentiation routine, which uses $5n$ qubits to factor an $n$-bit
number, requiring $\sim 54n^3$ gate times to run the algorithm on
large numbers.  The four curves are for different logical clock rates
from 1~Hz to 1~GHz.  The performance scales linearly with clock speed.
Factoring a 576-bit number in one month of calendar time requires a
clock rate of 4~kHz.  A 1~MHz clock will solve the problem in about
three hours.  If the clock rate is only 1~Hz, the same factoring
problem will take more than three hundred years.

The execution time shown in Figure~\ref{fig:scaling-bcdp} can be
improved by understanding that relationship of architecture and
algorithm.  The performance of the VBE and BCDP carry-ripple adders,
and by extension their entire modular exponentiation algorithms, is
almost independent of architecture.  Carry-ripple adders, which use
only nearby qubits during their execution, do not take advantage of
long-distance gates even when the architecture supports them, so any
architectural analysis based solely on these algorithms is likely to
conclude that long-distance gates are not useful.  However, the
performance of most polynomial-time algorithms, including other types
of adder, varies noticeably depending on the system architecture.

\subsection{Mapping Adders to Architectures}
\label{sec:map-add}

Figures~\ref{fig:vbe8} and \ref{fig:look8} on pages~\pageref{fig:vbe8}
and \pageref{fig:look8} showed two types of quantum adder circuits,
the Vedral-Barenco-Ekert (VBE) carry-ripple
adder~\cite{vedral:quant-arith} and the Draper-Kutin-Rains-Svore
carry-lookahead adder~\cite{draper04:quant-carry-lookahead}.  The
first, most obvious difference between the two is how ``busy'' the
diagrams appear.  The carry-ripple adder shows that most of the qubits
sit idle during most of the computation, waiting for the carry to
ripple across the circuit (and back, as a cleanup operation).  The
carry-lookahead adder is much denser, accomplishing its work in fewer
time steps by executing more gates in parallel.

The second most prominent visual difference is the span of the gates
(vertical line segments).  Carry-ripple adders operate only on qubits
that are nearby, while the carry-lookahead adder leapfrogs long
distances.  This gives the carry-ripple adder $O(n)$ latency, compared
to $O(\log n)$ for the carry-lookahead --- if long-distance gates are
supported.

\begin{figure*}
\centerline{\hbox{
\includegraphics[width=14cm]{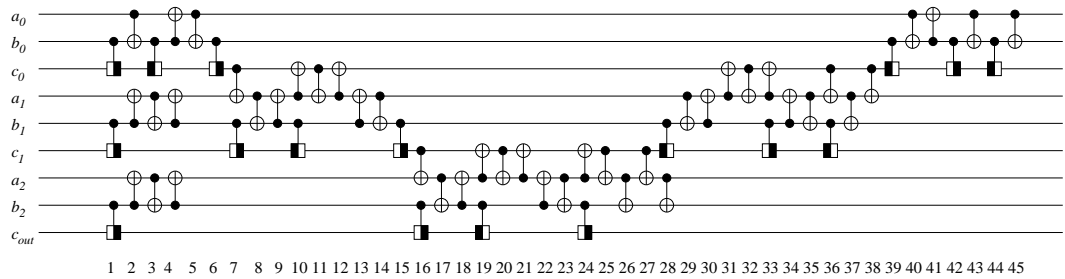}}}
\caption[Optimized, concurrent three bit VBE ADDER for
  \ntc]{Optimized, concurrent three bit VBE ADDER for the \ntc\
  architecture.  Numbers across the bottom are time steps.}
\label{fig:nmr-adder}
\end{figure*}

Figure~\ref{fig:nmr-adder} shows a fully optimized, concurrent, but
otherwise unmodified version of the VBE ADDER for three bits on a
neighbor-only machine (\ntc\ architecture).  The latency is
\begin{equation}
t_{ADD}^{NTC} = (20n-15;0)\#(2;\thinspace{}3n+1)
\label{eq:add-ntc}
\end{equation}
or 45 gate times for the three-bit adder.  A 128-bit adder will have a
latency of $(2545;0)$.  The diagram shows a concurrency level of
three, but simple adjustment of execution time slots can limit that to
two for any $n$, with no latency penalty.

\begin{table}
\centerline{
\begin{tabular}{|l|p{0.85in}|c|c|} \hline
technology & adder & conc. & latency \\ 
\hline
Si NMR 	& carry-ripple & 2 & $O(n)$ \\
solution NMR 	& carry-ripple & 2 & $O(n)$ \\
1-D quantum dot	& carry-ripple, Fourier & 2 or $n$ & $O(n)$ \\
1-D JJ charge 	& carry-ripple, Fourier & 2 or $n$ & $O(n)$ \\
1-D Kane model 	& carry-ripple, Fourier & 2 or $n$ & $O(n)$ \\
scalable ion trap 	& carry-lookahead, conditional-sum & $n$ or $2n$ & $O(\log n)$ \\
Oskin lattice 	& carry-lookahead, conditional-sum & $n$ or $2n$ & $O(\sqrt{n})$ \\
all-optical 	& carry-lookahead, conditional-sum & $n$ or $2n$ & $O(\log n)$ \\
\hline
\end{tabular}
}
\caption[Qubit technologies and recommended choice of adder]{Qubit
technologies and recommended choice of adder. conc., required
application-level concurrency}
\label{tab:adder-choice}
\end{table}

Table~\ref{tab:adder-choice} lists recommendations for adders that
match various technologies.  For example, the Fourier
adder~\cite{draper00:quant-addition} uses only $2n$ space, compared to
the $3n$ of standard carry-ripple
adders~\cite{vedral:quant-arith,beckman96:eff-net-quant-fact}.
Unfortunately, it requires $n$ concurrent gates to achieve the $O(n)$
time bound when performing the quantum Fourier transform (QFT)
required to move numbers into and out of the Fourier representation,
compared to concurrency of 2 for carry-ripple.  The Fourier adder also
requires precise rotations similar to those in the QFT, which may be
hard to implement accurately.  The newly designed CDKM carry-ripple
adder (Section~\ref{sec:cdkm}) uses only $2n$ space and small
concurrency, making it now the preferred choice in many
cases~\cite{cuccaro04:new-quant-ripple}.

Likewise, some entries recommend both the conditional-sum and
carry-lookahead adders, which have almost identical $O(\log n)$
latencies.  A conditional-sum adder requires more space and
concurrency than carry-lookahead.  However, it has different locality
characteristics which might make it map better to an irregular
architecture.

Irregular architectures, or those with regular but more complex
layouts, complicate the analysis.  In particular, the scalable ion
trap has limited concurrency, but the distance an ion must move may
have a factor of two or more performance impact, making locality
desirable.  Although the design of such a system is not yet advanced
enough to definitively choose between the two proposed types of
adders, Thaker et al. have begun analyzing the performance of the
carry-lookahead adder on one proposed system~\cite{thaker06:_cqla}.
In their analysis, the carry-lookahead adder is limited in performance
by available application-level concurrency, leading us to suggest that
the CDKM carry-ripple adder may provide similar performance while
using fewer qubits.  For the two-dimensional layout of the Kane
lattice, an ideal $O(\log n)$ adder can reach latency of only
$O(\sqrt{n})$ due to the communications cost of moving qubits.

For the Josephson-junction qubits, we recommend using long-distance
inductive or capacitive transfer structures only if concurrent
operations can be preserved for at least some qubits.  Alternating
cycles of a single long-distance interaction and many nearest-neighbor
interactions would be adequate.  Designs in which only some of the
qubits can transfer long distances while others execute concurrent
nearest-neighbor operations seem physically plausible, and would
result in intermediate performance, possibly using a carry-select or
conditional-sum adder.  Concrete performance analysis will depend on
the details of such a heterogeneous architecture.  Vartiainen has done
some analysis on such a structure~\cite{vartiainen04:_implem_josep}.

The common format of circuit diagram abstracts away the physical
layout of qubits, and for any layout other than linear nearest
neighbor, gives the wrong impression of ``nearby''.  Therefore, we
have begun animating the action of some circuits for more complex
topologies~\cite{van-meter:arith-webpage}.

\subsection{Acceleration}

This section presents an engineering tradeoff analysis of
parallelizing the multiplication steps, an improved modulo arithmetic
method, and a brief analysis of the indirection method of
Section~\ref{sec:trading}, in the context of Shor's algorithm.

\subsubsection{Concurrent Exponentiation}
\label{sec:conc-exp}

In Section~\ref{sec:modexp}, we discussed Cleve and Watrous' method
for parallelizing multiplication, as shown in Figure~\ref{fig:pmodexp}
on page~\pageref{fig:pmodexp}.  For $s$ multipliers, $s \le n$, each
multiplier must combine $r = \lfloor (2n+1)/s \rfloor$ or $r+1$
numbers, using $r-1$ or $r$ multiplications (the first number being
simply set into the running product register), where $\lfloor
x\rfloor$ indicates the largest integer not larger than $x$.  The
intermediate results from the multipliers are combined using $\lceil
\log_2 s\rceil$ quantum-quantum multiplication steps.

For a parallel version of VBE, the exact latency, including cases
where $rs \neq 2n + 1$, is
\begin{eqnarray}
\label{eq:vmults}
R_{V} &=&
2r  + 1 + \lceil \log_2 (\lceil(s- 2n - 1 + rs)/4\rceil\nonumber\\
&& + 2n + 1 - rs)\rceil
\end{eqnarray}
times the latency of our multiplier.  For small $s$, this is $O(n)$;
for larger $s$,
\begin{equation}
\lim_{s\rightarrow n} O(n/s + \log s) = O(\log n)
\end{equation}

\subsubsection{Reducing the Cost of Modulo Operations}
\label{sec:cut-mod}

The VBE algorithm does a trial subtraction of $N$ in each modulo
addition block; if that underflows, $N$ is added back in to the total.
This accounts for two of the five ADDER blocks and much of the extra
logic to compose a modulo adder.  The last two of the five blocks are
required to undo the overflow bit.

\begin{figure}
\centerline{\hbox{
\includegraphics[width=8.6cm]{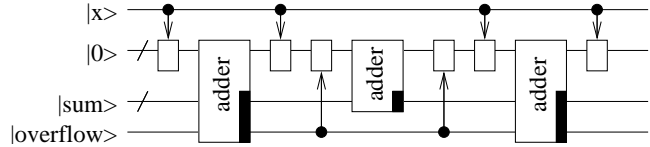}}}
\caption[More efficient modulo adder]{More efficient modulo adder.
  The blocks with arrows set the register contents based on the value
  of the control line.  The position of the black block indicates the
  running sum in our output.}
\label{fig:ema2}
\end{figure}

Figure~\ref{fig:ema2} shows a more efficient modulo adder than VBE,
based partly on ideas from BCDP and Gossett.  It requires only three
adder blocks, compared to five for VBE, to do one modulo addition.
The first adder adds $x^j$ to our running sum.  The second
conditionally adds $2^n-x^j-N$ or $2^n-x^j$, depending on the value of
the overflow bit, {\em without} affecting the overflow bit, arranging
it so that the third addition of $x^j$ will overflow and clear the
overflow bit if necessary.  The blocks pointed to by arrows are the
addend register, whose value is set depending on the control lines.
Figure~\ref{fig:ema2} uses $n$ fewer bits than VBE's modulo
arithmetic, as it does not require a register to hold $N$.

In a slightly different fashion, we can improve the performance of
VBE by adding a number of qubits, $p$, to our result register, and
postponing the modulo operation until later.  This works as long as we
don't allow the result register to overflow; we have a redundant
representation of modulo $N$ values, but that is not a problem at this
stage of the computation.

The largest number that doesn't overflow for $p$ extra qubits is
$2^{n+p}-1$; the largest number that doesn't result in subtraction is
$2^{n+p-1}-1$.  We want to guarantee that we always clear that
high-order bit, so if we subtract $bN$, the most iterations we can go
before the next subtraction is $b$.  The largest multiple of $N$ we
can subtract is $\lfloor2^{n+p-1}/N\rfloor$.  Since $2^{n-1} < N <
2^n$, the largest $b$ we can allow is, in general, $2^{p-1}$.  To
perform $b$ modular additions requires $2b+1$ ADDER calls.  For
example, adding three qubits, $p = 3$, allows $b = 4$, reducing the 20
ADDER calls VBE uses for four additions to 9 ADDER calls, a 55\%
performance improvement.

We must use $3p$ adder calls at the end of the calculation to perform
our final modulo operation.  As $p$ grows larger, the cost of the
adjustment at the end of the calculation also grows and the additional
gains are small.  Calculations suggest that $p$ of up to 10 or 11
continues to improve in speed.

This approach almost eliminates the penalty for doing modulo
arithmetic instead of ordinary integer arithmetic.  The number of
calls to our adder block necessary to make an $n$-bit modulo
multiplier is reduced from the $5n$ in VBE to $3n$ using
Figure~\ref{fig:ema2} to
\begin{equation}
\label{eq:adder-calls-for-modulo}
R_{M} = n(2b+1)/b
\end{equation}
for the overflow approach described in these last few paragraphs; this
last expression is only slightly above two adder calls per modulo
addition for reasonable values of $b$.

\subsubsection{Indirection}
\label{sec:modexp-indirect}

Adapting equation~\ref{eq:vmults} to both indirection and concurrent
multiplication, we have a total latency for our circuit, in multiplier
calls, of
\begin{equation}
\label{eq:imults}
R_{I} = 
2r + 1 + \lceil \log_2 (\lceil(s- 2n - 1 + rs)/4\rceil + 2n + 1 - rs)\rceil
\end{equation}
where $r = \lfloor \lceil (2n+1)/w\rceil / s \rfloor$.

\subsection{Example: Exponentiating a 128-bit Number}
\label{sec:example}


In this section, we combine these techniques into complete algorithms
and examine the performance of modular exponentiation of a 128-bit
number.  We assume the primary engineering constraint is the available
number of qubits.  In Section~\ref{sec:conc-exp} we showed that using
twice as much space can almost double our speed, essentially linearly
until the log term begins to kick in.  Thus, in managing space
tradeoffs, this will be our standard: any technique that raises
performance by more than a factor of $c$ in exchange for $c$ times as
much space will be used preferentially to parallel multiplication.
Carry-select adders (Sec.~\ref{sec:csla}) easily meet this criterion,
being perhaps six times faster for less than twice the space.

Because we are interested in systems with some realistic limitations,
in this section we have chosen to limit the space available to $100n$
qubits.  This is a large enough number to see the effects of
parallelism, but small enough to constrain the behavior of the
algorithm somewhat.  In later sections, we will relax this space
restriction to $2n^2$ qubits, the maximum number we have found to be
useful.

Algorithm {\bf D} uses $100n$ space and our conditional-sum adder
$CSUM$.  Algorithm {\bf E} uses $100n$ space and the carry-lookahead
adder $QCLA$.  Algorithms {\bf F} and {\bf G} use the Cuccaro adder
and $100n$ and minimal space, respectively.  Parameters for these
algorithms are shown in Table~\ref{tab:algorithms}.  We have included
detailed equations for concurrent VBE and {\bf D} below, and numeric
results for all of the algorithms in Table~\ref{tab:128-gate-count};
the detailed equations for the other algorithms are easily derived in
a similar fashion.  The performance ratios are based only on the
\textsc{ccnot} gate count for \ac, and only on the \textsc{cnot} gate
count for \ntc.

\begin{table*}
\centerline{
\begin{tabular}{|l|r|r|r|r|r|r|} \hline
algo. & adder & modulo & indirect & $s$ & space & concurrency \\ \hline
cVBE & VBE & VBE & N/A & 1 & 897 & 2 \\
{\bf D}
& CSUM($m = 4$) & $p=11, b=1024$ & $w=2$ & 12 & 11969 & $126\times 12=1512$	\\
{\bf E}
& QCLA & $p=10, b=512$ & $w=2$ & 16 & 12657 & $128\times 16 = 2048$	\\
{\bf F}
& CDKM & $p=10, b=512$ & $w=4$ & 20 & 11077 & $20\times 2 = 40$	\\
{\bf G}
& CDKM & fig.~\ref{fig:ema2} & $w=4$ & 1 & 660 & $2$	\\
\hline
\end{tabular}
}
\caption[Parameters for our algorithms, chosen for 128
  bits]{Parameters for our algorithms, chosen for 128 bits.  $s$,
  number of independent multiplier units.}
\label{tab:algorithms}
\end{table*}

\begin{table*}
\centerline{
\begin{tabular}{|l|r|r|r|r|r|} \hline
algo.	  	 & \multicolumn{2}{c|}{\ac} & \multicolumn{2}{c|}{\ntc} \\
 \hline
& gates & perf. & gates & perf. \\
\hline
cVBE &
 $(1.25\times 10^{8};\thinspace{}8.27\times
 10^{7};\thinspace{}0.00\times 10^{0})$ & 1.0 &
 $(8.32\times 10^{8};\thinspace{}0.00\times 10^{0})$      & 1.0 \\
{\bf D}             &
 $(2.19\times 10^{5};\thinspace{}2.57\times
 10^{4};\thinspace{}1.67\times 10^{5})$ & 570 &
 N/A    & N/A \\
{\bf E}             &
 $(1.71\times 10^{5};\thinspace{}1.96\times
 10^{4};\thinspace{}2.93\times 10^{4})$ & 727 &
 N/A    & N/A \\
{\bf F}             &
 $(7.84\times 10^{5};\thinspace{}1.30\times
 10^{4};\thinspace{}4.10\times 10^{4})$ & 159 &
 $(4.11\times 10^{6};\thinspace{}4.10\times 10^{4})$      & 203 \\
{\bf G}             &
 $(1.50\times 10^{7};\thinspace{}2.48\times
 10^{5};\thinspace{}7.93\times 10^{5})$ & 8.3 &
 $(7.87\times 10^{7};\thinspace{}7.93\times 10^{5})$      & 10.6 \\
\hline
\end{tabular}
}
\caption[Latency to factor a 128-bit number]{Latency to factor a
  128-bit number for various architectures and choices of algorithm.
  \ac, abstract concurrent architecture. \ntc\ neighbor-only,
  two-qubit gate, concurrent architecture.  perf, performance relative
  to VBE algorithm for that architecture, based on \textsc{ccnot}s for \ac
  and \textsc{cnot}s for \ntc.}
\label{tab:128-gate-count}
\end{table*}

\subsubsection{Concurrent VBE}

On \ac, the concurrent VBE ADDER is $(3n-3; 2n-3; 0) = (381;253;0)$
for 128 bits.  This is the value we use in the concurrent VBE line in
Table~\ref{tab:128-gate-count}.  This will serve as our best baseline
time for comparing the effectiveness of more drastic algorithmic
surgery.

The unmodified full VBE modular exponentiation algorithm, consists of
$20n^2-5n = 327040$ ADDER calls plus minor additional logic.  A
128-bit VBE adder, from Equation~\ref{eq:add-ntc}, will have a latency
of $(2545;0)$.  This gives a total latency of
\begin{eqnarray}
t_{V}^{NTC} &=& (20n^2-5n)t_{ADD}^{NTC} \nonumber\\
&=& (400n^3-400n^2+75n;\thinspace{}0)
\end{eqnarray}
for VBE.

\subsubsection{Algorithm {\bf D}}

The overall structure of algorithm {\bf D} is similar to VBE, with our
conditional-sum adders instead of the VBE carry-ripple, and our
improvements in indirection and modulo.  As we do not consider CSUM
to be a good candidate for an algorithm for \ntc, we evaluate only
for \ac.  Algorithm {\bf D} is the fastest algorithm for $n=8$ and
$n=16$.  The total latency is
\begin{eqnarray}
t_{D} &=& R_{I} R_{M}\nonumber\\
&& \times(t_{CSUM}+t_{ARG})\nonumber\\
&& + 3pt_{CSUM}.
\end{eqnarray}
Expanding the terms in this equation and letting $r = \lfloor \lceil
(2n+1)/w\rceil / s \rfloor$, the latency and space requirements for
algorithm {\bf D} are
\begin{eqnarray}
t_{D}^{AC} &=& 2r + 1 + \lceil \log_2 (\lceil(s- 2n - 1 +
 rs)/4\rceil\nonumber\\
&& + 2n + 1 - rs)\rceil n(2b+1)/b\nonumber\\
&& \times ((2m + 4\lceil\log_2 (g-1)\rceil + 2;\thinspace{}4;\nonumber\\
&&\thinspace{}4\lceil\log_2 (g-1)\rceil+2)
 +(4;\thinspace{}0;\thinspace{}4))\nonumber\\
&& + 3p(2m + 4\lceil\log_2 (g-1)\rceil + 2;\thinspace{}4;\nonumber\\
&&\thinspace{}4\lceil\log_2 (g-1)\rceil+2)
\end{eqnarray}
and
\begin{eqnarray}
S_{D} &=& s(S_{CSUM}\nonumber\\
&& + 2^w + 1 + p + n) + 2n + 1\nonumber\\
&=& s(7n - 3m - g + 2^w + p\nonumber\\
&& + \lceil3(g-1)/2 - 2 + (n-m)/2\rceil)\nonumber\\
&& + 2n + 1.
\end{eqnarray}

\subsubsection{Algorithm {\bf E}}

Algorithm {\bf E} uses the carry-lookahead adder QCLA in place of the
conditional-sum adder CSUM.  Although CSUM is slightly faster than
QCLA, its significantly larger space consumption means that in our
$100n$ fixed-space analysis, we can fit in 16 multipliers using QCLA,
compared to only 12 using CSUM, as listed in
Table~\ref{tab:algorithms}.  This allows the overall algorithm {\bf E}
to be 28\% faster than {\bf D} for 128 bits.

\subsubsection{Algorithms {\bf F} and {\bf G}}
\label{sec:algo-g}

The CDKM carry-rippler adder has a latency of $(10n+5;\thinspace{}0)$
for \ntc.  This is twice as fast as the VBE adder.  We use this in our
algorithms {\bf F} and {\bf G}.  Algorithm {\bf F} uses $100n$ space,
while {\bf G} is our attempt to produce the fastest algorithm possible
in the minimum amount of space.

\subsubsection{Smaller $n$ and Different Space}

\begin{figure}
\centerline{\hbox{
\includegraphics[width=12cm]{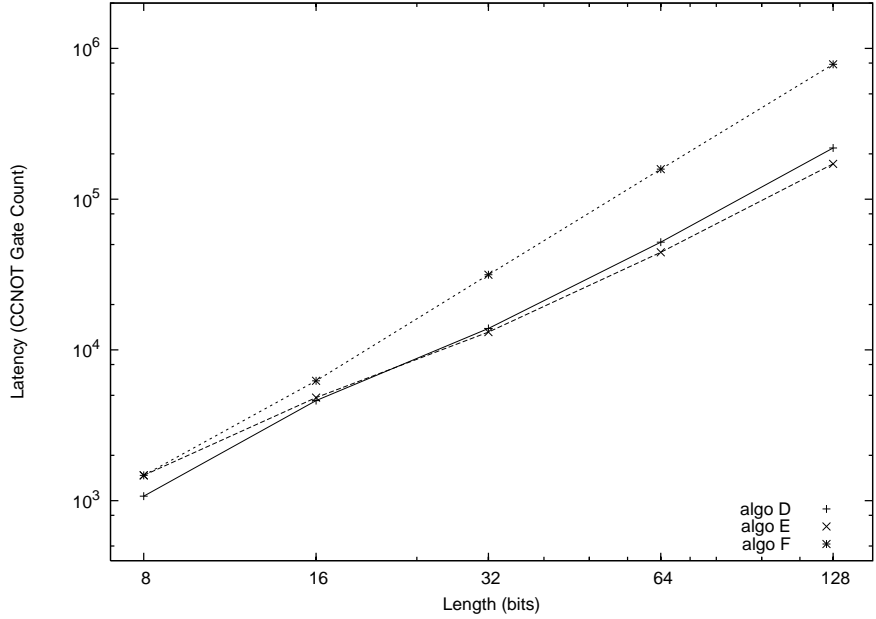}}}
\caption[Execution time for our algorithms for space $100n$]{Execution
  time for our algorithms for space $100n$ on the \ac\
  architecture, for varying value of $n$.}
\label{fig:exp-graph}
\end{figure}

Figure~\ref{fig:exp-graph} shows the execution times of our three
fastest algorithms for $n$ from eight to 128 bits.  Algorithm {\bf D},
using CSUM, is the fastest for eight and 16 bits, while {\bf E},
using QCLA, is fastest for larger values.  The latency of 1072 for $n
= 8$ bits is 32 times faster than concurrent VBE, achieved with $60n =
480$ qubits of space.

\begin{figure}
\centerline{\hbox{
\includegraphics[width=12cm]{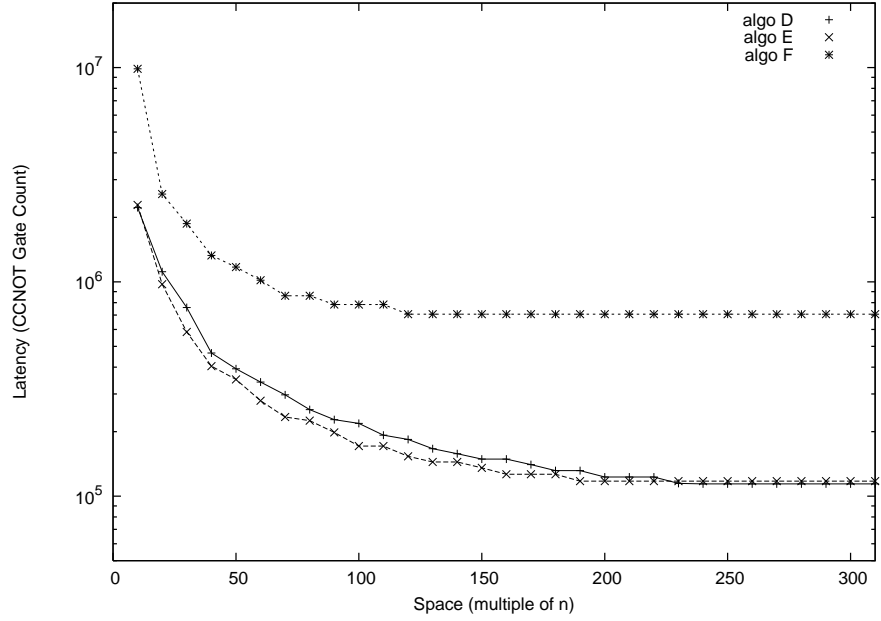}}}
\caption[Execution time v. space for our algorithms for 128
  bits]{Execution time for our algorithms for 128 bits on the \ac\
  architecture, for varying multiples of $n$ space available.}
\label{fig:space-graph}
\end{figure}

Figure~\ref{fig:space-graph} shows the execution times for $n = 128$
bits for various amounts of available space.  All of our algorithms
have reached a minimum by $240n$ space (roughly $1.9n^2$).

\subsection{Asymptotic Behavior}

The focus of this dissertation is the constant factors in modular
exponentiation for important problem sizes (up to a thousand bits or
so) and architectural characteristics.  However, let us look briefly
at the asymptotic behavior of our circuit depth, which will tell us
about the behavior of systems on very large problems.  As we have
mentioned before, the arbitrary-distance \ac\ model is not
physically realistic for very large systems; likewise, no one would
propose carrying \ntc\ to its extreme and building a
one-dimensional line of a million or more qubits.  Therefore, these
expressions should be treated as ``not to exceed'' upper and lower
bounds.

In Section~\ref{sec:conc-exp}, we showed that the latency of our
complete algorithm is
\begin{equation}
\label{eq:expcost}
 O(n/s + \log s)\times\text{(latency of multiplication)}
\end{equation}
as we parallelize the multiplication using $s$ multiplier blocks.  Our
multiplication algorithm is still
\begin{equation}
\label{eq:multcost}
 O(n)\times\text{(latency of addition)}.
\end{equation}

Algorithms {\bf D} and {\bf E} both use an $O(\log n)$-depth adder.
Combining equations \ref{eq:expcost} and \ref{eq:multcost} with the
adder cost, we have asymptotic circuit depth of
\begin{equation}
t_{D}^{AC} = t_{E}^{AC} = O((n\log n)(n/s + \log s))
\end{equation}
 for algorithms {\bf D} and {\bf E}.  As $s\rightarrow n$, these
approach $O(n \log^2 n)$ and space consumed approaches $O(n^2)$.

Algorithm {\bf F} uses an $O(n)$ adder, whose asymptotic behavior is
the same on both \ac\ and \ntc, giving
\begin{equation}
t_{F}^{AC} = t_{F}^{NTC} = O((n^2)(n/s + \log s))
\end{equation}
approaching $O(n^2 \log n)$ as space consumed approaches $O(n^2)$.

These results compare favorably to the asymptotic behavior of $O(n^3)$
for VBE, BCDP, and algorithm {\bf G}, each of which uses $O(n)$ space.
The asymptotic behavior of these three algorithms is independent of
whether the architecture is \ac\ or \ntc.

The ultimate limit of performance for \ac\ will be achieved using a
Gossett carry-save multiplier and large $s$.  The carry-save
multiplier consumes $O(n^2)$ space.  Gossett has shown that the
latency of a carry-save multiplier will be $O(\log n)$, using a tree
structure to combine partial results, and the latency of the entire
modular exponentiation algorithm will be $O(n\log n)$.  Parallelizing
the multiplication raises the space consumed to $O(n^3)$ and reduces
the latency to $O(\log^3 n)$.  The requirement for $n^3$ qubits
quickly moves into the billions as $n$ nears one thousand, and into
the trillions as $n$ nears ten thousand; none of the proposed
technologies we know of are likely to reach such levels of
scalability, though it is possible that nanotechnology will eventually
reach levels in which large numbers of individual atoms in bulk
materials are controllable.

For physically realizable systems, as we noted in
Section~\ref{sec:adder-ultimate}, an adder will ultimately be limited
to $O(\sqrt[3]{n})$ when $O(n)$ qubits are packed in three-dimensional
space, because all signal propagation methods are limited to be linear
in distance, and are subject to the final limit of the speed of light.
The complete modular exponentiation algorithm, using $O(n^2)$ adders
calls, is therefore limited to $O(n^2\sqrt[3]{n}) = O(n^{7/3})$
latency when using $O(n)$ qubits and a nominally $O(\log n)$ adder.
When using $O(n^2)$ qubits, the performance limit is $O(n^{5/3})$.
When using $O(n^3)$ qubits, the distance across the entire ensemble is
$O(n)$, and this turns out to be the limit of our performance, too.

Thus, we can say that modular exponentiation is ultimately limited to
$O(n)$ performance, where $n$ is limited only by the size (and age) of
the Universe and the availability of matter (or energy) to implement
the qubits.

\subsection{Results}

In this section, we extend our results by expanding the qubit space
available, and, at last, bringing clock speed into the picture.  On
the \ac\ architecture, our algorithms have shown a speed-up factor
ranging from 4,000 times for factoring a 576-bit number to nearly one
million for a 100,000-bit number, when using $100n$ space.  This is
about fifteen times the space consumption of the original VBE
algorithm, at $7n$, and twenty times the space of BCDP, at $5n$.
Using BCDP as our baseline, we compare the {\bf D} and {\bf F}
algorithms, with {\bf D} being the fastest algorithm on \ac\ and
{\bf F} being the fastest on \ntc.  The values reported here for
both algorithms are calculated using $2n^2$ qubits of storage to
exponentiate an $n$-bit number, the largest number of qubits our
algorithms can effectively use.  Algorithm {\bf D} with $2n^2$ qubits
on \ac\ is 13,000 times faster than BCDP at factoring a 576-bit
number, and one million times faster for a 6,000 bit number.
Algorithm {\bf F} on \ntc, by contrast, is only about 1,000
times faster than BCDP at factoring a 6,000-bit number.  For very
large $n$, the latency of {\bf D} is $\sim 9n\log_2^2(n)$.  The
latency of {\bf F} is $\sim 20n^2\log_2(n)$.

\begin{figure}
\centerline{\hbox{
\includegraphics[width=12cm]{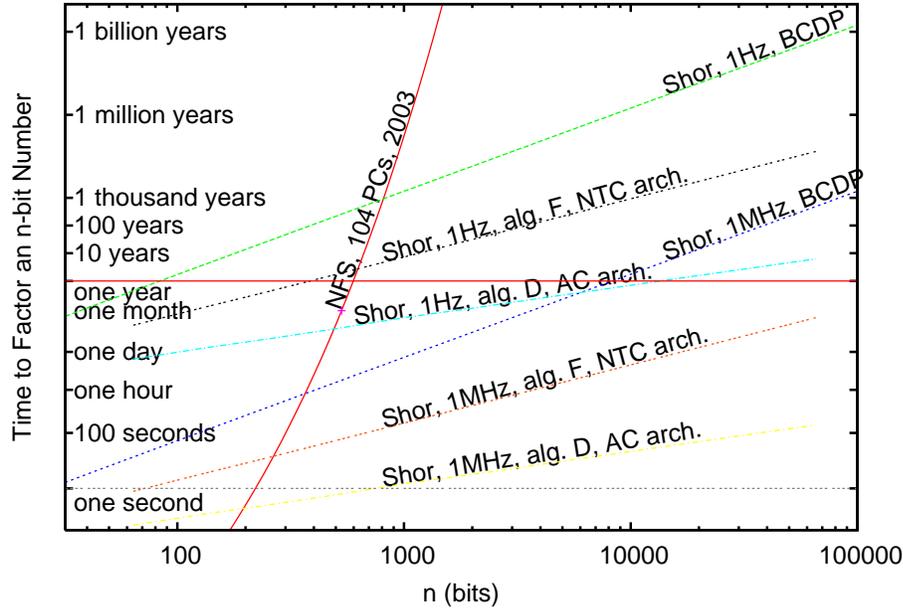}}}
\caption[Scaling of number field sieve (NFS) and Shor's
algorithms]{Scaling of number field sieve (NFS) and Shor's algorithms
for factoring, using faster modular exponentiation algorithms and
$2n^2$ space.}
\label{fig:scaling-fast}
\end{figure}

Figure~\ref{fig:scaling-fast} updates the performance shown in
Figure~\ref{fig:scaling-bcdp} on page~\pageref{fig:scaling-bcdp},
adding our fastest algorithms.  We have kept the 1~Hz and 1~MHz lines
for BCDP, and added matching lines for our fastest algorithms on the
\ac\ and \ntc\ architectures at the same clock speeds.  These speeds
are, of course, logical clock speeds, after accounting for the
overhead of fault tolerance and QEC.  The clock speed is for Toffoli
gates for BCDP and {\bf D}, and for two-qubit gates for {\bf F}.  For
\ac, our algorithm {\bf D} requires a clock rate of only about 0.3~Hz
to factor a 576-bit number in one month.  For \ntc, using our
algorithm {\bf F}, a clock rate of around 27~Hz is necessary.


\subsection{Error Correction Needs}

We saw in Section~\ref{sec:steane-models} that we can estimate the
required strength of error correction, roughly, by calculating $KQ$,
where $K$ is the number of qubits and $Q$ is the number of time steps.
$KQ$ represents the number of QEC cycles that must be performed
throughout the entire system during the course of the complete
computation.  This approach is predicated on the observation that QEC
consumes such a large percentage of the total operations in the system
that the effects of the logical gates are unimportant for this
analysis.  Steane's analysis treats $KQ$ somewhat abstractly; here we
show that $K$ varies over the course of the execution of an
algorithm~\cite{steane02:ft-qec-overhead}.

A carry-ripple adder to add two $n$-qubit numbers, whether VBE or
CDKM, uses $O(n)$ qubits and takes $O(n)$ time steps, giving a $KQ =
O(n^2)$.  The carry-lookahead and conditional-sum adders likewise use
$O(n)$ qubits, but run in $O(\log n)$ time steps, for $KQ = O(n\log
n)$.  Table~\ref{tab:adder-kq} shows approximate values of $KQ$ for
the different adders.  For $n = 1,024$, $KQ$ is about four million for
the CDKM adder, but only 160,000 for the conditional-sum adder, a
factor of twenty five better.  Of course, this analysis assumes the
\ac\ architecture's support for long-distance gates.  Thus, we see
that not only does \ac\ have a better error threshold, but the demands
of the application are lower.  This factor will result in
higher-fidelity calculations, or possibly even a reduction in the
necessary strength of QEC, saving space and time.

\begin{table}
\centerline{
\begin{tabular}{|l|c|c|c|} \hline
adder & $K$ & $Q$ & $KQ$ \\ 
\hline
VBE carry-ripple & $3n$ & $3n$ & $9n^2$ \\
CDKM carry-ripple & $2n$ & $2n$ & $4n^2$ \\
conditional-sum & $6n$ & $4\log_2 n$ & $24n\log_2 n$ \\
carry-lookahead & $4n$ & $4\log_2 n$ & $16n\log_2 n$ \\
\hline
\end{tabular}
}
\caption[Approximate $KQ$ for some different adder
  circuits]{Approximate $KQ$ to add two $n$-qubit numbers using some
  different adder circuits, in units of qubit-Toffoli times.}
\label{tab:adder-kq}
\end{table}

\begin{figure}
\centerline{\hbox{
\includegraphics[width=12cm]{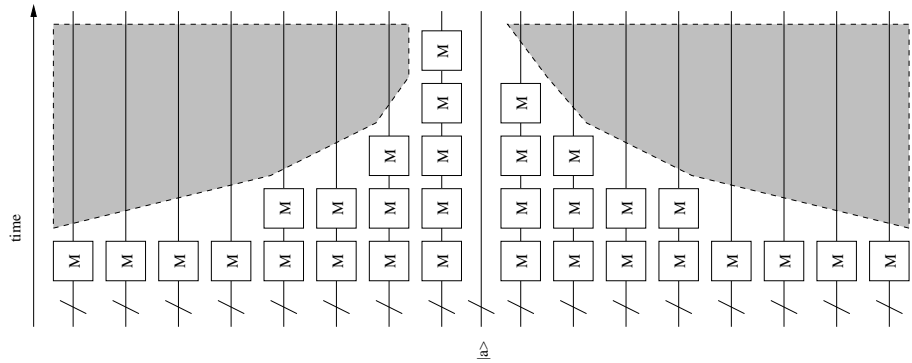}}}
\caption[Cleve-Watrous parallel multiplication]{Cleve-Watrous parallel
  multiplication (rotated ninety degrees relative to other graphs,
  with time flowing bottom to top).  Gray areas represent
  disentangled, unused qubits.}
\label{fig:cw-kq}
\end{figure}

In all of our proposed algorithms, modular multiplication consists of
$O(n)$ calls to the adder routine, giving $KQ = O(n^3)$ for a
multiplication when using carry-ripple adders and $KQ = O(n^2\log n)$
when using log-depth adders.  We have also proposed parallelizing
multiplication using the Cleve-Watrous method.  In its broadest form,
as in Figure~\ref{fig:cw-kq}, it uses $n$ multiplier units and
requires $\log_2 n$ steps.  This may appear to result in $KQ$ being
$n\log_2 n$ times the $KQ$ of a multiplication, which would be an
increase of a factor $\log_2 n$ over a simple linear string of
multiplications.  However, the gray areas in the figure are {\em
dis}entangled from the running computation.  They do not affect the
results, and should not be counted in the $KQ$ for the overall
computation.  (The unused resources ideally shouldn't go to waste, but
that's a different problem.)  Thus, regardless of the arrangement of
the multipliers, the total $KQ$ for modular exponentiation is $2n$
times the cost of a multiplier, or, when using the indirection of
Sections~\ref{sec:modexp-indirect} and~\ref{sec:trading}, $2l =
2\lceil n/w\rceil$ times the cost of a multiplier.  The one minor
complication is that our parallel multiplications keep only a single
copy of $|a\rangle$, rather than one for each multiplier unit.  For
algorithms {\bf D}, {\bf E}, {\bf F}, and {\bf G}, we ignore the cost
of the $|a\rangle$ register in the table, it being small compared to
the overall size of the system; for small values of $s$ this
approximation is not good, but the result is still within 40\% or so
at worst.  Recognizing from Equation~\ref{eq:adder-calls-for-modulo}
that even for modest values of $b$, the number of adder calls $R_{M}$
to make a modulo multiplier is $\sim 2n$, we can simplify our
expressions for $KQ$ and arrive at the values in
Table~\ref{tab:modexp-kq}.  The terms in the expressions in the table
are, in order, number of modulo multiplier calls; number of modulo
adder calls per modulo multiplier; adder calls per modulo adder; adder
depth; and first-order term in number of qubits.  Our algorithm {\bf
G} is an order of magnitude better than VBE, and {\bf F} is almost two
orders of magnitude better, on the \ntc\ architecture.  For \ac, we
can use {\bf D} and {\bf E} for further gains.  The asymptotic growth
is substantially slower; numerically, for $n = 1,024$, for VBE $KQ
\approx 2\times 10^{14}$, and {\bf E} is $\approx 2.4 \times 10^{11}$,
almost three orders of magnitude better.  All of these values are
for indirection (Section~\ref{sec:trading}) using $w = 2$ to $w = 4$,
as shown in Table~\ref{tab:algorithms}; an additional factor of 4 or
more seems quite plausible, as shown in Figure~\ref{fig:cost-plot} on
page~\pageref{fig:cost-plot}, when error correction becomes an
overriding concern.

\begin{table}
\centerline{
\begin{tabular}{|l|c|} \hline
algorithm & $KQ$ \\ 
\hline
cVBE & $2n \times n \times 5 \times 3n \times 7n = 210n^4$ \\
algo. {\bf D} & $2l \times n \times 2 \times 4\log_2 n \times 5n \approx
40n^3\log_2 n$ \\
algo. {\bf E} & $2l \times n \times 2 \times 4\log_2 n \times 3n \approx
24n^3\log_2 n$ \\
algo. {\bf F} &  $2l \times n \times 2 \times 2n \times 3n
\approx 6n^4 $ \\
algo. {\bf G} & $2l \times n \times 3 \times 2n \times 6n \approx 18n^4$ \\
\hline
\end{tabular}
}
\caption[Approximate $KQ$ for complete modular exponentiation
  circuits]{Approximate $KQ$ for our complete modular exponentiation
  circuits, in units of qubit-Toffoli times.}
\label{tab:modexp-kq}
\end{table}

Steane calculated that, for a physical gate error rate of $\sim
10^{-5}$ and a memory error rate of $\sim 10^{-6}$ on an \ac-like
architecture, $KQ$ of $10^{15}$ can be achieved using only about a
factor of twelve increase in storage, via the BCH [[127,43,13]]
code~\cite{steane02:ft-qec-overhead}\footnote{Steane uses extra
  ancillae for measurement and fault tolerance, resulting in a total
  consumption of $\sim 4n$ physical qubits to store $k$ logical qubits
  in an [[$n$,$k$,$d$]] code.}.

\section{Summary}

This chapter opened with a discussion of the performance of Shor's
algorithm on a quantum computer, showing in
Figure~\ref{fig:scaling-bcdp} that logical clock speed has an
important impact on the utility of a quantum computer, despite the
apparent gains in computational class compared to classical
computers.  This fact is often under-appreciated by physicists, who
tend to assume that the gain in class will prove decisive.

It is possible to significantly accelerate quantum modular
exponentiation using a stable of techniques, culminating in the
much-improved performance shown in Figure~\ref{fig:scaling-fast}.  I
have provided exact gate counts, rather than asymptotic behavior, for
the $n = 128$ case, showing algorithms that are faster by a factor of
200 to 700, depending on architectural features, when $100n$ qubits of
storage are available.  For $n = 1024$, this advantage grows to more
than a factor of 5,000 for non-neighbor machines (\ac).  Neighbor-only
(\ntc) machines can run algorithms such as addition in $O(n)$ time at
best, when non-neighbor machines (\ac) can achieve $O(\log{n})$
performance.

Our contribution has focused on parallelizing execution of the
arithmetic through improved adders, concurrent gate execution, and
overall algorithmic structure.  We have also made improvements that
resulted in the reduction of modulo operations, and traded some
classical for quantum computation to reduce the number of quantum
operations.  It seems likely that further improvements can be found in
the overall structure and by more closely examining the construction
of multipliers from adders~\cite{ercegovac-lang:dig-arith}.  We also
intend to pursue multipliers built from hybrid carry-save adders.

The three factors which most heavily influence performance of modular
exponentiation are, in order, concurrency, the availability of large
numbers of application-level qubits, and the topology of the
interconnection between qubits.  Without concurrency, it is of course
impossible to parallelize the execution of any algorithm.  Our
algorithms can use up to $\sim2n^2$ application-level qubits to
execute the multiplications in parallel, executing $O(n)$
multiplications in $O(\log n)$ time steps.  Finally, if any two qubits
can be operands to a quantum gate, regardless of location, the
propagation of information about the carry allows an addition to be
completed in $O(\log n)$ time steps instead of $O(n)$.  We expect that
these three factors will influence the performance of other algorithms
in similar fashion.

As we alluded to in Section~\ref{sec:map-add}, not all physically
realizable architectures map cleanly to one of our models.  A full
two-dimensional mesh, such as neutral atoms in an optical
lattice~\cite{brennen99:_optical_lattice}, and a loose trellis
topology~\cite{oskin:quantum-wires} probably fall between \ac\ and
\ntc.  The behavior of the scalable ion
trap~\cite{kielpinski:large-scale} is not immediately clear, but will
be controlled by ion movement times and realizable concurrency.

In this chapter, we have analyzed the performance of the modular
exponentiation step of Shor's factoring algorithm for some abstract
architectural models, and shown how to dramatically improve that
performance.  Depending on the post-quantum error correction,
application-level effective clock rate for a specific technology,
choice of exponentiation algorithm may be the difference between hours
of computation time and weeks, or between seconds and hours.  This
difference, in turn, feeds back into the system requirements for the
necessary strength of error correction and coherence time.  The next
chapter will develop a design for a machine we call a {\em quantum
multicomputer}, designed to run Shor's algorithm in a distributed
fashion, and show optimized forms of arithmetic to run on it.

\chapter{The Quantum Multicomputer}
\label{ch:qmc}
\section{System Overview}
\label{ch:arch-over}

\cq{The scientist describes what is; the engineer creates what never was.}
{Theodore Von K\'arm\'an}

\cq{Music is your own experience, your own thoughts, your wisdom. If you
don't live it, it won't come out of your horn. They teach you there's
a boundary line to music. But, man, there's no boundary line to art.}
{Charlie Parker}

\cq{Plan to throw one away. You will do that, anyway. Your only choice is
whether to try to sell the throwaway to customers.}{Frederick P. Brooks}

At long last, we reach the objective of our pilgrimage: the design and
analysis of a distributed quantum computer, or {\em quantum
multicomputer}.  A multicomputer is a constrained form of distributed
system~\cite{athas:multicomputer}.  It is composed of nodes, each of
which is an independent quantum computer, and an interconnect network
of links connecting the nodes.  As we noted in
Section~\ref{sec:intro-dist-qc}, distributed quantum computation
requires shared entanglement; in Yepez's terminology, our quantum
multicomputer is a type I system~\cite{yepez01:_type_ii}.  The network
is used to create EPR pairs shared between pairs of nodes, and those
EPR pairs are then used to teleport qubits (teledata) or quantum gates
(telegate).  Our goal with such a system is to increase both the {\em
storage} and {\em performance} of the total system well beyond what a
single, monolithic quantum computer is capable of; we want our
multicomputer to be {\em scalable}.  This chapter provides an overview
of the entire system, including the node and network hardware and
software.  The first section will justify our decision to explore
distributed quantum computer architectures.  Succeeding sections will
go into more detail on the impact of quantum error correction and
finally a performance analysis of adder circuits run on our system.

\section{An Engineer's Definition of Scalability}
\label{ch:scalability-defn}

What will constrain our ability to build a quantum computing system as
large as we care to attempt?  In this section, we discuss the
practical aspects of scaling up the size (in qubits) of a quantum
computer.  We also reason that technological limitations on most
proposed technologies make it necessary to plan to use multiple
machines to solve large problems, laying the foundation for our
quantum multicomputer work.

Chuang has defined scalability to mean that the combination of fault
tolerant methods and a particular technology, including its base error
rate, meet the threshold criterion.  Combinations that meet this
criterion are scalable; those that do not, are not.  However, the term
``scalable'' has different meanings in different contexts.  I am
interested in building a complete, practical quantum computing system.
In this context, Chuang's definition is a necessary, but not
sufficient, condition.  Instead, I offer the following, broader but
less formal, definition.

\begin{quotation}
Above all, it must be possible, physically and economically, to grow
the {\em system} through the region of interest.  Addition of physical
resources must raise the performance of the system by a useful amount
(for all important metrics of performance, such as calculation speed
or storage capacity), without excessive increases in negative features
(e.g., failure probability).
\end{quotation}

This definition refers to several important criteria, summarizing our
taxonomy from Chapter~\ref{ch:taxonomy}.  It also points out that
scalability is never indefinite in the real world; there are always
limits, and we must begin by deciding what those limits are.  No one
would say that a system that costs a hundred thousand dollars per
qubit or that covers an optical lab bench for each gate is scalable in
any practical sense.  Thus, good engineers say, ``This scales to...''
and name a level, metric, and what part of the system constrains the
scalability.  (Better engineers tell you why, and great engineers find
a way around the limitations.)  In this section, we provide a
qualitative look at some of these issues.

\subsection{Economics}

My estimate of the price at which the first production quantum
computer will be sold is four hundred U.S. dollars per qubit.  The
definition of ``production'' in this case is a machine that is bought
and installed for the purpose of solving real problems. That is, it
has to solve a problem for which there is not a comparable classical
solution.

To arrive at this estimate, I assume that the machine will be built to
run Shor's factoring algorithm on a 1,024-bit number. That takes about
five kilobits of application-level qubit space; we will multiply by
fifty to support two levels of QEC.  This gives a total requirement of
a quarter of a million physical qubits.

One hundred million U.S. dollars is a reasonable price for a machine
with unique capabilities.  The U.S. government clearly spends that
much on cluster supercomputers today. BlueGene, for example, built by
IBM, has 131,072 processors (65,536 dual-core chips).  Counting
packaging, power, cooling, memory, storage, and networking, the price
of such a system undoubtedly exceeds a thousand dollars per processor
(all of these prices are ignoring physical plant, including the
building).

Our price point, then, is $\$100$M$/250$K qubits $ = \$400/$qubit.
This estimate might easily be one or two orders of magnitude high or
low; other applications, such as physical simulations, may require
fewer qubits for a production machine (indeed, one estimate is that as
few as 30 qubits might be enough to be
useful~\cite{aspuru-guzik05:_simul_quant_comput_molec_energ}), or a
high error rate may demand more error correction and more physical
qubits.

The dollar cost is a real-world constraint that must be satisfied; a
large system will not get built until it justifies itself
economically.

\subsection{Infrastructure Needs}
\label{sec:infra-needs}

Each technology has its own physical infrastructure requirements.
Packaging, cooling, and housing a semiconductor-based quantum computer
may be non-trivial. Even though a quantum computer manipulates
individual quanta, the space, power, thermal, and helium budgets for
such a system are large.  In Section~\ref{sec:manufop}, we discussed
the size and cooling capacity of dilution refrigerators; this will be
one limit on the number of qubits we can support in each such dil
fridge.  For our quantum multicomputer, we plan to connect many dil
fridges together into a complete system.  We will call a setup of a
dil fridge and the electronics to support the qubits inside a ``pod''.
We will examine what constitutes a ``node'' in our multicomputer in
Section~\ref{ch:arch-over}.

Thermal engineering and packaging are serious problems.  In
Section~\ref{sec:node-hw-con}, we will discuss this issue; here we
assert that this issue will limit us to only a few logical qubits per
pod, which in turn requires us to have a large number of pods.  For
the moment, we assume one node per pod, and again set our target at a
machine for factoring a 1,024-bit number.  We must have clearance
around the dil fridge for operators and rack-mount equipment to move
equipment down the aisles.  Quite a bit of space, power, and money are
required for each such setup.  If each pod requires an area three
meters square, we need an area approximately 100 meters by 100 meters
for our total machine, a large but certainly achievable amount of
floor space.  However, growing an order of magnitude beyond this size
seems impractical.

With dilution refrigerator prices of about \$100,000 per pod, one
thousand dil fridges would consume our entire budget, leaving no money
for support electronics or the qubits themselves.  This clearly shows
that thermal engineering and packaging will be key issues in building
large-scale production systems based on quantum dot or
Josephson-junction devices; we need to fit more than one node into
each pod, or more qubits into each node.

This linear extrapolation from the current state of research is
unlikely to be the way production systems will really be
built~\footnote{It's also worth noting that NMR, ion trap, optical
lattice, and atom chip systems would require a completely different
analysis.}. However, this brief discussion should illustrate the
problems that must be solved.  Without solutions, we do not have a
system that scales to reach our desired performance target.

\subsection{Performance}

We introduced performance as an issue in quantum computing back in
Chapter~\ref{ch:shor}.  A system running an $O(n^3)$ algorithm that
requires a year to solve a problem of size $n$ is unlikely to be
considered a viable choice to solve a problem of size $10n$, even if
the hardware can be scaled to an appropriate level, as there are few
solutions for which funders and researchers are willing to wait 1,000
years.

\subsection{Single-Device Physical Limitations}

Before accepting the need to build a quantum multicomputer, we should
look at the scalability of a single, large, monolithic machine.
Thaker et al. estimated the size of an ion trap system to factor a
1,024-bit number to be about a tenth of a square meter of ion
traps~\cite{thaker06:_cqla}; a single device of this scale is
difficult to construct and operate, suggesting that smaller devices
interconnected via teleportation channels will be required.

We are most interested in VLSI-based qubits. In particular, let us
look at the superconducting Josephson-junction flux qubit from
Dr. Semba's group at NTT~\cite{kutsuzawa:_coher}. Their qubit is a
loop about 10$\mu$m square. This area is determined by the desired
physics of the device, not limited by achievable VLSI feature size;
the size of the loop determines the size of the flux quantum, which in
turn determines control frequencies and gate speed.  Dr. Semba's group
is working on connecting qubits via an LC oscillator which includes an
on-chip capacitor~\cite{johansson05:_vacuum_rabi_lc}.

Once they have demonstrated interconnection among multiple qubits
connected to the bus, will that meet DiVincenzo's criterion for a
scalable set of qubits?  In this case, we are looking for up to a
quarter of a million physical qubits. At first glance, it would seem
easy to fit that many qubits on a chip. Even a small 10mm square chip
would fit a million 10-micron square structures.  However, that
estimate ignores the need for I/O pads. Equally important, the
capacitor in the LC circuit is huge compared to a qubit (though only
one of those is required per bus that connects a modest-sized group of
qubits, and it may be possible to build the capacitor in some more
space-efficient manner, or maybe even put it off-chip). Still more
important, these qubits are magnetic, not charge; place them too close
together, and they'll interfere.  The strength of the interaction
could be a problem if the qubits are only a micron apart, but at
10$\mu$m spacing, the interaction drops to order of kHz, low enough
not to worry about much~\cite{semba05:private}. Control is achieved
with a microwave line run past the qubit; obviously, this line cannot
run that too close to other qubits.  Thus, there is a lot of physics
to be done even before the mundane engineering of floor-planning.
Above, we discussed the need for control lines to move into/out of the
dil fridge, crossing the thermal boundary.  The I/O requirement
applies directly to the chip, as well; now we need roughly a pin per
qubit. Without major advances in integration or some form of
multiplexing of control, we are probably limited to about a thousand
qubits per chip, simply because of the required pin count, and each
pin will conduct heat into the chip, affecting our overall thermal
budget.

With an estimated limit to the number of qubits of two orders of
magnitude or more below our total system requirements, we see the need
to connect multiple nodes together into a quantum multicomputer. We
need to create an entangled state that crosses node boundaries.  The
quantum I/O mechanisms discussed in Chapter~\ref{ch:taxonomy}
therefore become critical.  Having a quantum I/O mechanism allows us
to circumvent one entire set of scalability constraints.  The
governing constraints are likely to be overall ability to suppress
errors, performance, or cost.

\section{System Overview}
\subsection{Hardware Overview}

We constrain all parts of the system to be geographically collocated.
Short travel distances (up to a few tens of meters) between nodes
reduce latency, simplify coordinated control of the system, and
increase signal fidelity and reduce losses, freeing us from the need
to consider placing quantum repeaters~\cite{briegel98:_quant_repeater}
in the network.  We may wish, however, to use hardware proposed for
quantum repeaters as our local node and interconnect
technologies~\cite{childress05:_ft-quant-repeater}.

Figure~\ref{fig:multicomp-hw-block} on
page~\pageref{fig:multicomp-hw-block} showed the quantum multicomputer
architecture at a high level.  Here we deal only with the quantum
network and the nodes' interaction with it.  We choose a regular
network topology, assume a dedicated network environment, and set a
goal of scalability to thousands of nodes.  The dedicated network
assumption allows us to ignore security and contention for resources
beyond the instructions we schedule, and to assume in-order delivery
of data.  The links may be directly connected between a pair of nodes,
connected to a shared network medium, or switched at some lower
physical level.  Although the QEP protocol in theory supports EPR pair
creation over many kilometers, our design goal is a scalable quantum
computer in one location (such as a single lab).  We consider a 10
nanosecond classical communication latency, corresponding roughly to 2
meters' distance between nodes.  The performance figures found are
insensitive to this number.  The links in the multicomputer are
serial; Section~\ref{sec:qec-qubus} shows that parallel links would
have only a modest impact on performance and reliability, so we choose
to avoid the additional complexity.

We concentrate on a homogeneous node technology based on solid-state
qubits, with a qubus interconnect, though our results apply to
essentially any choice of node and interconnect technologies, such as
ion-trap nodes and single photon-based qubit transfer
interconnects~\cite{steane:ion-atom-light,wallraff04:_strong-coupling,matsukevich:matter-light-xfer}.
Each node has many qubits which are private to the node, and a few
transceiver qubits that can communicate with the outside world.  Node
size is limited by the number of elements that can practically be
built into a single device, considering control structures, external
signaling, packaging, cooling, and shielding constraints.

One or more nodes will be placed inside a dilution refrigerator, or
dil fridge.  Various rack-mount signal generators and measurement
devices, classical computing and control equipment, etc. must
accompany each node.  We will call such a setup a ``pod''.  For the
moment, we assume one node per pod.  The exact number of nodes and
qubits that can be placed in a pod will depend on volume, heat
extraction, and the cabling that must cross temperature boundaries.
This is perhaps {\em the} primary driver of system economics.  A dil
fridge includes multiple temperature stages, and different parts of
the system will be held at different levels.  The innermost,
millikelvin fridge can dissipate only a few hundred microwatts.
Unless the extraction rate of the dil fridge is raised substantially,
each transmission line crossing the inner temperature boundary is
limited to about a microwatt of thermal load, even if the device
itself dissipates no energy.

Finally, economics must be considered.  To be able to scale the system
to 1,024 nodes, we cannot exceed about US\$100,000 per node, almost
all of which will be consumed by the dil fridge if we have only one
node per pod.  Both cost and floor space can be reduced if more than
one node can be fit into a pod, but doubling or quadrupling the number
of coaxes and the heat budget is a daunting proposition on an already
extremely aggressive engineering challenge.  However, some researchers
have begun working on these problems and expect to make dramatic
improvements.  We will see in this and succeeding sections that such
progress is necessary to make the system viable.

These assumptions of a regular network topology and homogeneous nodes
will certainly hold for the first, small-scale systems that will be
built.  However, as the size of systems and our experience with them
grow, it is quite likely that a multi-stage network composed of
heterogeneous nodes will come to be the commonly-accepted
architecture.

\subsection{Node Architecture}
\label{sec:node-arch}

The basic architectural principles described in this dissertation are
largely independent of the technology on which the nodes are built.  A
node built on a semiconducting or superconducting base technology
serves as a useful model for evaluating performance.

\subsubsection{Technology-Independent Characteristics}

First, let us examine the roles each node must fulfill, regardless of
the implementation:

\begin{itemize}
\item Each node must include enough physical qubits to represent
  several logical qubits, once error correction is taken into account
  (we will vary our expectation of the exact number in later
  section).  The qubits must meet DiVincenzo's criteria, including
    adequately fast and accurate gates and measurements.
\item Each node must support one or more {\em transceiver qubits} that
  can connect to the qubus.  Because links are serial, only one
  transceiver qubit per link is required.
\item Qubus operations must be fast enough, relative to memory and
  gate times, and high enough fidelity that state transfer of logical
  qubits is possible, and basic performance constraints are met.
\item The technology and node implementation, including supporting
  equipment, must meet the physical, economic and operational
  constraints identified in Section~\ref{ch:scalability-defn}.
\end{itemize}

Expanding on the first criterion, if we assume, for the moment, that
each node contains three application-level qubits per node, and we use
one level of Steane [[7,1,3]] code and one level of [[23,1,7]] code,
then each node must contain about 500 physical qubits~\footnote{This
estimate ignores Steane's multiplier for multiple, concurrent QEC
syndrome extraction, which would raise the number by a factor of four
or so.  This factor depends on the cycle time of a measurement device,
which will be different for solid-state systems than ion traps.}.

\subsubsection{Hardware Constraints}
\label{sec:node-hw-con}

Solid-state qubits, including both semiconducting quantum dot and
superconducting Josephson junction-based devices, are operationally
challenging due to the millikelvin temperatures required and the large
number of sources of decoherence.  However, they are very attractive
for two reasons: among experimentally advanced technologies, they are
the fastest, with gate times in the low nanoseconds, and several
decades' collective experience with semiconductor design and
fabrication makes it possible that physical scalability will come more
easily to these technologies than some others, once the fundamental
hurdles of coherence and manipulation are cleared.  Josephson
junction-based devices may also support node-internal interconnects,
using various forms of resonators, that will transfer qubits long
distances and make them algorithmically more efficient.

In general, a node will be a single chip, with off-chip quantum
communication performed using the qubus protocol and teleportation.
More precisely, a node consists of the set of qubits that are under
unified control and clocking, and that can interact directly either as
neighbors or using resonator-based interconnects.  If the
communication between two qubits must be mediated by an EPR pair
created using the qubus protocol, those two qubits will be said to be
in different nodes.  Some hardware implementations may make the
boundary of a node fuzzier, using teleportation
internally~\cite{thaker06:_cqla,oskin:quantum-wires} or other methods
externally, but we will use these simplifying assumptions.

Each qubit requires certain control structures and lines; generally,
two to five signals each, including bias voltage, gate signals,
measurement devices, and qubit-qubit or qubit-resonator coupling
control.  Some of these signals can be shared among a small group of
qubits, potentially allowing an average of one to two signals per
qubit.  If the control structures remain off-chip, as is common today,
each signal requires an I/O pad and a line to the outside.  For the
chip package, ball grid array packages of more than 2,000 pins exist,
and the maximum number of package pins is predicted to reach 7,000 by
the year 2016~\cite{ITRS2005}.  At 250 qubits per chip, then, we may
not be pin-limited, though the I/O pads will still demand substantial
die space.  For a thousand qubits or more, once system demands such as
ground plane pins are met, it seems likely that packaging constraints
will come into play.  The engineering challenges of a bus consisting
of several thousand microcoaxial cables suitable to reach external
equipment are also large.  These pedestrian engineering issues suggest
that low-level qubit control must reside inside the dil fridge.  A
node may consist of several dice in a multi-chip module, or the
control structures may be integrated directly into the chip.  On-chip
demultiplexers may reduce the width of the bus to the outside world,
at the price of leaving qubits to fend for themselves for long periods
of time as control is multiplexed among a group of qubits.

This linear extrapolation from the current state of research
prototypes should be viewed as a strawman proposal demonstrating the
range of prosaic implementation problems that must be solved to build
production systems, rather than an actual proposal to implement.  It
is clear that, in addition to electrical and VLSI engineers for the
chip itself, packaging, thermal, and cabling engineers are needed to
create a production system.

\subsection{Network Topologies}
\label{sec:multi-net-topo}

\begin{figure}
\centerline{\hbox{
\includegraphics[width=15cm]{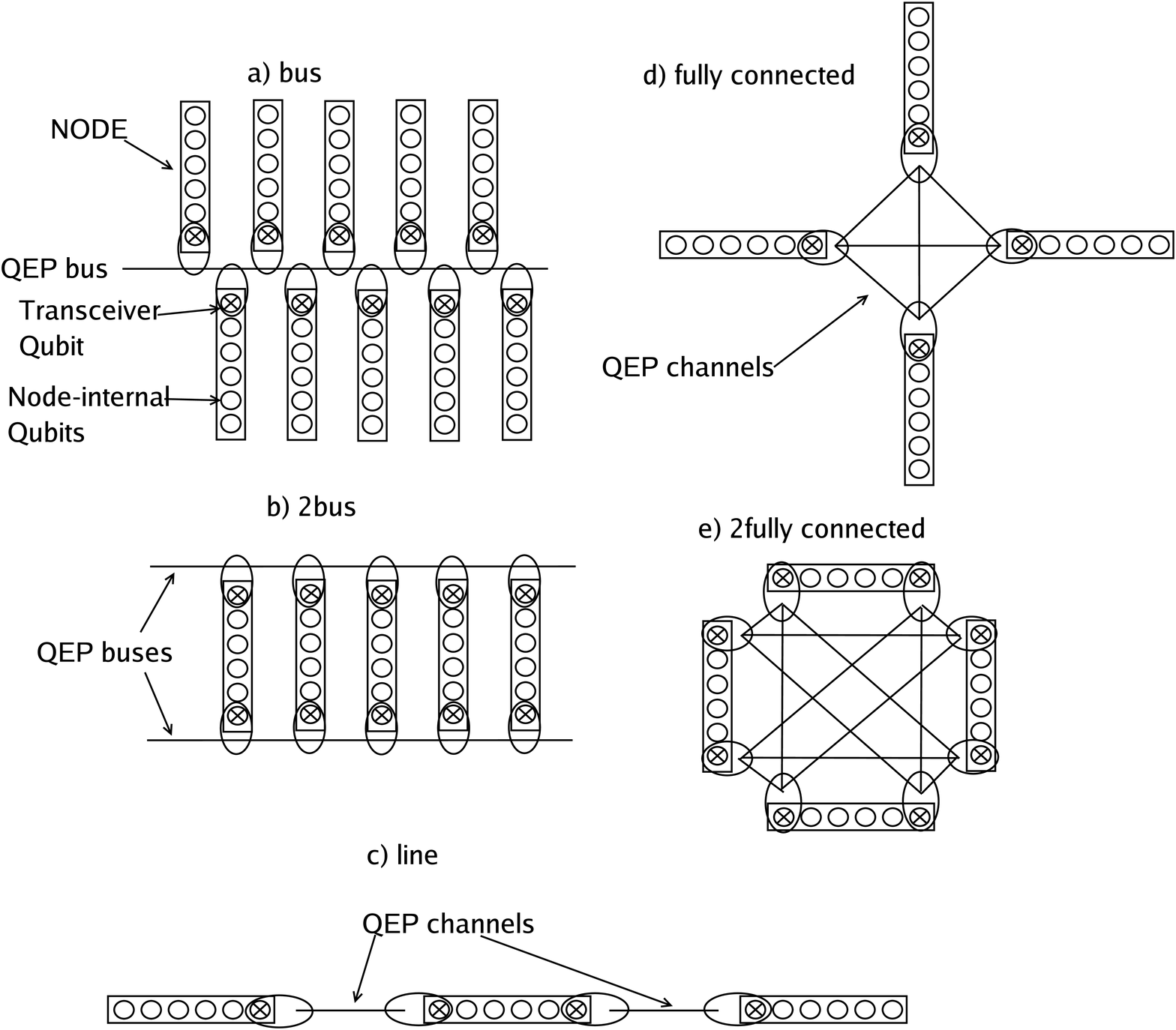}}}
\caption{The five physical topologies analyzed in this thesis.}
\label{fig:five-topos}
\end{figure}

For our proposed multicomputer, we have analyzed five network
topologies, as shown in Figure~\ref{fig:five-topos} and described in
Table~\ref{tab:topo-chars}, where the ``label'' column corresponds to
the label in the figure.  The bus, line, and fully connected
topologies were shown in Section~\ref{sec:multi-net}.  To these we
have added the 2bus and 2fully topologies.  In the 2bus and 2fully
topologies, each node is connected to two separate networks.  This set
of topologies explores whether the bottleneck in performance is the
network itself, or the ability to move data into and out of the nodes.
The network switching elements are integrated directly into the
computational nodes, except for the possibility of physical-layer
switching in the fully-connected networks.  There are no
store-and-forward routers or other intelligent elements in the
network.

For the shared bus, all nodes are connected to a single bus.  Any two
nodes may use the bus to communicate, but it supports only a single
transaction at a time.  In the line topology, each node uses two
transceiver qubits, one to connect to its left-hand neighbor and one
to connect to its right-hand neighbor.  Each link operates
independently, and all links can be utilized at the same time,
depending on the algorithm.  For the fully-connected network, a full
set of links creating a true fully-connected network would require
$n-1$ transceiver qubits at each node; obviously this number is
impractical.  We assume that each node has only a single transceiver
qubit, and that it can connect to any other node without penalty via
some form of classical, switched network such as a micromirror
device~\cite{aksyuk03:_microm_optic_cross_connec}.  Each transceiver
qubit can be involved in only one transaction at a time.  2bus and
2fully utilize two transceiver qubits per node for concurrent
transfers.

The effective topology may be different from the physical topology,
depending on the details of a bus transaction.  For example, even if
the physical topology is a bus, the system may behave as if it is
fully connected if the actions {\em internal} to a node to complete a
bus transaction are much longer than the activities on the bus itself,
allowing the bus to be reallocated quickly to another transaction.
Some technologies may support frequency division multiplexing on the
bus, allowing multiple concurrent transactions.

\begin{table*}
\centerline{
\begin{tabular}{|l|l|c|c|c|c|c|}\hline
label & name & degree & diameter & avg. dist. & bisection & total \\
\hline
a & bus & 1 & 1 & 1 & 1 & 1 \\
b & 2bus & 2 & 1 & 1 & 2 & 2 \\
c & line & 2 & $n-1$ & $(n+1)/3$ & 1 & $n-1$ \\
d & fully & 1 & 1 & 1 & $n-1$ & $n(n-1)/2$ \\
e & 2fully & 2 & 1 & 1 & $2(n-1)$ & $n(n-1)$ \\
\hline
\end{tabular}
}
\caption{Characteristics of our five network topologies.}
\label{tab:topo-chars}
\end{table*}

\subsection{Software}

Previous chapters have discussed the entire quantum modular
exponentiation that forms the most computationally intensive portion
of Shor's factoring algorithm, but here we will concentrate on the
adder algorithms that are the core arithmetic routines.
Section~\ref{sec:dist-shor} evaluates the VBE
(Sec.~\ref{sec:vbe})~\cite{vedral:quant-arith} and CDKM carry-ripple
adders (Sec.~\ref{sec:cdkm}) \cite{cuccaro04:new-quant-ripple}, and
the carry-lookahead adder
(Sec.~\ref{sec:qcla})~\cite{draper04:quant-carry-lookahead}.

As in general-purpose classical multicomputers, distribution of
software functionality and synchronization primitives are important
for both correctness and performance.  In the quantum multicomputer,
the distribution of functionality is at the level of a few gates,
simplifying the synchronization problem; we need not concern ourselves
with interrupt handlers and packet headers and the like.  Although
each node executes instructions (gates) independently on its qubits,
overall coordination requires that the nodes are in sync to within a
fraction of a gate, or on the order of a few nanoseconds.  This level
of synchronization can only be achieved through the real-time
classical network.  Small amounts of asynchrony must be tolerated as
propagation delays between nodes are significant compared to the clock
cycle time for individual gates.

Finally, although only application algorithms are presented here, it
is interesting to note that Magniez et al. have already discussed a
boot-time quantum self-test~\cite{magniez05:quantum-self-test}.

\subsection{Summary}

\begin{table*}
\centerline{
\begin{tabular}{|ll|}\hline
Node hardware: & $\sim 500$ physical qubits\\
& 2 transceiver qubits \\
QEC: & [[23,1,7]]$^i$+[[7,1,3]]$^o$ \\
logical capacity: & 3 qubits \\
Network: & Linear \\
& serial links \\
adder algorithm: & CDKM carry-ripple \\
\hline
\end{tabular}
}
\caption[Strawman system summary]{Summary of the strawman system proposal.}
\label{tab:hw-specs}
\end{table*}

We have already tipped our hand on one critical architecture issue:
the choice of serial links.  This decision will be justified in the
next section, along with analysis showing that the [[23,1,7]] Steane
code is the preferred bottom-level quantum error correction code.  The
following chapter will show that CDKM is the preferred adder circuit,
and that two-transceiver nodes with about 500 physical qubits and a
linear network will be adequate to scale systems up to hundreds of
nodes.  Table~\ref{tab:hw-specs} summarizes our strawman system
proposal.  Details of clock speed and the node-internal interconnect
are not specified because they are subject to technological
development.

The theme of the next two sections is the optimization of algorithms
that require qubits stored in separate nodes to interact.  The
engineering choice of performing gates via teleportation, as discussed
in Sec.~\ref{sec:telegate}, or teleporting data first, then executing
the desired gates locally (Sec.~\ref{sec:teledata}), is examined.  We
will see that teledata generally outperforms telegate for both QEC (in
Section~\ref{sec:qec-qubus}) and adder algorithms (in
Section~\ref{sec:dist-shor}).

\section{Distributed QEC and Bus Design}
\label{sec:qec-qubus}




We now take up the question of how to perform quantum error correction
(QEC) in our quantum multicomputer.  We show that it is possible to
execute QEC on a logical state where the physical qubits that make up
a QEC code block are distributed across multiple nodes.  We must also
determine how to utilize QEC to best protect logical states as they
are teleported from one node to another, and we show that the simplest
approach is best.

The performance of error correction influences an important hardware
design decision: should our network links be serial or parallel?  We
argue that the difference in both reliability and performance is
likely to be small, assuming that the reliability of teleportation is
less than that of quantum memory and that teleportation times are
reasonable compared to the cycle time of locally-executed QEC.

Teleportation, as we saw in Chapter~\ref{ch:travel}, is composed of
EPR pair creation, local gates, measurements, and classical
communication, and of course requires high-fidelity memory.  Until we
take up the issue of link design in Section~\ref{sec:link-design}, we
will assume that local gates, memory, and measurements are perfect, or
at least much better than EPR pair creation.  Therefore, when we talk
about limits on the failure rate of teleportation, we are really
referring to the quality of the EPR pair.  The quality can be improved
via purification, which has a cost logarithmic in the starting
fidelity; in this dissertation, we will not pursue further the best
way to achieve EPR pairs of the necessary quality.  We denote the
failure probability of a single teleportation as $p_t$.

First, let us briefly consider the failure probability assuming no
error correction on our qubits.  The probability of success of the
entire computation, then, rests on the success of {\em all} of the
individual teleportation operations.  If $t$ is the total number of
teleportations we must execute for the complete computation, our
success probability is
\begin{equation}
p_s = (1-p_t)^t = 1 - \binom{t}{1}p_t +
\binom{t}{2}(-p_t)^2 \cdots \approx 1 - tp_t
\label{eq:p_s}
\end{equation}
for small $p_t$.  Our failure probability grows linearly with the
number of teleportations we must execute, requiring $p_t \ll 1/t$.
Obviously, we need to do better than that, so we quickly conclude that
error correction on the logical states being transferred is necessary.

The argument here falls much along the lines of the threshold argument
for quantum computation in general, as discussed in
Section~\ref{sec:threshold}.  Because we are dealing with a small
number of levels of concatenation and a finite computation, we are
less interested in the threshold itself than a specific calculation of
the success probability for a chosen arrangement.  A detailed estimate
would differ slightly because we have three separate error sources in
memory, local gates, and teleportation, along the lines of Steane's
simulations~\cite{steane02:ft-qec-overhead}; here we restrict
ourselves to a simple analysis.  Table~\ref{tab:teleport-count} shows
rough teleportation counts for the complete modular exponentiation for
Shor's factoring algorithm, based on Table~\ref{tab:algorithms}
(page~\pageref{tab:algorithms}) and the teledata entries of
Table~\ref{tab:lat-topo-baseline}
(page~\pageref{tab:lat-topo-baseline}).  The number of multiplier
blocks has no significant impact on the number of teleportations we
must execute.  The choice of node size and adder are important; the
carry-lookahead adder requires ten to fifteen times as many
teleportations (for 16 to 1,024 bits), but may be faster under some
circumstances, as we will show in Section~\ref{sec:dist-shor}; this
accounts for the range of values in Table~\ref{tab:teleport-count}.

\begin{table*}
\centerline{
\begin{tabular}{|r|r|}\hline
length & teleportations ($t$) \\
\hline
16 & $14000$--$125000$\\
128 & $8\times 10^6$--$10^8$\\
1024 & $4\times 10^9$--$6\times 10^{10}$\\
\hline
\end{tabular}
}
\caption[Number of teleportations necessary for modular
exponentiation]{Number of teleportations necessary to execute the full
modular exponentiation for different problem sizes.}
\label{tab:teleport-count}
\end{table*}

\subsection{Distributed Logical Zeroes}

In Equation~\ref{eq:713-zero} (p.~\pageref{eq:713-zero}) and
Figure~\ref{fig:713-zero} (p.~\pageref{fig:713-zero}), we showed the
logical zero state ($|0_L\rangle$) for the Steane [[7,1,3]] quantum
error correcting code and a circuit to create the state.  This state
is used in the fault-tolerant construction of quantum error
correction.  In the multicomputer, we may need to perform QEC on
states that span two (or more) nodes, when moving data between nodes
in a quantum multicomputer, or simply trying to maintain the integrity
of a static state that spans multiple nodes.  Thus, we must find a way
to either
\begin{enumerate}
\item create a distributed $|0_L\rangle$ state;
\item do four-qubit parity (error syndrome) measurement using only
  weak nonlinearity on four qubits; or
\item find some other way to do syndrome measurement without the full,
  distributed $|0_L\rangle$ state.
\end{enumerate}

Of these three options, we have chosen the first.  We have also
invested some effort in looking for a way to calculate the parity of
$n$ qubits using the weak nonlinearity, but all of the schemes we have
found so far for more than three qubits scale poorly in terms of
noise; Yamaguchi et al. have designed a method that works for three
qubits but not more~\cite{yamaguchi05:_weak-nonlin-qec}.  Bacon has
developed a new method for creating self-correcting memories, using
the original Shor [[9,1,3]] code, that may not require the creation of
logical zeroes; its implications for actual implementation are
exciting but still poorly
understood~\cite{bacon05:_operator-self-qec,thaker06:_cqla}.  Thus,
$|0_L\rangle$ states must be created, and this chapter discusses the
performance and error characteristics of the creation process.

\begin{figure*}
\centerline{\hbox{
\input{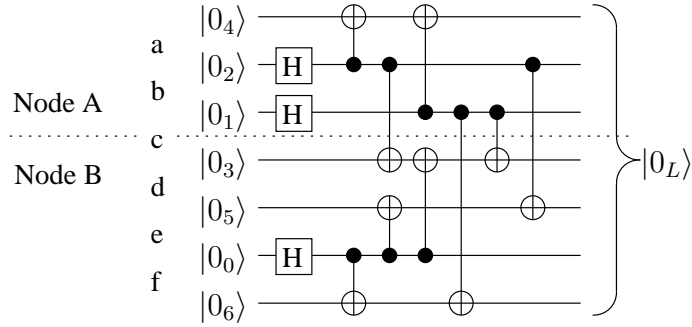}}}
\caption{Distributed circuit to create the $|0_L\rangle$ state for the
  Steane [[7,1,3]] code.}
\label{fig:dist-713-zero}
\end{figure*}

The logical $|0_L\rangle$ can be created using the same two methods as
any other distributed quantum computation: we can directly create the
state in a distributed fashion, using teleported gates (telegate), or
we can create the state within a single node and teleport several of
the qubits to the remote node before using the state in our QEC
(teledata).  First, consider the use of teleported gates to create the
$|0_L\rangle$ state.  Figure~\ref{fig:dist-713-zero} shows that
splitting the $|0_L\rangle$ state across two nodes, as at the line
labeled ``c'', forces the execution of four teleported CNOTs,
consuming four EPR pairs; breaking at ``d'' would require only three.
In the figure, the subscripts again represent the bit number in the
QEC block; the qubits have been reordered compared to
Figure~\ref{fig:713-zero} for efficiency.  Our second alternative is
to teleport portions of a locally-created $|0_L\rangle$ state.  If
enough qubits and computational resources are available at both nodes,
we are free to create the state in either location and teleport some
of the qubits; thus, the maximum number of qubits that must be
teleported is $\lfloor n/2\rfloor$, or 3 for the 7-bit Steane code.
Table~\ref{tab:713-breakpoints} shows the number of gate or data
teleportations necessary, depending on the breakdown of qubits to
nodes, showing that teledata requires the same or fewer EPR pairs, and
so is preferred.

\begin{table*}
\centerline{
\begin{tabular}{|c|c|c|}\hline
breakpoint & telegate & teledata \\
\hline
a & 2 & 1 ($B\rightarrow A$) \\
b & 3 & 2 ($B\rightarrow A$) \\
c & 4 & 3 ($B\rightarrow A$) \\
d & 3 & 3 ($A\rightarrow B$) \\
e & 3 & 2 ($A\rightarrow B$) \\
f & 2 & 1 ($A\rightarrow B$) \\
\hline
\end{tabular}
}
\caption[Breakpoints for the Steane {[[7,1,3]]} code]{Breakpoints
  (corresponding to Figure~\ref{fig:dist-713-zero}) and the cost of
  telegate v. teledata to create a logical zero state for the Steane
  [[7,1,3]] code, in EPR pairs consumed.  Also shown is the direction
  qubits must be teleported.}
\label{tab:713-breakpoints}
\end{table*}

\subsection{Distributed Data}

\subsubsection{Static Distributed States}

If a logical data qubit $|\psi_L\rangle$ is split between nodes A and
B in the same fashion as Figure~\ref{fig:dist-713-zero}, we will use
the $|0_L\rangle$ states to calculate the syndromes for the error
correction.  Each syndrome calculation consumes one $|0_L\rangle$
state, first executing some gates to entangle it with the logical data
qubit, then measuring the zero state.  The [[7,1,3]] code requires six
syndrome measurements (three value and three phase), and Steane
recommends measuring each syndrome at least twice, so each QEC cycle
consumes at least a dozen logical zero states.  With $|\psi_L\rangle$
divided at the ``d'' point, each $|0_L\rangle$ requires three
teleportations, for a total of $3\times 12 = 36$ EPR pairs destroyed
to execute a single, full cycle of QEC.

The split described here allows a single logical qubit plus its QEC
ancillae, a total of fourteen physical qubits, to be split between two
nodes.  The same principles apply to states split among a larger
number of nodes, potentially allowing significantly smaller nodes to
be useful, or allowing larger logical encoding blocks to used, spread
out among small, fixed-size nodes.  More importantly for our immediate
purposes, this analysis serves as a basis for considering the movement
of logical states from node to node.

\subsubsection{States in Motion}

When considering the teleportation of logical qubits and their error
correction needs, two general approaches are possible:
\begin{enumerate}
\item Transfer the entire QEC block, then perform QEC locally at the
  destination; or
\item use one of the methods described above for distributed QEC {\em
  between} the teleportations of the component qubits.
\end{enumerate}

The first approach is conceptually simpler; does the second offer any
advantages in either performance or failure probability?

We will examine one-level QEC and two-level concatenated QEC.  Steane
prefers the [[23,1,7]] code as the lowest layer of a multi-layer
code~\cite{steane02:ft-qec-overhead}.  This code can defend against
three errors, so we are interested in the probability of four errors.
All of the one- and two-layer combinations of [[7,1,3]] and
[[23,1,7]] are examined.


\begin{figure*}
\centerline{\hbox{
\input{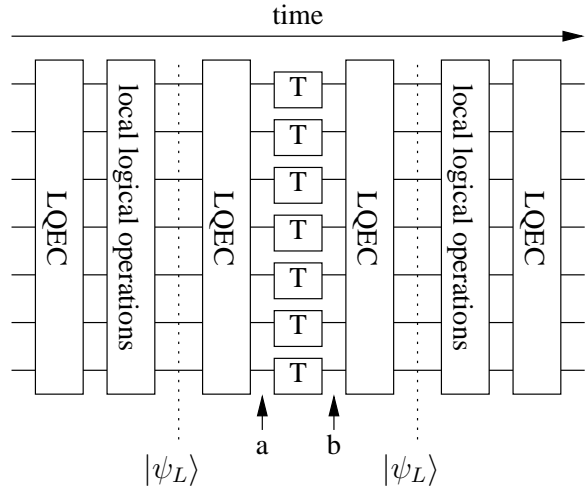}}}
\caption[Teleporting logical state using local QEC only]{Teleporting
logical state using local QEC only, no intermediate QEC.  The box
holding a ``T'' is the teleportation circuit.  Each line represents a
qubit variable, independent of its location, so that the teleportation
operation does not explicitly show the movement of the qubit from one
node to another.}
\label{fig:lqec-parallel}
\end{figure*}

The first approach, illustrated in Figure~\ref{fig:lqec-parallel},
obviously consumes seven EPR pairs to transfer the seven-qubit code
word from one node to the other.  Assume, for the moment, that local
gates and quantum memory are perfect, so that our only source of
errors is teleportation.  As we saw in Chapter~\ref{ch:err-mgmt}, for
an [[$n$,$k$,$d$]]-qubit error correction code, we use $n$ physical
qubits to hold $k$ logical qubits, and can correct up to $(d-1)/2$
errors.  If $p_t$ is the probability of an error occurring during the
teleportation of a single qubit, then the probability of $m$ errors
occurring is
\begin{equation}
p_e(n,m) = \binom{n}{m}(1-p_t)^{n-m}p_t^m \approx \binom{n}{m}p_t^m
\label{eq:p_e}
\end{equation}
for small $p_t$.  For $p_t \ll 1$, most failures will occur in the
lowest failure mode, $((d-1)/2)+1 = (d+1)/2$ errors.  We will
approximate our total failure probability as the probability of
$(d+1)/2$ errors occurring.

If $p_a$ is the failure probability of our total algorithm and $t$ is
the {\em total} number of {\em logical} qubit teleportations we use in
the computation, then
\begin{equation}
p_a = 1 - (1 - p_e)^t \approx \binom{t}{1}p_e\approx tp_e.
\label{eq:p_a}
\end{equation}

For the [[7,1,3]] code,
\begin{equation}
p_e(7,2) = \binom{7}{2} (1-p_t)^5 p_t^2 \approx 21p_t^2
\label{eq:p_e_713}
\end{equation}
is the probability of two errors occurring in our block of seven
qubits.  Two qubit errors, of course, is more than the [[7,1,3]] code
can correct.  Our probability of algorithm failure becomes
\begin{equation}
p_a \approx tp_e = 21tp_t^2.
\label{eq:p_a_713}
\end{equation}
Thus, we can say that, to have a reasonable probability of success, we
should have $p_t \ll 1/\sqrt{21t}$.  This is a significant improvement
over the case with no error correction seen at the beginning of this
chapter.

Using a two-level concatenated code, the picture is less grim.  For
two levels of the [[7,1,3]] code, our total encoding will consist of
seven blocks of seven qubits each, and the computation will fail only
if {\em two} or more of those blocks fail.

Of course, the two codes need not be the same.  Adapting Steane's
terminology and notation, will refer to the physical-level code as the
``inner'' code, and the code built on top of that as the ``outer''
code~\cite{steane02:ft-qec-overhead}.  [[$n^i$,$k^i$,$d^i$]] or
[[$n$,$k$,$d$]]$^i$ is the inner code, and [[$n^o$,$k^o$,$d^o$]] or
[[$n$,$k$,$d$]]$^o$ is the outer code.  Approximating the error
probability according to Equations~\ref{eq:p_e} and \ref{eq:p_a}, we
have
\begin{equation}
p_a \approx t\binom{n^o}{m^o}\left(\binom{n^i}{m^i}p_t^{m^i}\right)^{m^o}
\end{equation}
where $m^i = (d^i+1)/2$ and likewise for $m^o$.

Table~\ref{tab:teleport-code-strengths} shows the estimates for the
teleportation failure probability $p_t$ that will give us a total
algorithm failure probability of $p_a < 0.1$.  Although
[[23,1,7]]$^i$+[[7,1,3]]$^o$ and [[7,1,3]]$^i$+[[23,1,7]]$^o$ are
different, by coincidence, their failure probabilities are almost
identical, so they are listed together.  Note that [[23,1,7]] offers
essentially the same error protection as [[7,1,3]]+[[7,1,3]], despite
using half the number of qubits and being conceptually simpler.

\begin{table*}
\centerline{
\begin{tabular}{|l|c|c|c|}\hline
error-correcting code & scale-up & teleportations & $p_t$ for $p_a < 0.1$\\
\hline
(none) & 1 & $10^{5}$ & $  0.1/t = 10^{-6}$ \\
& & $10^{8}$ & $ 0.1/t = 10^{-9}$  \\
& & $10^{11}$ & $ 0.1/t = 10^{-12}$  \\
\hline
[[7,1,3]] & 7 & $10^{5}$ & $1/\sqrt{21t} = 7\times 10^{-4}$ \\
& & $10^{8}$ & $1/\sqrt{21t} = 2\times 10^{-5}$ \\
& & $10^{11}$ & $1/\sqrt{21t} = 7\times 10^{-7}$ \\
\hline
[[23,1,7]] & 23 & $10^{5}$ & $ 1/(17t^{1/4})\approx 3\times 10^{-3}$ \\
& & $10^{8}$ & $ 1/(17t^{1/4})\approx 6\times 10^{-4}$ \\
& & $10^{11}$ & $ 1/(17t^{1/4})\approx 1\times 10^{-4}$ \\
\hline
[[7,1,3]]$^i$+[[7,1,3]]$^o$ & 49 & $10^{5}$ & $ 1/(17t^{1/4})
\approx 3\times 10^{-3}$ \\
& & $10^{8}$ &  $ 1/(17t^{1/4}) \approx 6\times 10^{-4}$ \\
& & $10^{11}$ &  $ 1/(17t^{1/4}) \approx 1\times 10^{-4}$ \\
\hline
[[23,1,7]]$^i$+[[7,1,3]]$^o$ & 161 & $10^{5}$ & $1/(19t^{1/8}) \approx 0.012$ \\
and [[7,1,3]]$^i$+[[23,1,7]]$^o$ & & $10^{8}$ & $1/(19t^{1/8}) \approx 5\times 10^{-3}$ \\
& & $10^{11}$ & $1(19t^{1/8}) \approx 2\times 10^{-3}$ \\
\hline
[[23,1,7]]$^i$+[[23,1,7]]$^o$ & 529 & $10^{5}$ & $1/(20t^{1/16}) \approx 0.025$ \\
& & $10^{8}$ & $1/(20t^{1/16}) \approx 0.016$ \\
& & $10^{11}$ & $1/(20t^{1/16}) \approx 0.010$ \\
\hline
\end{tabular}
}
\caption[An estimate of the necessary teleportation error rate]{An
estimate of the necessary error rate of teleportation ($p_t$) to
achieve a specific number of logical teleportations with 90\%
probability of success, for different error-correction schemes.}
\label{tab:teleport-code-strengths}
\end{table*}

\begin{figure*}
\centerline{\hbox{
\resizebox{16cm}{!}{\input{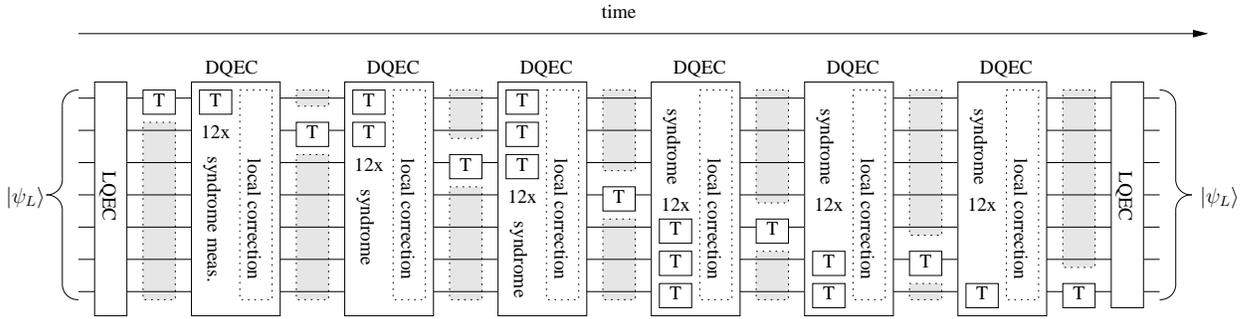}}}}
\caption{Teleporting logical state using intermediate, teledata
distributed QEC.}
\label{fig:dqec}
\end{figure*}

The second approach described above, doing error correction after
serially sending each qubit, is shown in Figure~\ref{fig:dqec}.  Using
this approach, we attempt to reduce the overall error probability by
incrementally correcting the logical state as it is teleported; to
teleport the seven-bit state we perform local QEC before beginning,
then do distributed QEC after each of the first six teleportations,
then local QEC again after the seventh teleportation.  Each
distributed QEC (DQEC) block performs twelve distributed syndrome
measurements.  We can again choose telegate or teledata for the
$|0_L\rangle$ state creation; the figure illustrates teledata.  Using
telegate, we would need the sum of the telegate column in
Table~\ref{tab:713-breakpoints}, or $2+3+4+3+3+2=17$, inter-node
gates, for each syndrome that must be measured.  To perform twelve
measurements we consume a total of $12\times 17 = 204$ EPR pairs.
Using teledata, we would need only $1+2+3+3+2+1 = 12$ per syndrome, or
144 EPR pairs for the full twelve syndromes in a cycle.  The
worst-case DQEC block is $3\times 12 = 36$ teleportations.  Obviously,
the probability of error is higher for 36 teleportations than for
seven.  Therefore, unless someone develops a means of measuring
syndromes without using the $|0_L\rangle$ states, this second approach
does not achieve its goal of reducing the total error probability.
Performance-wise, the penalty for doing step-wise QEC is also stiff;
we conclude that this approach is not useful.

\subsection{Implications for Link Design}
\label{sec:link-design}

\begin{figure*}
\centerline{\hbox{
\input{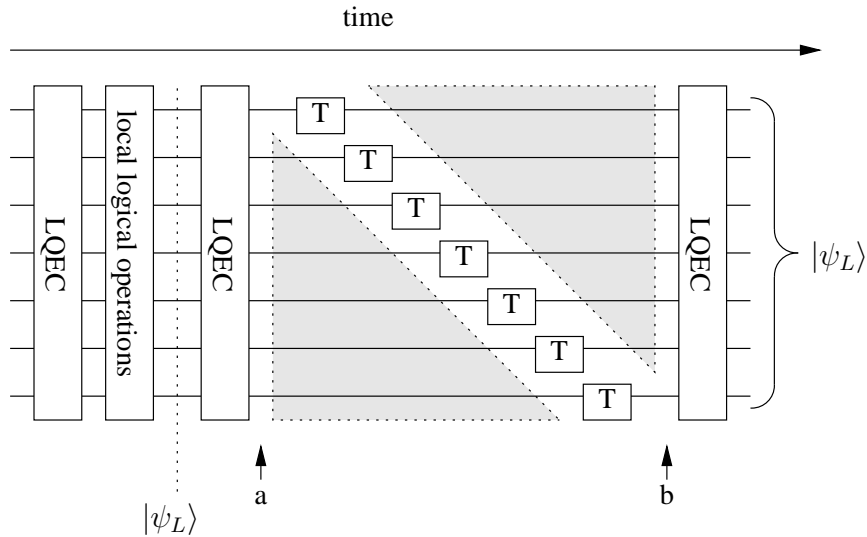}}}
\caption{Local QEC only, no intermediate QEC, serial interface.}
\label{fig:lqec-serial}
\end{figure*}

Figure~\ref{fig:lqec-parallel} shows a [[7,1,3]] state being
transferred in parallel and Figure~\ref{fig:lqec-serial} shows the
serial equivalent.  In these diagrams, each line represents a qubit
that is a member of a code block, essentially following the variable
rather than the storage locations; at a $T$ block, representing
teleportation, of course the qubit moves from one node to the other.
If the transfer is done serially, the wait to {\em start} the QEC
sequence is seven times as long, but the {\em total} time for transfer
plus QEC (that is, time from the start of one QEC cycle to the next,
from the first $|\psi_L\rangle$ to the point marked ``b'' in the
figures) won't grow by nearly as large a factor if local QEC requires
significant time compared to a teleportation.  Thus, we need to
determine if the increase in wait time caused by the lengthening of
the interval the point marked ``a'' to the point marked ``b'' in
Figures~\ref{fig:lqec-parallel} and \ref{fig:lqec-serial} has an
unacceptably large impact on our overall failure rate.

The gray areas in the serial figure indicate increased wait time for
the qubits.  Each qubit spends one cycle teleporting, and six waiting
for the other teleportations.  If $p_m$ is the probability of error
for a single qubit during the time to execute a single teleportation,
then the probability of no error on one bit during that time is
$(1-p_m)^6$ for a [[7,1,3]] code.  For an [[$n$,$k$,$d$]] code, the
failure probability of that qubit during the serial transfer waiting
time is $p'_m = 1 - (1-p_m)^{n-1}$.  The probability of $m$ memory
errors is
\begin{equation}
p_M(n,m) = \binom{n}{m}{p'_m}^m(1-p'_m)^{n-m}\approx \binom{n}{m}{p'_m}^m
\approx \binom{n}{m}(n-1)p_m^m.
\label{eq:tele-mem-prob}
\end{equation}

Combining Equations~\ref{eq:tele-mem-prob} and \ref{eq:p_e}, we need
the two error sources together to generate less than $m = (d+1)/2$
errors.  We will constrain the final combined memory and teleportation
error rate $p_f$ for the serial link to be similar to the
teleportation errors for the parallel link,
\begin{equation}
p_f(n,m) = \sum_{i=0}^{m}p_M(n,i)p_e(n,m-i) \sim p_e(n,m).
\label{eq:tele-mem-comp}
\end{equation}

For the error codes we are considering, [[7,1,3]] and [[23,1,7]],
numeric evaluation for $p_m = p_t/10(n-1)$ gives 25\% and 50\%
increase in failure probability compared to the $p_m = 0$ (perfect
memory) case.  Thus, we can say, very roughly, that a memory failure
probability two orders of magnitude less than the failure probability
of the teleportation operation will mean that the choice of serial or
parallel buses has minimal impact on the overall system error rate.

Although this section has focused on reliability rather than
performance, the choice of serial or parallel links also affects
performance.  It is easy to see that choosing a serial link does not
result in a factor of $n$ degradation in system performance when QEC
is taken into account.  Let $t_t$ be our teleportation time, and
$t_{LQEC}$ be the time to perform local error correction.  $t_t$ is
related to the qubus detector time and $t_{LQEC}$ is related to the
local qubit measurement time.

If $nt_t \ll t_{LQEC}$, then in accordance with Amdahl's Law the
choice also has minimal impact on our overall
performance~\cite{amdahl67}.  Of course, if the resources available at
each node are large enough, teleportation and error correction can be
pipelined, but the growth in resources is significant if $t_{LQEC}$ is
large and the performance gains are small if $t_{LQEC}$ is small.
In addition, as we will see in the next section, arithmetic algorithms
rarely have enough data waiting for teleportation that pipelining will
be effective, so pipelining here would be a second-order effect on
overall system performance.  Therefore, we recommend using serial
links without pipelining, if the qubus detector time is reasonable.

\subsection{Summary}

I originally believed that the issues of serial v. parallel and
intermediate QEC v. block transfer were tied together.  However, it
is now clear that the two are separate issues, and that, unless a
better method for creating logical zeroes is found or Bacon's method
of calculating syndromes without using logical zeroes proves to be
practical, intermediate QEC offers no benefit.  I therefore recommend
block-wise error correction, shipping the entire QEC block from source
to destination before performing QEC.

The results in Table~\ref{tab:teleport-code-strengths} show that a
teleportation error rate (really, EPR pair infidelity) of $\sim 1\%$
will allow computations as large as the factoring of a 1,024-bit
number to proceed with a high probability of success.  This estimate
is for a data encoding of [[23,1,7]]$^i$+[[23,1,7]]$^o$ on the link
and a memory error rate in the time it takes to perform a
teleportation at least two orders of magnitude better than the
teleportation failure rate.  Our analysis, though somewhat simpler
than Steane's, supports his recommendation of the [[23,1,7]] code.
Replacing one level with the [[7,1,3]] code still allows an error rate
of one part in a thousand or better, with a noticeable savings in
storage requirements.  Of course, we do not have to compute or store
data using the same encoded states that we use during data
transport~\cite{thaker06:_cqla}.  In this dissertation, for
simplicity, we have assumed that the system uses only a single choice
of encoding.

This section has argued that the difference in both performance and
reliability between serial and parallel network links will be small
for a reasonable set of assumptions.  Serial links will dramatically
simplify our hardware design by reducing the number of required
transceiver qubits in each node, and eliminating concerns such as
jitter and skew between pairs of conductors or wave guides.  Moreover,
if we do choose to have multiple transceiver qubits in each node,
system performance on some workloads may be boosted more by creating a
richer node-to-node interconnect topology than by creating parallel
channels between pairs of nodes in a simpler topology, as we will see
in the next section.

\section{Distributed Form of Shor's Algorithm}
\label{sec:dist-shor}

This section evaluates the performance of quantum arithmetic
algorithms run on a quantum multicomputer.  We vary the node capacity
and I/O capabilities, and the network topology.  The tradeoff of
choosing between telegate and teledata is examined.  We show that the
teledata approach performs better, and that carry-ripple adders
perform well when the teleportation block is decomposed so that the
key quantum operations can be parallelized.  A node size of only a few
logical qubits performs adequately, provided that the nodes have two
transceiver qubits.  A linear network topology performs acceptably for
a broad range of system sizes and performance parameters.  We
therefore recommend pursuing small, high-I/O bandwidth nodes and a
simple network, as described at the end of Section~\ref{ch:arch-over}.


The first question in considering a multicomputer is whether the
system performance will be acceptable {\em if} the implementation
problems can be solved.  Chapter~\ref{ch:large-perf} provided the
tools and algorithms for this analysis; here they are applied.  Our
evaluation criterion is the latency to complete one addition.  The
goal is to achieve ``reasonable'' performance for Shor's factoring
algorithm for numbers up to a thousand bits.  This analysis is done
attempting to minimize the required number of qubits in a node while
retaining reasonable performance; we investigate node sizes of one to
five logical qubits per node.

This section shows that:
\begin{itemize}
\item teleportation of data is better than teleportation of gates;
\item decomposition of teleportation into a series of smaller
  operations brings big benefits in performance, making a carry-ripple
  adder effective even for large problems;
\item a linear topology is an adequate network for the foreseeable
  future; and
\item small nodes (only a few logical qubits) perform acceptably, but
  I/O bandwidth is critical.
\end{itemize}
A multicomputer built around these principles and based on solid-state
qubit technology will perform well on Shor's algorithm.  These results
collectively represent a large step in the design and performance
analysis of distributed quantum computation.

Next, we discuss the mapping of arithmetic algorithms to our system.
The bulk of this section progressively refines performance estimates,
including decomposing the teleportation operation to make the
performance of carry-ripple adders competitive with the
carry-lookahead adder, with a simpler network and smaller nodes.

\subsection{Algorithm}
\label{sec:algorithm}

We evaluate three different addition algorithms: the
Vedral-Barenco-Ekert (VBE) style of carry-ripple adder
(Sec.~\ref{sec:vbe}) ~\cite{vedral:quant-arith}, the faster, smaller
Cuccaro-Draper-Kutin-Moulton (CDKM) carry-ripple adder
(Sec.~\ref{sec:cdkm}) ~\cite{cuccaro04:new-quant-ripple}, and the
carry-lookahead adder (Sec.~\ref{sec:qcla})
~\cite{draper04:quant-carry-lookahead}.  In this section, we discuss
the adders without regard to the network topology; the following
section presents numeric values for different topologies and gate
timings.

\subsubsection{Carry-Ripple Adders}

\begin{figure*}
\includegraphics[width=16cm]{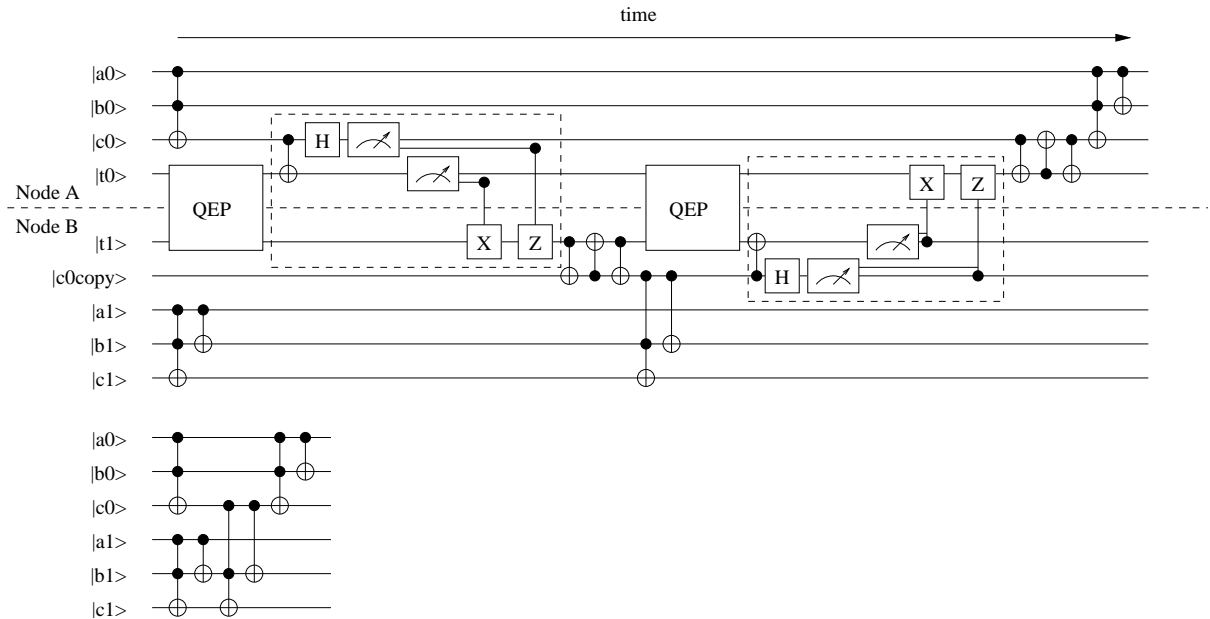}
\caption[Details of a distributed 2-qubit VBE adder]{Details of a
distributed 2-qubit VBE adder.  The top circuit is the distributed
form using the teledata method; the bottom circuit is the monolithic
equivalent.  The solid box (QEP) is the qubus EPR pair generator; the
circuits in dashed boxes are standard quantum teleportation circuits.
Graphical notation as in Fig.~\ref{fig:basic-gates} on
page~\pageref{fig:basic-gates}.}
\label{fig:teleport-2bit-adder}
\end{figure*}

Figure~\ref{fig:teleport-2bit-adder} shows a two-qubit VBE
carry-ripple adder in its monolithic (bottom) and distributed (top)
forms.  The QEP block creates an EPR pair using the qubus entanglement
protocol described in Sec.~\ref{sec:qubus}.  The dashed boxes
delineate the teleportation circuit (which is assumed to be perfect)
that moves the qubit $c0$ from node A to node B.  $c0$ is used in
computation at node B, then moved back to node A via a similar
teleportation to complete the computation.  The two qubits $t0$ and
$t1$ are used as transceiver qubits, and are reinitialized as part of
the QEP sub-circuit.

\begin{figure*}
\centerline{\hbox{
\includegraphics[width=13cm]{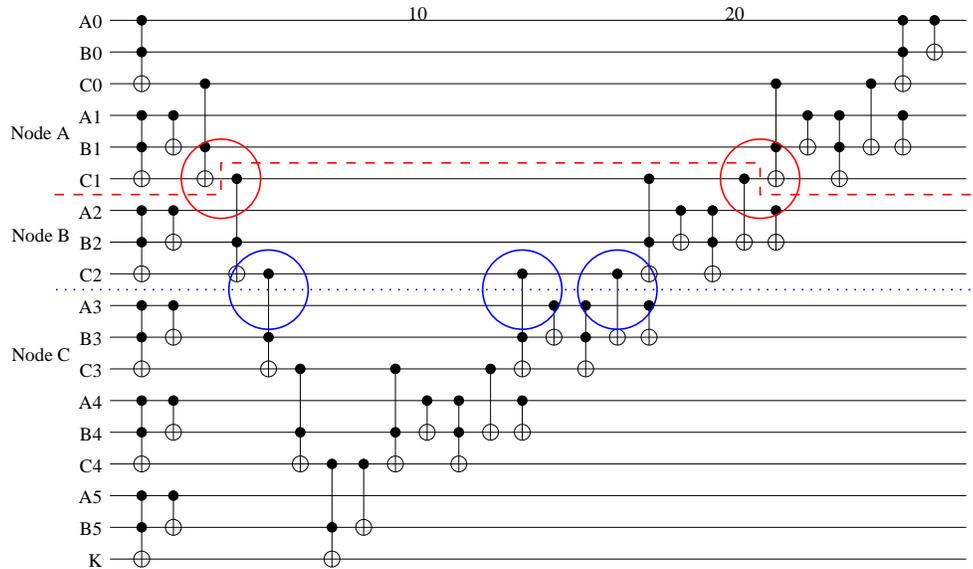}}}
\caption[Visual approach to determining cost for distributed VBE
  adder]{Visual approach to determining relative cost of teleporting
  data versus teleporting gates for a VBE adder.  The upper, dashed
  (red) line shows the division between two nodes (A and B) using data
  teleportation.  The circles show where the algorithm will need to
  teleport data.  The lower, dotted line (blue) shows the division
  using gate teleportation (nodes B and C).  The circles show where
  teleported gates must occur.  Note that two of these three are CCNOT
  gates, which may entail multiple two-qubit gates in actual
  implementation.  The numbers at the top are clock cycles.}
\label{fig:vbe-tradeoff}
\end{figure*}

Figure~\ref{fig:vbe-tradeoff} shows a larger VBE adder circuit and
illustrates a visual method for comparing telegate and teledata.  For
telegate, we can draw a line across the circuit, with the number of
gates (vertical line segments) crossed showing our cost.  For
teledata, the line must {\em not} cross gates, instead crossing the
qubit lines.  The number of such crossings is the number of
teleportations required.  This approach works well for analyzing the
VBE and CDKM adders, but care must be taken with the carry-lookahead
adder, because it uses long-distance gates that may be between
e.g. nodes 1 and 3.

The VBE adder latency to add two $n$-qubit numbers on an $m$-node
machine using the teledata method is $2m-2$ teleportations plus the
circuit cost.  For the telegate approach, using the five-gate
breakdown for CCNOT built from $\sqrt{X}$ gates and CNOTs, as in
Figure~\ref{fig:5gcc} on page~\pageref{fig:5gcc}, would require three
teleported two-qubit gates to form a CCNOT.  Therefore, implementing
telegate, the latency is $7m-7$ gate teleportations, or 3.5x the cost.

\begin{figure}
\centerline{\hbox{
\includegraphics[width=10cm]{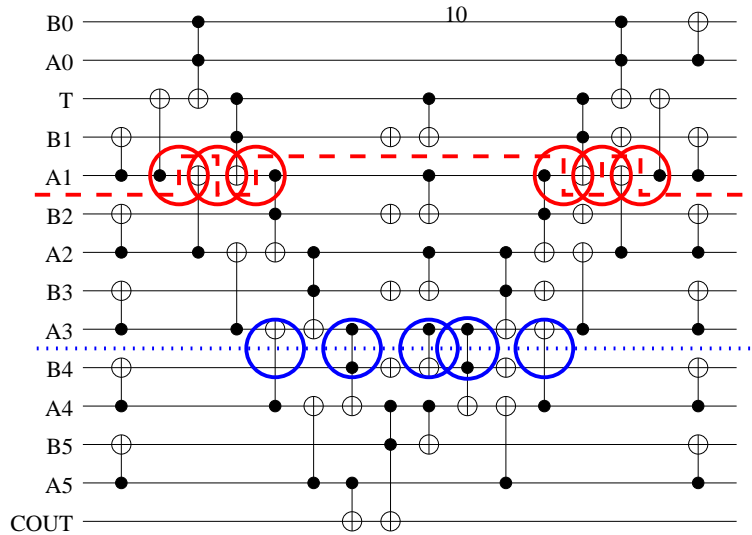}}}
\caption[Visual approach to determining cost for distributed CDKM
  adder]{Visual approach to determining relative cost of teleporting
  data versus teleporting gates for a CDKM adder.  The upper, dashed
  (red) line shows the division between two nodes using data
  teleportation.  The circles show where the algorithm will need to
  teleport data.  The lower, dotted line (blue) shows the division
  using gate teleportation.  The circles show where teleported gates
  must occur.  Note that two of these five are CCNOT gates, which may
  entail multiple two-qubit gates in actual implementation.}
\label{fig:cuca-tradeoff}
\end{figure}

For the CDKM carry-ripple adder, which more aggressively reuses data
space, teledata requires a minimum of six movements, whereas telegate
requires two CCNOTs and three CNOTs, or a total of nine two-qubit
gates, as shown in figure~\ref{fig:cuca-tradeoff}.  The CDKM adder
pipelines extremely well, so the actual latency penalty for more than
two nodes is only $2m+2$ data teleportations, or $6m$ gate
teleportations, when there is no contention for the inter-node links,
as in our line and fully-connected topologies.  The bus topology
performance is limited by contention for access to the interconnect.

\subsubsection{Carry Lookahead}

Analyzing the carry-lookahead adder is more complex, as its structure
is not regular, but grows more intertwined toward the middle bits.
Gate scheduling is also variable, and the required concurrency level
is high.  The latency is $O(\log n)$, making it one of the fastest
forms of adder for large
numbers~\cite{draper04:quant-carry-lookahead,van-meter04:fast-modexp,ercegovac-lang:dig-arith}.

Let us look at the performance in a monolithic quantum computer, for
$n$ a power of two.  Based on table 1 from Draper et
al.~\cite{draper04:quant-carry-lookahead}, for $n = 2^k$, the circuit
depth of $4k+3$ Toffoli gates is 19, 31, and 43 Toffoli gates, for 16,
128, and 1,024 bits, respectively.  We assume a straightforward
mapping of the circuit to the distributed architecture.  Most nodes
are assigned four logical qubits ($A_i$, $B_i$, $C_i$, and one
temporary qubit used as part of the carry propagation).  In the next
subsection, we see that the transceiver qubits are the bottleneck; we
cannot actually achieve this $4k+3$ latency.

\subsection{Performance}
\label{sec:perf}

The modular exponentiation to run Shor's factoring algorithm on a
1,024-bit number requires approximately 2.1 million calls to the
integer adder~\cite{van-meter04:fast-modexp}.  With a 100 $\mu$sec
adder, one run of the algorithm will require less than five minutes;
with a 1 msec adder, it will take just over half an hour, allowing
about twelve hundred ``runs'' per month.  Even a system two to three
orders of magnitude slower than this will have attractive performance,
provided that error correction can sustain the system state for that
long, and that the system can be built and operated economically.
This section presents numerical estimates of performance which show
that this criterion is easily met by a quantum multicomputer under a
variety of assumptions about logical operation times, providing plenty
of headroom for quantum error correction.

\subsubsection{Initial Estimate}
\label{sec:iniperf}

Our initial results are shown in table~\ref{tab:lat-topo-baseline}.
Units are in number of complete teleportations, treating teleportation
and EPR pair generation as a single block, and assuming zero cost for
local gates.  In the following subsections these assumptions are
revisited.  We show three approaches (baseline, telegate, and
teledata) and three adder algorithms (VBE, CDKM, carry-lookahead) for
five networks (bus, 2bus, line, fully, 2fully) and three problem sizes
(16, 128, and 1024 bits).  In the baseline case, each node contains
only a single logical qubit; gates are therefore executed using the
telegate approach.  For the telegate and teledata columns, we chose
node sizes to suit the algorithms: two, three, and four qubits per
node for the CDKM, VBE, and carry-lookahead adders, respectively, when
using telegate, and three, four and five qubits when using teledata.

The VBE adder, although larger than CDKM and slower on a monolithic
computer, is faster in a distributed environment.  The VBE adder
exhibits a large (3.5x) performance gain by using the teledata method
instead of telegate.  For teledata, the performance is independent of
the network topology, because only a single operation is required at a
time, moving a qubit to a neighboring node.  The CDKM adder also
communicates only with nearest neighbors, but performs more transfers.
The single bus configuration is almost 3x slower than the line
topology.  On a line, in most time slots, three concurrent transfers
are conducted (e.g., between nodes $1\rightarrow 2$, $3\rightarrow 2$,
and $3\rightarrow 4$).

An unanticipated but intuitive result is that the performance of the
carry-lookahead adder is better in the baseline case than the telegate
case, for the fully-connected network.  This is due to the limitation
of having a single transceiver qubit per node.  Putting more qubits in
a node increases contention for the transceiver qubit, and reduces
performance even though the absolute number of gates that must be
executed via teleportation has been reduced.  Our numbers also show
that the carry-lookahead adder is not a good match for a bus
architecture, despite the favorable long-distance transport, again
because of excessive contention for the bus.

The carry-lookahead adder is easily seen to be inappropriate for the
line architecture, since the carry-lookahead requires long-distance
gates to propagate carry information quickly.  Using the linear
network naturally degenerates to linear cost to share data over a long
distance.  Using nested purification techniques, as with quantum
repeaters~\cite{childress05:_ft-quant-repeater,briegel98:_quant_repeater},
it might be possible to reduce the linear time to $O(\log n)$ time,
but even the factor of ten introduced for a 1,024-bit number will make
the carry-lookahead adder slower than the carry-ripple adders.  If the
required resources on the line are spatially overlapping, the penalty
might actually exceed ten times, exacerbating the problem.  Therefore,
we have ruled out using the carry-lookahead adder on a linear network,
and do not analyze it further.

For telegate, performing some adjustments to eliminate intra-node
gates, we find $8n - 9k - 8$ total Toffoli gates that need arguments
that are originally stored on three separate nodes, plus $n-2$
two-node CNOTs.  For the bus case, which allows no concurrency, this
is our final cost.  For the fully-connected network, we find a depth
of $8k-10$ three-node CCNOTs, 8 two-node CCNOTs, and 1 CNOT.  These
numbers must be multiplied by the appropriate CCNOT breakdown.  For
the teledata fully-connected case, each three-node Toffoli gate
requires four teleportations (in and out for each of two variables).
For the 2fully network, the latency of the three-node Toffolis is
halved, but the two-node Toffolis do not benefit, giving us a final
cost of slightly over half the fully network cost.
\begin{landscape}
\begin{table*}
\centerline{\hbox{
\begin{tabular}{||r|r||r|r|r||r|r|r|r|r||r|r|r|r|r||}\hline
algo. & size & \multicolumn{3}{c||}{Baseline} &
\multicolumn{5}{c||}{Telegate} &
\multicolumn{5}{c||}{Teledata} \\
\hline
& & bus & line & fully & bus & 2bus & line & fully & 2fully & bus & 2bus & line & fully & 2fully \\ 
\hline
VBE & 16 & 360 & 305  & 182 & 105 & 105 & 105 & 105 & 105
& 30 & 30 & 30 & 30 & 30 \\
& 128 & 3048 & 2545  & 1526 & 889 & 889 & 889 & 889 & 889
& 254 & 254 & 254 & 254 & 254 \\
& 1024 & 24552 & 20465 & 12278 & 7161 & 7161 & 7161 &
7161 & 7161 & 2046 & 2046 & 2046 & 2046 & 2046 \\
\hline
CDKM & 16 & 232 & 160   & 160 & 138 & 96 & 96 & 97 & 96
& 90 & 60 & 34 & 90 & 34 \\
& 128 & 1912 & 1280 & 1280 & 1146 & 768 & 768 & 768 & 768
& 762 & 508 & 258 & 762 & 258 \\
& 1024 & 15352 & 10240 & 10240 & 9210 & 6144 & 6144 &
6145 & 6144 & 6138 & 4092 & 2050 & 6138 & 2050 \\
\hline
Carry- & 16 & 644 & N/A & 99 & 444 & 222 & N/A & 136 & 135
& 260 & 178 & N/A & 96 & 56 \\
look- & 128 & 6557 & N/A & 159 & 4901 & 2451 & N/A & 256 & 255
& 3176 & 2028 & N/A & 192 & 104 \\
ahead & 1024 & 54806 & N/A & 219 & 41502 & 20751 & N/A & 376 & 375
& 27260 & 17206 & N/A & 288 & 152 \\
\hline
\end{tabular}
}}
\caption[Estimate of adder latency using monolithic teleportation
  blocks]{Estimate of latency necessary to execute various adder
  circuits on different topologies of quantum multicomputer, assuming
  monolithic teleportation blocks (Sec.~\ref{sec:iniperf}). Units are
  in number of teleportation blocks, including EPR pair creation (bus
  transaction), local gates and classical communication.  Size, length
  of the numbers to be added, in bits.  Lower numbers are faster
  (better).}
\label{tab:lat-topo-baseline}
\end{table*}
\end{landscape}

\subsubsection{Improved Performance}
\label{sec:pipeperf}

The analysis in Section~\ref{sec:iniperf} assumed that a teleportation
operation is a monolithic unit.  However,
Figure~\ref{fig:teleport-2bit-adder} makes it clear that a
teleportation actually consists of several phases.  The first portion
is the creation of the entangled EPR pair via the qubus.  The second
portion is local computation and measurement at the sending node,
followed by classical communication between nodes, then local
operations at the receiving node.  The EPR pair creation is not
data-dependent; it can be done in advance, as resources (bus time
slots, qubits) become available, for both telegate and teledata.  With
these assumptions, we are free to reduce the entire performance
problem to making all needed EPR pairs as quickly as possible.

Our initial execution time model treats local gates and classical
communication as zero cost, assuming that EPR pair creation is the
most expensive portion of the computation.  For example, for the
teledata VBE adder on a linear topology, all of the EPR pairs needed
can be created in two time steps at the beginning of the computation.
The execution time would therefore be 2, constant for all $n$ and $m$.
Table~\ref{tab:lat-topo-pipeline} shows the performance under this
assumption.  The performance of the carry-lookahead adder does not
change compared to the initial estimate, as the bottleneck link is
busy full-time creating EPR pairs.

This model gives a misleading picture of performance once EPR pair
creation is decoupled from the teleportation sequence.  When the cost
of the teleportation itself or of local gates exceeds $\sim 1/n$ of
the cost of the EPR pair generation, the simplistic model breaks down;
in the next subsection, we examine the performance with a more
realistic model.
\begin{landscape}
\begin{table*}
\centerline{\hbox{
\begin{tabular}{||r|r||r|r|r||r|r|r|r|r||r|r|r|r|r||}\hline
algo. & size & \multicolumn{3}{c||}{Baseline} &
\multicolumn{5}{c||}{Telegate} &
\multicolumn{5}{c||}{Teledata} \\
\hline
& & bus & line & fully & bus & 2bus & line & fully &2fully & bus & 2bus & line & fully & 2fully \\ 
\hline
VBE & 16 & 360 & 16  & 16 & 105 & 53 & 7 & 14 & 7
& 30 & 15 & 2 & 4 & 2 \\
& 128 & 3048 & 16  & 16 & 889 & 445 & 7 & 14 & 7
& 254 & 127 & 2 & 4 & 2 \\
& 1024 & 24552 & 16 & 16 & 7161 & 3581 & 7 &
14 & 7 & 2046 & 1023 & 2 & 4 & 2 \\
\hline
CDKM & 16 & 232 & 21   & 19 & 135 & 68 & 11 & 18 & 9
& 90 & 60 & 6 & 12 & 6 \\
& 128 & 1912 & 21 & 19 & 1146 & 573 & 11 & 18 & 9
& 762 & 508 & 6 & 12 & 6 \\
& 1024 & 15352 & 21 & 19 & 9210 & 4605 & 11 &
18 & 9 & 6138 & 4092 & 6 & 12 & 6 \\
\hline
Carry- & 16 & 644 & N/A & 99 & 444 & 222 & N/A & 89 & 45
& 260 & 178 & N/A & 96 & 56 \\
look- & 128 & 6557 & N/A & 159 & 4901 & 2451 & N/A & 149 & 75
& 3176 & 2028 & N/A & 192 & 104 \\
ahead & 1024 & 54806 & N/A & 219 & 41502 & 20751 & N/A & 209 & 105
& 27260 & 17206 & N/A & 288 & 152 \\
\hline
\end{tabular}
}}
\caption[Estimated latency for adders using decomposed teleportation
  blocks]{Estimated latency to execute various adders on different
  topologies, for decomposed teleportation blocks
  (sec.~\ref{sec:pipeperf}), assuming classical communication and
  local gates have zero cost.  Units are in EPR pair creation times.
  Size, length of the numbers to be added, in bits.  Lower numbers are
  faster (better).}
\label{tab:lat-topo-pipeline}
\end{table*}
\end{landscape}

\subsubsection{Detailed Estimate}

\begin{figure*}
\centering
\subfigure{
\includegraphics[width=.45\textwidth]{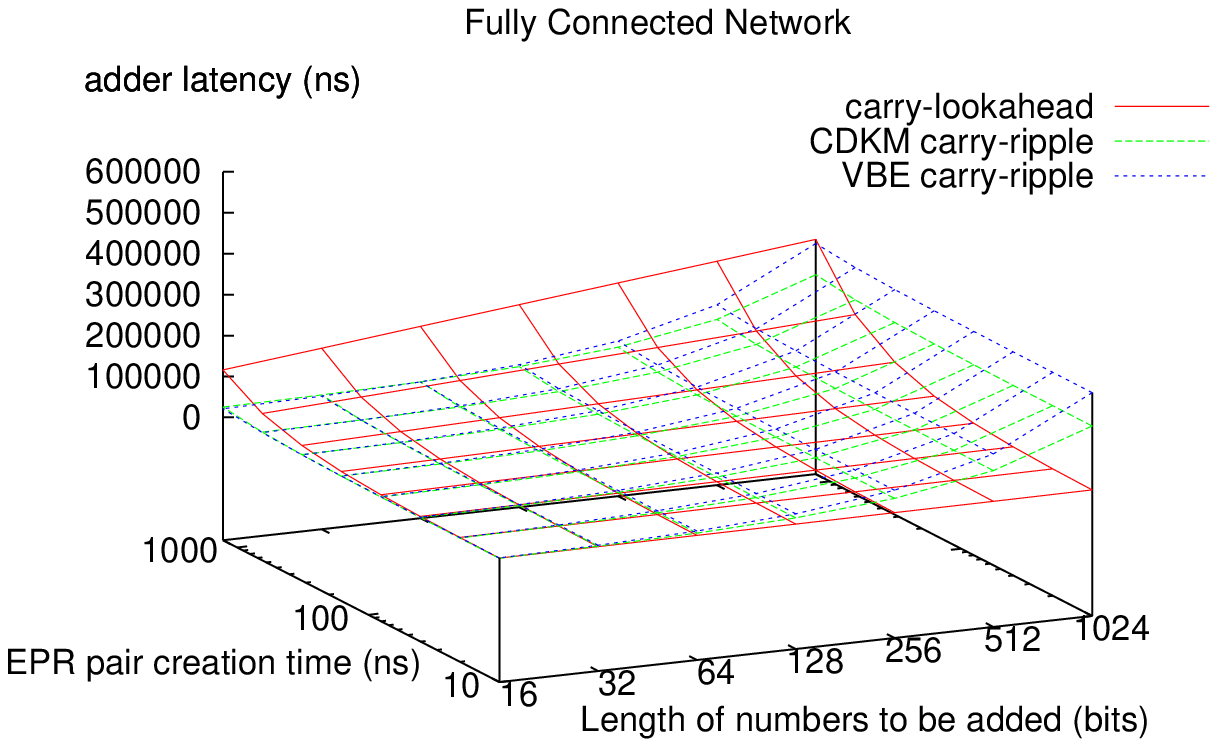}}
\hspace{.3in}
\subfigure{
\includegraphics[width=.45\textwidth]{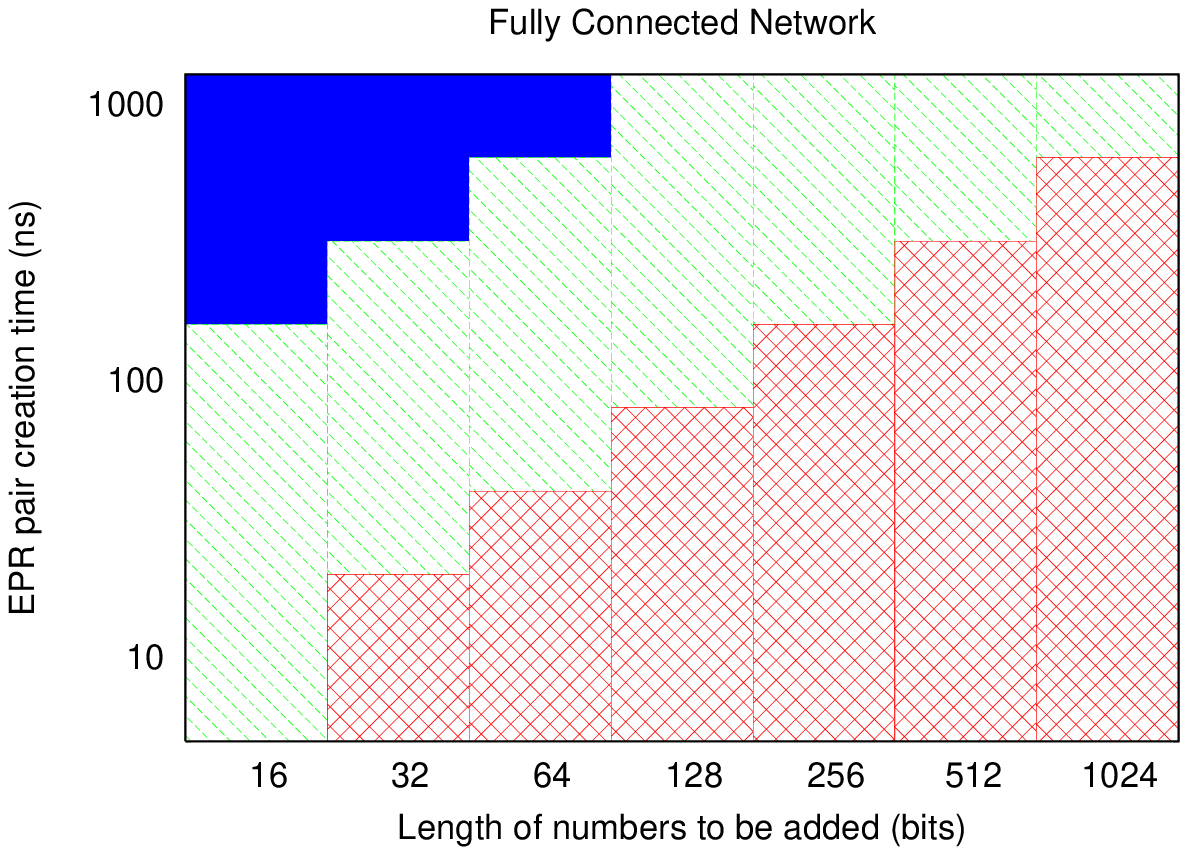}}
\hspace{.3in}
\subfigure{
\includegraphics[width=.45\textwidth]{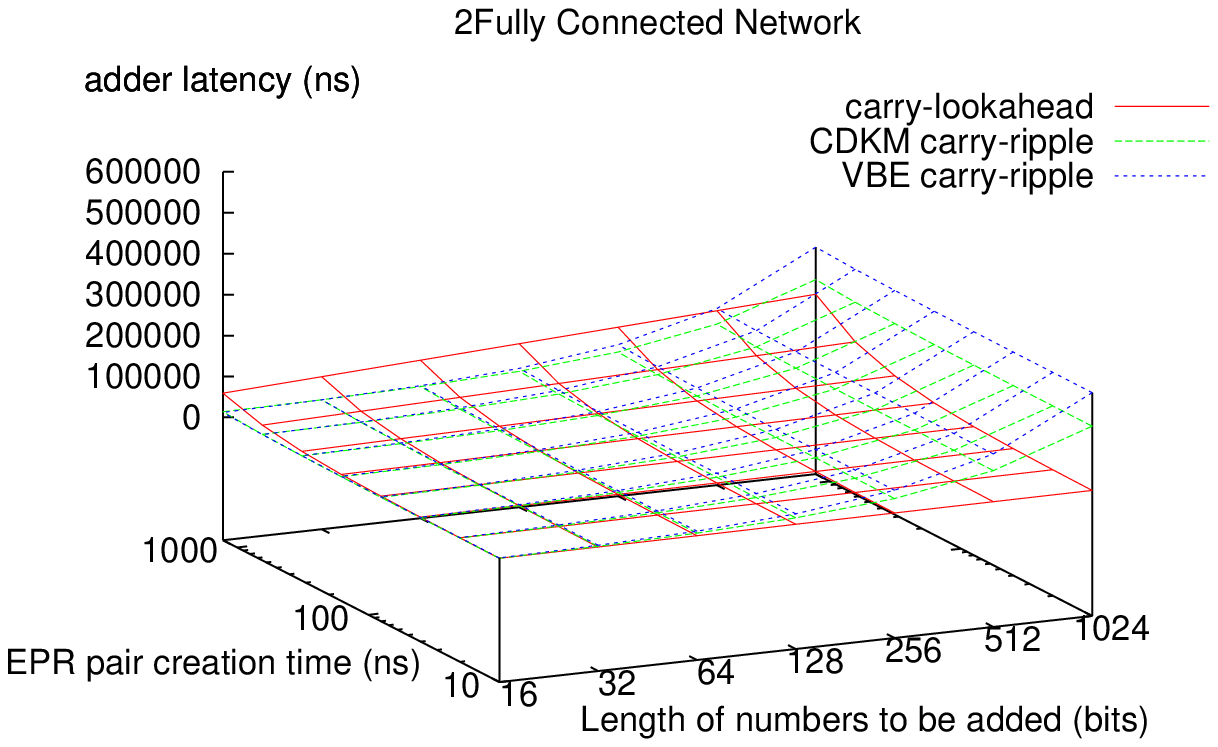}}
\hspace{.3in}
\subfigure{
\includegraphics[width=.45\textwidth]{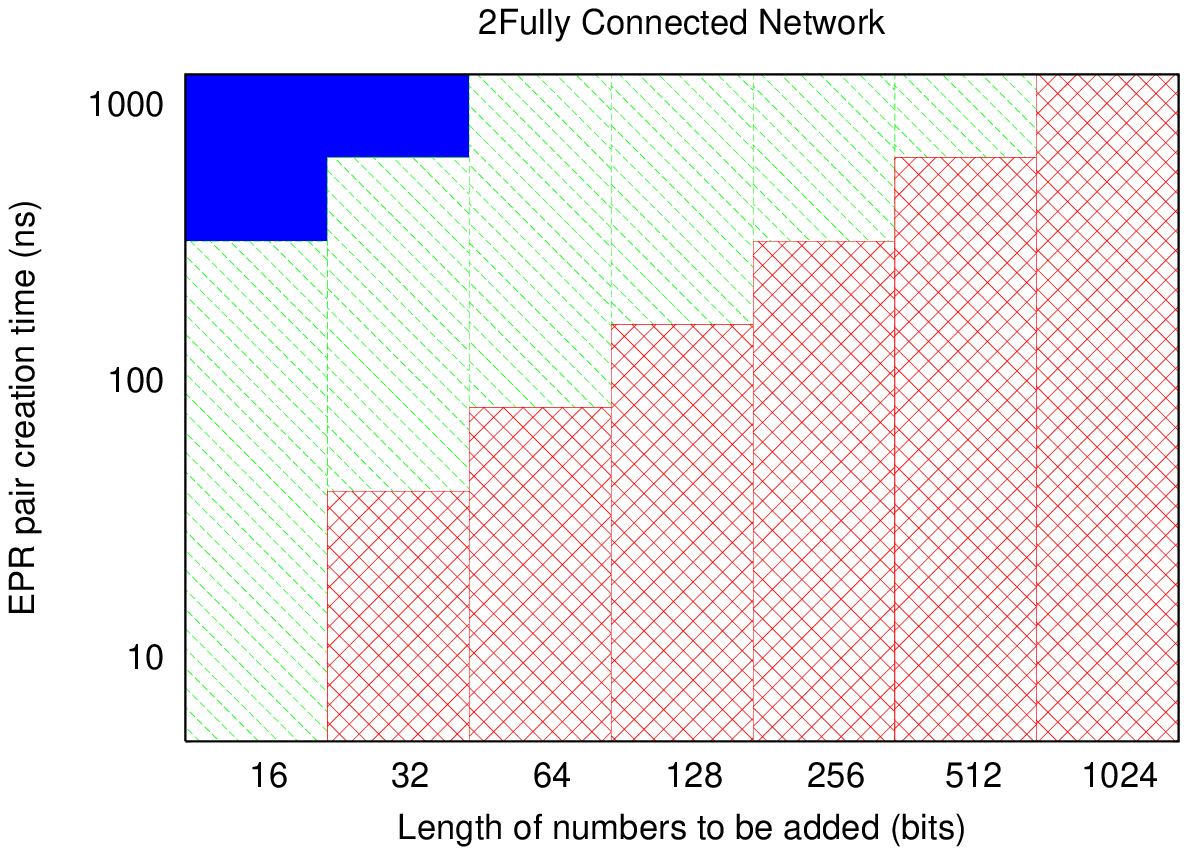}}
\hspace{.3in}
\subfigure{
\includegraphics[width=.45\textwidth]{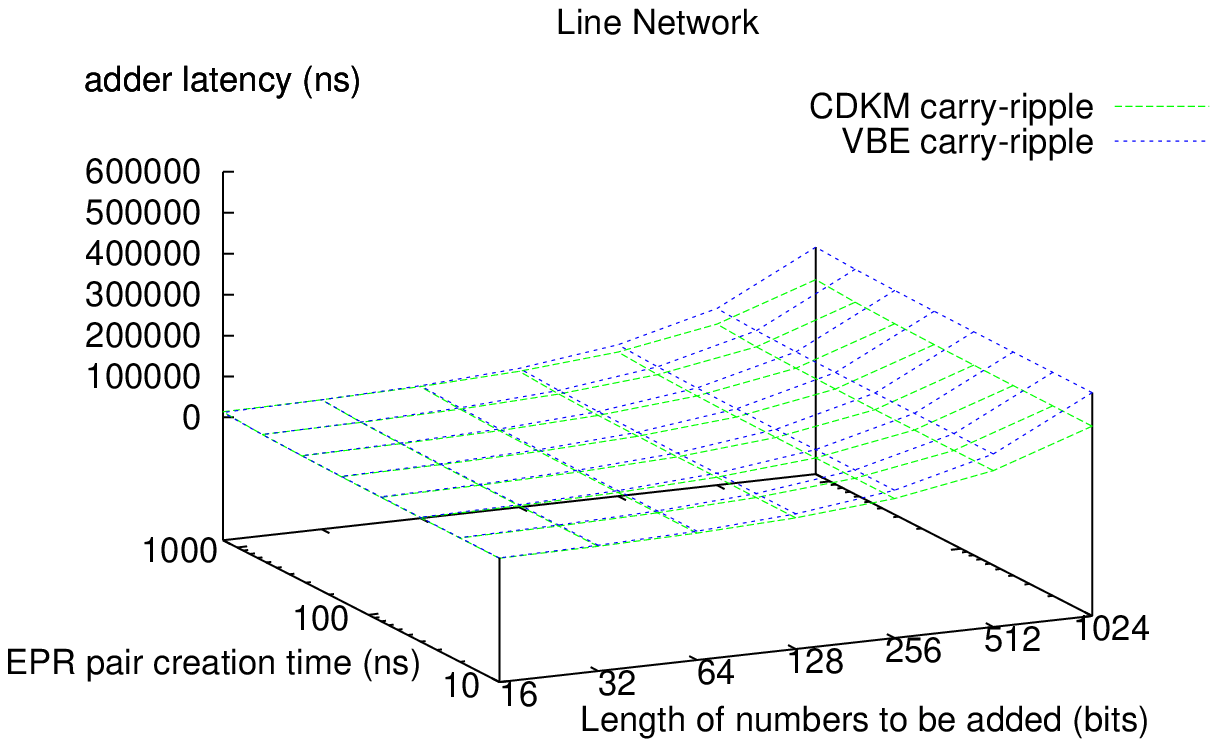}}
\hspace{.3in}
\subfigure{
\includegraphics[width=.45\textwidth]{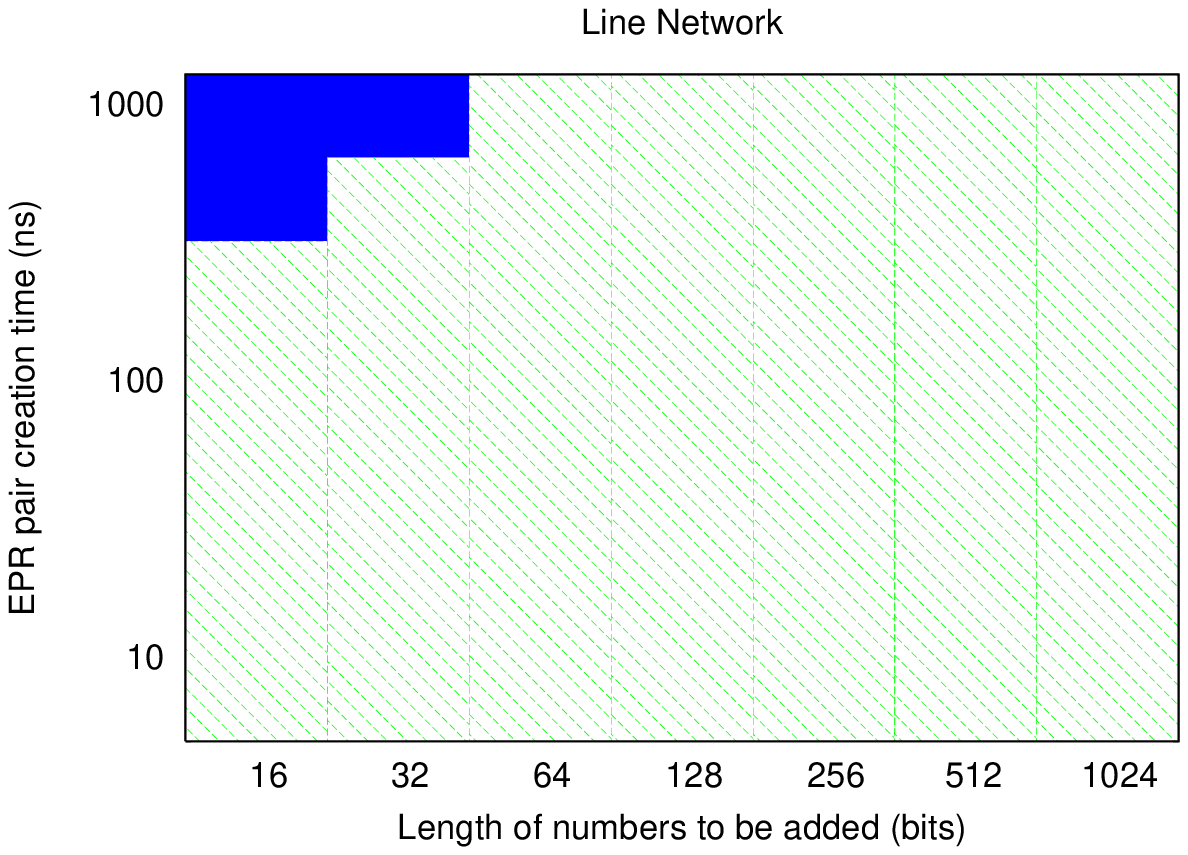}}
\caption[(Telegate) Performance of different adders on three different
  networks]{(Telegate) Performance of different adders on three
  different networks, one fully-connected with a single link and one
  with two links per node (2fully), and one line configuration.  In
  this graph, we vary the latency to create a high-quality EPR pair
  and the length of the numbers we are adding.  Classical
  communication time is assumed to be 10nsec, Toffoli gate time
  50nsec, CNOT gate time 10nsec.  The left hand graph of each pair
  plots adder execution time (vertical axis) against EPR pair creation
  time and number length.  In the right hand graph of each pair, the
  hatched red area indicates areas where carry-lookahead is the
  fastest, the diagonally lined green area indicates CDKM
  carry-ripple, and solid blue indicates VBE carry-ripple.  The
  performance of the carry-lookahead adder is very sensitive to the
  EPR pair creation time.  If EPR pair creation time is low, the
  carry-lookahead adder is very fast; if creation time is high, the
  adder is very slow.}
\label{fig:fully}
\end{figure*}

\begin{figure*}
\centering
\subfigure{
\includegraphics[width=.45\textwidth]{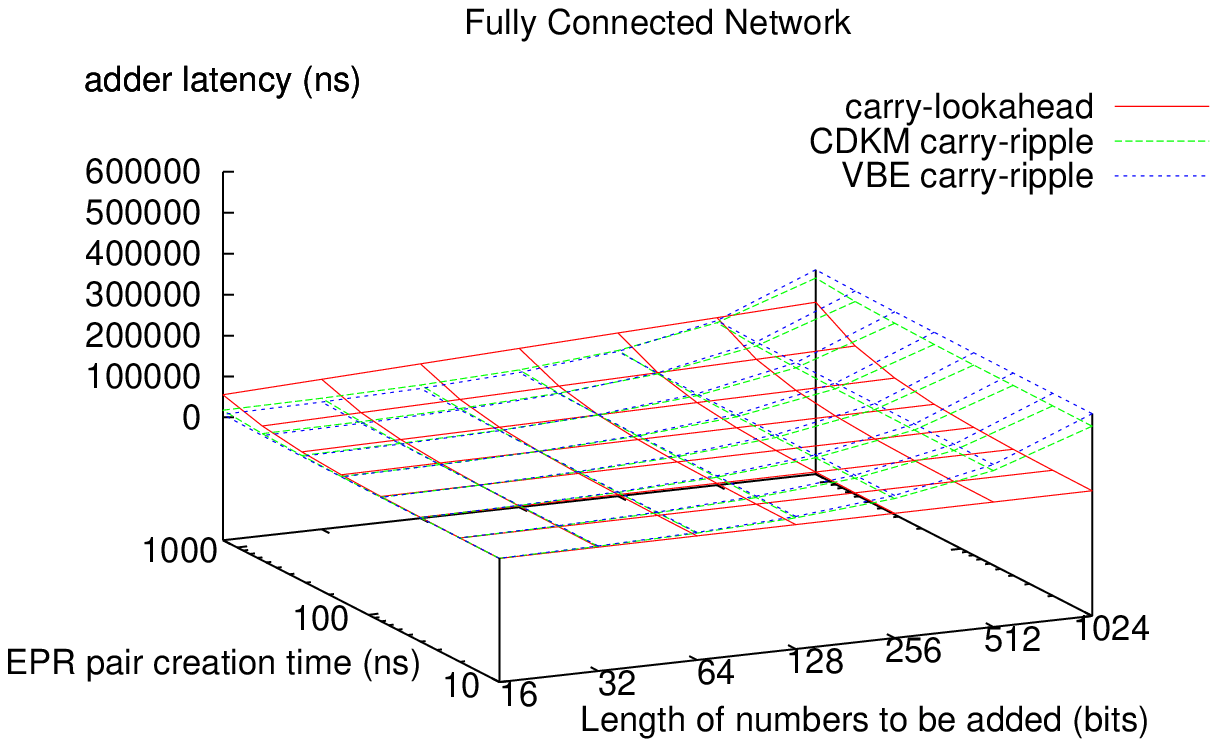}}
\hspace{.3in}
\subfigure{
\includegraphics[width=.45\textwidth]{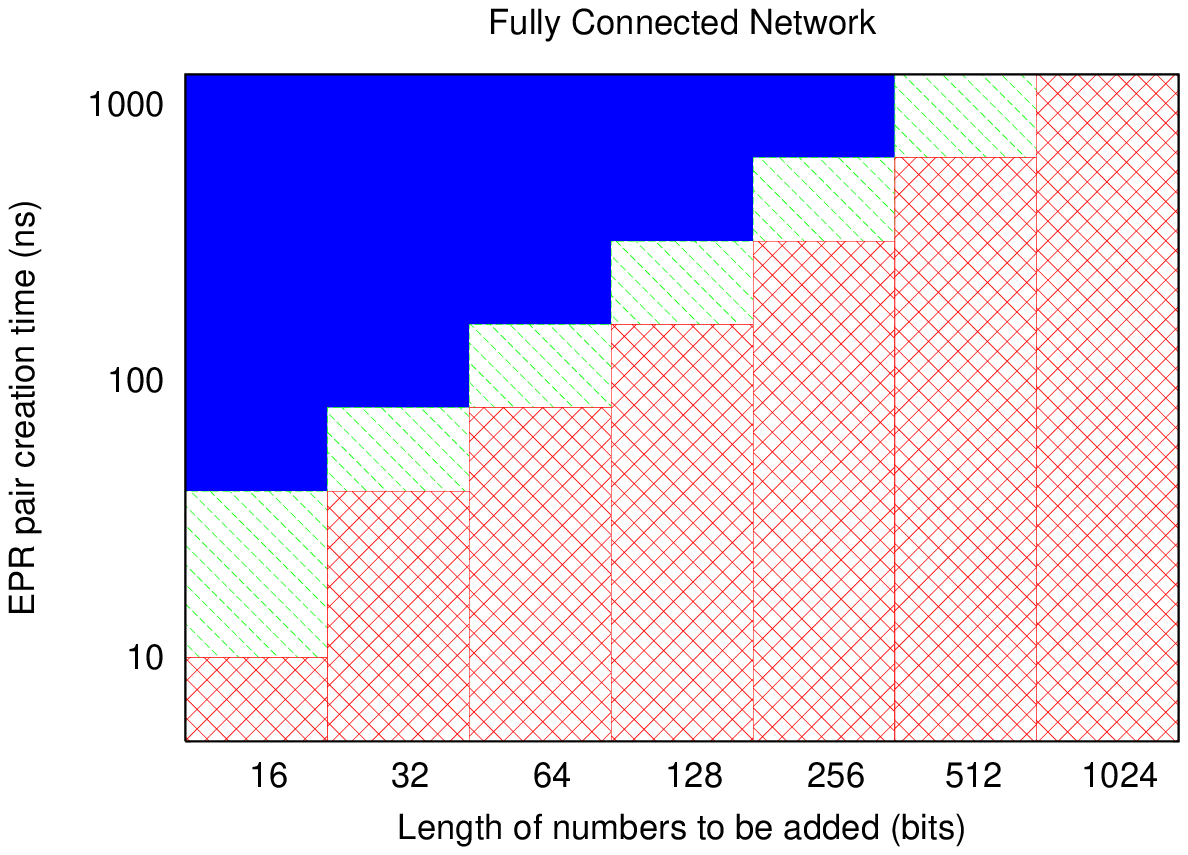}}
\hspace{.3in}
\subfigure{
\includegraphics[width=.45\textwidth]{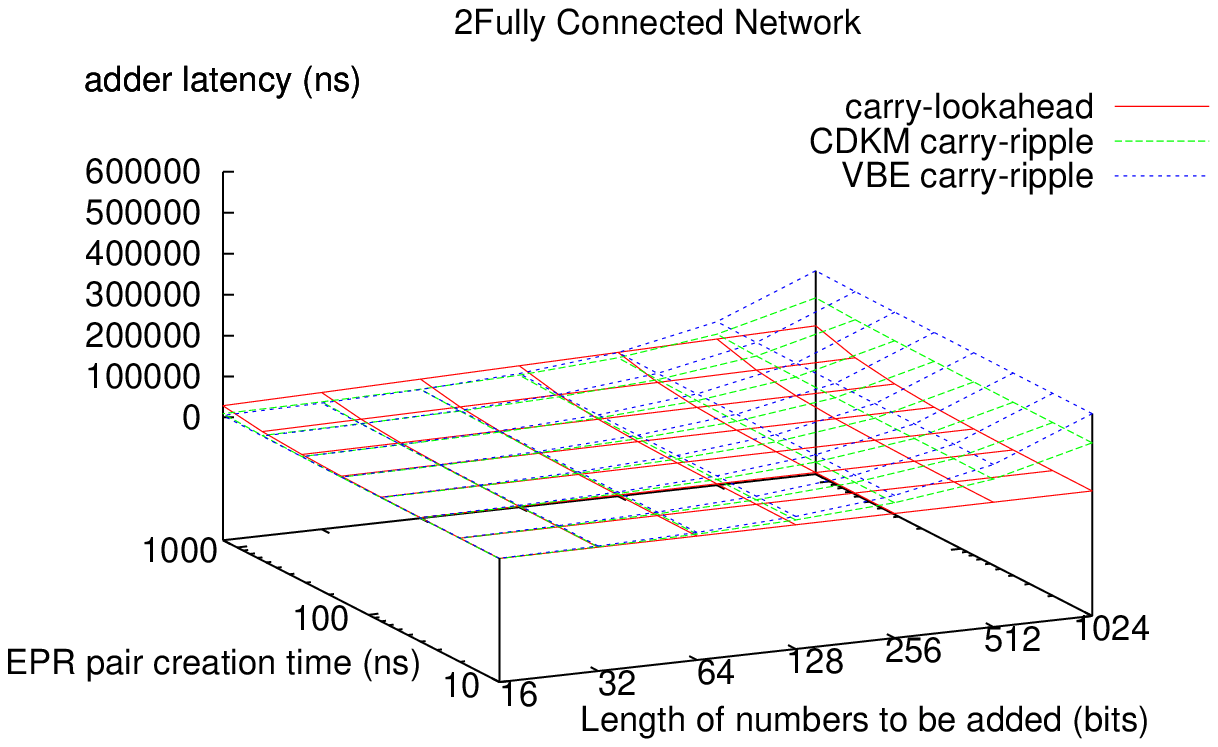}}
\hspace{.3in}
\subfigure{
\includegraphics[width=.45\textwidth]{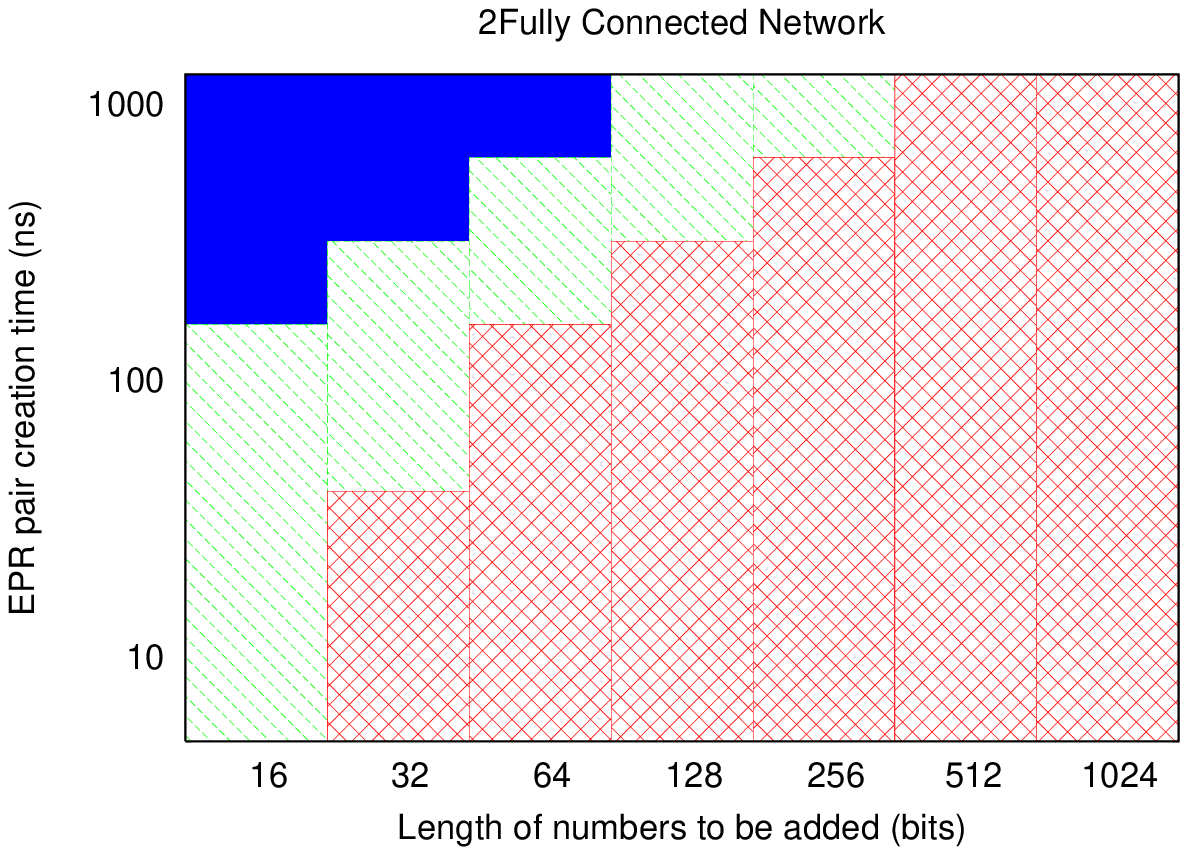}}
\hspace{.3in}
\subfigure{
\includegraphics[width=.45\textwidth]{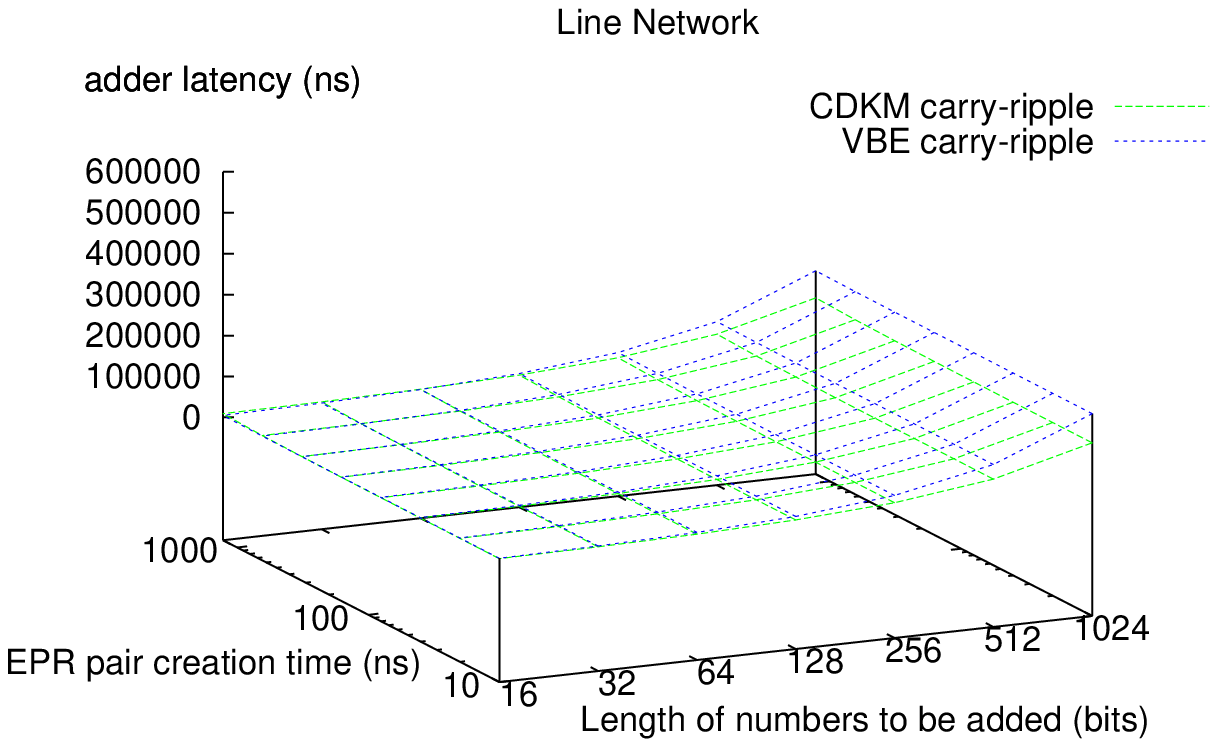}}
\hspace{.3in}
\subfigure{
\includegraphics[width=.45\textwidth]{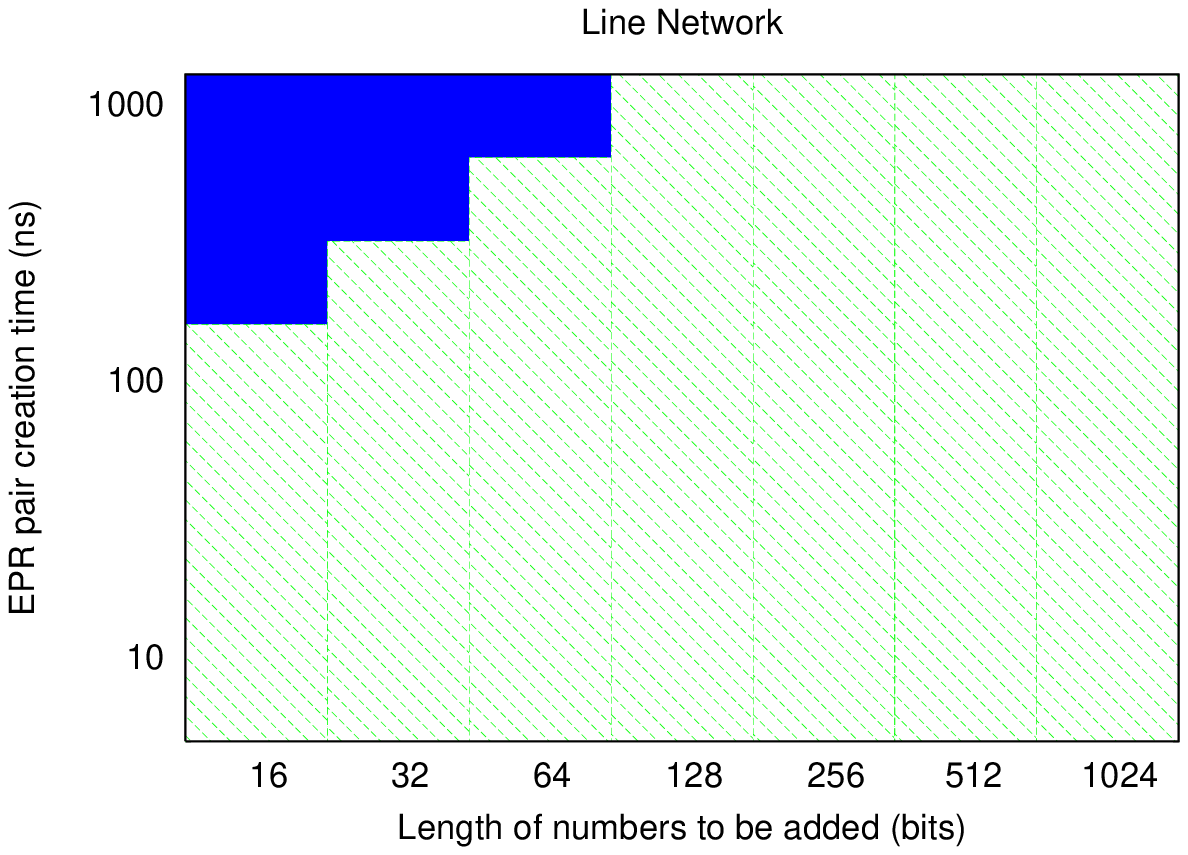}}
\caption[(Teledata) Performance of different adders on three different
  networks]{(Teledata) Performance of different adders on three
  different networks, one fully-connected with a single link and one
  with two links per node (2fully), and one line configuration.  In
  this graph, we vary the latency to create a high-quality EPR pair
  and the length of the numbers we are adding.  Classical
  communication time is assumed to be 10nsec, Toffoli gate time
  50nsec, CNOT gate time 10nsec.  In the right hand graph of each
  pair, the hatched red area indicates areas where carry-lookahead is
  the fastest, the diagonally lined green indicates CDKM carry-ripple,
  and solid blue indicates VBE carry-ripple.  The performance of the
  carry-lookahead adder is very sensitive to the EPR pair creation
  time.  If EPR pair creation time is low, the carry-lookahead adder
  is very fast; if creation time is high, the adder is very slow.}
\label{fig:fully-td}
\end{figure*}

\begin{figure*}
\centering
\subfigure{
\includegraphics[width=.45\textwidth]{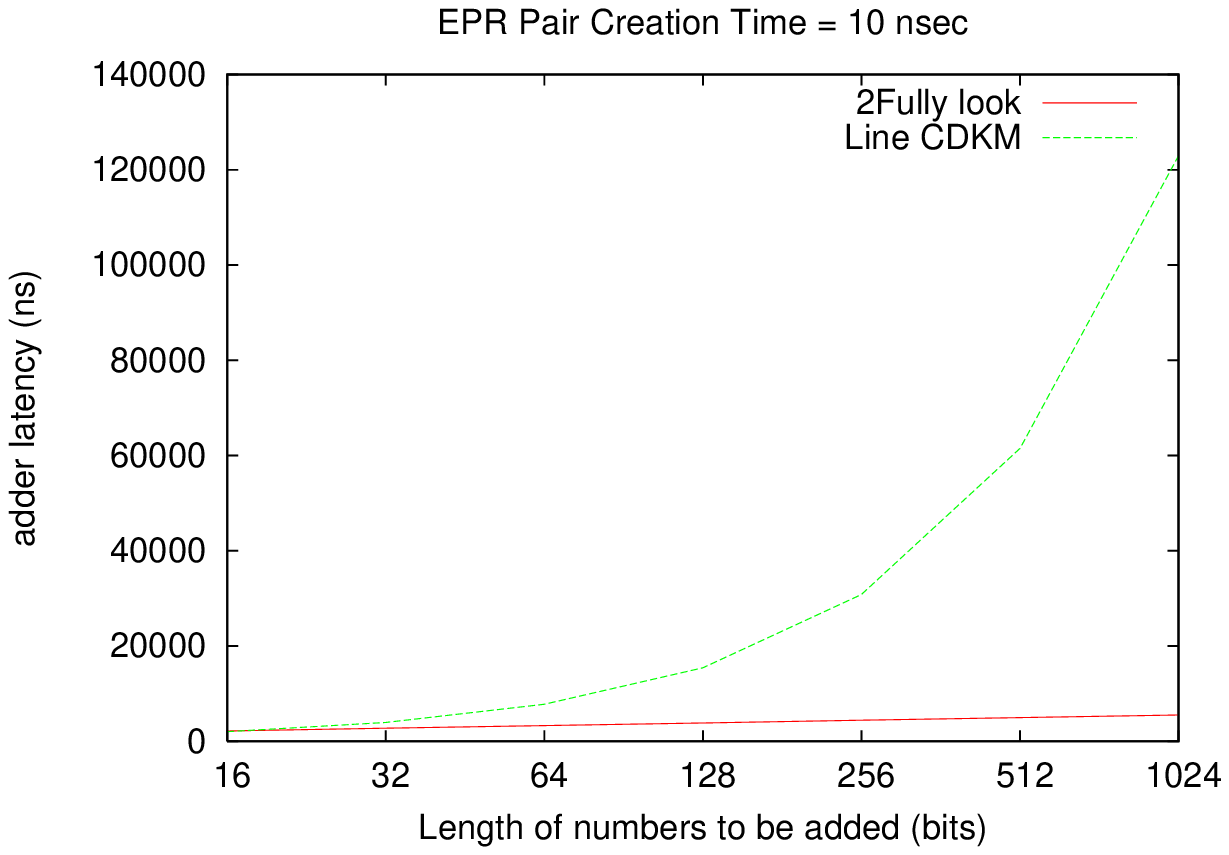}}
\hspace{.3in}
\subfigure{
\includegraphics[width=.45\textwidth]{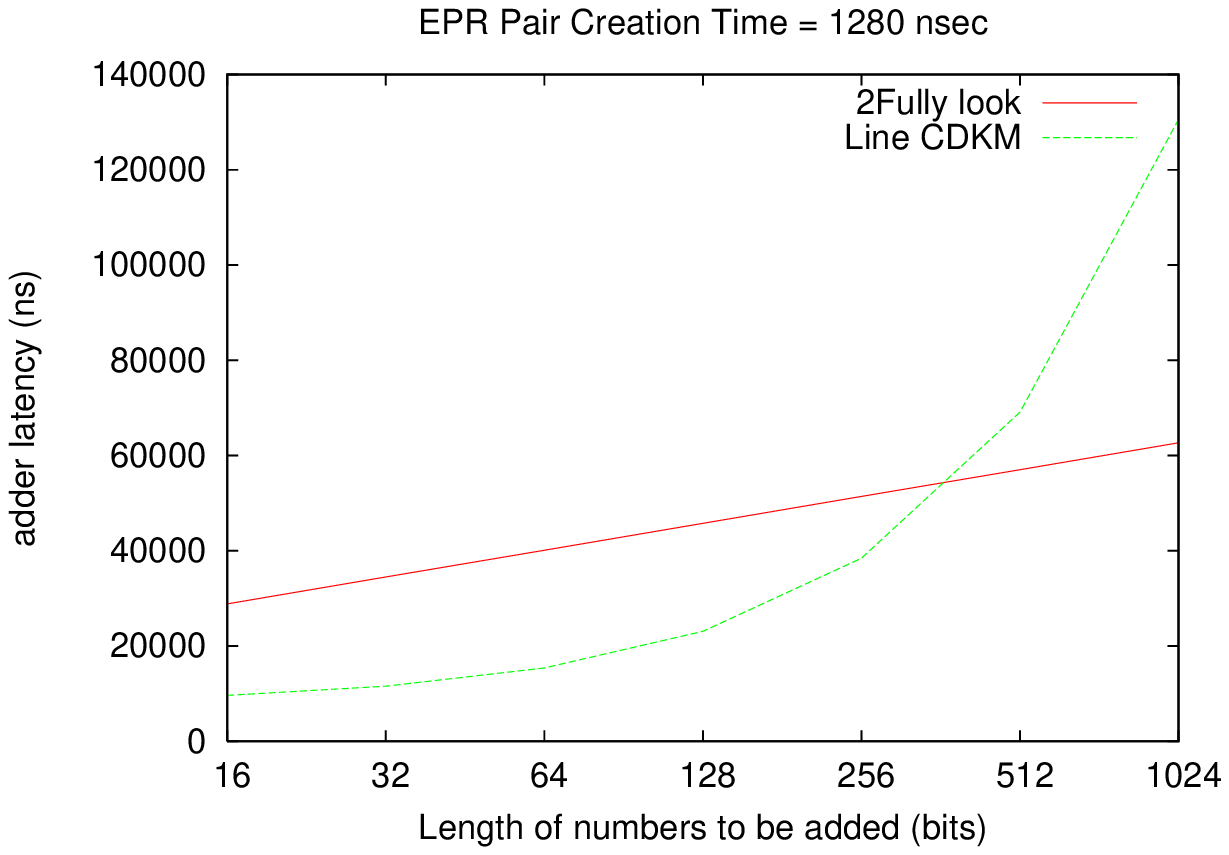}}
\caption[(Teledata) CDKM+line to carry-lookahead+2fully
  comparison]{(Teledata) Comparison of CDKM on a line network with
  carry-lookahead on a 2fully network.  These are the ``front'' and
  ``back'' cross-sections of figure~\ref{fig:fully-td}.}
\label{fig:fully-xsec}
\end{figure*}

To create Figures~\ref{fig:fully}-\ref{fig:fully-xsec}, we make
assumptions about the execution time of various operations.  Classical
communication between nodes is 10nsec.  A CCNOT (Toffoli) gate on
encoded qubits takes 50nsec, CNOT 10nsec, and NOT 1nsec.  These
numbers can be considered realistic but optimistic for a technology
with physical gate times in the low nanoseconds.  For quantum error
correction-encoded solid-state systems, the bottleneck is likely to be
the time for qubit initialization or reliable single-shot measurement,
which is still being designed, so actual performance may be one to two
orders of magnitude slower.

We vary the EPR pair creation time from 10nsec to 1280nsec.  This
creation process is influenced by the choice of parallel or serial bus
and the cycle time of an optical homodyne detector, as discussed in
the last section.  Photodetectors may be inherently fast, but their
performance is limited by surrounding
electronics~\cite{armen02:_adaptive-homodyne,stockton02:_fpga-homodyne}.
Final performance may be faster or slower than our model, but the
range of values we have analyzed is broad enough to demonstrate
clearly the important trends.

Figures~\ref{fig:fully} and \ref{fig:fully-td} show, top to bottom,
the fully, 2fully, and line networks for the telegate and teledata
methods.  The graphs plot adder time against EPR pair creation time
and the length of the numbers to be added.  The left hand plot shows
the shape of the surfaces, with the $z$ axis being latency to complete
the addition.  The right hand plot, with the same $x$ and $y$ axes,
shows the region in which each type of adder is the fastest.

These figures show that the teledata method is faster than telegate.
They also show that the carry-lookahead adder is very dependent on EPR
pair creation time, while neither type of carry-ripple adder is. In
Figure~\ref{fig:fully-xsec} we show this in more detail.  For fast
(10nsec) EPR pair creation, the carry-lookahead adder is faster for
all problem sizes.  For slow (1280nsec) EPR pair creation time,
carry-lookahead is not faster until we reach 512 bits.

Although I have not includes graphs, we have also varied the time for
classical communication and the other types of gates.  The performance
of an adder is fairly insensitive to these changes; it is dominated by
the relationship between CCNOT and EPR pair creation times.

\section{Summary}

This chapter has covered the overall quantum multicomputer
architecture, including justifying the need for distributed quantum
computation, investigating distributed quantum error correction and
network link design, and ended by evaluating the performance of
arithmetic circuits on a quantum multicomputer for different problem
sizes, interconnect topologies, and gate timings.  Although we have
assumed that the interconnect is based on the qubus entanglement
protocol creation of EPR pairs, our analysis, especially
Table~\ref{tab:lat-topo-baseline}, applies equally well to any
two-level structure with low-latency local operations and high-latency
long-distance operations.  The details of the cost depend on the
interconnect topology, number of transceiver qubits, and the chosen
breakdown for CCNOT.  Gate time ratios are more important than actual
gate times for this analysis.  The time values presented here are
reasonable for solid-state qubits under optimistic assumptions about
advances in the underlying technology.  Applying our results to slower
technologies (or the same technology using more layers of quantum
error correction) is a simple matter of scaling by the appropriate
clock speed and storage requirements.

We found that the teledata method is faster than the telegate method,
that separating the actual data teleportation from the necessary EPR
pair creation allows a carry-ripple adder to be efficient for large
problems, and that a linear network topology is adequate for up to a
hundred nodes or more, depending on the cost ratio of EPR pair
creation to local gates.  For very large systems, switching
interconnects, which are well understood in the optical
domain~\cite{kim03:_1100_port_mems,marchand:opt-aug-3d-computer,szymanski:_design-terabit-photonic-backplane},
may become necessary, though we recommend deferring adding switching
due to the complexity and the inherent signal loss; switching time in
such systems also must be considered.

These results show that node size, interconnect topology, distributed
gate approach (teledata v. telegate), and choice of adder affect
overall performance in sometimes unexpected ways.  Increasing the
number of logical qubits per node, for example, reduces the total
number of interconnect transfers but concentrates them in fewer
places, causing contention for access.  Therefore, increasing node
size is not favorable {\em unless node I/O bandwidth increases
proportionally}; we recommend keeping the node size small and fixed
for the foreseeable future.

This data presents a clear path forward.  I recommend pursuing a node
architecture consisting of only a few logical qubits and initially two
transceiver (quantum I/O) qubits.  This will allow construction of a
linear network, which will perform adequately with a carry-ripple
adder up to moderately large systems.  Engineering emphasis should be
placed on supporting more transceiver qubits in each node, which can
be used to parallelize transfers, decrease the network diameter, and
provide fault tolerance.  Significant effort is warranted on
minimizing the key parameter of EPR pair creation time.  Only once
these avenues have been exhausted should the node size be increased
and a switched optical network introduced.  This approach should lead
to the design of a viable quantum multicomputer.

\chapter{Conclusion}

\cq{Now this is not the end.  It is not even the beginning of the end.
But it is, perhaps, the end of the beginning.}  {Winston Churchill,
November 1942}

This dissertation has described the architecture of a quantum
multicomputer and the structure of the algorithms to run on it.
Shor's factoring algorithm has served as a convenient, concrete
benchmark, but the overall architecture, building blocks and analysis
methods are general.  Although small-scale quantum computers exist,
prospects for large-scale ones remain uncertain.  The physicists have
many problems to solve, of course, including decoherence time and gate
quality, both of which are affected by many physical sources.  The
engineers, as well, have many problems to solve.  At the highest
levels, the process of balancing performance, reliability, physical
feasibility and system cost has just begun.  Utilization of
heterogeneous structures, continued progress in error management, and
further optimization of application algorithms for particular
architectures continue to be promising areas of research.  At lower
levels, integration of system components, thermal engineering, and
packaging remain issues.  Once these problems are solved, a quantum
multicomputer built on many nodes based on solid-state qubits is a
viable, highly scalable, high-performance architecture.

The creation of the quantum multicomputer began with the optimization
of quantum modular exponentiation for Shor's factoring algorithm,
first in an architecture-independent fashion, then considering two
specific architectural models, \ac\ and \ntc.  The primary difference
is that \ac\ allows two qubits anywhere in the system to interact
without penalty, while \ntc\ allows only nearest neighbors in a line
topology to interact.  Both models are somewhat simplistic, but serve
as useful upper and lower bounds.  Classical computation can be traded
for quantum; increasing the classical computation by a factor of $2^w$
allows a factor of $w$ decrease in quantum, a good trade for small
values of $w$.  Two new adder algorithms, the carry-select and
conditional-sum adders, were developed.  The carry-select adder runs
in $O(\sqrt{n})$ time to add two $n$-bit numbers, and the
conditional-sum adder, which is similar but uses a more complex
demultiplexer, runs in $O(\log n)$ time.  These techniques, as well as
the fast, efficient CDKM carry-ripple adder, the $O(\log n)$-depth
carry-lookahead adder, Cleve-Watrous parallel multiplication, and some
original optimizations, are used to create complete modular
exponentiation algorithms.  The algorithms presented here will reduce
wall-clock time by a factor of one million for a six-thousand bit
number on the \ac\ architecture, or a factor of 13,000 on \ntc.  These
circuits are $O(nlog^2 n)$ and $O(n^2\log n)$ in circuit depth,
respectively, and demonstrate the paramount importance of architecture
when planning for performance.  The primary architectural features of
interest are the ability to execute multiple gates concurrently, the
number of application-level qubits available, and the interconnection
network of qubits.

The quantum multicomputer transcends the physical limitations of an
individual quantum computer by combining the power of multiple quantum
computers, in direct analogy to classical, distributed-memory
multicomputers.  It is obvious that a multicomputer can store more
data than any individual quantum computer; what was less certain
before this research was done was the performance of such a system.
Extracting performance improvements, as in classical distributed
systems, depends on finding parallelism in the algorithms and on
minimizing the costs of communication.  This research has shown that
application-level parallelism is plentiful, and that the communication
costs are reasonable.  A linear network of nodes, each containing just
a few logical qubits and two transceiver qubits for the quantum links,
performs well up to several hundred nodes.  Subdividing quantum
teleportation of the data into the EPR pair creation and the later
teleportation act allows high levels of parallelism in the EPR pair
creation to be used, and a simple carry-ripple adder performs well.
As the problem size approaches a thousand bits, the linear costs of
the carry-ripple adder begin to dominate, and the logarithmic depth
carry-lookahead adder becomes attractive.  Efficient implementation of
distributed carry-lookahead requires a more complex network.
Increasing the size of individual nodes risks turning I/O into the
system bottleneck, making it necessary to increase the number of
transceiver qubits as node size grows.

With this summary, the detailed technical work of this thesis draws to
a close.  The remainder of this final chapter of the dissertation is
more speculative: first, some rough estimates of the wall clock time
that will actually be required to execute modular exponentiation on
the quantum multicomputer are presented, then future work and some
thoughts on the prospects for quantum computation, and the
dissertation ends with some final, personal comments.

\section{Complete Performance Estimates}

Table~\ref{tab:teleport-adder-calls} shows the number of adder calls
for the complete modular exponentiation.  These values assume that $w
= 4$ and that $p$ is large enough for the modulo arithmetic to have no
impact, giving a required $2n^2$ calls to the adder routine.  These
numbers are combined with the data presented in the previous chapter
to create total teleportation counts; the range of numbers is due to
the difference between carry-ripple and carry-lookahead adders, with
the carry-lookahead adder being more expensive.  These total numbers
were used in Section~\ref{sec:qec-qubus} to derive the necessary
reliability of teleportation operations.

\begin{table*}
\centerline{
\begin{tabular}{|r|r|r|}\hline
length & adder calls & tot. teleportations ($t$) \\
\hline
16 & 481 & $14000$--$125000$\\
128 & 32544 & $8\times 10^6$--$10^8$\\
1024 & $2.1\times 10^6$ & $4\times 10^9$--$6\times 10^{10}$\\
\hline
\end{tabular}
}
\caption[Number of teleportations and adder calls for modular
exponentiation]{Number of teleportations and adder calls necessary to
execute the full modular exponentiation for different problem sizes.}
\label{tab:teleport-adder-calls}
\end{table*}

Because of the manner in which EPR pair creation and the actual gates
are composed, it is now no longer possible to talk about performance
strictly in units of ``gate times''; we must now talk in terms of
clock time for certain operations.  Table~\ref{tab:dist-perf} shows
performance estimates derived from the figures and extrapolated for
the complete algorithm.

These EPR pair creation times are for enough high-quality EPR pairs to
transfer an entire logical qubit.  Using the
[[23,1,7]]$^i$+[[7,1,3]]$^o$ error correction code, we must transfer
161 physical qubits for a single logical qubit.  Using a serial link,
performing 161 transfers in 1280nsec (the upper end of the graphs
shown) requires a physical EPR pair creation time of about 8nsec.
Although this time is faster than what has been achieved
experimentally, much of the time in adaptive homodyne measurements is
spent on (classical digital) calculations, usually carried out on
FPGAs~\cite{stockton02:_fpga-homodyne,armen02:_adaptive-homodyne}.
The qubus measurement time therefore seems amenable to significant
improvement as technology advances.

Likewise, the gate times we have chosen, such as 50nsec for a Toffoli
gate, must be seen in the light of fault tolerance and error
correcting techniques; the [[23,1,7]] code requires about three dozen
time steps to measure and correct, while using significant concurrent
gate execution~\cite{steane02:ft-qec-overhead}.  The exact performance
when combined with the upper-layer [[7,1,3]] code is unclear, and the
implementation of both codes is very different for \ac\ and \ntc, but
the total performance penalty is likely around two orders of
magnitude.  A 50nsec logical Toffoli gate would therefore require
physical gates well under a nanosecond, significantly faster than
current physical implementations.

Thus, it is likely that the absolute performance numbers for the adder
circuits presented in Section~\ref{sec:dist-shor} are one to two
orders of magnitude too optimistic.  However, the basic analysis
depends primarily on the ratio of gate times to teleportation and
communication times, so the qualitative results are valid and the
numbers need only scaling by the appropriate factors, which remain
unclear.

\begin{table*}
\centerline{
\begin{tabular}{|r||r|r|r||r|r|r|}\hline
length & \multicolumn{3}{c||}{CDKM, linear} &
\multicolumn{3}{c|}{Lookahead, 2fully} \\
\hline
 & 10nsec & 160nsec & 1280nsec & 10nsec & 160nsec & 1280nsec \\
\hline
16 & $960\usec$ & $1.4$msec & 4.6msec & 1.0msec & 2.5msec & 14msec \\
128 & 500msec & 530msec & 750msec & 125msec & 290msec & 1.5sec \\
1024 & 260sec & 260sec & 270sec & 12sec & 26sec & 130sec \\
\hline
\end{tabular}
}
\caption[Estimated time to complete distributed modular
  exponentiation]{Estimated time to complete a single run of
  distributed modular exponentiation.  The data are for the CDKM adder
  on a linear network and a carry-lookahead adder on a 2fully network,
  each for three different {\em logical} EPR pair creation times, 10,
  160, and 1280nsec.  Other gate times as described in text.}
\label{tab:dist-perf}
\end{table*}

Moreover, the numbers presented here are for a {\em single} run of the
algorithm.  For a perfect quantum computer, it is known that the
probability of success with Shor's algorithm is $\ge 40\%$,
independent of $n$, meaning that a very small number of runs will
produce a good answer~\cite{shor:siam-factor,knill95:_shor_prob}.
However, for an imperfect quantum computer, decoherence and the
precision required in the gates for the QFT ($O(2^{-k})$ for bit $k$)
present problems.  The approximate QFT (AQFT) is a reduced-precision
form of the QFT~\cite{coppersmith:approx-qft}, which has been
investigated by various researchers who have produced differing
estimates of the success probability, based on differing sets of
assumptions~\cite{barenco:approx-qft,fowler04:_limited-rotation-shor}.
Resolving this discrepancy for real-world conditions is a very high
priority issue.

One final factor throws a large uncertainty into the wall-clock time
estimates: the number of concurrent multiplications ($s$) we
implement.  We saw in Section~\ref{sec:conc-exp} that $s = n$ units
will allow us to complete the full modular exponentiation in $\log_2
n$ times the latency for one multiplication.  With the full $s = 1024$
multiplier units, the modular exponentiation for a 1,024-bit number
would run one hundred times as fast as for $s = 1$.  For this approach
to be economically and physically viable, integration must increase
one hundred fold over that proposed in Chapter~\ref{ch:arch-over}, to
about 50,000 physical qubits per pod, whether in one node or multiple
nodes, or the cost and floor space per dilution refrigerator must
decline by a similar amount.

\section{Future Work}
\label{sec:future-work}

The pursuit of performance in computing systems is never-ending.  In
classical computing systems, we have half a century of experience; in
quantum computing, the race has just begun.  It could be said that, at
the moment, answering many questions about quantum computer
architecture requires a great deal of insight and only moderate
amounts of sweat.  In the classical world, deep insight is also
required, beginning with an understanding of where the bottlenecks in
existing systems lie; however, in a mature field such as classical
architecture, acting on that insight, first demonstrating that your
insight is useful in limited circumstances, then achieving wide-spread
adoption, often requires an {\em enormous} amount of
effort~\footnote{For example, the TRIPS microprocessor team is over
twenty-five faculty, staff and students, and in turn is only a small
fraction of the size of a microprocessor team in a major semiconductor
manufacturer~\cite{DBLP:journals/computer/BurgerKMDJLMBMY04}.}.  Over
the next decade or so, as quantum computer architecture matures, this
will no doubt become true in this field as well.

The future work presented here blends smoothly from specific,
low-level continuations of the research in this dissertation to a
research agenda for the larger quantum computer architecture
community.  Further refinement of the quantum multicomputer design
requires the selection of a node technology and improvement in the
detail of hardware design. Specifically, we must determine with some
precision the number of qubits that can fit on a single chip,
investigate on-chip demultiplexers for external control signals, and
move as much control as possible into the device.  Heterogeneous node
types and heterogeneous qubit types within a node need to be
investigated, as well as multi-level interconnect architectures.  QEC
optimized for ion trap is progressing rapidly; similar optimizations
for solid state are desirable.  And, of course, supporting
experimental implementation of qubus and multi-qubit nodes will
advance the architecture also.

Improving the accuracy of estimates for the number of runs of Shor's
algorithm on QEC-encoded states on machines with limited physical
accuracy, and the detailed cost of high-precision operations on the
encoded states, tops the list of follow-on work on algorithms.
Continued algorithmic improvements in arithmetic, such as the
completion of the smaller, faster conditional-sum adder mentioned in
Section~\ref{sec:csla}, is necessary.  Optimizations for \ntc\ and
more complex topologies and more work to balance quantum and classical
computation will also contribute to reduced run times for quantum
algorithms, with consequent improvements in reliability and economic
benefits.

Can technologies with disparate characteristics be combined into a
hybrid, heterogeneous quantum computer, much as CPU, cache, RAM, and
magnetic disks are combined into a classical computer?  This will
depend on development of the ability to transfer qubits from one
technology to another and back, e.g. nuclear spin $\leftrightarrow$
electron spin $\leftrightarrow$
photon~\cite{mehring03:_entan-electron-nuke,jelezko04:_observ,childress05:_ft-quant-repeater}.
It will also require development of algorithms capable of taking
advantage of such an architectural feature, presumably based on the
classical techniques of caching, virtual memory, and out-of-core
algorithms~\cite{kilburn62:_one-level-store,knuth:v3}.

For all quantum computing technologies, we are entering the era where
automatic and semi-automatic design tools are
needed~\cite{udrescu04:_quantum-hdls,svore06:_sw-arch-computer,cross05:masters-thesis}.
A primary theme of architecture research going forward will no doubt
be creating and utilizing heterogeneity in structures.  Optimizing the
choice of hardware structures, their layout and interconnections, and
the algorithms to be run on them is a complex problem that will
require powerful tools.  Even for algorithms as simple and regular as
arithmetic, many mappings of qubits to nodes (and gates to bus
time slots) are possible; I do not claim the arrangements presented
here are optimal.  We are investigating further layouts using
evolutionary algorithms, and expect to report those results at a
future date.  Other researchers have been doing excellent work on
tools for automatic generation of QEC algorithms and structures,
especially for ion traps; continued improvement in these tools holds
the key to fast, accurate research into quantum computer
architectures.

In the early 1980s, although chip layout was done {\em on} a computer,
it was mostly done {\em by} a human being --- including much of the
verification (at Caltech, it was common to post a plot of a chip
layout on the wall for visual inspection and correction by
passers-by).  A decade later, engineers often mused that it had become
impossible to design a computer without using one; the layout and
especially validation of the design, including design rule checking
and simulation at both logical and electrical levels, could only be
done by computer, and designs were far too complex to get right
without the validation.  Obviously, detailed simulation of a large
quantum computer requires a quantum computer; the first large-scale
quantum computers must be built without data from the most desirable
simulations.  When will a quantum computer first be used to design its
successor, and when will it become indispensable to do so?

\section{Prospects}

Few of the researchers working on implementations of quantum computing
will commit to a timetable for delivering a machine large enough,
reliable enough, and fast enough to solve classically intractable
problems.  Off the record, some are optimistic that ``step functions''
in total capabilities are on the horizon; others are pessimistic
enough to say, ``I'm not sure we will have a useful quantum computer
in my lifetime.''

Personally, I am optimistic.  I believe we are on the verge of
stepping onto a Moore's Law-like growth curve, with the number of
qubits entangled in a single state growing exponentially over a
sustained period.  Ion trap systems are generating enormous
excitement, and the technical problems surrounding them seem to be
well on their way to being solved; a Moore's Law-like curve seems very
plausible for this technology.  System architects have already begun
making serious contributions in this area.  Solid-state technologies
such as quantum dot and Josephson junction still have hurdles to clear
for individual qubits, including coherence time, gate quality and
fast, reliable single-shot measurement.  Once those problems are
solved, it seems possible that the number of qubits on a chip can grow
quite rapidly; when this step function happens, the need for
system-level architects will be immediate.  All technologies, as
integration levels grow, will need improved control systems.  The
existing rack-mount equipment will quickly become prohibitive in both
space and money.

Once any of these technologies becomes ``turn-key'' ready, so that
system design, fabrication and experimentation are available to lay
systems folk rather than the initiates of physics, interest in quantum
computation will explode and systems will develop rapidly.  When the
physical technology reaches the point that individual researchers can
create quantum computer designs and fabricate them without dedicated
facilities, as the MOSIS project did more than two decades ago for
VLSI, the base of capable researchers will broaden
dramatically~\cite{tomovich88:_mosis,pina01:_mosis}.  Putting these
systems in the hands of hackers may also result in useful algorithms.
We are, in effect, in the time of Babbage asking what Knuth, Lampson
and Torvalds will do with the machines we build.


The most prominent proposed use of quantum computers today is Shor's
algorithm for factoring large numbers, which has the potential to make
the widely used RSA public-key cryptosystem and Diffie-Hellman key
exchange protocol insecure.  The encrypting operations and the
execution of Shor's algorithm are, not coincidentally, both $O(n^3)$
for $n$-bit keys.  The number of qubits we can build in a quantum
system is much smaller than the number of classical bits we build in a
system, and both manufacturing and operating costs for qubits and
quantum gates will remain many orders of magnitude more expensive than
classical bits and gates for the foreseeable future.  Classical
systems can therefore afford to go to larger key lengths far more
easily than a quantum system, staying ahead in the cryptographic arms
race (although this cost must be borne by all users, not those
breaking the codes).  However, the known existence (or even imminent
delivery) of even a single large quantum computer may prompt a shift
away from cryptosystems perceived to be vulnerable~\footnote{We wish
to point out here that quantum key distribution does not solve the
problems that Shor's algorithm creates~\cite{paterson04:why-qkd}.}.
Thus, Shor's algorithm alone is unlikely to be adequate economic
incentive for the development and purchase of more than a handful of
large quantum computers.

Whether or not a specific quantum computing technology is useful
depends on the availability of important algorithms (e.g., Shor's
algorithm) and supporting algorithms or subroutines (e.g., the modular
exponentiation necessary to run Shor's algorithm) that map efficiently
to a system built on the technology.  Future developments in
algorithms, therefore, can make an architecture useful which had
earlier been dismissed due to lack of interesting, practical
applications.

The need for hardware/software co-design is very much in evidence here.
Because quantum computation in general, and architecture in
particular, is immature as a field, we start adrift on Lampson's Sea.
This thesis charts a course toward a particular goal, and maps out
some of the major shoals.  Course corrections, some major, are
inevitable, but our sails are full and we have a guide star to follow.
To be a complete system, many subsystems must be developed.  Indeed,
not just the subsystems themselves, but the development {\em tools}
must be built.  Chip layout tools must integrate smoothly with one or
more of the commercial successors to early VLSI tools such as the
Magic toolkit~\cite{ousterhout84:_magic}.  We need to develop the
quantum equivalent of classical design rules~\cite{mead79:_vlsi}, and
may ultimately wish to use direct silicon compilation to physical
circuits from programs~\cite{ayres83}.  Compilers that optimize a
circuit are already being developed; new back ends to create both
hardware and software will allow better optimization, at the expense
of tool complexity.

\section{Final Words}

When I began working on quantum computing three years ago, I was naive
about a great many of the technical aspects.  I wanted to focus on
software for quantum computers, and I was especially curious about how
our classical mechanisms for resource management (such as semaphores)
and naming --- two of the key functions of an operating system ---
would translate into the quantum world.  I quickly discovered that the
structure of the machines themselves was not yet advanced enough to
work seriously on such topics.  Surveying the state of hardware
proposals, it became clear that there was much room for
jacks-of-all-system-trades like me to contribute.  Each time I opened
one door, I found another.  Sometimes I found that someone had
unlocked the door before me, and I was happy to walk through on their
work.  Sometimes, I found the door locked, and faced the task of
picking the lock myself.  I am pleased with what I have accomplished,
but not satisfied; I imagine many, many productive years yet pushing
beyond what we currently know, though it is not always obvious exactly
what it {\em is} that we don't know.

I wish to close with two of my favorite quotes.  ``Life is either a
daring adventure or nothing,'' Heller Keller said.  Even when things
don't work out according to the original plan, you accomplish
something along the way, if you are flexible and work hard.  You must
let the path teach you, as much as you choose the path.

\begin{quotation}
{\sf Butter tea and wind pictures, the crystal mountain, and blue sheep
dancing on the snow --- it's quite enough!  Did you see the snow
leopard?  No!  Isn't that wonderful?}
\flushright{\sf\textbf{Peter Matthiessen, \textit{The Snow Leopard}}}
\end{quotation}

\appendix

\chapter{Glossary}
\label{ch:glossary}

In such an interdisciplinary thesis, a glossary would seem to be
essential.  The mathematical terms are defined here {\em extremely}
informally, for the benefit of newcomers to the field.

\begin{Lentry}
\item[ancilla] (plural ancillae) Bits holding temporary variables used
  during a reversible computation that must be returned to their
  initial state at the end of the computation.
\item[bisection] In a network, the number of links that must be cut to
  divide the network in half.
\item[bra] Dirac notation for a complex-conjugate row vector:
  $\langle\psi|$.  See also {\em ket}.
\item[cluster state computing] Also called {\em one-way computing} or
  {\em measurement-based
  computing}~\cite{raussendorf03:_cluster-state-qc,nielsen-2004}.  Has
  nothing to do with classical computing clusters; the cluster state
  is a very large entangled state which serves as a computing
  substrate.
\item[decoherence] The degradation of the state of a quantum system as
  it interacts with its environment in ways that are impossible to
  adequately characterize; causes errors in qubits.
\item[decoherence free subspace (DFS)] A form of error management in
  which logical states are encoded in the {\em relative} state of
  multiple qubits~\cite{lidar03:dfs-review,haeffner05:_robust,lidar98:dfs}.
\item[degree] The number of links, or connections to the network, at
  each node.
\item[density matrix] Describes the statistical state of a quantum
  system.  For an $n$-qubit system, a $2^n\times 2^n$ matrix.  Also
  called the {\em density operator}, and usually written $\rho$.  A
  valid density matrix has trace $\operatorname{Tr}(\rho) = 1$, and
  the diagonal elements are the probability of finding the system in
  the corresponding state when measured.
\item[diameter] The largest number of hops through the network to get
  from any node to any other.
\item[entanglement] The property of two or more qubits in which
  operations on one affect the state of the other.  For pure states,
  corresponds roughly to the qubits having dependent probabilities for
  their states.  {\em Karami-tsuki} in Japanese.
\item[full-duplex] A type of link in which data can be transferred in
  both directions at the same time.  Telephones are generally
  full-duplex.
\item[half-duplex] A type of link in which data can be transferred in
  either direction, but only in one direction at a time.  Many
  computer buses are half-duplex; push-to-talk walkie-talkies are
  half-duplex.
\item[ket] Dirac notation for a column vector: $|\psi\rangle$.  For an
  $n$-qubit system, consists of $2^n$ entries.  See also {\em bra}.
\item[link] A physical connection in a network between two nodes, or a
  node and a piece of dedicated networking equipment such as a
  router.  May be serial or parallel.
\item[mixed state] A state which has partially decohered due to
  interaction with its environment; must be represented by a density
  matrix $\rho$ which does not have $\operatorname{Tr}(\rho^2) = 1$
\item[mux] Multiplexer.
\item[network] In this dissertation, a collection of links that
  connect quantum computer nodes together.  Often used in the quantum
  computing literature to mean circuit or program.
\item[node] A computational element attached to a network.
\item[probe beam] For the qubus, the high-intensity beam that
  interacts with the qubits.
\item[pure state] A quantum state about which we have maximal
  knowledge; it is not entangled with the environment.  A pure state
  has $\rho = \rho^2$ and $\operatorname{Tr}(\rho^2) = 1$.  A pure
  state can be written in state-vector form as $|\psi\rangle$.
\item[qubit] A two-level quantum system that obeys DiVincenzo's
  criteria; the basic unit of quantum information.  A qubit may be in
  a superposition of its two states.  Qubits may be physical or
  logical.
\item[qubus] A system that uses a strong probe beam and weak
  nonlinearities to entangle two or more qubits over a distance.
\item[qubyte] Eight qubits.
\item[separable] Two quantum systems that are not entangled are
  separable.
\item[simplex] A unidirectional link.
\item[superposition] Two or more solutions to Schr\"odinger's equation
  added together to form a single state, with their weights adjusted
  so that the total weight is still one.  {\em Kasane-awase} in
  Japanese.
\item[trace] The sum of the diagonal of a matrix.
\item[transceiver qubit] A physical qubit that connects to a qubus.
\item[unitary transform] The most common mathematical representation
  of a quantum gate; for an $n$-qubit gate, a $2^n\times 2^n$ unitary
  matrix that effects a rotation in the appropriate space.  A unitary
  transform $U$ satisfies the condition that $U^{\dagger}U =
  UU^{\dagger} = I$.
\end{Lentry}

\chapter{List of Papers and Presentations}

\subsubsection{Peer-Reviewed Journals}

\begin{enumerate}
\item R.~Van{ }Meter and M.~Oskin.  Architectural implications of
quantum computing technologies.  {\em ACM J. Emerging Tech. in
Comp. Sys.}, 2(1), Jan. 2006.
\item R.~Van{ }Meter and K.~M. Itoh.  Fast quantum modular
exponentiation.  {\em Physical Review A}, 71(5):052320, May 2005.
\end{enumerate}

\subsubsection{International Conferences}

\begin{enumerate}
\item R.~Van{ }Meter, W.~J. Munro, K.~Nemoto, and K.~M. Itoh.
Distributed arithmetic on a quantum multicomputer.  In {\em
Proc. Int. Symp. on Computer Architecture (ISCA33)}, Jun. 2006.
\item R.~Van{ }Meter, K.~M. Itoh, and T.~D. Ladd.
Architecture-dependent execution time of {Shor's} algorithm, 
  In {\em Proc. Int. Symp. on Mesoscopic Superconductivity and
    Spintronics (MS+S2006)}, Feb. 2006.
\item R.~Van{ }Meter.  Trading classical for quantum computation using
indirection.  In {\em Realizing Controllable Quantum States:
Proc. Int. Symp. on Mesoscopic Superconductivity and Spintronics
(MS+S2004)}, Mar. 2004.
\end{enumerate}

\subsubsection{National Conferences and Workshops}

\begin{enumerate}
\item R.~Van{ }Meter.  Communications topology and distribution of the
 quantum {Fourier} transform.  In {\em Proc. Tenth Symposium on
 Quantum Information Technology (QIT10)}, pages 19--24, May 2004.
\end{enumerate}

\subsubsection{Teaching}

\begin{enumerate}
\item Jun. 2005: WIDE Project School of Internet, ``Introduction to
  Quantum Computing'', a 3-day intensive short course on quantum
  computing offered via satellite and Internet.  Attended by
  approximately fifty students from Nepal, Indonesia, Laos, Thailand,
  Japan, Malaysia, and Bangladesh.

\item Sept. 2004: U. Aizu, ``Introduction to Quantum Computing'', a
  3-day intensive short course on quantum computing offered to U. Aizu
  students for credit.
\end{enumerate}

\subsubsection{Other Presentations}

\begin{enumerate}
\item ``Fast Quantum Modular Exponentiation,'' Caltech Workshop on
Classical and Quantum Information Security (CQIS), Dec. 2005.

\item ``The Design of a Quantum Multicomputer,'' USC/ISI, Dec. 2005.

\item ``Fast Quantum Modular Exponentiation,'' BBN, Aug. 2005.

\item ``Quantum Computing {\em Systems}: State of the Art, Summer 2005,''
Carnegie Mellon University, Aug. 2005.

\item ``Fast Quantum Modular Exponentiation,'' HP Labs, Bristol, Jan. 2005.

\item ``Fast Quantum Modular Exponentiation,'' Oxford University, Jan. 2005.

\item ``Fast Quantum Modular Exponentiation,'' MIT, Nov. 2004.

\item ``Accelerating Shor's Algorithm Using Fast Quantum Modular
  Exponentiation,'' 2004 Workshop on Information Security Research
  (invited), Fukuoka, Japan, Oct. 2, 2004.

\item ``Introduction to Quantum Computing,'' Keio Shonan Fujisawa
  Campus, June 3, 2004 (in Japanese).

\item ``Trading Classical for Quantum Computation Using Indirection,''
  ERATO Kyoto, April 15, 2004 (in Japanese).

\item ``A Computer Systems Research Agenda for Quantum Computing,''
  Nara Institute of Science and Technology, April 16, 2004 (in
  Japanese).

\item ``Communications Topology and Distribution of the Quantum
  Fourier Transform,'' National Institute of Informatics, April 22,
  2004.

\item ``A Computer Systems Research Agenda for Quantum Computing,''
  NTT Basic Research Laboratory, October 7, 2003.
\end{enumerate}


\bibliography{paper-reviews}



\end{document}